\pdfoutput=1

\documentclass[a4paper,USenglish,autoref]{lipics-v2021}

\input{style-and-macros.sty} 
\usepackage[table]{xcolor}

\nolinenumbers
\hideLIPIcs

\title{Parameterized algorithm for replicated objects with local reads}
\titlerunning{Replicated objects with local reads}

\author{Changyu Bi}{Department of Computer Science, Stanford University, USA}{}{}{}

\author{Vassos Hadzilacos}{Department of Computer Science, University of Toronto, Canada}{}{}{}

\author{Sam Toueg}{Department of Computer Science, University of Toronto, Canada}{}{}{}

\authorrunning{C. Bi, V. Hadzilacos, and S. Toueg}



\category{} 

\relatedversion{} 

\supplement{}

\acknowledgements{We are grateful to Tushar Chandra who suggested the idea behind Algorithm~1.}

\begin{document}

\maketitle

\begin{abstract}
We consider the problem of implementing
	linearizable objects that support both read 
	and read-modify-write (RMW) operations in message-passing systems with process crashes.
Since in many systems read operations vastly outnumber RMW operations,
	we are interested in implementations that emphasize the efficiency of read operations.

We present a parametrized algorithm for partially synchronous systems
	where processes have access to external clocks that are synchronized within $\epsilon$.
With this algorithm, every read operation is local (intuitively, it does not trigger messages).
If a read is not concurrent with a conflicting RMW, it is performed immediately with no waiting;
	furthermore, even with a concurrent conflicting RMW,
	a read experiences very little delay in the worst-case.
For example, the algorithm's parameters can be set to ensure that every read takes $\epsilon$ time in the worst-case.
To the best of our knowledge this is the first algorithm
	to achieve this bound
	in the partially synchronous systems that we assume here.
Our parametrized algorithm generalizes the (non-parameterized) lease-based algorithm of Chandra \emph{et al.}~\cite{CHT16}
	where the worst-case time for reads is $3\delta$, where $\delta$ is the maximum message delay.

The algorithm's parameters can be used to trade-off the worst-case times for read and RMW operations.
They can also be used to take advantage of the fact that in many message-passing systems
	the delay of most messages is order of magnitudes smaller than the maximum message delay~$\delta$:
	for example,
	the parameters can be set so that,
	in ``nice'' periods where message delays are $\delta^* \ll  \delta$,
	reads take at most $\epsilon$ time
	while RMWs take at most $3 \delta^*$ time.
\end{abstract}

\newpage
\section{Overview}\label{sec-intro}

We consider the problem of implementing
	linearizable objects that support both read 
	and read-modify-write (RMW) operations in message-passing systems with process crashes.
Since in many systems read operations vastly outnumber RMW operations,
	we are interested in implementations that emphasize the efficiency of read operations.

We present a parametrized, leader-based algorithm for partially synchronous systems
	where processes have access to clocks that are synchronized within
	$\epsilon$;
	such clocks can be provided by external devices such as GPS~\cite{CD+12}
	which provide a very small $\epsilon$.
With this algorithm, every read operation is local (intuitively, it does not trigger messages).
If a read is not concurrent with a conflicting RMW, it is performed immediately with no waiting;
	furthermore, even with a concurrent conflicting RMW,
	a read experiences very little delay in the worst-case.
For example, the algorithm's parameters can be set to ensure that (after the system stabilizes) every read takes $\epsilon$ time in the worst-case.
If $\epsilon\le\delta/2$, where $\delta$ is the maximum message delay,
	this nearly matches a lower bound by Chandra \emph{et al.} (Theorem~4.1 in~\cite{CHT16}).
To~the best of our knowledge this is the first algorithm
	to achieve this for \emph{linearizable} object implementations in the partially synchronous systems that~we~assume~here.

The algorithm's parameters can be used to trade-off the worst-case times for read and RMW operations.
They can also be used to take advantage of the fact that in many message-passing systems
	the delay of most messages is orders of magnitude smaller than the maximum message delay~$\delta$:
	for example,
	the parameters can be set so that,
	in ``nice'' periods where message delays are $\delta^* \ll  \delta$,
	reads take at most $\epsilon$ time,
	while the RMWs issued by the leader take
	at most $3 \delta^*$.
	
Our parametrized algorithm generalizes the (non-parameterized) lease-based algorithm of~\cite{CHT16}
	(henceforth referred to as the ``CHT algorithm'')
	where the worst-case time for reads is~$3\delta$.
This generalization is achieved by adding two novel mechanisms,
	each of which is controlled by a parameter.
Roughly speaking, the first mechanism
	decreases the worst-case time for reads and
	enables
	a continuous trade-off between the worst-case times for read and RMW operations,
	and the second mechanism
	allows us to take advantage of ``nice'' periods when message delays are very short.
These mechanisms may be useful to achieve similar benefits in other lease-based algorithms.

We now describe our algorithm and the results in more detail.
To do so, we first explain our model, we then describe the CHT algorithm
	and the two mechanisms that we added to generalize it, and finally we
	compare the performance of the two algorithms for some parameter settings.
	
\textbf{Model sketch.} We consider message-passing systems where fewer than half of the processes may crash.\footnote{ If half of the processes or more crash,
	it is impossible to implement even linearizable registers, let alone objects that support arbitrary RMW operations,
	in our model of partial synchrony.
	This is easy to show using a standard partitioning argument.}

Initially, processes take steps at arbitrary speeds and
	messages take arbitrarily long and can even be lost.
There is, however, an \emph{unknown} time $\tau$ after which
	no process crash occurs, processes take steps at some known minimum speed,
	and every message that is sent is received
	within some known time bound $\delta$~\cite{DLS88}.
To simplify the exposition, we assume that after time $\tau$
	the time between consecutive steps of each nonfaulty process
	is negligible compared to $\delta$.
We use the terms ``after the system stabilizes'' and ``stable period''
	to refer to the time after~$\tau$.
When discussing the performance of an algorithm,
	\emph{we focus exclusively on the period after the system stabilizes}.
The correctness of our algorithms, however, is always preserved: in particular safety is never violated and all operations issued by correct processes, even those issued before the system stabilizes, terminate.

Processes have local clocks
	that are \emph{always} synchronized
	within some known $\epsilon\ge0$ of each other;
	such synchronized clocks can be provided
	by devices such as GPS~\cite{CD+12}.
To simplify the exposition, we first assume here that $\epsilon=0$.
In Section~\ref{sec:epsilon} we explain how to deal
	with an arbitrary clock skew $\epsilon > 0$, and how the
	clock skew affects the performance of our algorithms.

\textbf{The CHT algorithm}.
This algorithm 
	has the following desirable properties.
Every read operation is ``local'';
	furthermore, after the system stabilizes,
	(a) every read operation is ``non-blocking''
	unless it is concurrent with a RMW operation that conflicts with it, and
	(b) even if a read blocks, it completes in a bounded period of time.
We say that read operations are \emph{local} 
	if they do not result in messages being sent;
	more precisely, the number of messages sent during the execution of the algorithm
	does not depend on the number of read operations performed in the execution.
A read operation issued by process $p$
	is \emph{non-blocking}
	if it completes
	within a constant number of steps of $p$,
	without waiting for a message to arrive
	or for the process's clock to reach a certain value.\footnote{Because we do not assume a \emph{maximum} process speed, it is not possible to simulate waiting for a certain period of time by requiring the process to execute a minimum number of local steps.} 
A read operation $r$ \emph{conflicts with} a RMW operation $w$
	if there is an object state such that if we execute $r$ and $w$ starting from this state,
	$r$ reads different values
	depending on whether it executes before or after $w$.

Intuitively, the CHT algorithm works by combining two well-known mechanisms:
	(a) a consensus algorithm to process all RMW operations, and
	(b) a lease mechanism to allow local reads.
Both mechanisms rely on an eventual leader elector.
Roughly speaking, the (current) leader executes a ``two-phase commit'' algorithm
	to linearize all RMW operations across the object replicas.
The leader also issues \emph{read leases}: the holder
	of a read lease that expires at some time $t$ can read its local copy of the object until time $t$,
	unless it is aware of a concurrent conflicting operation.

The blocking time of an operation is the time that elapses
	from the moment a process issues this operation to the moment it completes it with a return value.
In the rest of this paper, we consider only the blocking time of RMW operations
	\emph{issued by the leader} when it is not currently processing other RMW operations.
Note that if a RMW operation is not issued by the leader,
	its blocking time may be longer by up to a round-trip delay
	($2\delta$ in the worst-case, and at most $2 \delta^*$ in the ``nice'' periods):
	this accounts for the time it takes for the issuer to send this operation to the leader
	and to learn from the leader that this operation was committed.

We now explain why and for how long operations block in the CHT algorithm,
	and we introduce the main ideas of our algorithm for decreasing
	the blocking time of reads
	with only a small or even no increase in the blocking time of RMW operations.
	
To see why operations may block with the CHT algorithm,
	suppose a process $p$ has a read lease that expires far in the future,
	but the leader $\ell$ wants to process a RMW operation that conflicts with the read.
 To do so, $\ell$ first sends prepare messages to notify processes of the impending operation;
 	then, when $\ell$ receives ``enough'' acknowledgements, it commits the operation
	(the state of the object is now changed);
	finally $\ell$ sends commit messages to notify processes that the operation
	was indeed committed.
Note that when $p$ receives the prepare message,
	it does not know whether the state of the object already changed or not.
So if $p$ wants to do a read now, it cannot read its local copy of the object (because it could be stale):
	it must wait until it gets the commit message from the leader.
Since
	messages take at most~$\delta$,
	it is clear that up to $3\delta$ time may elapse
	from the moment $p$ receives the prepare message
	to the moment $p$ receives the commit message;
	during that period the read of $p$ is blocked.
	
The blocking time of a RMW operation issued by the leader $\ell$ is the time that elapses from
	the moment $\ell$ starts processing the operation by sending prepare messages to
	the moment $\ell$ commits it
	having received enough acknowledgements.
This takes at most $2\delta$ time.

	In summary,
	with the CHT algorithm,
		a read operation that is concurrent with a conflicting RMW operation
		may block for up to $3 \delta$; and
		a RMW operation issued by the leader~may block~for~up~to~$2 \delta$.
		
	In this paper we introduce a parametrized algorithm
		that can reduce the blocking time of reads without 
		affecting the maximum blocking time of RMW operations;
		or can eliminate the blocking of reads altogether (more precisely,
		reduces the blocking time to just $\epsilon$,
	if clocks are not perfectly synchronized)
		at the cost of slightly increasing
		the maximum blocking time of RMW operations.
We do so by adding the two mechanisms described below.

\textbf{Two new mechanisms.}	Our algorithm generalizes
 	the CHT algorithm by adding two mechanisms.
	For pedagogical reasons we present our algorithm in two stages:
	``Algorithm~1'' incorporates only one of the mechanisms,
		and is parameterized by a quantity we denote $\alpha$.
	The CHT algorithm is the special case of this algorithm
		with $\alpha$ set to~0.
	``Algorithm~2'' adds to Algorithm~1 the second mechanism,
		and is parameterized by an additional quantity we denote $\beta$.
	Algorithm~1 is the special case of Algorithm~2
		with $\beta$ set to $\infty$.
	
\emph{Promise mechanism.}	Roughly speaking,
		the parameter $\alpha$ of Algorithm~1 is used as follows:
		when the leader $\ell$ starts processing a RMW operation $\op$
		at some time $t$,
		it sends prepare messages for $op$
		with the \emph{promise} not to commit $op$ before time $t+\alpha$,
		the expiration time of that promise.
	Now when a process $p$ (that has a valid lease) receives this message,
		it knows that the state of the object
		will not change before time $t+\alpha$,
		so it can read its local copy up to that time.
	We call this the \emph{promise mechanism}.
	Process
		$p$ will receive the commit message by time $t+3\delta$,
		and so the reads of $p$ are blocked only during the period
		$[t+\alpha, t+3\delta]$,
		i.e., for up to $3\delta-\alpha$ time.
	
	By setting $\alpha=3\delta$ we get an algorithm where
		\emph{all} reads are non-blocking.
	Note, however, that this setting also causes
		all RMW operations issued by $\ell$ to block for $3\delta$ time.
	Thus, with this setting of $\alpha$
		Algorithm~1 achieves the desirable goal of non-blocking reads,
		but at a considerable cost for RMW operations
		in comparison to the CHT algorithm:
	In the CHT algorithm a RMW operation blocks only for
		the \emph{actual} delay of a round-trip message
		while now \emph{all} RMW operations block for $3\delta$ time,
		even if messages flow fast.
	This is a problem because in many systems
		the \emph{worst-case} message delay $\delta$
		is orders of magnitude greater than the delay
		experienced by most messages.
	In particular,
		there can be long periods of time after the system stabilizes
		during which all messages take at most some $\delta^* \ll \delta$ time;
		we call these \emph{nice periods}.
	It is desirable to optimize the performance of algorithms
		during such periods.
	Here our goal is to decrease the maximum blocking time of reads without increasing
		(or increasing only by little) the maximum blocking time of RMWs in the nice periods.
	This is achieved by Algorithm~2, as we now explain.

\emph{Status mechanism.}	The main idea behind Algorithm~2 is to keep the promises short,
		and extend them as needed.
	Instead of sending prepare messages with a long promise,
		the leader $\ell$ sends ``status'' messages with a short promise~$\alpha$.
	If $\ell$ does not receive enough acknowledgements
		to commit an operation within a period $\beta$,
		it sends another round of status messages with a new promise of~$\alpha$.\footnote{Note that it is possible for a promise to expire before the next one is received, and this may occur \emph{even in the stable period}.  This is in contrast to the behaviour of read leases in the stable period.}
	This is repeated until $\ell$ receives enough acknowledgements,
		at which point it sends commit messages as before.
	We call this the \emph{status mechanism}.
	The cost of the status mechanism is the additional number of messages,
		but if we set $\beta \ge 2\delta^*$
		this cost is not incurred in nice periods,
		because the leader receives enough acknowledgements
		within $2\delta^*$ in these periods.
	Thus we focus on the behaviour of Algorithm~2 only for settings of $\alpha$ and $\beta$ where $\beta \ge 2 \delta^*$.
	
	With Algorithm~2,
		we can set $\alpha$ (the length of the promise) to a small value
		to reduce the blocking time of RMW operations in nice periods,
		and with a suitable setting of $\beta$
		(the time between successive status messages)
		we can also keep the blocking time of reads short.

\textbf{Performance and comparison with CHT.}	Tables~\ref{table1} and~\ref{table2} summarize
		the maximum blocking times of operations
		during the stable period and nice periods
		under our two algorithms for certain interesting settings of their parameters $\alpha$ and $\beta$.
		(The maximum blocking times of the two algorithms,
			expressed as a function of $\alpha$ and $\beta$,
			are given in Table~\ref{TableAlgo2Generic}.)
	The column labeled ``CHT'' 
		in both tables
		shows the maximum blocking times of the CHT algorithm,
		and serves as a baseline.


\begin{table}[h]
\centering
\begin{tabular}{ | c | c || c || c | c |}
\hline
    \multicolumn{2}{|c|}{}
    &\makecell{CHT\\ }
    &\makecell{Alg.~1 \\ $\alpha = 2\delta$}
    &\makecell{Alg.~2\\ $\alpha = \beta = 2\delta^*$}
\\
\hline

\hline
    \multirow{2}{*}{Stable Period} & 
    \RMW & $2\delta$ & $2\delta$ & $2\delta$
\\
    \cline{2-5}
    & \tRead & $3\delta$ & $\delta$ & $\delta$
\\
\hline

\hline
\multirow{2}{*}{Nice Periods} 
   & \RMW & $2\delta^*$ & $2\delta$ & $2\delta^*$
\\
    \cline{2-5}
    & \tRead & $3\delta^*$ & 0 & $\delta^*$
\\
\hline
\end{tabular}
\\[1pt]
\captionsetup{justification=centering}
\caption{Reducing the maximum blocking time of reads.}
\label{table1}
\end{table}

	Table~\ref{table1} shows
		parameter settings aimed at
		\emph{improving} the blocking of reads
		without increasing the blocking of RMW operations.
	By setting $\alpha=2\delta$ in Algorithm~1
		we reduce the blocking time of reads
		to one-third of the CHT algorithm during the stable period,
		and make all reads non-blocking during nice periods
		(provided $\delta^*\le 2\delta/3$,
		which holds because $\delta^*\ll\delta$).
	This setting, however,
		increases the maximum blocking of RMW operations during
		nice periods from $2\delta^*$ to $2\delta$.
	       We can avoid this drawback by using Algorithm~2 with parameters $\alpha=\beta=2\delta^*$.
	This decreases the maximum blocking time of reads
		to one-third of the CHT algorithm 
		during both the stable period and during the nice periods,
		without increasing the maximum blocking time of RMW operations
		during either type of period,
		and without incurring the overhead of additional status messages
		during nice periods.
	

\begin{table}[h]
\centering
\begin{tabular}{ | c | c || c || c | c | c |}
\hline
    \multicolumn{2}{|c|}{}
    &\makecell{CHT\\ }
    &\makecell{Alg.~1 \\ $\alpha = 3\delta$}
    &\makecell{Alg.~2 \\ $\alpha = \beta = 3\delta^*$}
    &\makecell{Alg.~2 \\ $\alpha = \delta + 3\delta^*$ \\ and $\beta = 3\delta^*$}
\\
\hline

\hline
    \multirow{2}{*}{Stable Period} 
    & \RMW & $2\delta$ & $3\delta$ & $2\delta$ & $3\delta$
\\
    \cline{2-6}
    & \tRead & $3\delta$ & $0$ & $\delta$ & $0$
\\
\hline

\hline
\multirow{2}{*}{Nice Periods} 
    & \RMW & $2\delta^*$ & $3\delta$ & $3\delta^*$ & $\delta + 3\delta^*$
\\
    \cline{2-6}
    & \tRead & $3\delta^*$ & 0 & 0 & 0
\\
\hline
\end{tabular}
\\[1pt]
\captionsetup{justification=centering}
\caption{Achieving non-blocking reads.}
\label{table2}
\end{table}
\vspace*{-6mm}

Table~\ref{table2} shows
	parameter settings aimed at
	\emph{eliminating} blocking of reads altogether,
	even if at the cost of some increase in the blocking time of RMWs.
As we have seen in our earlier discussion,
	by setting $\alpha = 3\delta$,
	Algorithm~1 ensures that read operations never block;
	but this setting increases the maximum
	blocking time of RMW operations to $3\delta$
	even during nice periods.
With a suitable choice of its two parameters,
	Algorithm~2 can do better.
For example, by setting $\alpha = \beta = 3\delta^*$:
	(1)~read operations block
	for at most $\delta$,
	and
	(2)~reads never block during nice periods;
	this is achieved at the cost of increasing
	the maximum blocking time of RMW operations only by~$\delta^*$,
	and only for nice periods.
Finally, the parameters can also be set so that
	\emph{all} reads are non-blocking;
	this is at the cost of an additional increase
	of the maximum blocking time of RMW operations
	by a single $\delta$
	(see last column of Table~\ref{table2}).

\textbf{Roadmap.} In Section~\ref{sec-algo} we describe our algorithm and its performance under the simplifying assumption that $\epsilon = 0$, and we consider the case where $\epsilon \ge 0$ in Section~\ref{sec:epsilon}. In Section~\ref{sec:discussion},
	we discuss our assumption of known message delays and the adaptiveness of the algorithm.
We briefly review some related work in Section~\ref{sec:related} and conclude the~paper~in~Section~\ref{sec:conclusion}.

\section{The algorithm}\label{sec-algo}

Algorithms~1 and~2 are described
	in sufficient detail but informally in English
	in Sections~\ref{subsec:promise} and~\ref{subsec:status}, respectively.	
The pseudocode of Algorithms~1 and~2 are given in Figures~\ref{ObjectAlgo-alg1-code} and~\ref{ObjectAlgo-alg2-code}
	(pages~\pageref{ObjectAlgo-alg1-code} and~\pageref{ObjectAlgo-alg2-code}), respectively.
Both algorithms use the same variables, so they are given only in Figure~\ref{ObjectAlgo-alg1-code}.
The code differences between Algorithm~1 and 2  are small and
	are highlighted in {\color{blue} blue} in Figure~\ref{ObjectAlgo-alg2-code}.
Reading the detailed pseudocode may be skipped, but our English description of the algorithms has line references
	to the pseudocode to help the reader who wishes
	to follow it.
A complete proof of the correctness of Algorithm~1 is given in
	Appendix~\ref{RO-Proof}.

\subsection{Eventual leader election}
Our algorithms use a leader election procedure $\Leader()$
	with the following property:
	there is a time after which every call to $\Leader()$ returns the \emph{same} correct process.
	This procedure is the failure detector $\Omega$ \cite{CHT96}; it can be implemented efficiently in partially synchronous systems
	(even without synchronized clocks)~\cite{ADFT08,ST08}.
Throughout the paper $\ell$ refers to this process.
Our algorithms also use the procedure $\AL(t,t')$,
	which can be implemented from $\Leader()$ in our model~\cite{CHT16}.
Intuitively,  $\AL(t,t')$ returns \textsc{True}
	if and only if the process that invoked it has been the
	leader \emph{continuously} during the entire time interval $[t, t']$;
	$\AL(-,-)$ also ensures that no two distinct processes can consider
	themselves to be leaders for two intersecting time intervals.
\begin{itemize}
\item
If the calls $\AL(t_1,t_2)$ and $\AL(t'_1,t'_2)$ by \emph{distinct} processes
	both return \textsc{True},
	then the time intervals $[t_1,t_2]$ and $[t'_1,t'_2]$ are disjoint.

\item
There is a time $t^*$ such that if
	$\ell$ calls $\AL(t_1,t_2)$ at time $t \ge t_2\ge t_1\ge t^*$,
	then this call returns \textsc{True},
	and if a process
	$q \neq \ell$ calls $\AL(t_1,t_2)$ with $t_2 \ge t^*$,
	then this call returns \textsc{False}.

\end{itemize}

Our algorithms use the procedure $\ML{t_1}{t_2}$
	to effectively divide time
	into a sequence
	of maximal non-overlapping intervals,
	during each of which at most one process is continuously the leader, 
	and the last of which is infinite and
	has a nonfaulty leader~$\ell$.
Intuitively, a leader has two functions:
	(i)~it linearizes the RMW operations using a consensus mechanism, and
	(ii)~it issues ``read leases'', which makes it possible
	to execute read operations efficiently.
We now describe how each of these functions work in our two algorithms.

\subsection{Algorithm~1: The promise mechanism}\label{subsec:promise}

For the first function,
	the leader collects into \emph{batches}
	the RMW operations submitted by processes
	(lines~\ref{rr1}--\ref{rr2}),\footnote{In this subsection line numbers refer to Figure~\ref{ObjectAlgo-alg1-code}.}
	and it uses the two-phase commit protocol
	outlined in the introduction as follows
	(lines~\ref{nextops}--\ref{doops2failed} and
	procedure $\DO$ in lines~\ref{acceptcheck1}--\ref{done},
	called in line~\ref{second-doops}).
To commit a batch, the leader first attaches to the batch a sequence number $j$ and
	a \emph{promise time} $t+\alpha$,
	where $t$ is the current time and $\alpha$ is the
	parameter of the algorithm
	(line~\ref{set-promisetime}).
Intuitively, the leader guarantees that this batch of operations ``will not take effect'' before the promise time $t+\alpha$.
The leader then sends prepare messages to notify processes of batch~$j$
	(line~\ref{prep-send}).
When a process receives this message we say that
	it \emph{becomes aware} of batch $j$,
	and it responds with an acknowledgment
	(lines~\ref{a3b}--\ref{sendPack}).
When the leader receives enough acknowledgements,
	it commits this batch $j$ and sends commit messages to all processes
	(lines~\ref{prep-condition}--\ref{sendcommit}).
Note that when a batch is committed, it does \emph{not} mean that the operations in this batch have taken effect:
	the algorithm ensures that these operations are not visible to users
	(and in particular they do not return) before the batch's promise time.
Roughly speaking, a batch of RMW operations takes effect when it has been committed \emph{and} its promise time has been reached.

Each process applies to its local replica the committed batches in sequence, and
	applies the operations of each batch in some pre-determined order,
	the same for all processes
	(procedure $\EB$, lines~\ref{EB-start}--\ref{EB-end}).
When a process applies one of its own RMW operations to its replica, 
	it determines the response of that operation,
	\emph{and then it waits until the promise time
	of the batch containing
	that operation before
	returning this response}
	(lines~\ref{rmw-until}--\ref{rmw-return}).
Since all processes apply the same sequence of RMW operations in the same order
	(which is consistent with the order of non-concurrent operations)
	the execution of RMW operations is linearizable.

	The second function of the leader is to periodically issue read leases to allow processes to read locally, as we now explain.
	Recall that the leader starts processing  batch $j$ at some time $t$ and commits this batch with promise time
		$t+\alpha$.
	After committing batch $j$,
		the leader issues the read lease $(j,s)$ with $s = t+\alpha$
		by sending a lease message to all processes;
		this message is combined with the commit message
		(line~\ref{sendcommit}).
	We say that the read lease $(j,s)$ \emph{starts} at time $s$ and
		\emph{expires} at time $s+\lambda$, 
		where $\lambda$ is the \emph{lease period};
		we also say that the lease $(j,s)$ is \emph{valid at time $t'$} if $t' < s+\lambda$.
	At some time $s'$ before the read lease $(j,s)$ expires,
		the leader renews the lease by issuing the lease $(j,s')$.
	Such lease renewals for batch $j$ occur periodically
		until the leader commits batch $j+1$
		(line~\ref{sendcommit2} within the main loop of the $\LW$ procedure,
		lines~\ref{mainwhile}--\ref{doops2failed}).\footnote{The lease period $\lambda$
		and the frequency of lease renewals are chosen so that after the system stabilizes
		all the correct processes always have valid leases.}
	Note that when the leader issues
		the \emph{first} read lease $(j,s)$ for batch~$j$
		(line~\ref{sendcommit}),
		the start time $s=t+\alpha$ of this lease \emph{can be in the future},
		but whenever the leader issues
		a lease \emph{renewal} $(j,s')$ for batch $j$
		(line~\ref{sendcommit2}),
		the start time $s'$ is when this lease is issued.
	
We now explain the semantics of read leases,
	and how they are used by processes to read from their local replicas.
If a process $p$ has a valid lease $(k^*,t^*)$ at time $t'$ then the following two lease properties hold:

\begin{enumerate}
	\item
	No batch $j > k^*$ takes effect before time $t^*$.
	
	This property is ensured as follows.
	If $(k^*,t^*)$ is the \emph{first} read lease
		that the leader issued for batch $k^*$ (line~\ref{sendcommit}),
	  then the leader ``promised'' that batch $k^*$ will not take effect before time $t^*$
	  (and the algorithm ensures this promise is kept);
	  this implies that no batch $j > k^*$ takes effect before time $t^*$.
	If $(k^*,t^*)$ is a read lease renewal (line~\ref{sendcommit2}),
		then when the leader issues it at time $t^*$
		it has not yet committed any batch $j > k^*$.

	\item
	No batch $j >k^*$ takes effect during the interval $[t^*,t^* +\lambda)$
		before $p$ is aware of batch $j$.
		
	Intuitively, this property is ensured as follows.
	The leader keeps track of the processes that may hold a valid read lease on 
		the last batch it committed
		(these are the $\LeaseHolders$);
		before the leader commits a new batch $j$
		it waits until all the $\LeaseHolders$
		acknowledge
		the prepare messages for
		this batch (so they are now aware of batch $j$);
		if some of them do not acknowledge batch $j$
		then the leader waits until time $t^* +\lambda$, i.e., until all read leases expire
		(lines~\ref{wait2}--\ref{lh2})
		before committing the new batch $j$.
	
	\end{enumerate}

Now suppose that a process $p$ wants to read the object at some time $t'$
	(lines~\ref{read-invoke}--\ref{read-respond}).
To do so, intuitively $p$ needs to determine the maximum number $\hat{k}$
	such that batch $\hat{k}$
	took
	effect by time $t'$:
	$p$ can then read the state of the object after batch $\hat{k}$,
	i.e., after applying all the operations in batches $0$ to $\hat{k}$ to its local replica.
If $p$ holds a valid lease $(k^*,t^*)$ at the time $t'$ when it wants to read, 	it can determine this $\hat{k}$ by using
	the lease properties and the promise mechanism as follows:
	
\begin{case}
\item $t' < t^*$.
By the first lease property, only batches with
	sequence number at most $k^*$ can take effect by time $t' < t^*$.
By the promise mechanism, only batches with a promise time at most $t'$ can take effect by time $t'$.
Process $p$ determines the maximum batch number $\hat{k}$ such that $\hat{k} \le k^*$
	\emph{and} the promise time of batch $\hat{k}$ is at most~$t'$.
Note that batch $\hat{k}$ took effect by time $t'$: this is because it was committed by time $t'$\footnote{Since no leader can issue the lease $(k^*,-)$ before batches $0$, $1$, $2, \ldots , \hat{k} , \ldots , k^*$ have been committed.}
	\emph{and} the promise time of batch $\hat{k}$ is at most $t'$.
Thus $\hat{k}$ is the maximum batch number such that batch $\hat{k}$ took effect by time~$t'$.

Our algorithm ensures that
	because $p$ holds a lease $(k^*,t^*)$ at time $t'$,
	it has already received all the batches up to and including $k^*$
	by time $t'$.
After determining $\hat{k}$, process $p$ just reads the state of the object after batch $\hat{k}$
	at time $t'$ without any waiting.

\item $t' \ge t^*$.
First note that batch $k^*$ took effect by time $t'$:
	this is because $k^*$ was committed by time $t^* \le t'$
	\emph{and} the promise time of batch $k^*$ is at most $t^* \le t'$.
Thus $\hat{k} \ge k^*$.
Since the lease $(k^*,t^*)$ is valid at time $t'$, we have $t^* \le t' < t^* +\lambda$.
By the second lease property, the only batches with sequence number $j > k^*$ that can take effect by time $t'$ are those
	that $p$ is aware of at time $t'$.
By the promise mechanism, the only batches that can take effect by time~$t'$
	are those with a promise time at most $t'$.
Process $p$ determines the set $B$ of batches
	with sequence numbers $j > k^*$ such that:
	(a) $p$ is aware of batch $j$ at time~$t'$, and
	(b) the promise time of batch $j$ is at most $t'$.
From the above, $B$ consists of \emph{all} the batches with a sequence number greater than $k^*$ that could have taken effect by time~$t'$.
Thus, process $p$ can now compute
	$\hat{k}$ to be the maximum batch number in $B$ if $B$ is not empty,
	and $\hat{k} = k^*$ otherwise.
From the above, $\hat{k}$ is the maximum number such that
	batch $\hat{k}$ could have taken effect by time~$t'$.

After computing $\hat{k}$, process $p$ first waits until it has all batches up to $\hat{k}$
	and until the promise time of batch $\hat{k}$ has passed.\footnote{The promise time of batch $\hat{k}$ can change (and increase) since the time $p$ determined the set $B$ if and only if the leader trying to commit batch $\hat{k}$ changes.  As an optimization, it turns out that waiting for the promise time of $\hat{k}$ to pass is not necessary!}
It then reads the state of the object after batch $\hat{k}$.\footnote{Like the CHT algorithm, our algorithm
	incorporates
	a further optimization that ensures no read blocks
	unless it is concurrent with a conflicting RMW operation:
	to determine $\hat{k}$, $p$ eliminates from the set $B$ every batch that contains only RMW operations that do \emph{not}
	conflict with its read operation.
It can do so because the operations in these batches do not affect the value that it reads.}

\end{case}

Having explained how the read operations work
	with the new semantics of read leases under the promise mechanism,
	we now point out a subtelty with how promise times
	must be handled when a new leader takes over.
Note that the leaders must ensure that,
	\emph{even across leadership changes},
	all nonfaulty processes agree
	on the same sequence of batches,
	and
	that each RMW operation is included
	in exactly one batch.
To do so, the first thing that a new leader does
	is to wait long enough for all leases issued by previous leaders
	to expire (line~\ref {wait-lease-expire}).
It then commits or recommits the last batch $j$ that the previous leader
	attempted to commit but may have left half-done
	(lines~\ref{est-request}--\ref{first-doops}).
The new leader should not give a future promise time to batch $j$
	because doing so would allow processes
	to read the state of the object
	before the operations of batch~$j$ have been applied to it,
	even though batch $j$ could have already taken effect
	under the previous leader.
So, to be safe, the new leader uses the promise time~0 for batch $j$;
	effectively giving no promise for batch~$j$
	(line~\ref{first-doops}).

\textbf{Maximum blocking time analysis.}
The column of Table \ref{TableAlgo2Generic}
	labeled ``Algorithm~1''
	gives the maximum blocking times of RMW and read operations
	during the stable period (where all messages take at most $\delta$) and during nice periods (where all messages take at most $\delta^* \ll \delta$) for arbitrary values of $\alpha \le 3 \delta$.
Setting $\alpha > 3 \delta$ only increases the blocking of RMW
	operations without any benefit for the reads.
We now justify the entries of that column.

Consider the system in the stable period.
Suppose that a process $p$ wants to read at time~$t'$ and
	holds a valid lease $(k^*,t^*)$ at time $t'$.
If $t' < t^*$, then by Case 1 above this read does not block.
If $t' \ge t^*$, then by Case 2 above the read may block because $p$ waits until it knows all batches up to $\hat{k}$
	and until the promise time of batch $\hat{k}$ has passed.
If $\hat{k} = k^*$ then the read does not block since these two conditions are already met by time $t'$:
	this is because $p$ has the read lease $(k^*,t^*)$ at time $t'$.
Now assume that $\hat{k} > k^*$, so $\hat{k} \in B$.
Let $t$ be the time when the leader sent the prepare messages for batch $\hat{k}$;
	so the promise time of batch $\hat{k}$ is $t +\alpha$.
Since batch $\hat{k}$ is in the set $B$, $p$ is aware of batch $\hat{k}$ and the promise time of $\hat{k}$ 
	is at most $t'$, i.e., $t +\alpha \le t'$.
Because the system is in the stable period, $p$ will receive all batches up to $\hat{k}$ by time $t + 3\delta$.
So $p$ blocks from time $t' \ge t +\alpha$ to at most time $t + 3\delta$, i.e., for at most $3\delta - \alpha$.

Now suppose the leader wants to issue a RMW operation at time $t$.
	To process this operation, the leader
	waits for acknowledgments for the batch that contains the RMW operation; this will be done by time $t+2\delta$.
	It must also wait until the promise time $t +\alpha$  before it returns the response to the RMW operation.
	So the RMW completes by time $\max(t+2\delta , t +\alpha)$, i.e., it blocks for $\max(2\delta , \alpha)$

The analysis for the nice periods is similar.

\subsection{Algorithm~2: The status mechanism}\label{subsec:status}

Recall that in Algorithm~1 each batch $j$ has a promise time,
	which is a lower bound on the time when the batch takes effect.
In Algorithm~2, a batch does not have a fixed promise time but
	a sequence of increasing promise times,
	and thus a sequence of increasing lower bounds
	on the time when it takes effect.
To accomplish this,
	when the leader wants to commit a new batch $j$
	it does not send prepare messages that notify processes of
	the batch~$j$ and its associated promise time, as in Algorithm~1.
Instead, every $\beta$ time units the leader sends
	a new round of so-called \emph{status} messages for batch~$j$
	with promise time $t+\alpha$,
	where $t$ is the time when this round of status messages is sent
	(lines~\ref{algo2-dop-repeat}--\ref{algo2-dop-sendstatus-d}).\footnote{In this subsection line numbers refer to Figure~\ref{ObjectAlgo-alg2-code}.}
The leader stops sending status messages for batch~$j$
	as soon as it receives enough acknowledgements
	(line~\ref{algo2-dop-condition}).
It then sends commit messages for batch~$j$ to all processes,
	just as in Algorithm~1.
By choosing the parameter $\beta \ge 2 \delta^*$,
	in nice periods only one round of status messages is sent per batch.
This round replaces the prepare messages of Algorithm~1,
	and so the algorithm does not incur extra messages during nice periods.
In fact, with such a $\beta$, Algorithm~2 behaves exactly as Algorithm~1 during nice periods.

The leader also sends read leases:
The \emph{first} lease $(j,s)$ for batch $j$ 
	is sent alongside the commit message for that batch
	with a start time
	equal to
	the promise time of the \emph{last} status message for batch $j$
	that the leader sent --- i.e., a time that could be in the future
	(line~\ref{algo2-dop-sl2}).
As in Algorithm~1, the start time of each lease \emph{renewal} for batch $j$ is the time when it is sent
	(lines~\ref{algo2-taketime1}--\ref{algo2-sl1}).
Read leases have the same two properties as in Algorithm~1.

A subtlety that concerns the initialization of a new leader
	is worth pointing out.
As with Algorithm~1, the new leader first commits or recommits the last batch $j$ that the previous leader
	attempted to commit but may have left half-done,
	and to be safe the new leader uses the promise time 0 for batch $j$.
So Algorithm~2 uses the exact same procedure as Algorithm~1
	to commit batch $j$ during its initialization
	(see procedure $\DO$).
To commit subsequent batches, Algorithm~2
	uses the procedure described above,
	which sends successive rounds of status messages
	with increasing promise times
	(see procedure $\DO'$ in Figure~\ref{ObjectAlgo-alg2-code}).

\textbf{Maximum blocking time analysis.}
We now analyse the maximum blocking time of reads after the system stabilizes.
This
	analysis
	also shows how the ``status mechanism'' unblocks certain read operations
	that would remain blocked for a longer period under Algorithm~1.
Suppose that a process $p$ holding a valid lease $(k^*,t^*)$ at time $t'$
	wishes to perform a read at time $t'$ and is blocked.
As with Algorithm~1, this blocking can occur only in Case~2,
	i.e., when $t' \ge t^*$
	and the read is blocked
	because $p$ is aware of a batch $j>k^*$
	that has promise time at most $t'$.
Under Algorithm~1, such a read will remain blocked until
	$p$ has all batches up to $j$
	which may take $3\delta - \alpha$
	(see the first column of Table~\ref{TableAlgo2Generic}).
Consider now the same scenario under Algorithm~2.
Every $\beta$ units of time
	the leader sends a status message (with a new promise) for batch $j$,
	or it has already sent a commit message for batch $j$.
If it sends a status message after time $t'-\alpha$,
	the associated promise time is greater than $t'$.
So by time $t'-\alpha+\beta$
	the leader sends a status message with a promise time greater than $t'$,
	or it has already sent a commit message, for batch $j$.
Process $p$ receives that message by time $t'-\alpha+\beta+\delta$,
	and this unblocks the read:
	if it is a status message with a promise time greater than $t'$, then $p$ can read before batch $j$;
	if it is a commit message, $p$ can read after batch $j$.
Therefore, under Algorithm~2
	$p$'s read operation is blocked only
	during the interval $[t',t'-\alpha+\beta+\delta]$,
	i.e., for at most $\delta+\beta-\alpha$ units of time.


\begin{table}[h]
\centering
\begin{tabular}{ | c | c | c | c |}
\hline
    \multicolumn{2}{|c|}{}
    &\makecell{Algorithm 1 \\ $\alpha \leq 3\delta$}
    &\makecell{Algorithm 2\\ $\alpha \leq \delta + \beta$ \\ $2\delta^* \leq \beta \leq 2\delta$} 
\\
\hline
    \multirow{2}{*}{Stable Period}
    & RMW & $\max(2\delta, \alpha)$ & $\max(2\delta, 2\delta - \beta + \alpha)$
\\
    \cline{2-4}
    & Read & $3\delta - \alpha$ & $\delta + \beta - \alpha$
\\
\hline
\multirow{2}{*}{Nice periods} 
    & RMW & $\max(2\delta^*, \alpha)$ & $\max(2\delta^*, \alpha)$
\\
    \cline{2-4}
    & Read & $\max(3\delta^* - \alpha, 0)$ & $\max(3\delta^* - \alpha, 0)$
\\
\hline
\end{tabular}
\\[1pt]
\captionsetup{justification=centering}
\caption{Maximum blocking times under Algorithms~1 and~2 ($\epsilon=0$).}
\label{TableAlgo2Generic}
\end{table}

\vspace*{-6mm}
For the analysis of the maximum blocking time of RMW operations,
	it is convenient to assume that $\beta$ divides $2 \delta$.
Suppose the leader wants to issue a RMW operation at time $t$.
Before it returns the response to this RMW operation,
	the leader waits for acknowledgments
	for the batch that contains the RMW operation;
	this will be done by time $t+2\delta$.
It must also wait until the promise time of the \emph{last} status message 
	that it sent for that batch;
	since $\beta$ divides $2\delta$,
	that sending occurs by time $t+2\delta - \beta$,
	and so the promise time of that status message is
	at most $t+2\delta - \beta +\alpha$.
So the RMW completes by time $\max(t+2\delta, t+2\delta - \beta +\alpha)$,
	i.e., it blocks for $\max(2\delta, 2\delta - \beta +\alpha)$.
	
Since we assume that $\beta \ge 2\delta^*$,
	and in this case
	Algorithm~2 behaves exactly as Algorithm~1 during nice periods,
	the blocking times during these periods are the same as in Algorithm~1.
The maximum blocking times with Algorithm~2 are shown
	in the second column of Table~\ref{TableAlgo2Generic}.

	\section{Approximately Synchronized Clocks}\label{sec:epsilon}

	Recall that in our model all local clocks are always synchronized within $\epsilon$ with each other.
	To simplify the presentation,
		so far we have been assuming that $\epsilon=0$.
	In this section we explain how to modify our algorithms so that they work
		even when local clocks are not perfectly synchronized,
		i.e., when $\epsilon > 0$, and give their performance.
	We refer to the values of local (process) clocks as \emph{local time}
		to distinguish it from \emph{real time}. 
	
	The main challenge when $\epsilon>0$ is that 
		processes may not agree whether, at some real time, a batch has taken effect yet, and 
		they may execute operations that violate linearizability.
	For example, suppose that at every real time
		the clock of process $p^-$ shows local time $\epsilon/2$ less than real time
		while the clock of process $p^+$ shows local time $\epsilon/2$ more than real time.
	Suppose now that batch $j$ has promise time $s$.
	At real time~$s$,
		when the clock of $p^+$ shows $s+\epsilon/2>s$,
		$p^+$ reads the state of the object after batch $j$.
	At the later real time $s+\epsilon/4$,
		when the clock of $p^-$ shows $s+\epsilon/4-\epsilon/2<s$,
		$p^-$ reads the state of the object before batch $j$.
	This violates linearizability.
	
	We address this problem in the same way in both Algorithms 1 and 2 as follows.
	Whenever a process $p$ waits for the promise time $s$ of some batch $j$ to expire,
		we require $p$ to wait for an extra $\epsilon$,
		i.e., until its clock reaches $s+\epsilon$.
	Thus, if a process $p$ wants
		to read the state of the object after batch~$j$
		(line~\ref{wait-promise-2})
		or to return the response from a RMW operation contained in batch $j$
		(line~\ref{rmw-wait-promise}),
		$p$ now waits until its clock shows time $s+\epsilon$.
		(Throughout this section,
		line numbers refer to the pseudocode of Algorithm~1.)

Perhaps surprisingly, the computation of $\hat{k}$
	(lines~\ref{get-k-hat-then} and~\ref{get-k-hat-else}--\ref{get-k-hat-5})
	does not change when $\epsilon>0$.
To see this suppose that
	process $p$ wishes to perform a read operation
	at real time $\tau$ and local time $t'$, and
	$p$ is aware of a batch $j$ with promise time $s > t'$.
At real time $\tau$,
	the local clock of every process is at most $t' +\epsilon$.
Since $t'+\epsilon < s+\epsilon$,
	and each process $q$ waits until its local clock is
	at least $s+\epsilon$ before the promise of batch $j$ expires at $q$,
	by real time $\tau$
	no process could have read the state of the object
	after the operations of batch~$j$ have been applied,
	and no process could have returned the response from a RMW operation
	contained in batch~$j$.
So at real time $\tau$,
	$p$ can safely read the state of the object
	before the operations of batch~$j$ are applied,
	without violating linearizability.
This shows that process $p$ can compute $\hat k$
	in the same way as with $\epsilon = 0$,
	i.e., by considering only the batches~$j$ with promise $s \le t'$
	(as opposed to those with $s \le t' +\epsilon$).
To retain the property that $p$'s read does not block if there are no conflicting concurrent RMW operations,
	$p$ actually considers only the batches~$j$ with promise $s \le t'$ that
	contain RMW operations that conflict with $p$'s read.
(This is already done when computing $\hat{k}$ in lines \ref{get-k-hat-else}--\ref{get-k-hat-5},
	and the same must be done now also in line~\ref{get-k-hat-then}.)


\begin{table}[t]
\centering
\begin{tabular}{ | c | c | c | c |}
\hline
    \multicolumn{2}{|c|}{}
    &\makecell{Algorithm 1 \\ $\alpha \leq 3\delta$}
    &\makecell{Algorithm 2\\ $\alpha \leq \delta + \beta$ \\ $2\delta^* \leq \beta \leq 2\delta$} 
\\
\hline
    \multirow{2}{*}{Stable Period}
    & RMW & $\max(2\delta, \alpha + \epsilon)$ & $\max(2\delta, 2\delta - \beta + \alpha + \epsilon)$
\\
    \cline{2-4}
    & Read & $\max(3\delta - \alpha, \epsilon)$ & $\max(\delta + \beta - \alpha, \epsilon)$
\\
\hline
\multirow{2}{*}{Nice periods} 
    & RMW & $\max(2\delta^*, \alpha + \epsilon)$ & $\max(2\delta^*, \alpha + \epsilon)$
\\
    \cline{2-4}
    & Read & $\max(3\delta^* - \alpha, \epsilon)$ & $\max(3\delta^* - \alpha, \epsilon)$
\\
\hline
\end{tabular}
\\[1pt]
\captionsetup{justification=centering}
\caption{Maximum blocking times under Algorithms~1 and~2 (any $\epsilon\ge0$).}
\label{TableWithEpsilon}
\vspace*{-6mm}
\end{table}

There is a similar problem, and a similar solution,
	with the lease mechanism when $\epsilon > 0$.
To see the problem
	suppose all processes except $p^-$ (a process that is not the leader)
	have clocks that show real time,
	and process $p^-$ has a clock that shows $\epsilon$ less than real time.
Suppose that $p^-$ holds a lease $(j,t_j)$,
	and the leader that issued that lease
	wishes to commit a new batch $j+1$
	with a promise time of $t_j+\lambda-\epsilon$.
If $p^-$ does not receive the prepare message for batch~$j+1$
	(and therefore does not send an acknowledgement to the leader),
	the leader waits until the lease $(j,t_j)$ expires at real time $t_j+\lambda$.
At that real time the leader commits batch~$j+1$,
	issues a lease for that batch,
	and reads the state of the object after batch~$j+1$.
The lease $(j,t_j)$ that $p^-$ holds
	is valid at $p^-$ until local time $t_j+\lambda$,
	i.e., until real time $t_j+\lambda+\epsilon$.
So, $p^-$ can read the state of the object before batch~$j+1$
	during the real time interval $(t_j+\lambda,t_j+\lambda+\epsilon)$,
	which follows the time when
	the leader has read the state of the object after batch~$j+1$.
This violates linearizability.

The solution to this problem is similar to
	the solution for the corresponding problem with promises:
Whenever
	the leader waits for a lease $(j,t_j)$ to expire
	(lines~\ref{wait-lease-expire} and~\ref{wait-alg1}),
	we require it to wait for an extra $\epsilon$,
	i.e., until its clock reaches $t_j+\lambda+\epsilon$.
This implies that when the leader stops waiting,
	the lease $(j,t_j)$ has expired at all processes
	and thus it cannot be used to read.

With the above modifications to handle the case that $\epsilon\ge0$,
	the worst-case blocking times of our algorithms are shown in Table~\ref{TableWithEpsilon}.
As shown in this table, the maximum blocking times of RMW and read operations increase by at most $\epsilon$ compared to
	the special case that $\epsilon =0$.
As with~\cite{CHT16}, however, with our algorithms every read operation that does not conflict
	with a concurrent RMW operation remains non-blocking.

From Table~\ref{TableWithEpsilon} it is clear that we can set the algorithms' parameters
	so that the maximum blocking time for read operations is $\epsilon$;
	for example, $\epsilon$ is achieved by setting $\alpha=3\delta$ in Algorithm~1 or
	$\alpha=\delta+\beta$ in Algorithm~2.
If $\epsilon\le\delta/2$, this nearly matches a lower bound by Chandra \emph{et al.} (Theorem~4.1 in ~\cite{CHT16}).
Note that $\epsilon \le \delta/2$ holds in geo-distributed systems
	where, with present technology, clock skew can be under 10msec~\cite{CD+12}
	and message delays (say between data centres located in different continents)
	can be in the order of~100msec~or~more~\cite{KP+12}.

\section{Discussion}\label{sec:discussion}

\textbf{Knowing $\delta$ and $\delta^*$.}
Recall that our algorithms use two message delay estimates:
	 $\delta$ (the maximum message delay after the system stabilizes)
	 and $\delta^*$ (the maximum message delay during nice periods).
The reader may wonder whether it is reasonable to assume that
	$\delta$ and $\delta^*$ are known,
	and what happens if their assumed values are incorrect.

We first note that the assumption of a known $\delta$ is made routinely. 
For example, distributed algorithms that use timeouts on remote machines
	(say for detecting whether they are still alive)
	include an estimate of $\delta$ to determine the timeout period.
Also, many practical \emph{lease-based} distributed algorithms (e.g.,~\cite{B06})
	also use a known $\delta$ to calculate the length~of~the~lease.

What is the effect of assuming the wrong $\delta$?
In our algorithms, safety does not depend on having a correct estimate on $\delta$;
	it is always preserved.
\emph{Underestimating} $\delta$ can affect liveness:
	during ``bad'' periods where some messages take more than $\delta$
	it is possible that no progress is made.
\emph{Overestimating} $\delta$ may increase worst-case blocking times.

What is the effect of assuming the wrong $\delta^*$?
It turns out that neither safety \emph{nor} liveness depends
	on having a correct estimate on $\delta^*$.
The only consequence of \emph{underestimating} $\delta^*$ is
	that nice periods would be less frequent and shorter,
	so the maximum blocking times that we achieve for nice periods
	would be less useful.
The consequence of \emph{overestimating} $\delta^*$ is
	a possible increase in the worst-case blocking times.
But since safety and liveness do not depend on the choice of $\delta^*$,
	one can easily readjust the estimate of $\delta^*$ dynamically
	to match the ``current'' state of the system.

\textbf{Adaptiveness.}
Related to the question of the algorithm making use of $\delta$ and $\delta^*$
	is the property of ``adaptiveness'', in the following sense:
One of the advantages of the (completely) asynchronous model is that,
	because there are no known bounds on message delays,
	algorithms designed to work in that model
	tend to adapt to the actual operating conditions without
	making worst-case assumptions:
	if messages flow fast, such algorithms are correspondingly fast;
	if messages slow down, so does the algorithm.
This is a desirable property because, in practice, operating conditions
	are often favourable.
Unfortunately there are limits to implementing fault-tolerant objects
	in completely asynchronous systems;
	in particular,
	it is not possible to implement objects with arbitrary RMW operations as we do here~\cite{FLP83,Herlihy91}.

Note that in our algorithm \emph{all the read operations are adaptive},
	regardless of the parameter settings.
For RMW operations, our algorithm exhibits the flexibility of
	trading off their adaptivity with the worst-case blocking time of reads:
	if we set the parameter $\alpha$ to 0
	(i.e., the special case that is the CHT algorithm),
	the RMW operations are also adaptive;
	but in that case the (adaptive) reads may block for up to $3\delta$ time.
If, on the other hand, we prefer to optimize reads,
	we can set the parameters to reduce their worst-case blocking time
	at the cost of decreasing the adaptivity of the RMWs.
The best parameter setting for this trade-off depends
	on the relative frequency of read and RMW operations and
	on what one wants to achieve.
An advantage of our algorithm is that it allows for parameter settings
	that best fit different operating conditions and user objectives.

\section{Related work}\label{sec:related}

\textbf{Lower bounds.}
Attiya and Welch have shown some lower bounds on the time
	to read and write for linearizable implementations of registers~\cite{AW94}.
These bounds apply to systems where processes
	have clocks that run at the same rate as real time
	and \emph{all} the message delays are in the range $[\delta-u,\delta]$ for some known $\delta$ and message uncertainty $u$,
	where $0 \le u \le \delta$.
For $u = 0$, they prove that
	the \emph{sum} of the times to do a read and a write operation is
	at least $\delta$ (Theorem 4.1 in~\cite{AW94}).
For $u > 0$, they prove that
	a read operation requires at least $u/4$ time
	and a write operation requires at least $u/2$ time
	(Theorems 3.1 and 3.2 in~\cite{AW94}).

These bounds do not apply to the algorithms that we presented here
	because our model is incomparable
	to the model in~\cite{AW94}.
On one hand, our model is weaker because the maximum message delay applies only to messages sent after (an unknown)
	stabilization time.
On the other hand, it is also stronger because we assume that processes are equipped with external clocks
	that are synchronized within some $\epsilon \ge 0$.
In our model, after stabilization time we have $u = \delta$.
Note that for some parameter settings,
	reads in our algorithm take at most $\epsilon$ time which could be less than the $u/4$ lower bound of~\cite{AW94}
	\emph{if} the clocks are highly synchronized (e.g., via special devices such as atomic clocks and GPS signals,
	such as in the Spanner system~\cite{CD+12},
	or via special high priority messages).
This demonstrates a benefit of adding highly synchronized external clocks to partially synchronous systems.

\textbf{Algorithms.}
Replication is used extensively in distributed systems
	ranging from synchronous, tightly coupled ones,
	to asynchronous, geographically dispersed ones.
Below we highlight the main points of some replication algorithms that are most closely related~to~our~work.

Megastore~\cite{BB+11} is an early Google system designed
	to support distributed transaction processing with efficient reads.
Megastore implements a replicated log
	that can be written (by appending entries to it) and read.
Write operations are linearized using
	a version of the Paxos algorithm~\cite{LL98,LL01paxos},
	and read operations are local and non-blocking
	when there are no concurrent write operations.
To write the log
	Megastore requires the leader to receive acknowledgements
	from \emph{all} processes, or
	for crashed or disconnected processes to time out.
Thus, a process that crashes or becomes disconnected
	delays \emph{all} write operations issued while it is unresponsive.
In contrast, in our algorithms
	the leader keeps track of the current leaseholders, i.e.,
	the processes that acknowledged the last RMW operation,
	and in subsequent RMW operations it waits for acknowledgements
	only from them: so a process that crashes can
	delay at most one write operation.
As noted in~\cite{BB+11}, an asymmetric network partition
	can cause write operations to block indefinitely
	because of Megastore's reliance on the Chubby lock service
	(another Google system~\cite{B06}) for failure detection,
	a problem that requires operator intervention to resolve.

Paxos Quorum Leases (PQL)~\cite{MAK14} is an algorithm that addresses
	the above-mentioned problems with Megastore.
Similar to our algorithms,
	in PQL the leader keeps track of the current leaseholders
	and waits for acknowledgements to RMW operations only from them.
Lease renewals, however, are more expensive in PQL than
	in our algorithms:
Leases are granted not by the leader but by a majority of processes
	called ``lease grantors''.
Each lease renewal requires a quadratic number of messages
	in the number of participating processes
	(compared to linear, in our algorithms),
	and two message delays
	(compared to one, in our algorithm).
Furthermore, in PQL each change in the set of leaseholders triggers
	the use of a consensus algorithm (specifically of Paxos)
	among the lease grantors,
	whereas in our algorithm the leader manages this set on its own
	simply by noting the processes that acknowledge the last RMW operation.
Finally, in PQL a RMW operation revokes the current leases,
	and so a steady stream of RMW operations can disable
	local reads for arbitrarily long.
In our algorithms, all reads are local and block only for a bounded time.
	
Spanner~\cite{CD+12} is another Google system that,
	like its predecessor Megastore,
	supports distributed transactions and implements replicated objects.
Spanner is the first system we know of that uses the model
	we adopted in our paper:
	a partially asynchronous message-passing system
	equipped with accurately synchronized clocks.
Spanner	uses Google's TrueTime service, which maintains synchronized clocks,
	to attach timestamps to read and write operations,
	and executes these operations in timestamp order
	at each of the processes that manage a replicated object.
Thus, to execute a read operation with timestamp $t$,
	a process must know
	the write operation with the maximum timestamp $t'$ such that $t'<t$.
A process cannot determine this locally 
	unless it blocks until it receives a write operation
	with timestamp $t''>t$.
Thus a read operation either
	must involve communication with other processes and
	is therefore not local, or
	it may block indefinitely to wait for a write with a higher timestamp, or
	it may risk reading a stale value.

Hermes~\cite{KG+20} is a more recent system that supports replicated objects,
	designed with the express purpose of reducing the latency of operations.
To achieve this, Hermes allows any process to initiate a RMW operation,
	rather than channeling all such operations through the leader,
	as in our algorithms.
By doing so, RMW operations that are not issued by the leader
	save the round-trip delay of being sent to the leader and
	receiving the commit message.
To also achieve local reads, Hermes requires all processes to acknowledge
	each RMW operation, like Megastore.
If some process does not do so in a timely manner,
	a relatively expensive reconfiguration operation is triggered
	for a majority of processes to agree
	on the new set of processes that manage the replicated object.
This is done using a variant of Paxos called Vertical Paxos~\cite{LMZ09}.
In contrast, our algorithms weather permanent or transient
	disconnections of processes from the leader 
	using the more lightweight leaseholder mechanism.
As noted in~\cite{KG+20}, due to the lack of coordination by a leader,
	concurrent RMW operations in Hermes may abort,
	and thus they do not have a bounded blocking time.
Finally, as in PQL, a steady stream of write operations
	can disable local reads for arbitrarily~long.

\section{Conclusion}\label{sec:conclusion}

We presented a parameterized algorithm that works
	in partially synchronous systems where processes are equipped
	with clocks that are synchronized within $\epsilon$.
This algorithm generalizes the (non-parameterized) CHT algorithm,
	and for some settings of its parameters
	it ensures that no read takes more than $\epsilon$ time
	\emph{even in the presence of concurrent conflicting operations}.

A novel feature of our algorithm is that its parameters can be used for two benefits:
	They enable a continuous trade-off
	between the maximum blocking times of read and RMW operations,
	and they can be used to reduce these blocking times during ``nice'' periods where messages delays are smaller than the maximum message delay.
This is achieved by leveraging two new ideas,
	the promise mechanism and the status mechanism,
	which modify the semantics of leases.
Leases are used in a variety of settings in distributed computing,
	and we believe that our promise and status mechanisms can be used
        to achieve similar benefits in other lease-based algorithms.

\vfill    


\begin{figure}
	\hrule
\begin{tiny}
	\begin{multicols}{2}
		\begin{tabbing}
			bbb\=bbb\=bbb\=bbb\=bbb\=bbb\=bbb\=bbb\=bbb\=bbb\=bbb\=bbb\=  \kill\\
			
			\textsc{Code for process $p$:} \\ [2mm]
			
			\textbf{variables:} \\[1mm]
			
			$\maxT:= -1$	\cmmnt{max $t$ s.t. $p$ sent $\langle \EstReply,t,-,-,- \rangle$}\\
			$(\EstTuple,ts,k) := ({\bluetored}\emptyset{\bluetored},-1,0)$ \cmmnt{current estimate}\\[1mm]
			$\Batch[{\bluetoblack -1},0,1,2,..] := [\InitTuple,{\bluetored(}\emptyset{\bluetored,0)}, \InitTuple, \InitTuple, \ldots]$\\
			\cmmnt{currently known batches}\\
			\cmmnt{each batch has two fields: \textit{ops, promise}}\\

			$\sstate[-1, 0, 1, 2, \ldots] := [\sigma_0, \sigma_0, \bot, \bot, \ldots]$\\
			\cmmnt{object state after each batch; $\sigma_0=\text{init state}$}\\
			
			$\Reply{\op} := \bot$ \cmmnt{response to RMW operation $\op$}\\
			{\bluetored $\TakesEffect{\op} := \infty$} \bluecomment{promise time of the batch that $\op$ is in}\\
			
			$\cntr := 0$ 		\cmmnt{number of operations issued by $p$}\\
			$\OpsRequested := \emptyset$	\cmmnt{RMW operations requested}\\
			$\OpsDone := \emptyset$		\cmmnt{RMW operations committed}\\
			$\MBD := 0$	\cmmnt{max batch number up to which}\\
			\cmmnt{all RMW operations have been executed}\\
			
			$\replied[t] := \emptyset$  \cmmnt{responders to $\langle\EstRequest,t\rangle$}\\
			$\REPLIES[t] := \emptyset$  \cmmnt{responses to $\langle\EstRequest,t\rangle$}\\
			$\PACKED[t,j] := \emptyset$  \cmmnt{responders to $\langle \Prepare,-,t,j,- \rangle$}\\[2mm]

			{\color{black} $\PendingOps[0,1,\ldots] := [\InitTuple, \InitTuple, \ldots]$} \redcomment{pending batches}\\
			{\color{black} $\MaxPendingIndex := 0$}\redcomment{max pending batch number}\\
			{\color{black} $\LeaseHolders :=$ }{\bluetored$\emptyset$}  \bluetoredcomment{initially, no process holds a valid lease}\\
			\textcolor{black}{$\LP := \lambda$}
				  \redcomment{duration of the read lease period}\\
			\textcolor{black}{$\LRP := \Ptwo$} 	\redcomment{time between read lease renewals}\\
			\textcolor{black}{$\NextSendTime := 0$} 	\redcomment{time when next read lease is to be sent}\\
			\textcolor{black}{$\lease := (0, -\infty)$}		\redcomment{current lease held by $p$}\\
			\redcomment{$\lease$ has two fields: $\leasebatch$  and \leasetime}\\
			{\bluetored $\PP := \alpha$}
			\bluetoredcomment{duration of the promise period}\\[2mm]

		   \newcounter{algoonelinenumber}
			\setcounter{algoonelinenumber}{0}
			
			\textbf{cobegin}\\[1mm]
			
			// \textsc{Thread 1:}   \cmmnt{issue RMW or read operations}\\[1mm]
			
			\nnll\label{rmw1}\>\textbf{while} \textsc{True} \textbf{do} \\
			\nnll\label{rmw2}\>\>	\textbf{if} $p$ wants to execute a RMW operation $o$ \textbf{then}\\
			\nnll\label{rmw3}\>\>\>		$\cntr := \cntr +1$\\
			\nnll\label{rmw4}\>\>\>		$\operation := (o, (p,\cntr))$\\
			\nnll\label{rmw-send}\>\>\>  {\bluetoblack \textbf{periodically send} $\langle \OpRequest, \operation \rangle$ \textbf{to} $\textit{leader}$()}\\
			\nnll\label{rmw-until}\>\>\> {\bluetoblack \textbf{until} $\Reply{\operation} \neq \bot$}\\
			\nnll\label{rmw-wait-promise}\>\>\> {\bluetored \textbf{wait until} $\CT \geq \TakesEffect{\operation}$}\\
			\nnll\label{rmw-return}\>\>\>		\textbf{return} $\Reply{\operation}$\\
			
			\nnll\label{read-invoke}\>\> {\color{black} \textbf{if} $p$ wants to execute a read operation $o$ \textbf{then}}\\
			\nnll\label{read1}\>\>\>		{\color{black} $\cntr := \cntr +1$ }\\
			\nnll\label{read2}\>\>\>		{\color{black} $\operation := (o, (p,\cntr))$ }\\
			
			\nnll\label{getvalidlease-start}\>\>\>	{\color{black} \textbf{repeat}}\\
			\nnll\label{getleasetime}\>\>\>{\color{black}} \>	{\color{black} $t' := \CT$}\\
			\nnll\label{getlease}\>\>\>\> {\color{black} $(k^*,t^*) := \lease$} {\color{black}}\\
			\nnll\label{getvalidlease-end}\>\>\>	{\color{black} \textbf{until} $t' < t^* +\LP$}\\
			\nnll\label{get-k-hat-1}\>\>\> {\bluetored \textbf{if} $t'  < t^*$ \textbf{then}}\\
			\nnll\label{get-k-hat-then} \>\>\>\> {\bluetored $\hat{k} := \max \{j ~|~ 0 \le j \leq k^*$ \textbf{and}} {\bluetored $\BatchPromise{j} \leq t'\}$}\\
			\nnll\label{get-k-hat-3} \>\>\> {\bluetored \textbf{else}} \:\:\bluelcomment{$t^* \le t' < t^*+\LP$}\\
			\nnll\label{get-MPB} \>\>\>\> {\color{black} $u := \MaxPendingIndex$}\\
			\nnll\label{get-k-hat-else}\>\>\>\>{\color{black} $\hat{k} := \max \{j ~|~ j = k^*$ \textbf{or} $(k^* < j \leq u$ \textbf{and}}\\
			\nnll\label{get-k-hat-else-b}\>\>\>\>\>{\color{black} $o$ conflicts with an operation in}\\
			\nnll\>\>\>\>\>{$\PendingBatchOps{j}$ \textbf{and}}\\
			\nnll\label{get-k-hat-5}\>\>\>\>\>{\bluetored $\PendingBatchPromise{j} \leq t'$)\}}\\
			\nnll\label{FG3}\>\>\>\>{\color{black} 
				\textbf{wait for} $(\text{for all }j, k^* < j \leq \hat{k}, \Batch[j] \neq \InitTuple)$}\\
			\nnll\label{wait-promise-2}\>\>\>{\bluetored \textbf{wait until} $\CT \geq \BatchPromise{\hat{k}}$}\\
			
			\nnll\label{fill-gaps-to-k-hat}\>\>\>	{\color{black} $\ExecuteOpsUpToBatch{\hat{k}}$}\\
			\nnll\label{client-read-end}\>\>\> {\color{black} $(-,\reply) := \Apply(\sstate[\hat{k}],o)$}\\
			\nnll\label{read-respond}\>\>\> {\color{black} \textbf{return} $\reply$}\\[1mm]

			//  \textsc{Thread 2:}\\[1mm]
			
			\nnll\label{whileo}\> \textbf{while} \textsc{True} \textbf{do}\\
			\lcomment{determine whether to act as leader or client}\\
			\nnll\label{gt}\>\>		$t := \CT$\\
			\nnll\label{check}\>\>\textbf{if} $\ML{t}{t} = \textsc{True}$ \textbf{then} $\LeaderWork{t}$  \\
			\nnll\label{whilee}\>\> 
				$\PCM()$\\[1mm]
			
			//  \textsc{Thread 3:}\\[1mm]
			
			\nnll\label{thread3}\> 
			$\ReplyToMessages{}$ \cmmnt{reply to messages}\\[1mm]
			
			\textbf{coend}\\[1mm]

\>\\

			\textbf{procedure} $\LeaderWork{t}$:\\[1mm]

\> \lcomment{New leader initialization: find latest batch and (re)do}\\

\nnll\label{wait-lease-expire}\> {\bluetored \textbf{wait until} $\PP + \LP$ time has elapsed}\\
\nnll\label{lh1}\>{\color{black}		$\LeaseHolders := $ {\bluetored $\emptyset$}}\\
\nnll\label{est-request}\>			{\bluetoblack\textbf{periodically send}} $\langle \EstRequest,t \rangle$ \textbf{to} $\text{all processes} - \{p\}$ \\
\nnll\label{est-condition}\>		{\bluetoblack\textbf{until}}  $ | \replied[t] | \ge \lfloor n/2 \rfloor$ 
 	\textbf{or} $\ML{t}{\CT} = \textsc{False}$\\
\nnll\label{est-condition2}\>	{\bluetoblack\textbf{if} $| \replied[t] | < \lfloor n/2 \rfloor$ \textbf{then} \textbf{return}} \\

\nnll\label{selection}\>     $(\EstTupleStar,ts^*,k^*) := $ tuple with maximum $(ts^*,k^*)$\\
\>\> in $\REPLIES[t] \cup \{ (\EstTuple,ts,k) \} $ \\
\nnll\label{Egathercrumbs}\>  \textbf{if} $ts^* \ge  t$ \textbf{then}  \textbf{return}\\

\nnll\label{FG}\>	$\FillGaps{k^* -2}$\\
\nnll\label{first-doops}\> 	$\result := \DoOps{{\bluetored(}\EstTupleStar{\bluetored,0)}}{t}{k^*}$\\
\nnll\label{doopsfailed}\>  \textbf{if} $\result = \textsc{Failed} $ \textbf{then} \textbf{return}\\
\nnll\label{Do-a-NoOp}\>  initiate a $\NoOp$ as a RMW operation via Thread 1\\[1mm]

\>\\

\columnbreak

\>\lcomment{{\color{black}{Grant read leases and}} process new batches}\\
\nnll\label{mainwhile}\>	\textbf{while} \textsc{True} \textbf{do}\\
\nnll\label{taketime1}\>{\color{black}}\>		$t' := \CT$\\	
\nnll\label{recheck1}\>\>	\textbf{if} $\ML{t}{t'} = \textsc{False}$ \textbf{then return}\\
 \nnll\label{checksendtime}\>\>{\color{black}		\textbf{if} $t'  \ge \NextSendTime$ \textbf{then}}\\
 \nnll\label{setlease}\>\>\>{\color{black} $\lease := (k, t')$}\\
\nnll\label{sl1}\label{sendcommit2}\>\>\> {\bluetoblack \textbf{send} $\langle \ColoredCommitLease,\Batch[k],k {\bluetored,\lease, \LeaseHolders} \rangle$ \textbf{to} $\text{all processes} - \{p\}$}\\
\nnll\label{nst}\>\>\>{\color{black}			$\NextSendTime := t' + \LRP$}\\

\nnll\label{lh3}\>\>{\color{black} \textbf{if received} $\langle \RequestLease \rangle$ \textbf{from} a process $q$ \textbf{then} $\LeaseHolders := \LeaseHolders \cup \{q\} $}\\

\nnll\label{nextops}\>\>		$\NextOps := \OpsRequested - \OpsDone$\\
\nnll\label{nonempty2}\>\>		\textbf{if} $\NextOps \neq \emptyset$ \textbf{then}\\ 

\nnll\label{set-promisetime}\>\>\> ${\bluetored s := t'+\PP}$\\
\nnll\label{second-doops}\>\>\> $\result := \DoOps{{\bluetored (}\NextOps{\bluetored, s)}}{t}{k+1}$\\
\nnll\label{doops2failed}\label{endwhile}\>\>\>  \textbf{if} $\result = \textsc{Failed} $ \textbf{then}  \textbf{return}\\[2mm]

\textbf{procedure} $\DoOps{\DoOpsTuple}{t}{j}$:\\[1mm]
\bluelcomment{$O$ is the set of RMWs to be committed, $s$ is the promise time:}\\
\bluelcomment{$O$ will not be committed before time $s$}\\
\nnll\label{acceptcheck1}\> \textbf{if} $t < \maxT$ \textbf{then} $\textbf{return}$ \textsc{Failed}\\

\nnll\label{leader-accept}\> $(\EstTuple,ts,k) := ({\bluetored }O,t,j)$\\
\nnll\label{sendprep}\label{repeat}\label{prep-send}\>	{\bluetoblack \textbf{periodically send}} $\langle \Prepare,{\bluetored (}O{\bluetored, s)},t,j, \Batch[j-1] \rangle$ \textbf{to} $\text{all processes} - \{p\}$ \\
\nnll\label{condition}\label{endc}\label{prep-condition}\>		{\bluetoblack\textbf{until}}  $| \PACKED[t,j] | \ge  \lfloor n/2 \rfloor$  
	\textbf{or} $\ML{t}{\CT} = \textsc{False}$\\
\nnll\label{ack-condition}\label{prep-condition2}\> {\bluetoblack \textbf{if} $| \PACKED[t,j] | <  \lfloor n/2 \rfloor$ \textbf{then} \textbf{return} \textsc{Failed}}\\[1mm]

\nnll\label{wait2}\>{\color{black}		\textbf{wait}  \textbf{until} $\LeaseHolders \subseteq \PACKED[t,j]$
	\textbf{or} $2 \delta$ time has elapsed since $p$ first executed line~\ref{sendprep}}\\

\nnll\label{LHvsAcks}{\color{black} }\>{\color{black}		\textbf{if} $\neg (\LeaseHolders \subseteq \PACKED[t,j])$ 
{\bluetored \textbf{and} $s < \lease.start + \LP$}  \textbf{then}} \\
\nnll\label{wait-alg1}\>\>{\color{black}		\textbf{wait}  \textbf{until} $\CT \geq {\bluetored \lease.start} + \LP$}\\

\nnll\label{lh2}\>{\color{black}		$\LeaseHolders :=   \PACKED[t,j] $}\\

\nnll\label{batch1}\label{lease1}\>${\bluetored(}\Batch[j]{\bluetored,\lease)} := {\bluetored((}\Os{\bluetored, s),(j, s))}$\\

\nnll\label{ExecuteB2}\>		$\ExecuteOpsUpToBatch{j}$ \\

\nnll\label{sl2}\label{sendcommit}\> \textbf{send} $\langle \ColoredCommitLease, \Batch[j], j {\bluetored, \lease, \LeaseHolders} \rangle$ \textbf{to} $\text{all processes} - \{p\}$\\
\nnll\label{nst2}\>{\bluetored $\NextSendTime := s + \LRP$}\\
\nnll\label{done}\>	$\textbf{return}$ \textsc{Done}\\[2mm]

\textbf{procedure} $\FillGaps{k'}$:\\

\nnll\label{totoz8}\> \textbf{repeat} \\
\nnll\label{totoz9}\>\> $\Gaps := \{ j ~|~ 1 \le j \le k' \mbox{ and } \Batch[j] = {\color{black} (}\emptyset{\color{black}, \infty)} \}$\\

\nnll\label{totoz10}\>\>\textbf{if} $\Gaps \neq \emptyset$ \textbf{then} \textbf{send} $\langle \MyGaps,\Gaps \rangle$  \textbf{to} $\text{all processes} - \{p\}$\\
\nnll\label{totoz11}\> \textbf{until} $\Gaps = \emptyset$ \\

\nnll\label{totoz12}\> \textbf{return}\\[1mm]

\textbf{procedure} $\ExecuteBatch{j'}$:\\
\nnll\label{EB-start}\>	$\sigma := \sstate[j'-1]$\\
\nnll\label{EB-get-ops}\>		let $\op^1, \op^2, \ldots, \op^{m}$ be the operations in $\ColoredBatchOps{j'}$ listed in operation id order\\
\nnll\label{apply-op}\>               \textbf{for} $i = 1$ \textbf{to} $m$ \textbf{do} \\
\nnll\label{apply-op-1}\>\> $(\sigma,\Reply{\op^i}) :=  \Apply(\sigma,\op^i.\TYPE)$\\
\nnll\label{set-takesEffect}\>\> {\bluetored$\TakesEffect{\op^i} := \BatchPromise{j'}$}\\
\nnll\label{update-state}\>		$\sstate[j'] := \sigma$\\
\nnll\label{EB-end}\> \textbf{return}\\[1mm]

\textbf{procedure} $\ExecuteOpsUpToBatch{j'}$:\\

\nnll\label{EUTB}\> \textbf{for} $j = \MBD +1$ \textbf{to} $j'$ \textbf{do}\\
\nnll\label{EUTB2}\>\> $\ExecuteBatch{j}$\\
\nnll\label{addB1}\>\>		 $\OpsDone := \OpsDone \cup \ColoredBatchOps{j}$\\
\nnll\label{updateMBD1}\>\> 	$\MBD := \max (\MBD, j)$\\

\nnll\> \textbf{return}\\[1mm]

\textbf{procedure} $\PCM()$:\\[1mm]

\nnll\label{spcm}\label{a2b}\>	 \textbf{if received} $\langle \EstRequest, t \rangle$ \textbf{from} a process $q$ \textbf{then}\\
\nnll\label{incTmax}\>\>		$\maxT:= \max (\maxT, t)$\\
\nnll\label{a2e}\label{reply}\>\>		\textbf{send} $\langle \EstReply, t, \EstTuple,ts,k, Batch[k-1] \rangle$ \textbf{to} $q$\\

\nnll\label{a3b}\>	 \textbf{if received} $\langle \Prepare,\DoOpsTuple ,t,j, \BB \rangle$ \textbf{from} a process $q$ \textbf{then}\\
\nnll\label{setBatch3}\>\>	 $\Batch[j-1] := \BB$\\
\nnll\label{acceptcheck3}\>\>		\textbf{if}
			$t \ge \maxT$ \textbf{and} 
			$(t,j) > (ts,k)$ \textbf{then}\\
\nnll\label{client-accept}\>\>\>	$(\EstTuple,ts,k) := (O,t,j)$\\
\nnll\label{setPB}\>\>\>	{\color{black}	$\PendingOps[k] := \DoOpsTuple$}\\
\nnll\label{setMPB}\>\>\>	{\color{black}	$\MaxPendingIndex := \max (\MaxPendingIndex, k)$}\\

\nnll\label{sendPack}\>\> \textbf{if} $(\EstTuple,ts,k) = (O,t,j)$ \textbf{then} \textbf{send} $\langle \PACK,t,j \rangle$ \textbf{to} $q$\\

\nnll\label{lg0}\label{commitreceipt1}\> \textbf{if received} $\langle \ColoredCommitLease, \BB, j {\bluetored , \lease', \LeaseHolders'} \rangle$ \textbf{from} a process $q$ \textbf{then}\\
\nnll\label{batch2}\>\> $\Batch[j] := \BB$\\
\nnll\label{fill-lease-gap}\label{gmb2}\>\> $\FillGaps{j - 1}$\\
\nnll\label{gmb3}\>\> $\ExecuteOpsUpToBatch{j}$\\
\nnll\label{lg1}\>\>{\color{black}			\textbf{if} $p \in \LeaseHolders'$ \textbf{and} $\lease' > \lease$ \textbf{then}}
\\
\nnll\label{lg2}\>\>\> {\color{black}		$\lease := \lease'$}\\

\nnll\label{lg3}\>\> {\color{black}	 \textbf{else} \textbf{send} $\langle \RequestLease \rangle$ \textbf{to} $q$}\\

\nnll\label{epcm}\> \textbf{return}\\[2mm]

\textbf{procedure} $\ReplyToMessages{}$:\\[1mm]

\nnll\>	\textbf{while} \textsc{True} \textbf{do}\\

\nnll\label{rr1}\>\>	\textbf{if received} $\langle \OpRequest, \op \rangle$ \textbf{from} a process $q$ \textbf{then}\\

\nnll\label{rr2}\>\> \> 
	$\OpsRequested := \OpsRequested \cup \{\op\}$\\

\nnll\label{estrep1}\>\>	\textbf{if received} $\langle \EstReply,t,O',t',j', \BB'\rangle$ \textbf{from} a process $q$ \textbf{then}\\

\nnll\label{setBatch4}\>\>\> $\Batch[j'-1] := \BB'$ \\

\nnll\label{estrep2}\>\>\>		$\replied[t] := \replied[t] \cup \{ q \}$ \\

\nnll\label{estrep3}\>\>\>		$\REPLIES[t] := \REPLIES[t] \cup \{ (O',t',j') \}$ \\

\nnll\label{Rack1}\>\>	\textbf{if received} $\langle \PACK,t,j \rangle$ \textbf{from} a process $q$ \textbf{then}\\

\nnll\label{Rack2}\>\>\>		$\PACKED[t,j] := \PACKED[t,j] \cup \{ q \}$\\

\nnll\label{gb}\>\>\textbf{if received} $\langle \MyGaps,\Gaps' \rangle$ \textbf{from} a process $q$ \textbf{then}\\
\nnll\label{sendbatch}\>\>\>	\textbf{for all}  $j \in \Gaps'$ such that $\Batch[j] \neq \InitTuple$
		\textbf{send} $\langle \MyBatch, j,\Batch[j] \rangle$ \textbf{to} $q$\\

\nnll\label{batchreceipt}\>\> \textbf{if received} $\langle \MyBatch, j, \BB \rangle$ \textbf{from} a process $q$ \textbf{then}\\
\nnll\label{batch3}\label{ge}\>\>\> $\Batch[j] := \BB$\\

			\end{tabbing}

	\end{multicols}
\end{tiny}

\hrule
\caption{Algorithm 1}
\label{ObjectAlgo-alg1-code}
\end{figure}

\newcounter{ObjectAlgo-fig}  
\label{algo1page}


\newcounter{codelinenumber}
\setcounter{codelinenumber}{0}

\begin{figure}[H]
	\hrule

\begin{tiny}
	\begin{multicols}{2}
		\begin{tabbing}
			bbb\=bbb\=bbb\=bbb\=bbb\=bbb\=bbb\=bbb\=bbb\=bbb\=bbb\=bbb\=  \kill\\
\textsc{Code for process $p$:} \\
\textbf{cobegin}\\[1mm]

// \textsc{Thread 1:}   \cmmnt{issue RMW or read operations}\\

\nnlltwo\label{algo2-rmw1}\>\textbf{while} \textsc{True} \textbf{do} \\
\nnlltwo\label{algo2-rmw2}\>\>	\textbf{if} $p$ wants to execute a RMW operation $o$ \textbf{then}\\
\nnlltwo\label{algo2-rmw3}\>\>\>		$\cntr := \cntr +1$\\
\nnlltwo\label{algo2-rmw4}\>\>\>		$\operation := (o, (p,\cntr))$\\
\nnlltwo\label{algo2-rmw-send}\>\>\>  {\bluetoblack \textbf{periodically send} $\langle \OpRequest, \operation \rangle$ \textbf{to} $\textit{leader}$()}\\
\nnlltwo\label{algo2-rmw-until}\>\>\> {\bluetoblack \textbf{until} $\Reply{\operation} \neq \bot$}\\
\nnlltwo\label{algo2-rmw-wait-promise}\>\>\> {\bluetored \textbf{wait until} $\CT \geq \TakesEffect{operation}$}\\
\nnlltwo\label{algo2-rmw-return}\>\>\>		\textbf{return} $\Reply{\operation}$\\[1mm]

\nnlltwo\label{algo2-read-invoke}\>\> {\color{black} \textbf{if} $p$ wants to execute a read operation $o$ \textbf{then}}\\
\nnlltwo\label{algo2-read1}\>\>\>		{\color{black} $\cntr := \cntr +1$ }\\
\nnlltwo\label{algo2-read2}\>\>\>		{\color{black} $\operation := (o, (p,\cntr))$ }\\

\nnlltwo\label{algo2-getvalidlease-start}\>\>\>	{\color{black} \textbf{repeat}}\\
\nnlltwo\label{algo2-getleasetime}\>\>\>{\color{black}} \>	{\color{black} $t' := \CT$}\\
\nnlltwo\label{algo2-getlease}\>\>\>\> {\color{black} $(k^*,t^*) := \lease$} {\color{black}}\\
\nnlltwo\label{algo2-getvalidlease-end}\>\>\>	{\color{black} \textbf{until} $t' < t^* +\LP$}\\
\nnlltwo\label{algo2-get-k-hat-1}\>\>\> {\bluetored \textbf{if} $t'  < t^*$ \textbf{then}}\\
\nnlltwo\label{algo2-get-k-hat-then} \>\>\>\> {\bluetored $\hat{k} := \max \{j ~|~ 0 \le j \leq k^*$ \textbf{and}} {\bluetored $\BatchPromise{j} \leq t'\}$}\\
\nnlltwo\label{algo2-get-k-hat-3} \>\>\> {\bluetored \textbf{else}} \:\:\bluelcomment{$t^* \le t' < t^*+\LP$}\\
\nnlltwo\label{algo2-get-MPB} \>\>\>\> {\color{black} $u := \MaxPendingIndex$}\\

\nnlltwo\label{algo2-repeat-get-k-hat-else} \>\>\>\> {\color{blue} \textbf{repeat}} \\
\nnlltwo\label{algo2-get-k-hat-else}\>\>\>\>\>{\color{black} $\hat{k} := \max \{j ~|~ j = k^*$ \textbf{or} $(k^* < j \leq u$ \textbf{and}}\\
\nnlltwo\label{algo2-get-k-hat-else-b}\>\>\>\>\>\>{\color{black} $o$ conflicts with an operation in}\\
\nnlltwo\>\>\>\>\>\>$\PendingBatchOps{j}$ \textbf{and}\\
\nnlltwo\label{algo2-get-k-hat-5}\>\>\>\>\>\>{\bluetored $\PendingBatchPromise{j} \leq t'$)\}}\\
\nnlltwo\label{algo2-until-get-k-hat-else}\label{algo2-FG3} \>\>\>\> {\color{blue} \textbf{until} $(\text{for all }j, k^* < j \leq \hat{k}, \Batch[j] \neq \InitTuple)$}\\
\nnlltwo\label{algo2-wait-promise-2}\>\>\>{\bluetored \textbf{wait until} $\CT \geq \BatchPromise{\hat{k}}$}\\

\nnlltwo\label{algo2-fill-gaps-to-k-hat}\>\>\>	{\color{black} $\ExecuteOpsUpToBatch{\hat{k}}$}\\
\nnlltwo\label{algo2-client-read-end}\>\>\> {\color{black} $(-,\reply) := \Apply(\sstate[\hat{k}],o)$}\\
\nnlltwo\label{algo2-read-respond}\>\>\> {\color{black} \textbf{return} $\reply$}\\[1mm]

//  \textsc{Thread 2:}\\[1mm]

\nnlltwo\label{algo2-whileo}\> \textbf{while} \textsc{True} \textbf{do} \cmmnt{determine whether to act as leader or client}\\
\nnlltwo\label{algo2-gt}\>\>		$t := \CT$\\
\nnlltwo\label{algo2-check}\>\>\textbf{if} $\ML{t}{t} = \textsc{True}$ \textbf{then} $\LeaderWork{t}$  \\
\nnlltwo\label{algo2-whilee}\>\> 
	$\PCM()$\\[2mm]

//  \textsc{Thread 3:}\\[1mm]

\nnlltwo\label{algo2-thread3}\> 
$\ReplyToMessages{}$ \cmmnt{reply to messages}\\[2mm]

\textbf{coend}\\[2mm]

\>\\

\textbf{procedure} $\LeaderWork{t}$:\\[1mm]

\> \lcomment{New leader initialization: find latest batch and (re)do}\\

\nnlltwo\label{algo2-wait-lease-expire}\> {\bluetored \textbf{wait until} $\PP + \LP$ time has elapsed}\\
\nnlltwo\label{algo2-lh1}\>{\color{black}		$\LeaseHolders := $ {\bluetored $\emptyset$}}\\
\nnlltwo\label{algo2-est-request}\>			{\bluetoblack\textbf{periodically send}} $\langle \EstRequest,t \rangle$ \textbf{to} $\text{all processes} - \{p\}$ \\
\nnlltwo\label{algo2-est-condition}\>		{\bluetoblack\textbf{until}}  $ | \replied[t] | \ge \lfloor n/2 \rfloor$ 
 	\textbf{or} $\ML{t}{\CT} = \textsc{False}$\\
\nnlltwo\label{algo2-est-condition2}\>	{\bluetoblack\textbf{if} $| \replied[t] | < \lfloor n/2 \rfloor$ \textbf{then} \textbf{return}} \\

\nnlltwo\label{algo2-selection}\>     $(\EstTupleStar,ts^*,k^*) := $ tuple with maximum $(ts^*,k^*)$\\
\>\>in $\REPLIES[t] \cup \{ (\EstTuple,ts,k ) \} $ \\
\nnlltwo\label{algo2-Egathercrumbs}\>  \textbf{if} $ts^* \ge  t$ \textbf{then}  \textbf{return}\\

\nnlltwo\label{algo2-FG}\>	$\FillGaps{k^* -2}$\\
\nnlltwo\label{algo2-first-doops}\> 	$\result := \DoOps{{\bluetored(}\EstTupleStar{\bluetored,0)}}{t}{k^*}$\\
\nnlltwo\label{algo2-doopsfailed}\>  \textbf{if} $\result = \textsc{Failed} $ \textbf{then} \textbf{return}\\
\nnlltwo\label{algo2-Do-a-NoOp}\>  initiate a $\NoOp$ as a RMW operation via Thread 1\\[1mm]

\>\lcomment{{\color{black}{Grant read leases and}} process new batches}\\
\nnlltwo\label{algo2-mainwhile}\>	\textbf{while} \textsc{True} \textbf{do}\\
\nnlltwo\label{algo2-taketime1}\>{\color{black}}\>		$t' := \CT$\\	
\nnlltwo\label{algo2-recheck1}\>\>	\textbf{if} $\ML{t}{t'} = \textsc{False}$ \textbf{then return}\\
 \nnlltwo\label{algo2-checksendtime}\>\>{\color{black}		\textbf{if} $t'  \ge \NextSendTime$ \textbf{then}}\\
 \nnlltwo\label{algo2-setlease}\>\>\>{\color{black} $\lease := (k, t')$}\\
\nnlltwo\label{algo2-sl1}\label{algo2-sendcommit2}\>\>\> {\bluetoblack \textbf{send} $\langle \ColoredCommitLease,\Batch[k],k {\bluetored,\lease, \LeaseHolders} \rangle$}\\
\>\>\>\>\textbf{to} $\text{all processes} - \{p\}$\\
\nnlltwo\label{algo2-nst}\>\>\>{\color{black}			$\NextSendTime := t' + \LRP$}\\

\nnlltwo\label{algo2-lh3}\>\>{\color{black} \textbf{if received} $\langle \RequestLease \rangle$ \textbf{from} a process $q$ \textbf{then}}\\
\>\>\>$\LeaseHolders := \LeaseHolders \cup \{q\} $\\

\nnlltwo\label{algo2-nextops}\>\>		$\NextOps := \OpsRequested - \OpsDone$\\
\nnlltwo\label{algo2-nonempty2}\>\>		\textbf{if} $\NextOps \neq \emptyset$ \textbf{then}\\ 

\nnlltwo\label{algo2-second-doops}\>\>\> {\color{blue} $\result := \DO'(\NextOps,t,k+1)$}\\
\nnlltwo\label{algo2-doops2failed}\label{algo2-endwhile}\>\>\>  \textbf{if} $\result = \textsc{Failed} $ \textbf{then}  \textbf{return}

\\[4mm]

{\color{blue} \textbf{procedure} $\DO'({O},{t},{j})$}:\\[1mm]
\nnlltwo\label{algo2-dop-acceptcheck1}\> \textbf{if} $t < \maxT$ \textbf{then} $\textbf{return}$ \textsc{Failed}\\

\nnlltwo\label{algo2-dop-leader-accept}\> $(\EstTuple,ts,k) := ({\bluetored }O,t,j)$\\

\nnlltwo\label{algo2-dop-sendprep}\label{algo2-dop-repeat}\label{algo2-dop-prep-send}\>	{\color{blue}\textbf{repeat every $\beta$}}\\
\nnlltwo\label{algo2-dop-sendstatus-a}\>\> {\color{blue} $t' := \CT$}\\
\nnlltwo\label{algo2-dop-sendstatus-b}\>\> {\color{blue} \textbf{if} $\AL(t,t') = \textsc{False}$ \textbf{then return} \textsc{Failed}}\\
\nnlltwo\label{algo2-dop-sendstatus-c}\>\> {\color{blue} $s := t' + \PP$}\\
\nnlltwo\label{algo2-dop-sendstatus-d}\>\> {\color{blue} \textbf{send} $\langle \Status,(O,s),t,j, \Batch[j-1] \rangle$ \textbf{to} $\text{all processes} - \{p\}$} \\
\nnlltwo\label{algo2-dop-condition}\> {\color{blue} \textbf{until}  $| \PACKED[t,j] | \ge  \lfloor n/2 \rfloor$}\\

\nnlltwo\label{algo2-dop-wait2}\>{\color{black}		\textbf{wait}  \textbf{until} $\LeaseHolders \subseteq \PACKED[t,j]$}\\
	\>\>\textbf{or} $2 \delta$ time has elapsed since $p$ first executed \color{blue} line~\ref{algo2-dop-sendstatus-d}\\

\nnlltwo\label{algo2-dop-LHvsAcks}{\color{black} }\>{\color{black}		\textbf{if} $\neg (\LeaseHolders \subseteq \PACKED[t,j])$ 
{\bluetored \textbf{and} $s < \lease.start + \LP$}  \textbf{then}} \\
\nnlltwo\label{algo2-dop-wait-alg1}\>\>{\color{black}		\textbf{wait}  \textbf{until} $\CT \geq {\bluetored \lease.start} + \LP$}\\

\nnlltwo\label{algo2-dop-lh2}\>{\color{black}		$\LeaseHolders :=   \PACKED[t,j] $}\\

\nnlltwo\label{algo2-dop-batch1}\label{algo2-dop-lease1}\>${\bluetored(}\Batch[j]{\bluetored,\lease)} := {\bluetored((}\Os{\bluetored, s),(j, s))}$\\

\nnlltwo\label{algo2-dop-ExecuteB2}\>		$\ExecuteOpsUpToBatch{j}$ \\

\nnlltwo\label{algo2-dop-sl2}\label{algo2-dop-sendcommit}\> \textbf{send} $\langle \ColoredCommitLease, \Batch[j], j {\bluetored, \lease, \LeaseHolders} \rangle$ \textbf{to} $\text{all processes} - \{p\}$\\
\nnlltwo\label{algo2-dop-nst2}\>{\bluetored $\NextSendTime := s + \LRP$}\\
\nnlltwo\label{algo2-dop-done}\>	$\textbf{return}$ \textsc{Done}\\[2mm]

\textbf{procedure} $\DoOps{\DoOpsTuple}{t}{j}$:\\[1mm]
\bluelcomment{$O$ is the set of RMWs to be committed, $s$ is the promise time:}\\
\bluelcomment{$O$ will not be committed before time $s$}\\
\nnlltwo\label{algo2-acceptcheck1}\> \textbf{if} $t < \maxT$ \textbf{then} $\textbf{return}$ \textsc{Failed}\\

\nnlltwo\label{algo2-leader-accept}\> $(\EstTuple,ts,k) := ({\bluetored }O,t,j)$\\
\nnlltwo\label{algo2-sendprep}\label{algo2-repeat}\label{algo2-prep-send}\>	{\bluetoblack \textbf{periodically send}} $\langle {\color{blue}\Status},{\bluetored (}O{\bluetored, s)},t,j, \Batch[j-1] \rangle$ \textbf{to} $\text{all processes} - \{p\}$ \\
\nnlltwo\label{algo2-condition}\label{algo2-endc}\label{algo2-prep-condition}\>		{\bluetoblack\textbf{until}}  $| \PACKED[t,j] | \ge  \lfloor n/2 \rfloor$  
	\textbf{or} $\ML{t}{\CT} = \textsc{False}$\\
\nnlltwo\label{algo2-ack-condition}\label{algo2-prep-condition2}\> {\bluetoblack \textbf{if} $| \PACKED[t,j] | <  \lfloor n/2 \rfloor$ \textbf{then} \textbf{return} \textsc{Failed}}\\[1mm]

\nnlltwo\label{algo2-wait2}\>{\color{black}		\textbf{wait}  \textbf{until} $\LeaseHolders \subseteq \PACKED[t,j]$
	\textbf{or} $2 \delta$ time has elapsed since $p$ first executed {\color{blue} line~\ref{algo2-sendprep}}}\\

\nnlltwo\label{algo2-LHvsAcks}{\color{black} }\>{\color{black}		\textbf{if} $\neg (\LeaseHolders \subseteq \PACKED[t,j])$ 
{\bluetored \textbf{and} $s < \lease.start + \LP$}  \textbf{then}} \\
\nnlltwo\label{algo2-wait-alg1}\>\>{\color{black}		\textbf{wait}  \textbf{until} $\CT \geq {\bluetored \lease.start} + \LP$}\\

\nnlltwo\label{algo2-lh2}\>{\color{black}		$\LeaseHolders :=   \PACKED[t,j] $}\\ [1mm]

\nnlltwo\label{algo2-batch1}\label{algo2-lease1}\>${\bluetored(}\Batch[j]{\bluetored,\lease)} := {\bluetored((}\Os{\bluetored, s),(j, s))}$\\

\nnlltwo\label{algo2-ExecuteB2}\>		$\ExecuteOpsUpToBatch{j}$ \\[1mm]

\nnlltwo\label{algo2-sl2}\label{algo2-sendcommit}\> \textbf{send} $\langle \ColoredCommitLease, \Batch[j], j {\bluetored, \lease, \LeaseHolders} \rangle$ \textbf{to} $\text{all processes} - \{p\}$\\
\nnlltwo\label{algo2-nst2}\>{\bluetored $\NextSendTime := s + \LRP$}\\
\nnlltwo\label{algo2-done}\>	$\textbf{return}$ \textsc{Done}\\[2mm]

\textbf{procedure} $\FillGaps{k'}$:\\[1mm]

\nnlltwo\label{algo2-totoz8}\> \textbf{repeat} \\
\nnlltwo\label{algo2-totoz9}\>\> $\Gaps := \{ j ~|~ 1 \le j \le k' \mbox{ and } \Batch[j] = {\color{black} (}\emptyset{\color{black}, \infty)} \}$\\

\nnlltwo\label{algo2-totoz10}\>\>\textbf{if} $\Gaps \neq \emptyset$ \textbf{then} \textbf{send} $\langle \MyGaps,\Gaps \rangle$  \textbf{to} $\text{all processes} - \{p\}$\\
\nnlltwo\label{algo2-totoz11}\> \textbf{until} $\Gaps = \emptyset$ \\

\nnlltwo\label{algo2-totoz12}\> \textbf{return}\\[2mm]

 \textbf{procedure} $\ExecuteBatch{j'}$:\\[1mm]
\nnlltwo\label{algo2-EB-start}\>	$\sigma := \sstate[j'-1]$\\
\nnlltwo\label{algo2-EB-get-ops}\>		let $\op^1, \op^2, \ldots, \op^{m}$ be the operations in $\ColoredBatchOps{j'}$ listed in operation id order\\
\nnlltwo\label{algo2-apply-op}\>               \textbf{for} $i = 1$ \textbf{to} $m$ \textbf{do} \\
\nnlltwo\label{algo2-apply-op-1}\>\> $(\sigma,\Reply{\op^i}) :=  \Apply(\sigma,\op^i.\TYPE)$\\
\nnlltwo\label{algo2-set-takesEffect}\>\> {\bluetored$\TakesEffect{\op^i} := \BatchPromise{j'}$}\\
\nnlltwo\label{algo2-update-state}\>		$\sstate[j'] := \sigma$\\
\nnlltwo\label{algo2-EB-end}\> \textbf{return}\\[2mm]

\textbf{procedure} $\ExecuteOpsUpToBatch{j'}$:\\[1mm]

\nnlltwo\label{algo2-EUTB}\> \textbf{for} $j = \MBD +1$ \textbf{to} $j'$ \textbf{do}\\
\nnlltwo\label{algo2-EUTB2}\>\> $\ExecuteBatch{j}$\\
\nnlltwo\label{algo2-addB1}\>\>		 $\OpsDone := \OpsDone \cup \ColoredBatchOps{j}$\\
\nnlltwo\label{algo2-updateMBD1}\>\> 	$\MBD := \max (\MBD, j)$\\

\nnlltwo\> \textbf{return}\\[2mm]

\textbf{procedure} $\PCM()$:\\[1mm]

\nnlltwo\label{algo2-spcm}\label{algo2-a2b}\>	 \textbf{if received} $\langle \EstRequest, t \rangle$ \textbf{from} a process $q$ \textbf{then}\\
\nnlltwo\label{algo2-incTmax}\>\>		$\maxT:= \max (\maxT, t)$\\
\nnlltwo\label{algo2-a2e}\label{algo2-reply}\>\>		\textbf{send} $\langle \EstReply, t, \EstTuple,ts,k, Batch[k-1] \rangle$ \textbf{to} $q$\\

\nnlltwo\label{algo2-a3b}\>	 \textbf{if received} $\langle {\color{blue} \Status},\DoOpsTuple ,t,j, \BB \rangle$ \textbf{from} a process $q$ \textbf{then}\\
\nnlltwo\label{algo2-setBatch3}\>\>	 $\Batch[j-1] := \BB$\\
\nnlltwo\label{algo2-acceptcheck3}\>\>		\textbf{if}
			$t \ge \maxT$ \textbf{and} 
			$(t,j) > (ts,k)$ \textbf{then}\\
\nnlltwo\label{algo2-client-accept}\>\>\>	$(\EstTuple,ts,k) := (O,t,j)$\\
\nnlltwo\label{algo2-setPB}\>\>\>	{\color{black}	$\PendingOps[k] := \DoOpsTuple$}\\
\nnlltwo\label{algo2-setMPB}\>\>\>	{\color{black}	$\MaxPendingIndex := \max (\MaxPendingIndex, k)$}\\

\nnlltwo\label{algo2-updatePBPromise}\>\> {\color{blue} $\PendingBatchPromise{j} := \max(\PendingBatchPromise{j},s)$}\\
\nnlltwo\label{algo2-sendPack}\>\> \textbf{if} $(\EstTuple,ts,k) = (O,t,j)$ \textbf{then} \textbf{send} $\langle \PACK,t,j \rangle$ \textbf{to} $q$\\

\nnlltwo\label{algo2-lg0}\label{algo2-commitreceipt1}\> \textbf{if received} $\langle \ColoredCommitLease, \BB, j {\bluetored , \lease', \LeaseHolders'} \rangle$ \textbf{from} a process $q$ \textbf{then}\\
\nnlltwo\label{algo2-batch2}\>\> $\Batch[j] := \BB$\\
\nnlltwo\label{algo2-fill-lease-gap}\label{algo2-gmb2}\>\> $\FillGaps{j - 1}$\\
\nnlltwo\label{algo2-gmb3}\>\> $\ExecuteOpsUpToBatch{j}$\\
\nnlltwo\label{algo2-lg1}\>\>{\color{black}			\textbf{if} $p \in \LeaseHolders'$ \textbf{and} $\lease' > \lease$ \textbf{then}}
\\
\nnlltwo\label{algo2-lg2}\>\>\> {\color{black}		$\lease := \lease'$}\\

\nnlltwo\label{algo2-lg3}\>\> {\color{black}	 \textbf{else} \textbf{send} $\langle \RequestLease \rangle$ \textbf{to} $q$}\\[1mm]

\nnlltwo\label{algo2-epcm}\> \textbf{return}

\\[2mm]

	\textbf{procedure} $\ReplyToMessages{}$:\\[1mm]

\nnlltwo\>	\textbf{while} \textsc{True} \textbf{do}\\

\nnlltwo\label{algo2-rr1}\>\>	\textbf{if received} $\langle \OpRequest, \op \rangle$ \textbf{from} a process $q$ \textbf{then}\\

\nnlltwo\label{algo2-rr2}\>\> \> 
	$\OpsRequested := \OpsRequested \cup \{\op\}$\\

\nnlltwo\label{algo2-estrep1}\>\>	\textbf{if received} $\langle \EstReply,t,O',t',j', \BB'\rangle$ \textbf{from} a process $q$ \textbf{then}\\

\nnlltwo\label{algo2-setBatch4}\>\>\> $\Batch[j'-1] := \BB'$ \\

\nnlltwo\label{algo2-estrep2}\>\>\>		$\replied[t] := \replied[t] \cup \{ q \}$ \\

\nnlltwo\label{algo2-estrep3}\>\>\>		$\REPLIES[t] := \REPLIES[t] \cup \{ (O',t',j') \}$ \\

\nnlltwo\label{algo2-Rack1}\>\>	\textbf{if received} $\langle \PACK,t,j \rangle$ \textbf{from} a process $q$ \textbf{then}\\

\nnlltwo\label{algo2-Rack2}\>\>\>		$\PACKED[t,j] := \PACKED[t,j] \cup \{ q \}$\\

\nnlltwo\label{algo2-gb}\>\>\textbf{if received} $\langle \MyGaps,\Gaps' \rangle$ \textbf{from} a process $q$ \textbf{then}\\
\nnlltwo\label{algo2-sendbatch}\>\>\>	\textbf{for all}  $j \in \Gaps'$ such that $\Batch[j] \neq \InitTuple$
		\textbf{send} $\langle \MyBatch, j,\Batch[j] \rangle$ \textbf{to} $q$\\

\nnlltwo\label{algo2-batchreceipt}\>\> \textbf{if received} $\langle \MyBatch, j, \BB \rangle$ \textbf{from} a process $q$ \textbf{then}\\
\nnlltwo\label{algo2-batch3}\label{algo2-ge}\>\>\> $\Batch[j] := \BB$\\

		\end{tabbing}
    \end{multicols}
\end{tiny}

\hrule
\caption{Algorithm 2 (differences from Algorithm~\ref{ObjectAlgo-alg1-code} are highlighted in {\color{blue}blue})}
\label{ObjectAlgo-alg2-code}
\end{figure}
\label{algo2page}

\newpage
\bibliography{ms}

\newpage
\appendix

\section{Proof of correctness of Algorithm~1}\label{RO-Proof}

In this appendix we give a detailed proof of correctness of
	the algorithm shown in Figure~\ref{ObjectAlgo-alg1-code}.
As we have seen, this algorithm is based on three mechanisms:
	a consensus mechanism to order the RMW operations,
	a read-lease mechanism to allow processes to read locally, and
	the promise mechanism that allows trading off
	the blocking time of read operations against
	the blocking time of RMW operations.
Although these mechanisms are intuitive at a high level,
	each has its subtleties
	(largely arising from the need to cope with asynchrony and failures);
	and their interaction increases the complexity of the proof.

In Section~\ref{Model-Fullpaper} we state the assumptions
	on which the correctness of our algorithm is based.
Then in Sections~\ref{cons-safety}--\ref{readleaseNonBlockingProps}
	we prove the correctness of the algorithm.

In Section~\ref{cons-safety} we prove some basic safety properties
	of the consensus mechanism.
Recall that each process commits a sequence of batches,
	where each batch contains a set of RMW operations submitted by processes.
The key properties proved in this section are that:
	(a)~processes agree on the sequence of batches they commit
	(Theorem~\ref{finalsafetyx}),
	(b)~different batches committed contain disjoint sets of RMW operations
	(Theorem~\ref{finalsafetyb}), and
	(c)~committed batches are not lost: if a process commits batch $j$,
	each of the previous batches $1,2,\ldots,j-1$ is stored in a majority of processes
	(Corollary~\ref{successivebatches2}).

In Section~\ref{cons-liveness} we prove the liveness of the consensus mechanism:
Every RMW operation submitted by a correct process eventually terminates
	(Theorem~\ref{finally}).
	
In Section~\ref{readleaseBasicProps} we prove some basic properties of the
	read-lease mechanism, which are needed for the proof of linearizability,
	and the liveness and blocking time of read operations.

In Section~\ref{sec:linearizability} we prove that our algorithm
	implements a linearizable object:
Every execution of operations submitted by processes is equivalent to
	a \emph{sequential} execution of operations that
	(a)~contains all completed operations and a subset of incomplete operations
	submitted by processes;
	(b)~respects the semantics of the object being implemented; and
	(c)~respects the order of non-concurrent operations:
	if operation $\op$ completed before operation $\op'$ started
	in the actual execution,
	then $\op$ appears before $\op'$ in the equivalent sequential execution
	(Theorem~\ref{linearizability}).
	
In Section~\ref{sec:read-liveness} we prove the liveness of read operations:
Every read operation submitted by a correct process eventually terminates
	(Theorem~\ref{bertolucci}).
	
Finally, in Section~\ref{readleaseNonBlockingProps}
	we prove properties of the algorithm related to blocking of reads.
Specifically, we prove that eventually:
	(a)~every read operation
	that does not conflict with any pending RMW operation, or issued by the leader,
	completes without blocking
	(Theorems~\ref{granfinale1} and~\ref{granfinale2}); and
	(b)~every read operation
	(that conflicts with a pending RMW operation and is not issued by the leader)
	blocks only for a bounded period of time
	(Theorem~\ref{granfinale3}).

\subsection{Model}\label{Model-Fullpaper}

\subsubsection{Objects and operations}\label{objects-ops}

An object of a given type $\mathcal{T}$ is defined by specifying a set of states $\Sigma$,
	a set of operations $\Ops$, a set of responses $\Res$,
	and a transition function
	$\Apply:\Sigma\times\Ops\rightarrow\Sigma\times\Res$.
The transition function describes the
	effect of applying an operation $o \in\Ops$ to a state $\sigma\in\Sigma$:
	if $\Apply(\sigma,o)=(\sigma',v)$ then the new state of the object is $\sigma'$
	and the response of the operation is $v$.	
An operation $o$ is a \emph{read} operation if,
	for every $\sigma\in\Sigma$, $\Apply(\sigma,o)=(\sigma,v)$ for some $v\in\Res$;
	$o$ is a \emph{read-modify-write (RMW)} operation
	if it is not a read operation.

\subsubsection{System assumptions}
We assume a partially synchronous system that is the same as in \cite{CHT16}
	except that clocks are perfectly-synchronized.

\smallskip\noindent
\emph{\textbf{$\bullet$~Clocks.}}
Each process $p$ has a local clock denoted $\CT_p$.
The value of $\CT_p$ at real time~$\tau$, denoted $\CT_p(\tau)$, is the \emph{local time of $p$ at real time~$\tau$}.
We assume that local clocks are non-negative integers that are monotonically increasing and perfectly synchronized.
More precisely:

\begin{assumption}\label{xclocks}[Perfectly synchronized clocks]
 For all processes $p$, for all real times $\tau$,
 \begin{enumerate}
 \item\label{xcl1} For all processes $p$, for all real times $\tau$, $\CT_p(\tau)$ is a non-negative integer.

\item\label{xcl2} For all processes $p$, for all real times $\tau$ and $\tau'$ such that $\tau \le \tau'$, $\CT_p(\tau) \le \CT_p(\tau')$.

\item\label{xcl3} For all processes $p$, for all local times $t \ge 0$, there is a real-time $\tau$ such that $\CT_p(\tau)~\ge~t$.

\item\label{xcl4} For all processes $p$, the clock $\CT_p$ of $p$ increases by at least one time unit between any two successive readings of this clock by $p$.

\item\label{xcl5} For all processes $p$ and $q$, for all real times $\tau$, $\CT_p(\tau) = \CT_q(\tau)$.

\end{enumerate}
\end{assumption}

Assumption~\ref{xclocks}(\ref{xcl4}) can be enforced by delaying each clock reading until its value exceeds the previously read value.

\smallskip\noindent
\emph{\textbf{$\bullet$~Processes.}}
A majority of the processes are non-faulty, i.e., are \emph{correct}.
More precisely:

\begin{assumption}\label{majority-correct} [Process failures]
There are $n$ processes, they may fail only by crashing, and fewer than $n/2$ of them can crash.
\end{assumption}

We assume that there is a \emph{known} lower bound on the speed of processes
	that \emph{eventually} holds forever.
More precisely:

\begin{assumption}\label{process-speed} [Minimum process speed]
There is a known constant $C$ and an unknown real time $\tauc$ such that the following holds:
For all correct processes $p$, and all real time intervals $[\tau, \tau']$ such that
	$\tau' > \tau \ge \tauc$ and $| \CT_p(\tau') - \CT_p(\tau)| \ge C$,
	$p$ takes at least one step during interval $[\tau, \tau']$.
\end{assumption}

\smallskip\noindent
\emph{\textbf{$\bullet$~Messages.}}
We assume that there is a \emph{known} upper bound on message delays that \emph{eventually} holds forever.
More precisely:

\begin{assumption}\label{message-delay} [Maximum message delay]
There is a known constant~$\delta$ and an unknown time $\taum$
	after which the following holds:
	For all correct processes $p$ and $q$,
	if $p$ sends a message $m$ to $q$
	then $q$ receives $m$ within $\delta$ local time units from when it was sent, as measured on $p$'s or $q$'s clock.
\end{assumption}

\medskip
\noindent
Note that the clock properties are \emph{perpetual},
	while the process speed and message delay properties are \emph{eventual}.
Before these eventual properties hold,
	processes can be arbitrarily slow,
	and messages can take arbitrarily long to arrive and can even be lost.

\subsubsection{Leader election}\label{leader-election}

We assume that processes have access to an eventual leader election procedure $leader()$ that satisfies the following property:

\begin{assumption}\label{Omega}
There is a correct process $\ell$ and a real time $\tau_{\ell}$ after which every call to $leader()$ by any correct process returns $\ell$.
\end{assumption}

Throughout the paper ``\emph{(eventual) stable leader}'' refers to the process $\ell$ of the above assumption.

\cite{CHT16} describes a leader election enhancer algorithm that 
	transforms any implementation of $leader()$ as described above, 
	into a procedure $\ML{t_1}{t_2}$ that satisfies the following properties:

\begin{theorem}\label{leader-safety}
\textsc{[Safety]} For all processes $p \neq p'$ and
	all local times $t_1,t_2,t'_1,t'_2$ such that $t_1\le t_2$ and $t'_1\le t'_2$,
	if $p$ calls $\ML{t_1}{t_2}$
	and $p'$ calls $\ML{t'_1}{t'_2}$,
	and both calls return \textsc{True},
	then the intervals $[t_1,t_2]$ and $[t'_1,t'_2]$ do not intersect.
\end{theorem}

\begin{theorem}\label{leader-liveness-safety1}
\textsc{[Liveness]} There is an unknown time $\Tz$ such that for all $t' \ge t \geq \Tz$:
\begin{enumerate}
\item If $\ell$ calls $\ML{t}{t'}$ at a time $\tu$ where $\tu \ge t' \ge t \ge \Tz$
	then this call returns \textsc{True}.
\item If a process $q \neq \ell$ calls $\ML{t}{t'}$ with $t' \ge  \Tz$,
	and this call returns,
	then it returns \textsc{False}.
\end{enumerate}
\end{theorem}

\subsection{Consensus mechanism: safety properties}\label{cons-safety}
We first focus on the consensus mechanism (that processes RMW operations)
	and then on the read lease and the promise mechanism 
	(that enables local and non-blocking reads).

The consensus mechanism relies on the following assumptions:

\begin{enumerate}

\item Processes have access to the $\AL$ procedure of Section~\ref{leader-election}.

\item Local clocks are non-negative integers that are monotonically increasing (Assumption~\ref{xclocks} (\ref{xcl1})--(\ref{xcl4})).

\item Processes may fail only by crashing, and a majority of them do not fail (Assumption~\ref{majority-correct}).

\item Links are lossy but \emph{fair} (a weakening of Assumption~\ref{message-delay}). More precisely:

\begin{assumption}\label{fl}
	The communication link between any two correct processes $p$ and $q$ is \emph{fair}:
	messages can get lost,
	but if $p$ sends a message $m$ to $q$ infinitely often
	then $q$ receives $m$ infinitely often.	
\end{assumption}

\end{enumerate}

We first show that there is agreement on the set of operations in each $\Batch[j]$, and that for $j \neq j'$, $\BatchOps{j} \cap \BatchOps{j'} = \emptyset$.

\subsubsection{On accepting and locking}

From the way some variables are initialized and maintained by the algorithm
	it is clear that they each contain a set of operations.
In particular:
	
\begin{observation}\label{sets}
The variables $\OpsRequested$, $\OpsDone$, $\NextOps$, $\Ops$, $\Ops^*$, $\Os$, and $\BatchOps{j}$ for any $j \ge -1$,
	 contain a set of operations.
\end{observation}

Consider the variables $\OpsRequested$ and $\OpsDone$ of a process.
From the way they are initialized and updated
	(in line~\ref{rr2} for $\OpsRequested$,
	and in line~\ref{addB1} for $\OpsDone$):

\begin{observation}\label{opsdoneismonotonic}
$\OpsRequested$ and $\OpsDone$ contain a non-decreasing set of operations.
\end{observation}

\begin{definition}\label{bcmleader}
A process $\ell$ \emph{becomes leader at local time $t$}
	if:

\begin{enumerate}

\item $\ell$ gets the value $t$ from its $\CT$ in line~\ref{gt}, and

\item $\ell$ calls $\ML{t}{t}$, finds that $\ML{t}{t} = True$, and calls $\LeaderWork{t}$ in line~\ref{check}.
\end{enumerate}
\end{definition}

\begin{observation}\label{LeaderWork1}
If a process calls $\LeaderWork{t}$, then it became leader at local time $t$.
\end{observation}

\begin{lemma}\label{unique-LeaderWork}
If processes $p$ and $q$ both call $\LeaderWork{t}$, then $p = q$.
\end{lemma}
\begin{proof}
Suppose that processes $p$ and $q$ both call $\LeaderWork{t}$ for some $t$.
Then $p$ and $q$ both called $\ML{t}{t}$ in line~\ref{check}, 
	and this call returned \textsc{True}.
By Theorem~\ref{leader-safety}, $p=q$.
\qedhere~$_\text{\autoref{unique-LeaderWork}}$
\end{proof}

\begin{lemma}\label{monoLW}
If a process $p$ calls $\LeaderWork{t}$ and later calls $\LeaderWork{t'}$, then $t' >t$.
\end{lemma}

\begin{proof}
This is because $p$'s local clock is non-decreasing and that
	local clocks increase between successive readings 
	(Assumptions~\ref{xclocks}(\ref{xcl1}) and (\ref{xcl4})).
\qedhere~$_\text{\autoref{monoLW}}$
\end{proof}

\begin{corollary}\label{LeaderWorkUnique}
For each $t \ge 0$, a process calls $\LeaderWork{t}$ at most once.
\end{corollary}

\begin{observation}\label{DoopsCalls}
If a process calls $\DoOps{(-,-)}{t}{-}$,
	then it does so in line~\ref{first-doops} or~\ref{second-doops}
	of $\LeaderWork{t}$.
Moreover, if a process calls $\DoOps{(-,-)}{-}{-}$ in $\LeaderWork{t}$,
	then this call is of the form $\DoOps{(-,-)}{t}{-}$.
\end{observation}

\begin{definition}\label{accepts-def}
A process $p$ \emph{accepts the tuple $(\Os,t,j)$} if it sets the variables $(\Ops,ts,k)$ to $(\Os,t,j)$ in lines
	\ref{leader-accept} or \ref{client-accept} of the algorithm.
If a process accepts $(\Os,t,j)$, we say that \emph{$(\Os,t,j)$ is accepted}.
\end{definition}

\begin{observation}\label{AcceptAndOps-ts-k}
If a process has $(\Ops,ts,k) = (\Os,t,j) \neq (\emptyset,-1,0)$,
	then it previously accepted $(\Os,t,j)$.
\end{observation}

\begin{observation}\label{Accept-Doops}
If a process accepts the tuple $(\Os,t,j)$, then some process (possibly the same process) previously called $\DoOps{(\Os,-)}{t}{j}$, 
	and accepted the tuple $(\Os,t,j)$ in line~\ref{leader-accept} in that $\DoOps{(\Os,-)}{t}{j}$.
\end{observation}

\begin{observation}\label{LeaderWorkAcceptances}
All the tuples that a process $p$ accepts in $\LeaderWork{t}$
	are of the form $(-,t,-)$.
\end{observation}

\begin{lemma}\label{T-k-increase}
If a process accepts
	$(\Os_1,t_1,j_1)$ before accepting $(\Os_2,t_2,j_2)$,
	then $(t_2,j_2) > (t_1,j_1)$.
\end{lemma}

\begin{proof}
Suppose $p$ accepts $(\Os_1,t_1,j_1)$ and later accepts $(\Os_2,t_2,j_2)$.
We will prove that if these are consecutive tuples accepted by $p$,
	then $(t_2,j_2) > (t_1, j_1)$.
Then, by induction it follows that the lemma holds for non-consecutive tuples accepted by $p$.

When $p$ accepts $(\Os_1,t_1,j_1)$ it sets its variables $(\Ops,ts,k)$ to $(\Os_1,t_1,j_1)$.
Since $p$ modifies $(\Ops,ts,k)$ only when it accepts a tuple, the following holds:
(*) $p$ has $(\Ops,ts,k) = (\Os_1,t_1,j_1)$ from the moment
	it accepts
	 $(\Os_1,t_1,j_1)$ up to (not including) the moment
	 that it accepts it next tuple, namely, $(\Os_2,t_2,j_2)$.
	
There are several cases, depending on where $p$ accepts $(\Os_2,t_2,j_2)$.

\begin{enumerate}
\item $p$ accepts $(\Os_2,t_2,j_2)$ in line~\ref{client-accept}.
Note that this occurs because $p$ received a $\langle \Prepare,(\Os_2,-),t_2,j_2,- \rangle$ message in line~\ref{a3b}.
By (*), when $p$ executes line~\ref{acceptcheck3}, it has $(\Ops,ts,k) = (\Os_1,t_1,j_1)$.
Since $p$ executes line~\ref{client-accept}, the condition of line~\ref{acceptcheck3} is satisfied,
	and so $(t_2,j_2) > (ts,k)$.
Thus $(t_2,j_2) > (t_1,j_1)$.

\item $p$ accepts $(\Os_2,t_2,j_2)$ in line~\ref{leader-accept} of the $\DoOps{(-,-)}{-}{-}$ procedure.
Note that this occurs during $p$'s execution of $\DoOps{(\Os_2,-)}{t_2}{j_2}$ in $\LeaderWork{t_2}$.
There are two subcases.

\begin{enumerate}

\item $p$ calls $\DoOps{(\Os_2,-)}{t_2}{j_2}$ in line~\ref{first-doops}.
By (*), $p$ has $(\Ops,ts,k) = (\Os_1,t_1,j_1)$ in line~\ref{selection}.
Since $p$ selects the tuple $(\Ops^*,ts^*,k^*)$ in line~\ref{selection}
	as a tuple with maximum $(ts^*,k^*)$ in $\REPLIES[t] \cup \{ (\Ops,ts,k) \} $,
	we have $(ts^*,k^*) \ge (ts,k)$, and so $(ts^*,k^*) \ge (t_1,j_1)$.
Since $p$ reaches line~\ref{first-doops},
	the condition $ts^* \ge t$ in line~\ref{Egathercrumbs} must be false,
	and so $t > ts^*$.
Thus, $(t,k^*) > (ts^*,k^*) \ge (t_1,j_1)$.
Since $p$ executes  $\DoOps{(\Ops^*,0)}{t}{k^*} = \DoOps{(\Os_2,-)}{t_2}{j_2}$ in line~\ref{first-doops},
	$(t,k^*) = (t_2,j_2)$.
So $(t_2,j_2) > (t_1,j_1)$.

\item $p$ calls $\DoOps{(\Os_2,-)}{t_2}{j_2}$ in line~\ref{second-doops}.
It is clear that $p$ called $\DoOps{(-,-)}{t_2}{-}$ at least once before in $\LeaderWork{t_2}$ (in line~\ref{first-doops} or~\ref{second-doops}).
Consider the last $\DoOps{(-,-)}{t_2}{-}$ that $p$ executed before calling \linebreak $\DoOps{(\Os_2,-)}{t_2}{j_2}$ in $\LeaderWork{t_2}$.
This $\DoOps{(-,-)}{t_2}{-}$ must have returned $\textsc{Done}$ (because $p$ did not exit $\LeaderWork{t_2}$:
	it continued on to execute $\DoOps{(\Os_2,-)}{t_2}{j_2}$).
Thus, during the execution of this $\DoOps{(-,-)}{t_2}{-}$, $p$ accepted a tuple $(-,t_2,-)$ in line~\ref{leader-accept}.
Since a process accepts a tuple in either line~\ref{client-accept} and line~\ref{leader-accept},
	and $p$ does not execute $\PCM()$, hence line~\ref{client-accept}, during the execution of $\LeaderWork{t_2}$,
	the tuple $(-,t_2,-)$ is the last tuple that $p$ accepted before accepting $(\Os_2,t_2,j_2)$.
Therefore, $(-,t_2,-) = (\Os_1,t_1,j_1)$, and so $t_2 = t_1$.
By~(*), when $p$ call $\DoOps{(\Os_2,-)}{t_2}{j_2}$ in line~\ref{second-doops},
	$p$ has $(\Ops,ts,k) = (\Os_1,t_1,j_1)$,
	i.e., $p$ has $k = j_1$ at that time.
Since $p$ calls $\DoOps{(\NextOps,-)}{t}{k+1} = \DoOps{(\Os_2,-)}{t_2}{j_2}$ in line~\ref{second-doops},
	$j_2 = k+1$.
We conclude that $(t_2,j_2) = (t_2 , k+1) > (t_2, k) = (t_2, j_1) = (t_1, j_1)$, and so $(t_2,j_2) > (t_1, j_1)$.
\end{enumerate}
\end{enumerate}
Thus, in all cases $(t_2,j_2) > (t_1, j_1)$.
\qedhere~$_\text{\autoref{T-k-increase}}$
\end{proof}

\begin{corollary}\label{accepts-only-once}
A process can accept a tuple $(\Os,t,j)$ at most once.
\end{corollary}

\begin{lemma}\label{leaderfirst}
If a tuple $(\Os,t,j)$ is accepted, then
	the \emph{first} process to accept $(\Os,t,j)$ is a process $p$
	that became leader at local time $t$:
	$p$ called $\LeaderWork{t}$
	and accepted $(\Os,t,j)$ while executing $\DoOps{(\Os,-)}{t}{j}$ in $\LeaderWork{t}$.
\end{lemma}
\begin{proof}
Suppose a tuple $(\Os,t,j)$ is accepted and $p$ is the first process to accept this tuple.
If $p$ accepted this tuple in line~\ref{client-accept},
	then by Observation~\ref{Accept-Doops},
	some process $q$ previously accepted this tuple in line~\ref{leader-accept} in $\DoOps{(\Os,-)}{t}{j}$.
By Corollary~\ref{accepts-only-once}, $q \neq p$. 
This contradicts the assumption that
	$p$ is the first process to accept this tuple.
So $p$ must accept this tuple in line~\ref{leader-accept},
	and it is clear that this happens in $\DoOps{(\Os,-)}{t}{j}$.
By Observation~\ref{DoopsCalls}, $p$ accepts $(\Os,t,j)$ while executing $\DoOps{(\Os,-)}{t}{j}$ in $\LeaderWork{t}$.
\qedhere~$_\text{\autoref{leaderfirst}}$
\end{proof}

\begin{lemma}\label{leader-accepts}
If a process $\ell$ that becomes leader at local time $t$ accepts a tuple of the form $(-,t,-)$,
	it does so in line~\ref{leader-accept} of the $\DoOps{(-,-)}{t}{-}$ procedure
	that $\ell$ calls in line~\ref{first-doops} or \ref{second-doops}.
Furthermore, $\ell$ accepts its \emph{first} $(-,t,-)$ tuple
	when $\ell$ executes $\DoOps{(-,-)}{t}{-}$ in line~\ref{first-doops},
	and
	any other $(-,t,-)$ tuple when $\ell$ executes $\DoOps{(-,-)}{t}{-}$ in line~\ref{second-doops}.

\end{lemma}
\begin{proof}
Suppose $\ell$ becomes leader at local time $t$ and accepts a tuple $(\Os,t,j)$.
We first show that $\ell$ accepts this tuple in line~\ref{leader-accept} in $\DoOps{(\Os,-)}{t}{j}$.
By Observation~\ref{Accept-Doops}, some process $p$ previously called $\DoOps{(\Os,-)}{t}{j}$,
	and accepted the tuple $(\Os,t,j)$ in line~\ref{leader-accept} in $\DoOps{(\Os,-)}{t}{j}$.
By Observation~\ref{DoopsCalls}, this happened in $\LeaderWork{t}$.
By Lemma~\ref{unique-LeaderWork}, $p = \ell$.
So $\ell$ accepted $(\Os,t,j)$ in line~\ref{leader-accept} in $\DoOps{(\Os,-)}{t}{j}$.
Thus, by Corollary~\ref{accepts-only-once}, if $\ell$ accepts a tuple of form $(-,t,-)$,
	it does so in line~\ref{leader-accept} of the $\DoOps{(-,-)}{t}{-}$ procedure.
The lemma now follows from Observation~\ref{DoopsCalls} and the fact that 
	$\ell$ first calls $\DoOps{(-,-)}{t}{-}$ in line~\ref{first-doops}, 
	and calls any other $\DoOps{(-,-)}{t}{-}$ in line~\ref{second-doops} in $\LeaderWork{t}$.
\qedhere~$_\text{\autoref{leader-accepts}}$
\end{proof}

\begin{lemma}\label{T-k-Doops}
If a process calls $\DoOps{(-,-)}{t}{j}$ and then $\DoOps{(-,-)}{t}{j'}$, consecutively,
	then $j' = j+1$.
\end{lemma}

\begin{proof}
Suppose a process $p$ calls $\DoOps{(-,-)}{t}{j}$ and then $\DoOps{(-,-)}{t}{j'}$, consecutively.
By Observation~\ref{DoopsCalls}, $p$ makes both calls while executing $\LeaderWork{t}$.
By Corollary~\ref{LeaderWorkUnique}, $p$ makes both calls in the same $\LeaderWork{t}$.
Since $\DoOps{(-,-)}{t}{j'}$ is not the first $\DoOps{(-,-)}{t}{-}$ call that $p$ makes in $\LeaderWork{t}$,
	from the code of $\LeaderWork{}$,
	$p$ calls $\DoOps{(-,-)}{t}{j'}$ in line~\ref{second-doops}.
Thus $\DoOps{(-,-)}{t}{j'} = \DoOps{(-,-)}{t}{k+1}$, i.e., $j'$ is the value of $k+1$ at $p$ in line~\ref{second-doops}.
Note that when $p$ previously executed $\DoOps{(-,-)}{t}{j}$, $p$~set its variable $k$ to $j$ in line~\ref{leader-accept}
	(because this $\DoOps{(-,-)}{t}{j}$ must have returned $\textsc{Done}$).
Since $p$ does not execute $\PCM()$ while it is executing $\LeaderWork{t}$,
	$p$ does not update its variable $k$
	before calling $\DoOps{(-,-)}{t}{-}$ again.
Since $\DoOps{(-,-)}{t}{j}$ and $\DoOps{(-,-)}{t}{j'}$ are \emph{successive} calls of $\DoOps{(-,-)}{-}{-}$ by $p$,
	when $p$ calls $\DoOps{(-,-)}{t}{j'}$ in line~\ref{second-doops}, $p$'s variable $k$ is still equal to $j$.
So when $p$ is in line~\ref{second-doops}, we have $j' = k+1 = j+1$.
\qedhere~$_\text{\autoref{T-k-Doops}}$
\end{proof}
The following is an immediate corollary to the above lemma.

\begin{corollary}\label{T-k-DoopsCreduced}
If a process $p$ calls $\DoOps{(-,-)}{t}{j}$ before calling $\DoOps{(-,-)}{t}{j'}$
	then $j' > j$.
\end{corollary}

\begin{lemma}\label{doops-simplecase}
Suppose a process $p$ calls $\DoOps{(\Os,\s)}{t}{j}$ and $\DoOps{(\Os',\s')}{t}{j'}$.
If $j' =j$ then $(\Os',\s') = (\Os,\s)$.
\end{lemma}

\begin{proof}
Suppose $p$ calls $\DoOps{(\Os,\s)}{t}{j}$ and $\DoOps{(\Os',\s')}{t}{j'}$.
If $j' = j$, then Corollary~\ref{T-k-DoopsCreduced} implies that 
	$\DoOps{(\Os,\s)}{t}{j}$ and $\DoOps{(\Os',\s')}{t}{j'}$ are the same call,
	and so $(\Os',\s') = (\Os,\s)$.
\qedhere~$_\text{\autoref{doops-simplecase}}$
\end{proof}

\begin{lemma}\label{simplecase}
Suppose tuples $(\Os,t,j)$ and $(\Os',t,j')$ are accepted.
If $j' =j$ then $\Os' = \Os$.
\end{lemma}

\begin{proof}
Suppose $(\Os,t,j)$ and $(\Os',t,j')$ are accepted.
By Observation~\ref{Accept-Doops}
	some process $p$ called $\DoOps{(\Os,-)}{t}{j}$ and some process $q$ called $\DoOps{(\Os',-)}{t}{j'}$.
By Observation~\ref{DoopsCalls}, $p$ and $q$ did so in $\LeaderWork{t}$.
By Lemma~\ref{unique-LeaderWork}, $p=q$.
The result now follows from Lemma~\ref{doops-simplecase}.
\qedhere~$_\text{\autoref{simplecase}}$
\end{proof}

\begin{lemma}\label{lmx3}
If a process $p$ has $(\Ops^*,ts^*,k^*) \neq (\emptyset, -1, 0)$ in line~\ref{selection},
	then some process previously accepted tuple $(\Ops^*,ts^*,k^*)$.
\end{lemma}

\begin{proof}
Suppose $p$ has $(\Ops^*,ts^*,k^*) \neq (\emptyset, -1, 0)$ in line~\ref{selection}
	in an execution of $\LeaderWork{t}$ for some $t$.
So $p$ has the tuple $(\Ops^*,ts^*,k^*)$ in $\REPLIES[t] \cup \{ (\Os,t,j) \}$ where $(\Os,t,j)$ is the value of
	$p$'s variables $(\Ops,ts,k)$ in line~\ref{selection}.
Note that $(\Ops^*,ts^*,k^*)$ is not the initial value of $(\Ops,ts,k)$ at any process.
There are two cases:

\begin{enumerate}
\item $(\Ops^*,ts^*,k^*) = (\Os,t,j)$.
	Since $(\Os,t,j)$ is not the initial value $(\emptyset, -1, 0)$ of $(\Ops,ts,k)$ at $p$,
	by Observation~\ref{AcceptAndOps-ts-k},
	$p$ previously accepted $(\Os,t,j)$, i.e., it previously accepted $(\Ops^*,ts^*,k^*)$.
	
\item $(\Ops^*,ts^*,k^*) \in \REPLIES[t]$.
From the code of the algorithm concerning $\REPLIES[t]$
	(lines~\ref{est-request}-\ref{selection}, \ref{a2b}-\ref{a2e}, and \ref{estrep1}-\ref{estrep3}), 
	it is clear that $p$ previously received a $\langle \EstReply,t,\Ops^*,ts^*,k^*,- \rangle$ message from some
	process $q^*$.
When $q^*$ sent this message (in line~\ref{reply}), it had $(\Ops,ts,k) = (\Ops^*,ts^*,k^*)$.
Since $(\Ops^*,ts^*,k^*)$ is not the initial value of $(\Ops,ts,k)$ at $q^*$,
	$q^*$ accepted the tuple $(\Ops^*,ts^*,k^*)$, 
	and it did so before sending $\langle \EstReply,t,\Ops^*,ts^*,k^*,-\rangle$ to~$p$.
\end{enumerate}
In all cases some process accepted $(\Ops^*,ts^*,k^*)$ before $p$ selected $(\Ops^*,ts^*,k^*)$ in line~\ref{selection}. 
\qedhere~$_\text{\autoref{lmx3}}$
\end{proof}

\begin{lemma}\label{ne-doops}
If a process calls $\DoOps{(\Os,-)}{-}{j}$, then
\begin{enumerate}
	\item $j \ge 0$,
	\item $\Os = \emptyset$ if and only if $j=0$.
\end{enumerate}
\end{lemma}

\begin{proof}
Suppose for contradiction that some call to $\DO$ fails to satisfy
	the conditions of the lemma, and
	let the \emph{first} call to do so be the call $\DoOps{(\Os,-)}{t}{j}$,
	for some $\Os$, $t$, and $j$,
	made by some process $p$.
There are two cases:

\begin{enumerate}
\item $p$ calls $\DoOps{(\Os,-)}{t}{j}$ in line~\ref{first-doops}.
Before this call, $p$ has $(\Ops^*,ts^*,k^*)$ with $\Ops^* = \Os$ 
	and $k^* = j$ in line~\ref{selection}.
Thus, $p$ has $(\Ops^*,ts^*,k^*) \neq (\emptyset, -1, 0)$ in line~\ref{selection}.
From Lemma~\ref{lmx3},
	some process $q^*$ accepted the tuple $(\Ops^*,ts^*,k^*)$,
	and this occurred
	before $p$ calls $\DoOps{(\Os,-)}{t}{j}$ in line~\ref{first-doops}.
By Observation~\ref{Accept-Doops},
	a process called $\DoOps{(\Ops^*,-)}{ts^*}{k^*}$ before
	$q^*$ accepted $(\Ops^*,ts^*,k^*)$, and so before $p$ calls $\DoOps{(\Os,-)}{t}{j}$.
That call also fails to satisfy the conditions of the lemma,
	contradicting that $p$'s call to $\DoOps{(\Os,-)}{t}{j}$
	is the first to do so.

\item $p$ calls $\DoOps{(\Os,-)}{t}{j}$ in line~\ref{second-doops}.
By the guard in line~\ref{nonempty2}, $\Os \neq \emptyset$.
Since the call fails to satisfy the conditions of the lemma,
	either $j < 0$ or $(j=0 \wedge O \neq \emptyset)$. 
Thus, $j \le 0$.
Since $p$ calls $\DoOps{(\Os,-)}{t}{j}$ in line~\ref{second-doops},
	$p$ has $k+1 =j$, and therefore $(\Ops,ts,k) = (-,-,j-1)$, at that time.
Since $j-1 <0$, the tuple $(-,-,j-1)$ is not the initial value of $(\Ops,ts,k)$ at $p$,
	and therefore $p$ accepted this $(-,-,j-1)$ 
	before calling $\DoOps{(\Os,-)}{t}{j}$ in line~\ref{second-doops}.
By Observation~\ref{Accept-Doops},
	a process called $\DoOps{(-,-)}{-}{j-1}$ before $p$ accepted $(-,-,j-1)$,
	and so before $p$ calls $\DoOps{(\Os,-)}{t}{j}$ in line~\ref{second-doops}.
Since $j-1<0$, that call also fails to satisfy the conditions of the lemma,
	contradicting that $p$'s call to $\DoOps{(\Os,-)}{t}{j}$
	is the first to do so.
\qedhere~$_\text{\autoref{ne-doops}}$
\end{enumerate}	 
\end{proof}

\begin{definition}\label{locking}
	A process $p$ \emph{locks a tuple $(\Os,t,j)$}
		if $p$ executes $\DoOps{(\Os, -)}{t}{j}$ up to line~\ref{batch1} (included). 
	If a process locks $(\Os,t,j)$, we say that 
		\emph{$(\Os,t,j)$ is locked}.
\end{definition}

\begin{observation}\label{acceptedfirst}
If a process locks a tuple $(\Os,t,j)$, then it previously accepted this tuple.
\end{observation}

\begin{observation}\label{aboutlocking}
If a process locks a tuple $(\Os,t,j)$, then it does so while executing $\LeaderWork{t}$.
\end{observation}

From Lemma~\ref{ne-doops}, we have:

\begin{corollary}\label{ne-lock}
If a process locks a tuple $(\Os,t,j)$, then
\begin{enumerate}
	\item  $j \ge 0$,
	\item  $\Os = \emptyset$ if and only if $j=0$, and
\end{enumerate}
\end{corollary}

\begin{lemma}\label{specialcase}
Suppose $(\Os,t,j)$ and $(\Os',t,j')$ are locked.
If $j' =j$ then $\Os' = \Os$.
\end{lemma}

\begin{proof}
If $(\Os,t,j)$ and $(\Os',t,j')$ are locked,
	by Observation~\ref{acceptedfirst}, 
	$(\Os,t,j)$ and $(\Os',t,j')$ are also accepted.
The result now follows directly from Lemma~\ref{simplecase}.
\qedhere~$_\text{\autoref{specialcase}}$
\end{proof}

\begin{theorem}\label{lockingcore}
Suppose a tuple $(\Os,t,j)$ is locked.
For all $t'>t$, if a process $\ell$ accepts a tuple $(\Os',t',j')$ in $\LeaderWork{t'}$
	then
	$\ell$ selects
	a tuple $(\Ops^*,ts^*,k^*)$
	in line~\ref{selection} of $\LeaderWork{t'}$ such that:
\begin{enumerate}
\item $(ts^*,k^*) \ge (t,j)$, and
\item some process $q^*$ previously accepted $(\Ops^*,ts^*,k^*)$.
\end{enumerate}
\end{theorem}

\begin{proof}
Suppose a process $p$ locks $(\Os,t,j)$,
	and
	a process $\ell$ accepts a tuple $(\Os',t',j')$ in $\LeaderWork{t'}$ for some $t'>t$.
From the code of $\LeaderWork{t'}$,
	it is clear that
	before accepting $(\Os',t',j')$,
	$\ell$ selects
	a tuple $(\Ops^*,ts^*,k^*)$
	in line~\ref{selection} of $\LeaderWork{t'}$.

Since $p$ locks $(\Os,t,j)$,
	by Observation~\ref{aboutlocking} and
	Definition~\ref{locking},
	$p$ becomes leader at local time $t$ and
	it executes $\DoOps{(\Os,-)}{t}{j}$ up to line~\ref{lease1}.
So $p$ found $| \PACKED[t,j] | \ge  \lfloor n/2 \rfloor$ in line~\ref{ack-condition}.
Let $M_1$ be the set consisting of $p$ and the processes that sent a $\langle \PACK,t,j \rangle$ message to $p$.
Note that $|M_1| > n/2$.

\begin{claim}\label{claim-hacks}
Every process in $M_1$ accepts $(\Os,t,j)$.
\end{claim}

\begin{proof}
First note that $p$ accepts $(\Os,t,j)$ in line~\ref{leader-accept}.
Now let $p' \in M_1$ where $p' \neq p$.
So $p'$ sent a $\langle \PACK,t,j \rangle$ message to $p$ in line~\ref{sendPack}.
From lines~\ref{a3b}-\ref{sendPack} of the algorithm, it is clear that $p'$
	received some $\langle \Prepare,(\Os',-),t,j,-\rangle$ message from $p$,
	and that $p'$ has $(\Ops,ts,k) = (\Os',t,j)$ in line~\ref{sendPack}.
We claim that $\Os' = \Os$.
To see this, note that $p$ sent $\langle \Prepare,(\Os',-),t,j,- \rangle$ during an execution of $\DoOps{(\Os',-)}{t}{j}$.
Since $p$ calls $\DoOps{(\Os,-)}{t}{j}$ and $\DoOps{(\Os',-)}{t}{j}$,
	by Lemma~\ref{doops-simplecase}, $\Os' = \Os$.
So, $p'$ has $(\Ops,ts,k) = (\Os,t,j)$ in line~\ref{sendPack}.
Since $p$ became leader at local time $t$, $t \neq -1$.
Thus $(\Os,t,j)$ is not the initial value of $(\Ops,ts,k)$ at $p'$.
Therefore $p'$ accepted $(\Os,t,j)$ 
	before sending $\langle \PACK,t,j \rangle$ to $p$.
\qedhere~$_\text{\autoref{claim-hacks}}$
\end{proof}

Note that $\ell$ selects the tuple $(\Ops^*,ts^*,k^*)$ in line~\ref{selection}
	as a tuple with maximum $(ts^*,k^*)$ in $\REPLIES[t'] \cup \{ (\Ops,ts,k) \}$.
From the code in lines~\ref{estrep1}-\ref{estrep3},
	it is clear that:
	
\begin{itemize}
\item $\REPLIES[t']$ at $\ell$ is the set $\{ (\Os_q,t_q,j_q) ~|~ \ell \mbox{ received a}$
	$\langle \EstReply,t',\Os_q,t_q,j_q,- \rangle \mbox{ message} \}$, 
	and
\item $\replied[t']$ at $\ell$ is the set 
	$\{ q ~|~ \ell \mbox{ received some }
	\langle \EstReply,t',\Os_q,t_q,j_q,-\rangle$ $\mbox{message from } $q$ \}$.
\end{itemize}

In line~\ref{est-condition2}, $\ell$ finds $ | \replied[t'] | \ge  \lfloor n/2 \rfloor$.
So the set $\REPLIES[t']$ that $\ell$ uses to select $(\Ops^*,ts^*,k^*)$ in line~\ref{selection},
	contains tuples from at least $\lfloor n/2 \rfloor$ distinct processes in $\replied[t']$.
Since $\ell \not \in \replied[t']$ (because $\ell$ does not send a $\langle \EstReply,t',-,-,-,- \rangle$ to itself)
	these distinct processes are different than $\ell$.
Let $M_2$ be the set consisting of $\ell$ and the processes that are in $\replied[t']$
	at the time
	when $\ell$ selects $(\Ops^*,ts^*,k^*)$ in line~\ref{selection}.
Note that $|M_2| > n/2$.

Since $|M_1| > n/2$, $|M_2| > n/2$, and there are $n$ processes, the intersection of $M_1$ and $M_2$ is not empty.
Let $q$ be a process in $M_1 \cap M_2$.
There are two possible cases, namely, $q \neq \ell$ and $q = \ell$.
We now prove that $(ts^*,k^*) \ge (t,j)$ in both cases:

\begin{enumerate}

\item $q \neq \ell$.
Since $q \in M_2$, $q \in\replied[t']$
	when $\ell$ selects $(\Ops^*,ts^*,k^*)$ in line~\ref{selection}.
So $q$ sent some $\langle \EstReply,t',\Os_q,t_q,j_q,-\rangle$ message to $\ell$
	such that $(\Os_q,t_q,j_q) \in \REPLIES[t']$ when $\ell$ selects $(\Ops^*,ts^*,k^*)$ in line~\ref{selection}.
Consider the following two events:

\begin{enumerate}
\item\label{e1} $q$ accepts $(\Os,t,j)$ (this occurs because $q \in M_1$, see Claim~\ref{claim-hacks}).

\item\label{e2} $q$ sends the above $\langle \EstReply,t',\Os_q,t_q,j_q,-\rangle$ message to $\ell$.
\end{enumerate}

\begin{claim}\label{claim-paxos}
Event (a) occurred before event (b).

\end{claim}

\begin{proof}
Suppose, for contradiction,
	that (b) occurred before (a).
From the code in lines~\ref{a2b}-\ref{a2e},
	it is clear that before sending this $\langle \EstReply,t',\Os_q,t_q,j_q,-\rangle$
	message to $\ell$,
	process $q$ sets $\maxT$ to $max(\maxT, t')$;
	since $t' >t$, this means $q$ has $\maxT > t$
	before sending $\langle \EstReply,t',\Os_q,t_q,j_q,-\rangle$
	to~$\ell$.
Note that $\maxT$ is non-decreasing at $q$
	(because line~\ref{incTmax} is the only statement that modifies $\maxT$ in the algorithm).
So from the time when
	$q$ sent this $\langle \EstReply,t',\Os_q,t_q,j_q,-\rangle$ to $\ell$,
	process $q$ has $\maxT > t$ forever.

Note that $q$ accepts $(\Os,t,j)$ in line~\ref{leader-accept} or in line~\ref{client-accept}.
Since $q$ sent $\langle \EstReply,t',\Os_q,t_q,j_q,-\rangle$ to $\ell$ before accepting $(\Os,t,j)$,
	when $q$ compares $t$ with $\maxT$ in line~\ref{acceptcheck1} or line~\ref{acceptcheck3}
	just before accepting $(\Os,t,j)$,
	$q$ finds that $\maxT > t$.
So $q$ does not accept $(\Os,t,j)$ in line~\ref{leader-accept} or in line~\ref{client-accept}
	 --- a contradiction.
\qedhere~$_\text{\autoref{claim-paxos}}$
\end{proof}

By Claim~\ref{claim-paxos}, $q$ accepted $(\Os,t,j)$ before sending the above $\langle \EstReply,t',\Os_q,t_q,k_q,-\rangle$
	message to~$\ell$ (recall that $(\Os_q,t_q,j_q) \in \REPLIES[t']$ when $\ell$ selected $(\Ops^*,ts^*,k^*)$ in line~\ref{selection}).
Note that when $q$ sent this message, $q$'s variables tuple $(\Ops,ts,k)$ contained $(\Os_q,t_q,j_q)$,
	and so $(\Os_q,t_q,j_q)$ was the \emph{last} tuple that $q$ accepted before sending the message.
Thus, either $(\Os_q,t_q,j_q) = (\Os,t,j)$,
	or $q$ accepted $(\Os,t,j)$ before accepting $(\Os_q,t_q,j_q)$.
In the first case, $(t_q,j_q) = (t,j)$.
In the second case, by Lemma~\ref{T-k-increase}, $(t_q,j_q) > (t,j)$.
So $(t_q,j_q) \ge (t,j)$.
Since $\ell$ has $(\Os_q,t_q,j_q) \in \REPLIES[t']$ when it selects $(\Ops^*,ts^*,k^*)$
	as a tuple with maximum $(ts^*,k^*)$ in $\REPLIES[t'] \cup \{ (\Ops,ts,k) \}$ in line~\ref{selection},
	$(ts^*,k^*) \ge (t_q,j_q)$.
So $(ts^*,k^*) \ge (t,j)$.

\item $q = \ell$.
Thus, $\ell$ accepts the tuple $(\Os,t,j)$.
Since $\ell$ also accepts $(\Os',t',j')$ and $(t',j') > (t,j)$ (because $t' > t$),
	from Lemma~\ref{T-k-increase},
	$\ell$ accepts $(\Os,t,j)$ before accepting $(\Os',t',j')$.
By Observation~\ref{LeaderWorkAcceptances},
	from the instant
	$\ell$ calls $\LeaderWork{t'}$
	to the instant
	$\ell$ accepts $(\Os',t',j')$ in $\LeaderWork{t'}$,
	$\ell$ does not accept any tuple $(-,t,-)$ with $t \neq t'$.
Thus, $\ell$ accepts $(\Os,t,j)$ \emph{before} calling $\LeaderWork{t'}$.
Let $(\Os_\ell,t_\ell,j_\ell)$ be the \emph{last} tuple that $\ell$ accepts before calling $\LeaderWork{t'}$
	(it is possible that $(\Os_\ell,t_\ell,j_\ell) = (\Os,t,j)$).
From Lemma~\ref{T-k-increase}, $(t_\ell,j_\ell) \ge (t,j)$.
Note that $\ell$ has $(\Ops,ts,k) = (\Os_\ell,t_\ell,j_\ell)$
	from the instant
	it accepts $(\Os_\ell,t_\ell,j_\ell)$
	to the instant
	it selects $(\Ops^*,ts^*,k^*)$
	as a tuple with maximum $(ts^*,k^*)$ in $\REPLIES[t'] \cup \{ (\Ops,ts,k) \}$
	in line~\ref{selection} of $\LeaderWork{t'}$.
So $(ts^*,k^*) \ge (ts,k) =(t_\ell,j_\ell)$.
Since $(t_\ell,j_\ell) \ge (t,j)$,
	we have $(ts^*,k^*) \ge (t,j)$.
\end{enumerate}

So in all cases $(ts^*,k^*) \ge (t,j)$,
	proving part~(1) of the theorem.

Since the process $p$ that locked $(\Os,t,j)$ became leader at local time $t$, we have $t\ge0$.
Since $(ts^*,k^*) \ge (t,j)$ we have  $ts^* \ge t \ge 0$.
Thus $\ell$ has $(\Ops^*,ts^*,k^*) \neq (\emptyset, -1, 0)$ in line~\ref{selection}.
By Lemma~\ref{lmx3},
	some process $q^*$ previously accepted $(\Ops^*,ts^*,k^*)$,
	proving part~(2) of the theorem.
\qedhere~$_\text{\autoref{lockingcore}}$
\end{proof}

\begin{theorem}\label{generalcase1a}
Suppose a tuple $(\Os,t,j)$ is locked.
For all $t'>t$, if a tuple $(\Os',t',j')$ is accepted then:
\begin{enumerate}
\item\label{t1} $j' \ge j$, and
\item\label{t2} if $j' = j$ then $\Os' = \Os$.
\end{enumerate}
\end{theorem}

\begin{proof}
The proof is by contradiction.
Suppose that some $(\Os,t,j)$ is locked, and:

\smallskip\noindent
(*) there is a $t'>t$, $\Os'$, and $j'$, such that $(\Os',t',j')$ is accepted but:
\begin{enumerate}
\item[(a)] $j' < j$, or
\item[(b)] $j' = j$ and $\Os' \neq \Os$.
\end{enumerate}

Without loss of generality, assume that $t'$ is the smallest time $t'>t$
	for which there is a ``bad'' accepted tuple $(\Os',t',j')$.
From this assumption, we have:

\smallskip\noindent
(**) for all $\hat{t}$ such that $t <\hat{t}<t'$, if a tuple $(\widehat{\Os},\hat{t},\hat{j})$
	is accepted then:
\begin{enumerate}
\item\label{try} $\hat{j} \ge j$, and
\item if $\hat{j} = j$ then $\widehat{\Os} = \Os$.

\end{enumerate}

Consider the accepted tuple $(\Os',t',j')$.
By Lemma~\ref{leaderfirst}, the first process that accepts $(\Os',t',j')$
	is a process $\ell$ that becomes leader at time~$t'$ and accepts $(\Os',t',j')$ in $\LeaderWork{t'}$.
Since  $(\Os,t,j)$ is locked and $t' >t$, by Theorem~\ref{lockingcore},
	process $\ell$ selected a tuple $(\Ops^*,ts^*,k^*)$ in line~\ref{selection} of $\LeaderWork{t'}$ such that:

\begin{enumerate}
\item $(ts^*,k^*) \ge (t,j)$, and
\item some process $q^*$ previously accepted $(\Ops^*,ts^*,k^*)$.
\end{enumerate}

After selecting $(\Ops^*,ts^*,k^*)$ in line~\ref{selection},
	process $\ell$ first verified that $ts^* < t'$ in line~\ref{Egathercrumbs},
	and then
	$\ell$ executed lines~\ref{FG}-\ref{first-doops}.
	In particular, $\ell$ called $\DoOps{(\Ops^*,0)}{t'}{k^*}$ in line~\ref{first-doops}
	and $\ell$ accepted $(\Ops^*,t',k^*)$ during this execution.
Note that $(\Ops^*,t',k^*)$ is first tuple of the form $(-,t',-)$ that $\ell$ accepts.\footnote{The tuples $(\Ops^*,t',k^*)$ and $(\Os',t',j')$ that $\ell$ accepts are not necessarily distinct.}

\begin{claim}\label{about-star-tple}
Consider $(\Ops^*,ts^*,k^*)$:

\begin{enumerate}
\item $k^* \ge j$, and
\item if $k^* = j$ then $\Ops^* = \Os$.
\end{enumerate}
\end{claim}

\begin{proof}
Since $(ts^*,k^*) \ge (t,j)$, we have $ts^* \ge t$.
There are two possible cases:
\begin{enumerate}
\item $ts^* = t$.
So $k^* \ge j$, and
	$(\Ops^*,ts^*,k^*)$ is $(\Ops^*,t,k^*)$.
Since both $(\Os,t,j)$ and $(\Ops^*,t,k^*)$ are accepted, by Lemma~\ref{simplecase},
	if $k^* = j$ then $\Ops^* = \Os$.
	
\item $ts^* > t$.
Recall that before calling $\DoOps{(\Ops^*,0)}{t'}{k^*}$ in line~\ref{first-doops},
	process $\ell$ verified that $ts^* < t'$ holds (in line~\ref{Egathercrumbs}).
Since $ t < ts^* < t'$, and $(\Ops^*,ts^*,k^*)$ was accepted by some process,
	by~(**)
	we have
	$k^* \ge j$, and if $k^* = j$ then $\Ops^* = \Os$.
\end{enumerate}

So in all possible cases, the claim holds.
\qedhere~$_\text{\autoref{about-star-tple}}$
\end{proof}

Now consider $(\Os',t',j')$.
Recall that $(\Ops^*,t',k^*)$ is the \emph{first} tuple of the form $(-,t',-)$ that $\ell$ accepts.
Since $\ell$ accepts $(\Os',t',j')$,
	there are two possible cases:

\begin{enumerate}
\item $(\Os',t',j')$ is $(\Ops^*,t',k^*)$.
So $\Ops^* = \Os'$, and $k^* = j'$.
By Claim~\ref{about-star-tple}, $j' \ge j$ and if $j' = j$ then $\Os' = (\Os,\s')$.
	
\item $\ell$ accepts $(\Ops^*,t',k^*)$ before it accepts $(\Os',t',j')$.
By Lemma~\ref{T-k-increase}, $j' > k^*$.
By Claim~\ref{about-star-tple}, $k^* \ge j$.
Thus, $j' > j$.
\end{enumerate}

So in all cases we have $j' \ge j$, and
	if $j' = j$ then $\Os' = \Os$.
This contradicts the assumption~(*) about $(\Os',t',j')$.
\qedhere~$_\text{\autoref{generalcase1a}}$
\end{proof}

\begin{theorem}\label{generalcase2}
If tuples $(\Os,t,j)$ and $(\Os',t',j)$ are locked,
	then $\Os=\Os'$.
\end{theorem}

\begin{proof}
Suppose $(\Os,t,j)$ and $(\Os',t',j)$ are locked.
By Observation~\ref{acceptedfirst}, tuples $(\Os,t,j)$ and $(\Os',t',j)$ are also accepted.
If $t =t'$ then, by Lemma~\ref{specialcase}, $\Os = \Os'$.
If $t' >t$ or $t > t'$ then, by Theorem~\ref{generalcase1a}(\ref{t2}), $\Os = \Os'$.
So in all cases $\Os = \Os'$.
\qedhere~$_\text{\autoref{generalcase2}}$
\end{proof}

\subsubsection{Batch properties}

\begin{lemma}\label{acceptsB}
For all $j \ge 1$, if a process $p$ accepts a tuple $(-,-,j)$, then $p$ previously set $\Batch[j-1]$ to $(\Os,-)$ for some possibly empty set $\Os$.

\end{lemma}

\begin{proof}
Suppose, for contradiction, that there is a $j \ge1$ and a process $p$ such that
	$p$ accepts a tuple $(-,-,j)$, but it did not previously set $\Batch[j-1]$ to any pair.
Let $(\Os,t,j)$ be the \emph{first} $(-,-,j)$ tuple that $p$ accepts such that
	$p$ did not previously set $\Batch[j-1]$ to any pair.
Clearly, $(\Os,t,j)$ is the first $(-,-,j)$ tuple that $p$ accepts.
There are two cases, depending on where $p$ accepts $(\Os,t,j)$:

\begin{enumerate}
\item $p$ accepts $(\Os,t,j)$ in line~\ref{client-accept}.
	Then $p$ previously set $\Batch[j-1]$ to some pair in line~\ref{setBatch3} --- a contradiction.
	
\item $p$ accepts $(\Os,t,j)$ in line~\ref{leader-accept}.
This occurs during $p$'s execution of $\DoOps{(\Os,-)}{t}{j}$ in $\LeaderWork{t}$.
There are two cases, depending on where $p$ called $\DoOps{(\Os,-)}{t}{j}$.
\begin{enumerate}
\item $p$ called $\DoOps{(\Os,-)}{t}{j}$ in line~\ref{second-doops}.
From the code of $\LeaderWork{t}$, it is clear that $p$ called $\DoOps{(-,-)}{t}{-}$ at least once
	before calling $\DoOps{(\Os,-)}{t}{j}$ in line~\ref{second-doops}.
Let $\DoOps{(\Os',-)}{t}{j'}$ be the last call to \linebreak
	$\DoOps{(-,-)}{t}{-}$ that $p$ makes before calling $\DoOps{(\Os,-)}{t}{j}$.
By Lemma~\ref{T-k-Doops}, $j' = j-1$.
Since \linebreak
	$\DoOps{(\Os',-)}{t}{j-1}$ must have returned $\textsc{Done}$,
	$p$ set $\Batch[j-1]$ to $(\Os',-)$ in line~\ref{batch1} of $\DoOps{(\Os',-)}{t}{j-1}$.
Since this occurs before $p$ accepts $(\Os,t,j)$, it is a contradiction.

\item $p$ called $\DoOps{(\Os,-)}{t}{j}$ in line~\ref{first-doops}.
Let  $(\Ops^*,ts^*,k^*)$ be the tuple with maximum $(ts^*,k^*)$
	in $\REPLIES[t] \cup \{ (\Ops,ts,k) \}$ that $p$ selects in line~\ref{selection}.
From the code of lines~\ref{selection}-\ref{first-doops},
	it is clear that $\Ops^* = \Os$ and $k^* = j$,
	so $(\Ops^*,ts^*,k^*) = (\Os,ts^*,j)$.
Furthermore, since $p$ does not return in line~\ref{Egathercrumbs},
	$t \neq ts^*$, so the tuples $(\Os,ts^*,j)$ and $(\Os,t,j)$ are distinct.
There are two cases depending on how $p$ selected $(\Os,ts^*,j)$ in line~\ref{selection}.
\begin{enumerate}
\item $(\Os,ts^*,j)$ is the value of $(\Ops,ts,k)$ at $p$ in line~\ref{selection}.
Since $j \ge 1$, $(\Os,ts^*,j)$ is not the initial value $(\emptyset, -1, 0)$ of $(\Ops,ts,k)$ at $p$.
So, by Observation~\ref{AcceptAndOps-ts-k},
	$p$ previously accepted $(\Os,ts^*,j)$.
Thus $p$ accepted $(\Os,ts^*,j)$ before calling $\DoOps{(\Os,-)}{t}{j}$ in line~\ref{first-doops},
	and so before accepting $(\Os,t,j)$ in line~\ref{leader-accept} --- a contradiction.

\item $(\Os,ts^*,j)$ is a tuple in $\REPLIES[t]$ at $p$ in line~\ref{selection}.
From the code in lines~\ref{estrep1}-\ref{estrep3},
	it is clear that $\REPLIES[t] = \{ (\Os_q,t_q,j_q) ~|~ p \mbox{ received some}$
	$\langle \EstReply,-,\Os_q,t_q,j_q, B_q \rangle \mbox{ message} \}$.
So the following events occurred at $p$
	before $p$ selected $(\Os,ts^*,j)$ from $\REPLIES[t]$ in line~\ref{selection}:
	$p$~received a
	$\langle \EstReply,-,\Os,ts^*,j, B \rangle$ message for some pair $B$ in line~\ref{estrep1},
	$p$ set $\Batch[j-1]$ to $B$ in line~\ref{setBatch4},
	and then $p$ inserted $(\Os,ts^*,j)$ into $\REPLIES[t]$ in line~\ref{estrep3}.
So $p$ set $\Batch[j-1]$ to a pair $B$ before executing line~\ref{selection},
	and so before calling $\DoOps{(\Os,-)}{t}{j}$ in line~\ref{first-doops},
	and thus before accepting $(\Os,t,j)$ --- a contradiction.
\qedhere~$_\text{\autoref{acceptsB}}$	 

\end{enumerate}

\end{enumerate}

\end{enumerate}

\end{proof}

From the definition of locking, we have:

\begin{observation}\label{lock-batch}
If a process locks a tuple $(\Os,t,j)$ then 
	it sets $\Batch[j]$ to $(\Os,-)$ in line~\ref{batch1}.
\end{observation}

\begin{lemma}\label{commit-doops}
If a process $q$ sends a $\langle \CommitLease, (-,-), j, -, - \rangle$
	message to $p$ in line~\ref{sendcommit2} of a $\LeaderWork{t}$ for some~$t$,
	then $q$ previously executed some $\DoOps{(-,-)}{t}{j}$ in $\LeaderWork{t}$.
\end{lemma}
\begin{proof}
Suppose $q$ sends a $\langle \CommitLease, (-,-), j, -, - \rangle$
	message to $p$ in line~\ref{sendcommit2} of some $\LeaderWork{t}$.
	From the code of $\LeaderWork{t}$, it is clear that
	$q$ called $\DoOps{(-,-)}{t}{-}$ at least once in $\LeaderWork{t}$ before executing line~\ref{sendcommit2}.
	
	Let $\DoOps{(\Os',-)}{t}{j'}$ be the \emph{last} $\DoOps{(-,-)}{t}{-}$ that $q$ calls before executing line~\ref{sendcommit2}.
	In this $\DoOps{(\Os',-)}{t}{j'}$, $q$ sets its variable $k$ to $j'$
	(line~\ref{leader-accept}).
	Note that while $q$ is executing in $\LeaderWork{t}$
		$q$ cannot be executing concurrently in the procedure $\PCM()$,
		so $q$ cannot modify $k$ in line~\ref{client-accept} of $\PCM()$,
		and thus $q$ can modify $k$ only inside a call to $\DoOps{(-,-)}{t}{-}$
		(in line~\ref{leader-accept}).
	Therefore, since $\DoOps{(\Os',-)}{t}{j'}$ is the \emph{last} $\DoOps{(-,-)}{t}{-}$ that $q$ calls before executing line~\ref{sendcommit2},
		when $q$ executes line~\ref{sendcommit2} the values of $k$ is still $j'$.
	Since $q$ sent $\langle \CommitLease, (-,-), k, -, - \rangle = \langle \CommitLease, (-,-), j, -, - \rangle$
		to $p$ in line~\ref{sendcommit2}, when $q$ executes line~\ref{sendcommit2} the value of $k$ is $j$.
	So $j'=j$.
	Therefore the \emph{last} $\DoOps{(-,-)}{t}{-}$ that $q$ calls before executing line~\ref{sendcommit2}
	is  $\DoOps{(-,-)}{t}{j'} = \DoOps{(-,-)}{t}{j}$.	
\qedhere~$_\text{\autoref{commit-doops}}$	 
\end{proof}
\begin{lemma}\label{IamOutOfLableNames}
For all $j \ge 0$, if a process sets $\Batch[j]$ to a pair $(\Os,-)$ at some real time $\tau$,
	then some process locks a tuple $(\Os,-,j)$ by real time $\tau$. 
\end{lemma}

\begin{proof}
Suppose, for contradiction, that this lemma does not hold.
Suppose that the \emph{first time} that the lemma is violated
	is when: (*) process $p$ sets $\Batch[j]$ to a $(\Os,-)$ for some set $\Os$ at real time $\tau$ (for some $j \ge 0$),
	while no process locks $(\Os,-,j)$ by real time $\tau$.
This definition implies that: (**) \emph{no} process sets $\Batch[j]$ to $(\Os,-)$ \emph{before} real time~$\tau$.
There are several cases, depending on where $p$ set $\Batch[j]$ to $(\Os,-)$ at real time $\tau$.
We now show that each case leads to a contradiction, and so the lemma holds.
\begin{enumerate}
\item\label{ehad} $p$ sets $\Batch[j]$ to $(\Os,-)$ at real time $\tau$ in line~\ref{batch1}.
	Note that, by the definition of locking, $p$ simultaneously locks a tuple $(\Os,-,j)$  in line~\ref{batch1}.
	Thus $p$ locks $(\Os,-,j)$ by real time $\tau$ --- a contradiction to (*).

\item $p$ sets $\Batch[j]$ to $(\Os,-)$ at real time $\tau$ in line~\ref{batch2}.
	Thus, $p$ received a $\langle \CommitLease, (\Os,-), j, -, - \rangle$ message from some process $q \neq p$
	in line~\ref{commitreceipt1}, and so before real time $\tau$.
	There are two cases:
	
	\begin{enumerate}
	\item $q$ sent $\langle \CommitLease, (\Os,-), j, -, - \rangle$ to $p$ in line~\ref{sendcommit}.
	Thus $q$ previously set $\Batch[j]$ in line~\ref{batch1} and has $\Batch[j] = (\Os,-)$ in line~\ref{sendcommit},
		which implies that $q$ previously set $\Batch[j]$ to $(\Os,-)$
		--- a contradiction to (**).

	\item $q$ sent $\langle \CommitLease, (\Os,-), j, -, - \rangle$ to $p$ in line~\ref{sendcommit2},
	during the execution of $\LeaderWork{t}$ for some $t$.
	By Lemma~\ref{commit-doops},
		$q$ previously executed $\DoOps{(-,-)}{t}{j}$ in $\LeaderWork{t}$ and
		this call must returned \textsc{Done} since $q$ continued to execute line~\ref{sendcommit2}.
	Thus $q$ previously set $\Batch[j]$ in line~\ref{batch1} of $\DoOps{(-,-)}{t}{j}$.
	Since $q$ has $\Batch[j] = (\Os,-)$ in line~\ref{sendcommit2},
		it must previously set $\Batch[j]$ to $(\Os,-)$ 
		--- a contradiction to (**).
	\end{enumerate}
	
\item $p$ sets $\Batch[j]$ to $(\Os,-)$ at real time $\tau$ in line~\ref{batch3}.
Process $p$ must have received a $\langle \MyBatch, j, (\Os,-) \rangle$
	message from some process $q \neq p$ in line~\ref{batchreceipt} 
	($q\neq p$ because $q$ only sends $\MyBatch$ messages to processes from which it received a $\MyGaps$ message in line~\ref{gb}, and $p$ does not send $\MyGaps$ messages to itself in line~\ref{totoz10}).
So $q$ sent $\langle \MyBatch, j,(\Os,-) \rangle$ to $p$ in line~\ref{sendbatch}.
From the code of line~\ref{sendbatch}, 
	it is clear that $q$ had $\Batch[j] = (\Os,-) \neq (\emptyset,\infty)$
	when $q$ sent that message.
Since $(\Os,-) \neq (\emptyset,\infty)$ is not the initial value of $\Batch[j]$ at $q$,
	$q$ must have previously set $\Batch[j]$ to $(\Os,-)$.
So $q$ set $\Batch[j]$ to $(\Os,-)$ before $p$ does --- a~contradiction to (**).

\item $p$ sets $\Batch[j]$ to $(\Os,-)$ at real time $\tau$ in line~\ref{setBatch4}.
So $p$ received some $\langle \EstReply,-,\Os',t',j+1, (\Os,-) \rangle$ message
	from some process $q \neq p$ in line~\ref{estrep1} 
	before setting $\Batch[j]$ to $(\Os,-)$ in line~\ref{setBatch4}.
Note that when $q$ sent this message in line~\ref{reply},
	$q$ had $(\Ops,ts,k) := (\Os',t',j+1)$ and $\Batch[k-1] = \Batch[j] = (\Os,-)$.

We now show
	that $q$ set $\Batch[j]$ to $(\Os,-)$ before executing line~\ref{reply}
	(note that this contradicts (**)).
Since $q$ had $(\Ops,ts,k) = (\Os',t',j+1)$ in line~\ref{reply},
	and $j+1 \ge 1$, the tuple $(\Os',t',j+1)$ is not the initial value $(\emptyset,-1,0)$ of $(\Ops,ts,k)$ at $q$.
	So $q$ accepted $(\Os',t',j+1)$ before executing line~\ref{reply} by Observation~\ref{AcceptAndOps-ts-k}.
 	By Lemma~\ref{acceptsB}, $q$ set its variable $\Batch[j]$ before accepting $(\Os',t',j+1)$,
	and therefore before executing line~\ref{reply}.
	Thus, since $q$ has $\Batch[j] = (\Os,-)$ in line~\ref{reply},
	it is now clear that $q$ set $\Batch[j]$ to $(\Os,-)$ before executing line~\ref{reply}.
This implies that $q$ set $\Batch[j]$ to $(\Os,-)$ before $p$ did so at real time $\tau$ --- a contradiction to (**).

\item $p$ sets $\Batch[j]$ to $(\Os,\s)$ at real time $\tau$ in line~\ref{setBatch3}.
So $p$ received a $\langle \Prepare,(\Os',-),t',j+1, (\Os,-) \rangle$ message
	from some process $q \neq p$ in line~\ref{a3b} before setting $\Batch[j]$ to $(\Os,-)$ in line~\ref{setBatch3}.
Note that when $q$ sent this message in line~\ref{sendprep},
	$q$ had $(\Ops,ts,k) = (\Os',t',j+1)$  and $\Batch[j] = (\Os,-)$.

We claim that
	$q$ set $\Batch[j]$ to $(\Os,-)$ before executing line~\ref{sendprep}
	(note that this contradicts (**)).
The proof is virtually identical to the one that we saw above.
Since $q$ had $(\Ops,ts,k) = (\Os',t',j+1)$ in line~\ref{sendprep},
	and $j+1 \ge 1$,
	$q$ accepted $(\Os',t',j+1)$ before executing line~\ref{sendprep} by Observation~\ref{AcceptAndOps-ts-k}.
 	By Lemma~\ref{acceptsB}, $q$ set its variable $\Batch[j]$ before accepting $(\Os',t',j+1)$,
	and therefore before executing line~\ref{sendprep}.
	Thus, since $q$ has $\Batch[j] = (\Os,-)$ in line~\ref{sendprep},
		$q$ set $\Batch[j]$ to $(\Os,-)$ before executing line~\ref{sendprep}.
Therefore $q$ set $\Batch[j]$ to $(\Os,-)$ before $p$ did so at real time $\tau$ --- a contradiction to (**).
\qedhere~$_\text{\autoref{IamOutOfLableNames}}$
\end{enumerate}
\end{proof}

Lemmas~\ref{acceptsB} and~\ref{IamOutOfLableNames} imply the following:

\begin{corollary}\label{accepts-to-locking}
For all $j \ge 1$, if a process accepts a tuple $(-,-,j)$
	then some process previously locked a tuple $(-,-,j-1)$.
\end{corollary}

By Lemma~\ref{IamOutOfLableNames} and Corollary~\ref{ne-lock}, we have:

\begin{corollary}\label{ne-batch}
For all $j \ge 0$, if a process sets $\Batch[j]$ to $(\Os,-)$ for some set $\Os$, then $\Os = \emptyset$ if and only if $j=0$. 
\end{corollary}

\begin{theorem}\label{finalsafetya}
For all $j\geq 0$,
	if processes $p$ and $p'$ set $\Batch[j]$ to $(\Os,-)$ and $(\Os',-)$, respectively, 
	then $\Os = \Os'$.
\end{theorem}

\begin{proof}
Suppose $p$ and $p'$ set $\Batch[j]$ to $(\Os,-)$ and $(\Os',-)$, respectively.
Then, by Lemma~\ref{IamOutOfLableNames}, there are $t$ and $t'$
	such that $(\Os,t,j)$ and $(\Os',t',j)$ are locked.
By Theorem~\ref{generalcase2}, $\Os = \Os'$.
\qedhere~$_\text{\autoref{finalsafetya}}$
\end{proof}

\begin{corollary}\label{finalsafetyac}
If a process $p$ has $\Batch[j] = (\Os,-)$ 
	for some non-empty set $\Os$ at some real time $\tau$,
	then $p$ has $\Batch[j] = (\Os,-)$ at all all real times $\tau' \ge \tau$.
\end{corollary}

\begin{proof}
Suppose $p$ has $\Batch[j] = (\Os,-)$ 
	for some non-empty set $\Os$ at some real time $\tau$.
Since initially $\Batch[j] = (\emptyset,-)$ at $p$,
	$p$ set $\Batch[j]$ to~$(\Os, -)$ by real time $\tau$.
To change $\Batch[j]$ after real time $\tau$, $p$ must set it again.
By Theorem~\ref{finalsafetya}, $p$ can set it only to~$(\Os,-)$.
\qedhere~$_\text{\autoref{finalsafetyac}}$
\end{proof}

\begin{theorem}\label{finalsafetyx}
For all processes $p$ and $p'$, all integers $j \ge 0$, all non-empty sets of operations $\Os$ and~$\Os'$
	and~all real times $\tau$ and $\tau'$:
	if $p$ and $p'$ have $\Batch[j] = (\Os,-)$ and $\Batch[j]=(\Os',-)$
	at times $\tau$ and $\tau'$, respectively, 
	then~$\Os = \Os'$.
\end{theorem}

\begin{proof}
Suppose $p$ and $p'$ have $\Batch[j] = (\Os,-)$ and $\Batch[j]=(\Os',-)$ 
	for some non-empty sets of operations~$\Os$~and~$\Os'$
	at times $\tau$~and~$\tau'$, respectively.
Since $p$ and $p'$ have $\Batch[j] = (\emptyset,\infty)$ initially, and $\Os$ and $\Os'$ are non-empty, 
	$(\Os,-)$ and ($\Os',-)$ are not the initial values of $\Batch[j]$ at $p$ and $p'$ respectively.
So $p$ and $p'$ must set 
	$\Batch[j]$ to $(\Os,-)$ and $(\Os',-)$, by the times~$\tau$~and~$\tau'$, respectively.
By Theorem~\ref{finalsafetya}, $\Os = \Os'$.
\qedhere~$_\text{\autoref{finalsafetyx}}$
\end{proof}

\begin{lemma}\label{DeSica0}
For all $j \ge 0$,
	if a process $p$ calls $\FillGaps{j}$ and this call returns,
	then before this call returns, 
	$p$ set $\Batch[i]$ to $(\Os_i,-)$ for some non-empty set $\Os_i$ for all $i$, $1 \le i \le j$.
\end{lemma}
\begin{proof}
Suppose a process $p$ calls $\FillGaps{j}$ for some $j \geq 0$.
If $j = 0$, then the lemma holds trivially. 
If $j \geq 1$, it is clear from the code of $\FillGaps{}$
	(lines~\ref{totoz8}-\ref{totoz12} and~\ref{gb}-\ref{batch3})
	that if this call returns, 
	$p$ must find $\Batch[i] \neq (\emptyset,\infty)$
	for all $i$, $1 \le i \le j$, 
	before it exits the repeat-until loop of lines~\ref{totoz8}-\ref{totoz11}.
Since the initial value of $\Batch[i]$ is $(\emptyset, \infty)$ for all $i\geq 1$,
	$p$ must set $\Batch[i]=(\Os_i,-)$ for some set $\Os_i$ for all $i$, $1\leq i\leq j$ before the call $\FillGaps{j}$ returns.
By Corollary~\ref{ne-batch}, $\Os_i \neq \emptyset$ for all $i$, $1\leq i\leq j$.
\qedhere~$_\text{\autoref{DeSica0}}$
\end{proof}

\begin{lemma}\label{set-batch-before-doops}
For all $j \geq 1$, if a process $p$ calls $\DoOps{(-,-)}{-}{j}$ at real time $\tau$,
	then for all $i$, $1 \le i < j$, 
	there is a set $\Os_i \neq \emptyset$ such that $p$~sets $\Batch[i]$ to $(\Os_i, -)$ before real time $\tau$.
\end{lemma}
\begin{proof}
The proof is by induction on $j$.
The basis is when $j=1$ and the lemma holds trivially.

For the induction step, consider any integer $j \geq 2$. 
Suppose the lemma holds for $j-1$; we prove that it also holds for $j$.
Suppose process $p$ calls $\DoOps{(-,-)}{-}{j}$ at real time $\tau$.
There are two cases depending on where $p$ calls $\DoOps{(-,-)}{-}{j}$:
\begin{enumerate}
	\item $p$ calls $\DoOps{(-,-)}{-}{j}$ in line~\ref{first-doops} at real time $\tau$.
	\begin{claim}\label{hhh}
	There is a set $\Os_{j-1} \neq \emptyset$ such that $p$ sets $\Batch[j-1]$ to $(\Os_{j-1},-)$ before real time $\tau$.
	\end{claim}
	\begin{proof}
	From the code of lines~\ref{selection}-\ref{first-doops},
		it is clear that $p$ set $(\Ops^*,ts^*,k^*)$ to $(-,-,j)$ in line~\ref{selection} before time $\tau$.
	From the way $p$ set $(\Ops^*,ts^*,k^*)$ to $(-,-,j)$ in line~\ref{selection},
	there are two cases:
	\begin{enumerate}
	\item $(-,-,j)$ is the value of $(\Ops,ts,k)$ at $p$ in line~\ref{selection}.
	Since $j\ge 2$, by Observation~\ref{AcceptAndOps-ts-k}, $p$ previously accepted the tuple $(-,-,j)$.
	By Lemma~\ref{acceptsB}, $p$ set $\Batch[j-1]$ to $(\Os_{j-1}, -)$ for some set $\Os_{j-1}$ before accepting $(-,-,j)$.
	Since $j\ge2$, by Corollary~\ref{ne-batch}, $\Os_{j-1} \neq \emptyset$.
So $p$ sets $\Batch[j-1]$ to $(\Os_{j-1},-)$ for some non-empty set $\Os_{j-1}$ before real time $\tau$.

	\item $(-,-,j)$ is in the set $\REPLIES[t]$ of $p$ in line~\ref{selection}.
From the code of the algorithm concerning $\REPLIES[t]$
	(lines~\ref{est-request}-\ref{est-condition}, \ref{a2b}-\ref{a2e}, and \ref{estrep1}-\ref{estrep3}), 
	it is clear that before executing line~\ref{selection}:
	(i) $p$ received an \linebreak 
	$\langle \EstReply,t,-,-,j, (\Os_{j-1},-) \rangle$ message for some set $\Os_{j-1}$
	in line~\ref{estrep1}, and
	(ii) $p$ set $\Batch[j-1]$ to $(\Os_{j-1},-)$ in line~\ref{setBatch4}.
	Since $j\ge2$, by Corollary~\ref{ne-batch}, $\Os_{j-1} \neq \emptyset$.
So $p$ sets $\Batch[j-1]$ to $(\Os_{j-1},-)$ for some non-empty set $\Os_{j-1} \neq \emptyset$ before real time~$\tau$.
\end{enumerate}
\qedhere~$_\text{\autoref{hhh}}$
\end{proof}

	Then, from the code of lines~\ref{FG}-\ref{first-doops},
		$p$ called $\FillGaps{j-2}$ in line~\ref{FG} before real time $\tau$.
	Thus, by Claim~\ref{hhh} and by Lemma~\ref{DeSica0},
		for all $i$, $1 \leq i < j$, there is a set $\Os_i \neq \emptyset$
		such that $p$ sets $\Batch[i]$ to $(\Os_i,-)$ before real time $\tau$.

	\item $p$ calls $\DoOps{(-,-)}{-}{j}$ in line~\ref{second-doops} at real time $\tau$.
	Suppose this call is of form $\DoOps{(-,-)}{t}{j}$ for some $t$.
	Then, it is clear that $p$ completed a call to $\DoOps{(\Os,-)}{t}{j-1}$ for some set $\Os$ 
		before calling $\DoOps{(-,-)}{t}{j}$, and 
		this $\DoOps{(\Os,-)}{t}{j-1}$ call returned \textsc{Done}.
	By the induction hypothesis, 
		for all $i$, $1 \leq i < j-1$, there is a set $\Os_i \neq \emptyset$ such that $p$ sets $\Batch[i]$ to $(\Os_i,-)$
		before real time $\tau$.
	Since this call to $\DoOps{(\Os,-)}{t}{j-1}$ returns \textsc{Done},
		$p$ set $\Batch[j-1]$ to $(\Os, -)$ in line~\ref{batch1} before it returns \textsc{Done} in line~\ref{done}, 
		which is before real time $\tau$.
		Since $j-1 \geq 1$, by Corollary~\ref{ne-batch}, 
		$\Os \neq \emptyset$.
	Thus, 
	for all $i$, $1 \leq i < j$, there is some set $\Os_i \neq \emptyset$ such that 
		$p$ sets $\Batch[i]$ to $(\Os_i,-)$ before real time $\tau$.
\qedhere~$_\text{\autoref{set-batch-before-doops}}$
 \end{enumerate}
\end{proof}

\begin{lemma}\label{set-batch-before-lease}
For all $j \geq 0$, if a process $p$ has $\lease = (j,-)$ at some real time $\tau$,
	then for all $i$, $1 \le i \leq j$,
	there is a set $\Os_i \neq \emptyset$ such that $p$~sets $\Batch[i]$ to $(\Os_i, -)$ by real time $\tau$.
\end{lemma}
\begin{proof}
Suppose that a process $p$ has $\lease=(j,-)$ at some real time $\tau$.
If $j < 1$, then the lemma holds trivially.
Henceforth we assume $j \geq 1$.
Since $j\geq 1$, $(j, -)$ is not the initial value of variable $\lease$ at $p$,
	so $p$ must have set $\lease$ to $(j,-)$ by real time $\tau$.
There are three cases depending on where $p$ sets $\lease$ to $(j,-)$:
\begin{enumerate}
	\item $p$ sets $\lease$ to $(j,-)$ in line~\ref{setlease}.
	Suppose $p$ executes line~\ref{setlease} in $\LeaderWork{t}$ for some $t$.
	Then, it is clear from the code that $p$ completed at least one call to $\DoOps{(-,-)}{t}{-}$ before executing line~\ref{setlease}, 
		and this $\DoOps{(-,-)}{t}{-}$ returned \textsc{Done} (since $p$ continued to execute line~\ref{setlease}).
	Consider the last $\DoOps{(\Os,-)}{t}{j'}$ that $p$ executed before it sets $\lease$ to $(j,-)$ in line~\ref{setlease}; we claim that $j' = j$.
	To see this, first note that $p$ sets $k=j'$ in line~\ref{leader-accept} in $\DoOps{(\Os,-)}{t}{j'}$. 
	Since (i)~$p$ changes the value of the variable $k$ only in line~\ref{leader-accept} and line~\ref{client-accept},
		(ii)~$p$ does not execute $\PCM()$ concurrently with $\LeaderWork{t}$, and
		(iii)~$\DoOps{(\Os,-)}{t}{j'}$ is the last $\DO$ that $p$ executes before line~\ref{setlease},
		$p$ has $k=j'$ in line~\ref{setlease}.
	Since $p$ sets $\lease$ to $(j,-)$ in line~\ref{setlease}, we have $j = k = j'$.
	Now, since $j\geq 1$, by Lemma~\ref{ne-doops}, $\Os \neq \emptyset$.
	Since this $\DoOps{(\Os,-)}{t}{j}$ returns \textsc{Done},
		$p$ set $\Batch[j]$ to $(\Os,-)$ in line~\ref{batch1} before it returns \textsc{Done} in line~\ref{done}, 
		which is before real time $\tau$ when $p$ sets $\lease$ to $(j,-)$ in line~\ref{setlease}.
	Since $p$ called $\DoOps{(\Os,-)}{t}{j}$ before real time $\tau$,
		by Lemma~\ref{set-batch-before-doops}, 
		for all $i$, $1 \leq i < j$,
		there is a set $\Os_i \neq \emptyset$ such that
		$p$ sets $\Batch[i]$ to $(\Os_i,-)$ by real time $\tau$.

	\item $p$ sets $\lease$ to $(j,-)$ in line~\ref{lease1}.
	The proof for this case is similar to the proof above.
	Suppose $p$ sets $\lease$ to $(j,-)$ in $\DoOps{(\Os,-)}{t}{j}$ for some set $\Os$ and some $t$. 
	Then, when $p$ executes line~\ref{lease1} at real time $\tau$, 
		it also sets $\Batch[j]$ to $(\Os,-)$.
	Since $j\geq 1$, by Lemma~\ref{ne-doops}, $\Os \neq \emptyset$.
	Since $p$ called $\DoOps{(\Os,-)}{t}{j}$ before real time $\tau$,
		by Lemma~\ref{set-batch-before-doops}, 
		for all $i$, $1 \leq i < j$,
		there is a set $\Os_i \neq \emptyset$ such that
		$p$ sets $\Batch[i]$ to $(\Os_i,-)$ by real time $\tau$.

	\item $p$ sets $\lease$ to $(j,-)$ in line~\ref{lg2}.
	It is clear from the code in lines~\ref{batch2}-\ref{lg2} that,
		before setting $\lease$ to $(j,-)$,
		$p$ set $\Batch[j]$ to $(\Os_j,-)$ for some set $\Os_j$ and completed a call to $\FillGaps{j-1}$ .
	Since $j \geq 1$, by Lemma~\ref{ne-batch}, $\Os_j \neq \emptyset$.
	By Lemma~\ref{DeSica0}, $p$~sets $\Batch[i]$ to $(\Os_i, -)$ for some $\Os_i \neq \emptyset$ 
		for all $i$, $1 \le i \le j-1$, 
		before the call to $\FillGaps{j-1}$ returns,
		which is before real time $\tau$.
\end{enumerate}

So, in all cases, if $p$ has $\lease = (j,-)$ at real time $\tau$,
	then for all $i$, $1 \leq i \leq j$,
	there is a set $\Os_i \neq \emptyset$ such that 
	$p$ sets $\Batch[i]$ to $(\Os_i,-)$ by real time $\tau$.
\qedhere~$_\text{\autoref{set-batch-before-lease}}$
\end{proof}

\begin{lemma}\label{aboutEUB}
For all $j \ge 1$, if a process $p$ calls $\EUB{j}$ at some real time $\tau$,
	then, for all~$i$, $1 \le i \le j$, there is a set $\Os_i \neq \emptyset$ such that:
\begin{enumerate}
\item\label{part-I} $p$ sets $\Batch[i]$ to $(\Os_i,-)$ before real time $\tau$, and
\item\label{part-II} $p$ has $\Batch[i] = (\Os_i,-)$ at all real times $\tau' \ge \tau$.
\end{enumerate}
\end{lemma}

\begin{proof}
Suppose $p$ calls $\EUB{j}$ with $j \ge 1$ at real time $\tau$. 
We first show~(\ref{part-I}): for all $i$, $1 \le i \le j$, 
	there is a set $\Os_i \neq \emptyset$ such that
	$p$ sets $\Batch[i]$ to $(\Os_i,-)$ before real time~$\tau$.

Note that $p$ calls $\EUB{j}$ in line~\ref{fill-gaps-to-k-hat}, \ref{ExecuteB2} or \ref{gmb3}.
So we consider three cases:

\begin{enumerate}

\item $p$ calls $\EUB{j}$ in line~\ref{fill-gaps-to-k-hat}.
Thus, $p$ sets $\hat k=j$ in either line~\ref{get-k-hat-then} or lines~\ref{get-k-hat-else}-\ref{get-k-hat-5}.
Suppose that $p$ records $(k^*,-)$ from its variable $\lease$ at real time $\tau^*$ before real time $\tau$
	during the last iteration of the loop 
	of lines~\ref{getvalidlease-start}-\ref{getvalidlease-end}.
Then, by Lemma~\ref{set-batch-before-lease}, 
	for all $i$, $1 \le i \le k^*$, 
	there is a set $\Os_i \neq \emptyset$ such that process $p$ set $\Batch[i]$ to $(\Os_i,-)$ by time $\tau^* < \tau$ (*).
If $p$ sets $\hat k$ in line~\ref{get-k-hat-then}, 
	then it is clear that $j=\hat k \leq k^*$,
	and (\ref{part-I}) follows from (*).
If $p$ sets $\hat k$ in line~\ref{get-k-hat-else},
	then $p$ completed the wait condition $[\mbox{for all } i, k^* < i \le j, \Batch[i] \neq (\emptyset, \infty)]$
	in line~\ref{FG3} before time $\tau$.
Since the initial value of $\Batch[i]$ for $k^* < i \leq j$ is $(\emptyset, \infty)$,
	$p$ set $\Batch[i]$
	before real time~$\tau$.
By Corollary~\ref{ne-batch}, for all i, $k^* < i \le j$
	there is a set $\Os_i\neq \emptyset$ such that
	$p$ set $\Batch[i]=(\Os_i,-)$ before time $\tau$ (**).
So (\ref{part-I}) follows from (*) and (**).

\item $p$ calls $\EUB{j}$ in line~\ref{ExecuteB2}.
This must happen in some $\DoOps{(\Os,-)}{-}{j}$ call.
Then, by Lemma~\ref{set-batch-before-doops},
	by the time $p$ calls $\DoOps{(\Os,-)}{-}{j}$,
	for all $i$, $1 \le i < j$,
	there is a set $\Os_i \neq \emptyset$ such that $p$ sets $\Batch[i]$ to $(\Os_i,-)$.
Since $j \geq 1$ and $p$ sets $\Batch[j]$ to $(\Os,-)$ in line~\ref{batch1}, 
	by Corollary~\ref{ne-batch}, $\Os \neq \emptyset$.
So before real time $\tau$ when $p$ executes line~\ref{ExecuteB2},
	for all $i$, $1 \le i \le j$,
	there is a set $\Os_i \neq \emptyset$ such that $p$ sets $\Batch[i]$ to $(\Os_i,-)$,
	and hence (\ref{part-I}) holds.

\item $p$ calls $\EUB{j}$ in line~\ref{gmb3}.
Before doing so, $p$ set $\Batch[j]=(\Os_j,-)$ for some set $\Os_j$ in line~\ref{batch2},
	and $p$ executed $\FillGaps{j-1}$ in line~\ref{fill-lease-gap}.
	Since $j \ge 1$, by Corollary~\ref{ne-batch}, $\Os_j \neq \emptyset$.
By Lemma~\ref{DeSica0},
	$p$ set $\Batch[i]=(\Os_i,-)$ for some set $\Os_i \neq \emptyset$ for all $i$, $1 \le i \le j-1$, before it returns from $\FillGaps{j-1}$.
Thus, for all $i$, $1 \le i \le j$, there is a set $\Os_i\neq \emptyset$ such that
	process $p$ set $\Batch[i]$ to $(\Os_i,-)$ before real time $\tau$,
	and hence (\ref{part-I}) holds.
\end{enumerate}

Since
	for all $i$, $1 \le i \le j$,
	process $p$ sets $\Batch[i]$ to $(\Os_i,-)$ for some $\Os_i \neq \emptyset$ before time $\tau$,
by Corollary~\ref{finalsafetyac},
	for all~$i$, $1 \le i \le j$,
	process $p$ has $\Batch[i] = (\Os_i,-)$ at all real times $\tau' \ge \tau$,
	and hence~(\ref{part-II}) holds as well.
\qedhere~$_\text{\autoref{aboutEUB}}$
\end{proof}

\begin{lemma}\label{aboutEB}\label{EB1}\label{before-EB-1}
For all $j \ge 1$, if a process $p$ calls $\ExecuteBatch{j}$ at some real time $\tau$,
	then there is a set $\Os_j \neq \emptyset$ such that:
\begin{enumerate}
\item $p$ sets $\Batch[j]$ to $(\Os_j, -)$ before real time $\tau$, and
\item $p$ has $\Batch[j] = (\Os_j, -)$ at all real times $\tau' \ge \tau$.
\end{enumerate}
\end{lemma}

\begin{proof}
Suppose a process $p$ calls $\ExecuteBatch{j}$ for some $j \ge 1$ at some real time $\tau$.
This happens when process $p$ calls $\ExecuteBatch{j}$ in line~\ref{EUTB2} of $\ExecuteOpsUpToBatch{h}$ with $h \ge j$.
Since $p$ called $\ExecuteOpsUpToBatch{h}$ before calling
	$\ExecuteBatch{j}$ at real time $\tau$, 
	by Lemma~\ref{aboutEUB},
	there is a non-empty set $\Os_j$ such that $p$ sets $\Batch[j]$ to $(\Os_j,-)$ before real time $\tau$,
	and $p$ has $\Batch[j] = (\Os_j,-)$ at all real times $\tau' \ge \tau$.
\qedhere~$_\text{\autoref{aboutEB}}$
\end{proof}

Since $\MBD$ is initialized to $0$,
	and a process updates $\MBD$ only by executing
	the statement $\MBD := \max (\MBD, j)$
	in line~\ref{updateMBD1}, we have:
	
\begin{observation}\label{MBD-O1}
At every process $p$, $\MBD \ge 0$ and $\MBD$ is non-decreasing.
\end{observation}

\begin{observation}\label{MBD-O2}
For all $j \ge 0$, after a process $p$ executes $\EUB{j}$,
	or after $p$ executes $\DoOps{(-,-)}{-}{j}$ and this execution returns \textsc{Done},
	$p$ has $\MBD \ge j$.
\end{observation}

\begin{lemma}\label{MBD-L}
For all $j \ge 1$, if a process $p$ has $\MBD = j$,
	then the following events previously occurred at $p$.
For all $i$, $1 \le i \le j$:

\begin{enumerate}
\item $p$ sets $\Batch[i]$ to $(\Os_i,-)$ for some non-empty set $\Os_i$,
\item $p$ executes $\ExecuteBatch{i}$, and
\item $p$ executes $\OpsDone := \OpsDone \cup \Os_i$,
\end{enumerate}
in this order.
\end{lemma}

\begin{proof}
First note that $p$ modifies the variable $\MBD$ only by executing the statement
	 $\MBD := \max (\MBD, i)$
	 in line~\ref{updateMBD1} of an $\EUB{}$ that 
	 $p$ calls in line~\ref{fill-gaps-to-k-hat}, \ref{ExecuteB2} or \ref{gmb3}.

We now prove the lemma by induction on $j$.
For the base case, let $j =1$, and consider the first time that $p$ sets $\MBD$ to $1$.
By Observation~\ref{MBD-O1}, before this occurs $p$ has $\MBD = 0$.
So $p$ sets $\MBD$ to $1$ by executing the statement $\MBD := \max (0, 1)$.
	 
This occurs in an execution of
	$\EUB{h}$ for some $h \ge 1$
	(because $p$ does not do anything in $\EUB{h}$ if $h<1$).
Note that before executing $\MBD := \max (0, 1)$ in line~\ref{updateMBD1} of $\EUB{h}$,
	$p$ does the following in the \emph{first} iteration of the for loop of $\EUB{h}$:

\begin{enumerate}
\item $p$ executes $\ExecuteBatch{1}$ in line~\ref{EUTB2}, and
\item $p$ executes $\OpsDone := \OpsDone \cup \BatchOps{1}$ in line~\ref{addB1}.
\end{enumerate}
By Lemma~\ref{aboutEB}, $p$ sets $\Batch[1]$ to $(\Os_1,-)$ for some non-empty set $\Os_1$ before
	executing $\ExecuteBatch{1}$ in line~\ref{EUTB2},
	and $p$ has $\BatchOps{1} = \Os_1$ in line~\ref{addB1}.

The above shows that the lemma holds for the base case of $j=1$.

For the induction step, suppose the lemma holds for every $i$, $1 \le i \le j$;
	we now prove that it also holds for $i=j+1$.
Consider the first time that $p$ sets $\MBD$ to $j+1$,
	and suppose this occurs at real time $\tau$.
By Observation~\ref{MBD-O1}, $p$ has $0 \le \MBD \le j$ before real time $\tau$.
So, at real time $\tau$, $p$ sets $\MBD$ to $j+1$
	by executing the statement $\MBD := \max (\MBD, j+1)$, where $0 \le \MBD \le j$.

This must occur in an execution of $\EUB{h}$ for some $h \ge j+1$
	(because if $h<j+1$ then $p$ does not execute $\MBD := \max (\MBD, j+1)$ in $\EUB{h}$).

Let $h'$ be the value of $\MBD$ when $p$ calls $\EUB{h}$.
Since this call occurs before real time $\tau$, from Observation~\ref{MBD-O1},
	$0\le h' \le j$.
Thus, by the induction hypothesis,\footnote{For $h' = 0$, the statement that follows is trivially true;
	we use the induction hypothesis only
	for the case that $1 \le h' \le j$.}
	the following events occurred before $p$ called $\EUB{h}$.
For all~$i$, $1 \le i \le h'$:
\begin{enumerate}
\item $p$ set $\Batch[i]$ to $(\Os_i,-)$ for some non-empty set $\Os_i$,
\item $p$ executed $\ExecuteBatch{i}$, and
\item $p$ executed $\OpsDone := \OpsDone \cup \Os_i$,
\end{enumerate}
in this order.

Since $p$ has $\MBD = h'$ when $p$ calls $\EUB{h}$ with $h \ge j+1$,
	from the code of $\EUB{h}$, 
	the following events occur at~$p$ before $p$
	executes the statement $\MBD := \max (\MBD, j+1)$ in line~\ref{updateMBD1}.
For all $i$, $h'+1 \le i \le j+1$:
	
\begin{enumerate}
\item $p$ executes $\ExecuteBatch{i}$ in line~\ref{EUTB2}, and
\item $p$ executes $\OpsDone := \OpsDone \cup \BatchOps{i}$ in line~\ref{addB1}.
\end{enumerate}
Note that by Lemma~\ref{aboutEB}, $p$ sets $\Batch[i]$ to $(\Os_i,-)$ for some non-empty set $\Os_i$ before
	executing $\ExecuteBatch{i}$ in line~\ref{EUTB2},
	and $p$ has $\BatchOps{i} = \Os_i$ when it executes $\OpsDone := \OpsDone \cup \BatchOps{i}$ in line~\ref{addB1}.
Thus, the following events occur at $p$ before $p$ first sets $\MBD$ to $j+1$ at real time~$\tau$.
For all $i$, $1 \le i \le j+1$:

\begin{enumerate}
\item $p$ sets $\Batch[i]$ to $(\Os_i,-)$ for some non-empty set $\Os_i$,
\item $p$ executes $\ExecuteBatch{i}$, and
\item $p$ executes $\OpsDone := \OpsDone \cup \Os_i$,
\end{enumerate}

in this order.
\qedhere~$_\text{\autoref{MBD-L}}$
\end{proof}

Observation~\ref{MBD-O2} and Lemma~\ref{MBD-L} immediately imply the following:

\begin{corollary}\label{MBD-L-C}
For all $j \ge 1$, if a process $p$ returns from $\EUB{j}$,
	or it returns from $\DoOps{(-,-)}{-}{j}$ with a \textsc{Done},
	then the following events previously occurred at $p$.
For all $i$, $1 \le i \le j$:

\begin{enumerate}
\item $p$ sets $\Batch[i]$ to $(\Os_i,-)$ for some non-empty set $\Os_i$,
\item $p$ executes $\ExecuteBatch{i}$, and
\item $p$ executes $\OpsDone := \OpsDone \cup \Os_i$,
\end{enumerate}
in this order.
\end{corollary}

\begin{lemma}\label{MBD3}
For all $j \ge 2$, if a process calls $\ExecuteBatch{j}$ then it has $\MBD \ge j-1$ before this call.
\end{lemma}

\begin{proof}
Suppose a process $p$ calls $\ExecuteBatch{j}$ for $j \ge 2$.
Then it does so in line~\ref{EUTB2} of an \linebreak$\EUB{h}$, for some $h \ge j$.
From the code of the for loop in lines~\ref{EUTB}--\ref{updateMBD1} of $\EUB{h}$,
	it is clear that $p$ has $\MBD = j-1$ just before it executes $\ExecuteBatch{j}$ in this loop.
So in all cases, $p$ has $\MBD \ge j-1$
	before it calls $\ExecuteBatch{j}$.
\qedhere~$_\text{\autoref{MBD3}}$
\end{proof}

Lemmas~\ref{MBD-L} and~\ref{MBD3} immediately imply the following:

\begin{corollary}\label{EB2}\label{before-EB-2}
For all $j \ge 2$, if a process calls $\ExecuteBatch{j}$ then it previously completed a call to $\ExecuteBatch{j-1}$.
\end{corollary}

\begin{lemma}\label{exactly-once-simple-doops}
Suppose a process $p$ calls $\DoOps{(\Os,-)}{t}{j}$ and $\DoOps{(\Os',-)}{t}{j'}$.
If $j' \neq j$ then $\Os' \cap \Os = \emptyset$.
\end{lemma}
\begin{proof}
Suppose $p$ calls $\DoOps{(\Os,-)}{t}{j}$ and $\DoOps{(\Os',-)}{t}{j'}$.
Assume, without loss of generality, that 
	$p$ calls $\DoOps{(\Os,-)}{t}{j}$ before calling $\DoOps{(\Os',-)}{t}{j'}$.
If $j=0$,
	then by Lemma~\ref{ne-doops},
	$\Os = \emptyset$, and hence $\Os \cap \Os' = \emptyset$.
Henceforth we assume that $j \geq 1$.
Since $p$ continues to call $\DoOps{(\Os',-)}{t}{j'}$ after $\DoOps{(\Os,-)}{t}{j}$,
	the call to $\DoOps{(\Os,-)}{t}{j}$ returns \textsc{Done}.
By Corollary~\ref{MBD-L-C},
	by the time when $p$ returns from $\DoOps{(\Os,-)}{t}{j}$,
	it set $\Batch[j]$ to $(\Os_j,-)$ for some non-empty set $\Os_j$
	and it executed $\OpsDone := \OpsDone \cup \Os_j$.
Since $p$ sets $\Batch[j]$ to $(\Os, -)$ in line~\ref{batch1} of $\DoOps{(\Os,-)}{t}{j}$, 
	by Theorem~\ref{finalsafetya}, $\Os_j = \Os$.
So, by the monotonicity of $\OpsDone$ (Observation~\ref{opsdoneismonotonic}),
	when $p$ computes $\NextOps := \OpsRequested - \OpsDone$ in line~\ref{nextops}
	(just before executing $\DoOps{(\Os',-)}{t}{j'}$ with $\Os' = \NextOps$ in line~\ref{second-doops})
	we have $\Os \subseteq \OpsDone$.
Therefore $\NextOps \cap \Os = \emptyset$, i.e., \mbox{$\Os' \cap \Os = \emptyset$}.
\qedhere~$_\text{\autoref{exactly-once-simple-doops}}$
\end{proof}

\begin{lemma}\label{exactly-once-simple-accept}
Suppose tuples $(\Os,t,j)$ and $(\Os',t,j')$ are accepted.
If $j' \neq j$\Ch 
	then $\Os' \cap \Os = \emptyset$.\
\end{lemma}
\begin{proof}
Suppose $(\Os,t,j)$ and $(\Os',t,j')$ are accepted.
By Observation~\ref{Accept-Doops}
	some process $p$ called $\DoOps{(\Os,-)}{t}{j}$ and some process $q$ called $\DoOps{(\Os',-)}{t}{j'}$.
By Observation~\ref{DoopsCalls}, $p$ and $q$ did so in $\LeaderWork{t}$.
By Observation~\ref{LeaderWork1}, $p$ and $q$ became leaders at time $t$,
	so both called $\ML{t}{t}$ and this call returned \textsc{True}.
By Theorem~\ref{leader-safety}, $p=q$.
The result now follows from Lemma~\ref{exactly-once-simple-doops}.
\qedhere~$_\text{\autoref{exactly-once-simple-accept}}$
\end{proof}

\begin{theorem}\label{generalcase1b}
Suppose a tuple $(\Os,t,j)$ is locked.
For all $t'>t$, if a tuple $(\Os',t',j')$ with $j' \neq j$\Ch 
	is accepted then $\Os' \cap \Os = \emptyset$.
\end{theorem}

\begin{proof}
The proof is by contradiction.
Suppose that some $(\Os,t,j)$ is locked, and:

\smallskip\noindent
(*) there is a $t'>t$ and a tuple $(\Os',t',j')$ with $j' \neq j$\Ch 
	that is accepted but $\Os' \cap \Os \neq \emptyset$.

Without loss of generality, assume that $t'$ is the smallest $t'>t$
	for which there is a ``bad'' accepted tuple $(\Os',t',j')$.
From this assumption, we have:

\smallskip\noindent
(**) for all $\hat{t}$ such that $t <\hat{t}<t'$, if a tuple $(\widehat{\Os},\hat{t},\hat{j})$
	with $\hat{j} \neq j$\Ch 
	is accepted then $\widehat{\Os} \cap \Os = \emptyset$.

Consider the accepted tuple $(\Os',t',j')$.
By Lemma~\ref{leaderfirst}, the first process that accepts $(\Os',t',j')$
	is a process $\ell$ that becomes leader at local time~$t'$ and accepts $(\Os',t',j')$ in $\LeaderWork{t'}$.
Since  $(\Os,t,j)$ is locked and $t' >t$, by Theorem~\ref{lockingcore},
	process $\ell$ selected a tuple $(\Ops^*,ts^*,k^*)$ in line~\ref{selection} of $\LeaderWork{t'}$ such that
	$(ts^*,k^*) \ge (t,j)$ and some process $q^*$ previously accepted $(\Ops^*,ts^*,k^*)$.

After selecting $(\Ops^*,ts^*,k^*)$ in line~\ref{selection},
	process $\ell$ first verified that $ts^* < t'$ in line~\ref{Egathercrumbs},
	and then
	$\ell$ executed lines~\ref{FG}-\ref{first-doops}.
	In particular, $\ell$ called $\DoOps{(\Ops^*,0)}{t'}{k^*}$ in line~\ref{first-doops}
	and $\ell$ accepted $(\Ops^*,t',k^*)$ during this execution.
Note that $(\Ops^*,t',k^*)$ is the first tuple of the form $(-,t',-)$ that $\ell$ accepts.\footnote{The tuples $(\Ops^*,t',k^*)$ and $(\Os',t',j')$ that $\ell$ accepts are not necessarily distinct.}

\begin{claim}\label{about-star-tple2}
If $k^* \neq j$\Ch 
	then $\Ops^* \cap \Os = \emptyset$.
\end{claim}

\begin{proof}
Since $(ts^*,k^*) \ge (t,j)$, we have $ts^* \ge t$.
There are two possible cases:
\begin{enumerate}
\item $ts^* = t$.
So $(\Ops^*,ts^*,k^*)$ is $(\Ops^*,t,k^*)$.
Since both $(\Os,t,j)$ and $(\Ops^*,t,k^*)$ are accepted, by Lemma~\ref{exactly-once-simple-accept},
	if $k^* \neq j$,\Ch 
	then $\Ops^* \cap \Os = \emptyset$.

\item $ts^* > t$.
Recall that $ts^* < t'$.
Since $ t < ts^* < t'$, and $(\Ops^*,ts^*,k^*)$ was accepted by some process,
	by~(**)
	we have
	if $k^* \neq j$,\Ch 
	then $\Ops^* \cap \Os = \emptyset$.
\end{enumerate}

So in all possible cases, the claim holds.
\qedhere~$_\text{\autoref{about-star-tple2}}$
\end{proof}

Now consider $(\Os',t',j')$.
Recall that $(\Ops^*,t',k^*)$ is the \emph{first} tuple of the form $(-,t',-)$ that $\ell$ accepts.
Since $\ell$ accepts $(\Os',t',j')$,
	there are two possible cases:

\begin{enumerate}
\item $(\Os',t',j')$ is $(\Ops^*,t',k^*)$.
So $\Ops^* = \Os'$, and $k^* = j'$.
By Claim~\ref{about-star-tple2}, if $j' \neq j$\Ch 
	then $\Os' \cap \Os = \emptyset$.
	
\item $\ell$ accepts $(\Ops^*,t',k^*)$ before it accepts $(\Os',t',j')$.
By Lemma~\ref{T-k-increase}, $j' > k^*$.
Since $(\Os,t,j)$ is locked and $(\Ops^*,t',k^*)$ is accepted and $t'>t$,
	By Theorem~\ref{generalcase1a}(\ref{t1}),
	$k^* \ge j$.
Thus, $j' > j$ (so $j' \neq j$).\Ch
We now show that $\Os' \cap \Os = \emptyset$.

Suppose first that $j=0$. 
In this case, by Corollary~\ref{ne-lock} $\Os = \emptyset$, and so $\Os' \cap \Os = \emptyset$ is obvious.
Henceforth we assume that $1 \le j$, and so we have $1 \le j \le k^*$.

Since $(\Os',t',j')$ is not the first tuple that $\ell$ accepts, by Lemma~\ref{leader-accepts},
	$\ell$ accepts $(\Os',t',j')$ during its execution of $\DoOps{(\Os',-)}{t'}{j'}$ in line~\ref{second-doops}
	(in the while loop of lines~\ref{mainwhile}-\ref{endwhile}).

\begin{claim}\label{PrevBatches}
$\ell$ has $\Os \subseteq \OpsDone$
	before executing the while loop of lines~\ref{mainwhile}-\ref{endwhile}.
\end{claim}

\begin{proof}
First note that since $(\Os,t,j)$ is locked, by Observation~\ref{lock-batch},
	some process $p$ sets $\Batch[j]$ to~$(\Os,-)$.
Before $\ell$ executes the while loop of lines~\ref{mainwhile}-\ref{endwhile}, 
	$\ell$ calls $\DoOps{\Ops^*}{t'}{k^*}$ 
	in line~\ref{first-doops},
	and this call returns $\textsc{Done}$
	(because $\ell$ later executes $\DoOps{(\Os',-)}{t'}{j'}$
	in the while loop of lines~\ref{mainwhile}-\ref{endwhile}).
Let $\tau$ be the real time when $\DoOps{(\Ops^*, -)}{t'}{k^*}$ returns $\textsc{Done}$.
By Corollary~\ref{MBD-L-C},
	for all $i$, $1 \le i \le k^*$,
	$\ell$ sets $\Batch[i]$ to $(\Os_i,-)$ for some non-empty set $\Os_i$ and
	then it executes $\OpsDone := \OpsDone \cup \Os_i$,
	by real time $\tau$.
Since $1 \le j \le k^*$ and the set $\OpsDone$ is non-decreasing
	(Observation~\ref{opsdoneismonotonic}),
	$\OpsDone$ contains $\Os_j$ by real time $\tau$ (and all real times thereafter).	
Since $p$ and $\ell$ set $\Batch[j]$ to $(\Os,-)$ and $(\Os_j,-)$, respectively,
	by Theorem~\ref{finalsafetya}, $\Os = \Os_j$.
So $\ell$ has $\Os \subseteq \OpsDone$ by real time $\tau$,
	i.e., before it executes the while loop of lines~\ref{mainwhile}-\ref{endwhile}.
\qedhere~$_\text{\autoref{PrevBatches}}$
\end{proof}

Note that just before calling $\DoOps{(\Os',-)}{t'}{j'}$ in line~\ref{second-doops},
	$\ell$ computes $\NextOps := \OpsRequested - \OpsDone$ in line~\ref{nextops},
	and $\Os'$ is the resulting $\NextOps$.
By Claim~\ref{PrevBatches} and the monotonicity of $\OpsDone$ (Observation~\ref{opsdoneismonotonic}),
	$\Os \subseteq \OpsDone$ in line~\ref{nextops},
	so $\NextOps \cap \Os = \emptyset$, i.e., $\Os' \cap \Os = \emptyset$.

\end{enumerate}
So in all cases we have if $j' \neq j$\Ch 
	then $\Os' \cap \Os = \emptyset$.
This contradicts the assumption~(*) about $(\Os',t',j')$.
\qedhere~$_\text{\autoref{generalcase1b}}$
\end{proof}

\begin{theorem}\label{intersect1}
If tuples $(\Os,t,j)$ and $(\Os',t',j')$ are locked and $j' \neq j$, then $\Os' \cap \Os = \emptyset$.
\end{theorem}

\begin{proof}
Suppose $(\Os,t,j)$ and $(\Os',t',j')$ are locked and $j' \neq j$.
By Observation~\ref{acceptedfirst}, tuples $(\Os,t,j)$ and $(\Os',t',j')$ are also accepted.
If $t' =t$ then, by Lemma~\ref{specialcase}, $\Os' \cap \Os = \emptyset$.
If $t' >t$ or $t > t'$ then, by Theorem~\ref{generalcase1b}, $\Os' \cap \Os = \emptyset$.
So in all cases $\Os' \cap \Os = \emptyset$.
\qedhere~$_\text{\autoref{intersect1}}$
\end{proof}

\begin{theorem}\label{finalsafetyb}
For all $j, j' \ge 0$, suppose processes $p$ and $p'$ set $\Batch[j]$ and $\Batch[j']$ to $(\Os,-)$ and $(\Os',-)$, respectively.
If $j' \neq j$ then $\Os' \cap \Os = \emptyset$.
\end{theorem}

\begin{proof}
Suppose $p$ and $p'$ set $\Batch[j]$ and $\Batch[j']$ to $(\Os,-)$ and $(\Os',-)$, respectively, with $j' \neq j$.
By Lemma~\ref{IamOutOfLableNames}, there are local times $t$ and $t'$
	such that $(\Os,t,j)$ and $(\Os',t',j')$ are locked.
By Theorem~\ref{intersect1}, $\Os' \cap \Os = \emptyset$.
\qedhere~$_\text{\autoref{finalsafetyb}}$
\end{proof}

\subsubsection{Each batch is recorded by a majority}
\begin{lemma}\label{memory}
For all $j \ge 1$,
	if a process sets $\Batch[j]$ to some pair $(\Os_{j},-)$ at some real time $\tau$,
	then more than $n/2$ processes set $\Batch[j-1]$ to $(\Os_{j-1},-)$ for some set $\Os_{j-1}$ before real time $\tau$.
\end{lemma}

\begin{proof}
Let $j \ge 1$ and suppose a process sets $\Batch[j]$ to some pair $(\Os_{j},-)$ at some real time $\tau$.
By Lemma~\ref{IamOutOfLableNames}, some process $p$ locks a tuple $(\Os_j,t,j)$ for some $t$ by real time $\tau$.
Note that $p$ did so in line~\ref{batch1} of $\DoOps{(\Os_j,-)}{t}{j}$,
	and that $p$ previously accepted $(\Os_j,t,j)$ in line~\ref{leader-accept} of that $\DoOps{(\Os_j,-)}{t}{j}$.
Since $j\ge1$, by Lemma~\ref{acceptsB},
	$p$ set $\Batch[j-1]$ to $(\Os_{j-1},-)$ for some set $\Os_{j-1}$ 
	before accepting $(\Os_j,t,j)$ in line~\ref{leader-accept}.
We claim that after setting $\Batch[j-1]$ to $(\Os_{j-1},-)$, 
	process $p$ has $\Batch[j-1]$ of form $(\Os_{j-1},-)$ forever.
To see this, note that:
	(1) if $j-1 = 0$, by Corollary~\ref{ne-batch}, $\Os_{j-1} =\emptyset$, and if $p$ later sets $\Batch[j-1]$, then it sets $\Batch[j-1]$ to $(\emptyset,-)$. 
	So $p$ has $\Batch[j-1]$ of form $(\Os_{j-1}, -)$ forever after setting $\Batch[j-1]$ to $(\Os_{j-1}, -)$; 
	and
	(2) if $j-1 > 0$, by Corollary~\ref{ne-batch}, $\Os_{j-1} \neq \emptyset$,
	and, by Corollary~\ref{finalsafetyac},
	$p$ has $\Batch[j-1] = (\Os_{j-1},-)$ forever after setting $\Batch[j-1]$ to $(\Os_{j-1},-)$.
	
After accepting $(\Os_j,t,j)$ in line~\ref{leader-accept},
	$p$ sent $\langle \Prepare,(\Os_j,-),t,j, \Batch[j-1] \rangle$ messages
	to all processes $q \neq p$ in lines~\ref{repeat}-\ref{condition},
	and it found $| \PACKED[t,j] | \ge  \lfloor n/2 \rfloor$ in line~\ref{ack-condition}.
Since $p$ set $\Batch[j-1]$ to $(\Os_{j-1},-)$ before accepting $(\Os_j,t,j)$ in line~\ref{leader-accept},
	by the above claim
	these $\Prepare$ messages have $\Batch[j-1] = (\Os_{j-1},-)$.
From the code of the algorithm concerning $\PACKED[t,j]$
	(lines~\ref{a3b}-\ref{sendPack} and lines~\ref{Rack1}-\ref{Rack2}),
	at least $\lfloor n/2 \rfloor$ processes different than $p$ executed the following events
	before $p$ found $| \PACKED[t,j] | \ge  \lfloor n/2 \rfloor$:
	(1)~they received the $\langle \Prepare,(\Os_j,-),t,j, (\Os_{j-1},-) \rangle$ message from $p$
	in line~\ref{a3b},
	(2)~they set their variable $\Batch[j-1]$ to $(\Os_{j-1},-)$ in line~\ref{setBatch3},
	and
	(3)~they sent a $\langle \PACK,t,j \rangle$ to $p$ in line~\ref{sendPack}.
Since $p$ also sets $\Batch[j-1]$ to $(\Os_{j-1},-)$,
	a total of more than $n/2$ processes set their $\Batch[j-1]$ to $(\Os_{j-1},-)$;
	note that they all do so before $p$ locks $(\Os_j,t,j)$ in line~\ref{batch1} of $\DoOps{(\Os_j,-)}{t}{j}$.
Thus, more than $n/2$ processes set $\Batch[j-1]$ to $(\Os_{j-1},-)$ before real time $\tau$.
\qedhere~$_\text{\autoref{memory}}$
\end{proof}

By Lemma~\ref{memory} and induction we have:

\begin{corollary}\label{successivebatches2}
For all $j \ge 1$,
	if a process sets $\Batch[j]$ to some pair $(\Os_{j},-)$ at some real time $\tau$,
	then for all~$i$, $0 \le i \le j-1$,
	more than $n/2$ processes set $\Batch[i]$ to $(\Os_{i},-)$ for some set $\Os_i$ before real time $\tau$.
\end{corollary}

\begin{theorem}\label{memory2}
For all $j \ge 2$,
	if a process accepts a tuple $(-,-,j)$ at some real time $\tau$,
	then for all~$i$, $0 \le i \le j-2$,
	more than $n/2$ processes set $\Batch[i]$ to $(\Os_{i},-)$ for some set $\Os_i$ before real time $\tau$.
\end{theorem}

\begin{proof}
Let $j \ge 2$, and suppose that some process accepts a tuple $(-,-,j)$ at some real time $\tau$.
By Lemma~\ref{acceptsB}, some process set $\Batch[j-1]$ to some pair $(\Os_{j-1},-)$, before real time $\tau$.
Since $j-1 \ge 1$, by Corollary~\ref{successivebatches2},
	for all~$i$, $0 \le i \le j-2$,
	more than $n/2$ processes set $\Batch[i]$ to $(\Os_{i},-)$ for some set $\Os_i$ before real time~$\tau$.
\qedhere~$_\text{\autoref{memory2}}$
\end{proof}

\subsection{Consensus mechanism: liveness properties}\label{cons-liveness}

Recall that $\ell$ is the process that becomes leader after local time $c_0$
	(see Theorem~\ref{leader-liveness-safety1} in Section~\ref{leader-election}).

\begin{lemma}\label{lm6.1}
For all $t' \ge t$:

\begin{enumerate}

\item\label{leader1} If $\ell$ calls $\ML{t}{t'}$ with $t \ge \Tz$,
	then this call returns \textsc{True}.
\item\label{leader2} If a process $q \neq \ell$ calls $\ML{t}{t'}$ with $t' \ge \Tz$,
	and this call returns,
	then it returns \textsc{False}.
\end{enumerate}

\end{lemma}

\begin{proof}
In our 
	algorithm,
	it is clear that if a process calls $\ML{t}{t'}$ at some local time $t''$, then $t'' \ge t' \ge t$.
The lemma now follows directly from Theorem~\ref{leader-liveness-safety1}.
\qedhere~$_\text{\autoref{lm6.1}}$
\end{proof}

\begin{assumption}\label{finite-LP}
The parameter $\LP=\Pone$ is positive and finite.
\end{assumption}
\begin{assumption}\label{promiseRange}
The parameter $\PP=\Pthree$ is non-negative and finite.
\end{assumption}

From these assumptions it follows that:

\begin{observation}\label{finite-first-lw-wait}
No correct process waits forever in line~\ref{wait-lease-expire}.
\end{observation}

\begin{lemma}\label{lm1w}
No process $q \neq \ell$ executes the loop of lines~\ref{est-request}-\ref{est-condition} forever.
\end{lemma}

\begin{proof}
Suppose, for contradiction, that a process $q \neq \ell$ executes the loop of lines~\ref{est-request}-\ref{est-condition} forever.
Suppose that this occurs when $q$ executes $\LeaderWork{t}$, so $q$ became leader at local time $t$.
Since $q$ executes the loop of lines~\ref{est-request}-\ref{est-condition} forever,
	there is a real time after which $q$ has $\CT \geq \Tz$
	(Assumptions~\ref{xclocks}(\ref{xcl2}) and (\ref{xcl3}))
	and $q$ calls $\ML{t}{\CT}$ in line~\ref{est-condition} of this loop.
By Lemma~\ref{lm6.1}(\ref{leader2}), this call returns \textsc{False}, and so $q$ exits the loop --- a contradiction.
\qedhere~$_\text{\autoref{lm1w}}$
\end{proof}

\begin{theorem}\label{DeSica}
For all $j \ge 0$,
	if for all $i$, $1 \le i \le j$,
	more than $n/2$ processes have $\BatchOps{i} \neq \emptyset$ at some real time $\tau$,
	and
	a correct process $p$ calls $\FillGaps{j}$ at some real time $\tau' \ge \tau$,
	then:
\begin{enumerate}
\item\label{Yo1} $p$ eventually returns from $\FillGaps{j}$, and
\item\label{Yo2} when $p$ returns from $\FillGaps{j}$ and thereafter,
	for all $i$, $1 \le i \le j$, $\BatchOps{i} \neq \emptyset$ at $p$.
\end{enumerate}
\end{theorem}

\begin{proof}
Let $j \ge 0$ be such that for all $i$, $1 \le i \le j$,
	more than $n/2$ processes have $\BatchOps{i} \neq \emptyset$ at some real time $\tau$.
Thus, for every~$i$, $1 \le i \le j$,
	at least one \emph{correct} process $q_i$ has $\BatchOps{i} \neq \emptyset$ at real time $\tau$;
	by Corollary~\ref{finalsafetyac}, $q_i$ has $\BatchOps{i} \neq \emptyset$ from time $\tau$ on.
	
Suppose a correct process $p$ calls $\FillGaps{j}$ at some real time $\tau' \ge \tau$.
Consider any $i$, $1 \le i \le j$,
	such that $p$ has $\Batch[i] = (\emptyset,\infty)$ when $p$ calls $\FillGaps{j}$.
From the above, some correct process $q_i \neq p$ has $\BatchOps{i} \neq \emptyset$ from real time $\tau$ on.
From lines~\ref{totoz9}-\ref{totoz11}
	and lines~\ref{gb}-\ref{ge} of the algorithm,
	and since the communication link between correct processes $p$ and $q_i$ is fair
	(Assumption~\ref{fl}),
	it is clear that	
	$p$ eventually receives a $\langle \MyBatch, i, \BB \rangle$
	message with $\BB.ops \neq \emptyset$ in line~\ref{batchreceipt}
	from some process, and $p$ then sets $\Batch[i] = \BB$ in line~\ref{batch3}.
By Corollary~\ref{ne-batch}, 
	$\BatchOps{i}$ remains not equal to $\emptyset$ thereafter.
Thus the set $\Gaps := \{ i ~|~ 1 \le i \le j \mbox{ and } \Batch[i] = (\emptyset,\infty) \}$ at $p$
	is eventually empty.
Since $(\emptyset,\infty)$ is the the initial value of $\Batch[i]$ at $p$ for all $i$, $1 \le i \le j$,
	$p$ must previously set $\Batch[i]$ for all $i$, $1 \le i \le j$.
By Corollary~\ref{ne-batch} and Corollary~\ref{finalsafetyac}, 
	$p$ has $\BatchOps{i} \neq \emptyset$ thereafter.
So $p$'s call to $\FillGaps{j}$ returns, and when it does and thereafter, 
	we have that 
	for all $i$, $1 \le i \le j$, $\BatchOps{i} \neq \emptyset$ at~$p$.
\qedhere~$_\text{\autoref{DeSica}}$
\end{proof}

\begin{lemma}\label{lm2}
If a correct process calls $\FillGaps{k^* -2}$
	in line~\ref{FG},
	then this call
	returns.
\end{lemma}

\begin{proof}
Suppose a correct process $p$ calls $\FillGaps{k^* -2}$
	in line~\ref{FG}.
First note that if $k^* \leq 2$, then from the code of $\FillGaps{}$
	it is easy to see that this call immediately returns.
Henceforth assume that $k^* > 2$.
So $p$ has $(\Ops^*,ts^*,k^*) \neq (\emptyset,-1,0)$ in line~\ref{selection}
	(before calling $\FillGaps{k^* -2}$
	in line~\ref{FG}).
Thus, from Lemma~\ref{lmx3},
	some process $q$ \emph{previously} accepted some tuple $(-,-,k^*)$.
So, by Theorem~\ref{memory2},
	for all~$i$, $1 \le i \le k^*-2$,
	more than $n/2$ processes set $\Batch[i]$ to $(\Os_{i},-)$ for some set $\Os_{i}$ before $q$ accepted $(-,-,k^*)$,
	and so before $p$ calls $\FillGaps{k^* -2}$
	in line~\ref{FG}.
By Corollary~\ref{ne-batch}, for all~$i$, $1 \le i \le k^*-2$, $\Os_{i} \neq \emptyset$.
Thus, by Corollary~\ref{finalsafetyac}, when $p$ calls $\FillGaps{k^* -2}$
	in line~\ref{FG} the following holds:
	for all~$i$, $1 \le i \le k^*-2$,
	more than $n/2$ processes have $\Batch = (\Os_{i},-)$ for some $\Os_{i} \neq \emptyset$.
By Theorem~\ref{DeSica}(\ref{Yo1}), this call returns.
 \qedhere~$_\text{\autoref{lm2}}$
\end{proof}

\begin{lemma}\label{lm3w}
No process $q \neq \ell$ executes the loop of lines~\ref{prep-send}-\ref{prep-condition} forever.
\end{lemma}

\begin{proof}
The proof is similar to the proof of Lemma~\ref{lm1w}.
Suppose, for contradiction, that a process $q \neq \ell$ executes the loop of lines~\ref{prep-send}-\ref{prep-condition} forever.
Suppose that this occurs when $q$ executes $\LeaderWork{t}$, so $q$ became leader at local time $t$.
Since $q$ executes the loop of lines~\ref{prep-send}-\ref{prep-condition} forever,
	there is a real time after which $q$ has $\CT \geq \Tz$
	and $q$ calls $\ML{t}{\CT}$ in line~\ref{prep-condition} of this loop.
By Lemma~\ref{lm6.1}(\ref{leader2}), this call returns \textsc{False}, and so $q$ exits the loop --- a contradiction.
\qedhere~$_\text{\autoref{lm3w}}$
\end{proof}

\begin{lemma}\label{lm4}
No correct process waits in line~\ref{wait2} or~\ref{wait-alg1} forever.
\end{lemma}

\begin{proof}
No correct process can wait in line~\ref{wait2} for more than $2 \delta$ local time units on its $\CT$.
Now consider a correct process $p$ that waits in line~\ref{wait-alg1}.
If $p$ has $\leasetime = -\infty$, i.e. the initial value of $\leasetime$, 
	then it is clear that $p$ does not execute line~\ref{wait-alg1} forever
	(in fact, $p$ does not wait in this line).
If $\leasetime \neq -\infty$, then it is clear that $\leasetime$ is finite, and by Assumption~\ref{finite-LP}, 
	$\leasetime + \LP = \leasetime + \Pone$ is finite in line~\ref{wait-alg1}.
So, by Assumptions~\ref{xclocks}(\ref{xcl2}) and (\ref{xcl3}),
	there is a real time after which $p$ has $\CT \geq \leasetime + \LP$
	(note that while $p$ waits in line~\ref{wait-alg1},
	it does not change the value of its variable $\leasetime$),
	and $p$ does not wait in line~\ref{wait-alg1} forever.
\qedhere~$_\text{\autoref{lm4}}$
\end{proof}

\begin{lemma}\label{lm2y}
If a correct process calls $\EUB{j}$
	in line~\ref{ExecuteB2},
	then this call
	returns.
\end{lemma}

\begin{proof}
The procedure $\EUB{}$ does not contain any unbounded loops.
\qedhere~$_\text{\autoref{lm2y}}$
\end{proof}

\begin{lemma}\label{lm6w}
If a correct process $q \neq \ell$ calls $\DoOps{(-,-)}{-}{-}$,
	then this call
	returns.
\end{lemma}

\begin{proof}
This is immediate from Lemmas~\ref{lm3w}, \ref{lm4}, and \ref{lm2y}, and the code of $\DoOps{(-,-)}{-}{-}$.
\qedhere~$_\text{\autoref{lm6w}}$
\end{proof}

\begin{lemma}\label{lm6.3w}
For all $t \ge 0$, no correct process $q \neq \ell$ executes forever in $\LeaderWork{t}$.
\end{lemma}

\begin{proof}
Suppose, for contradiction, that a correct process $q \neq \ell$ executes forever in $\LeaderWork{t}$ for some $t\ge0$.
By Observation~\ref{finite-first-lw-wait}, $q$ does not wait forever in line~\ref{wait-lease-expire}.
By~Lemma~\ref{lm1w},
	$q$ exits the loop of lines~\ref{est-request}-\ref{est-condition};
	by Lemma~\ref{lm2},
	$q$ returns from the call of $\FillGaps{k^* -2}$
	in line~\ref{FG};
	and
	by Lemma~\ref{lm6w},
	$q$ returns from the call of $\DoOps{(-,-)}{-}{-}$ in line~\ref{first-doops}.
Thus $q$ reaches line~\ref{mainwhile} of the ``\textbf{while} \textsc{True} \textbf{do}'' loop of lines~\ref{mainwhile}-\ref{endwhile}.
Since $q$ executes forever in $\LeaderWork{t}$,
	$q$ never returns in lines~\ref{recheck1} or \ref{doops2failed} of this while loop.
Moreover, by Lemma~\ref{lm6w},
	$q$ returns from every call of $\DoOps{(-,-)}{-}{-}$ in line~\ref{second-doops}.
Since $q$ is correct, it is now clear that $q$ executes infinitely many iterations of the while
	loop of lines~\ref{mainwhile}-\ref{endwhile}.
Since $\ell$ executes this loop forever,
	by Assumptions~\ref{xclocks}(\ref{xcl2}) and (\ref{xcl3}),
	there is a $t' > \Tz$ such that $q$ gets $t'$ from its $\CT$ in line~\ref{taketime1}
	and $q$ calls $\ML{t}{t'}$ in line~\ref{recheck1} of this loop.
By Lemma~\ref{lm6.1}(\ref{leader2}), this call returns \textsc{False},
	and so $q$ returns from $\LeaderWork{t}$ in line~\ref{recheck1} --- a contradiction.
\qedhere~$_\text{\autoref{lm6.3w}}$
\end{proof}

\begin{lemma}\label{lm7}
For all $t \ge \Tz$, no process $q \neq \ell$ calls $\LeaderWork{t}$.
\end{lemma}

\begin{proof}
Let $t \ge \Tz$ and $q \neq \ell$.
Suppose, for contradiction, that $q$ calls $\LeaderWork{t}$.
Thus, $q$ previously called $\ML{t}{t}$ in line~\ref{check}, and this call returned \textsc{True}.
This contradicts Lemma~\ref{lm6.1}(\ref{leader2}).
\qedhere~$_\text{\autoref{lm7}}$
\end{proof}

\begin{lemma}\label{lm6.8w}
There is a real time after which
 	no correct process $q \neq \ell$ executes inside the $\LeaderWork{}$ procedure.\footnote{For any property $\Phi$,
	``there is a real time after which $\Phi$'' means that there is a real time after which $\Phi$ holds forever; more precisely, it means that there is a real time $\tau$ such that for all $\tau' \ge \tau$ the property $\Phi$ holds at real time $\tau'$.}  
\end{lemma}

\begin{proof}
Suppose, for contradiction, that there is a correct process $q \neq \ell$ such that: for every real time $\tau$,
	there is a real time $\tau' > \tau$
	such that $q$ is executing in $\LW$ at real time $\tau'$.
Then, from Lemma~\ref{lm6.3w}, $q$ returns from $\LW$ infinitely often.
So $q$ calls $\LW$ infinitely often.
Since $q$'s local clock is non-decreasing and it eventually exceeds any given value
	(Assumptions~\ref{xclocks}(\ref{xcl2}-\ref{xcl3})),
	there is a real time after which $q$'s local clock is at least $\Tz$.
Since $q$ calls $\LW$ infinitely often,
	it will eventually call $\LW(t')$, with $t'\ge\Tz$
--- a contradiction to Lemma~\ref{lm7}.
\qedhere~$_\text{\autoref{lm6.8w}}$
\end{proof}

\begin{lemma}\label{zozo6}
For all $j \ge 0$, if a correct process $p$ receives a $\langle \CommitLease, (-,-), j,-,- \rangle$ message then:
\begin{enumerate}
\item $p$ calls $\FillGaps{j-1}$ in line~\ref{gmb2} and this call returns, and
\item $p$ calls $\ExecuteOpsUpToBatch{j}$ in line~\ref{gmb3} and this call returns.
\end{enumerate}
\end{lemma}

\begin{proof}
Let $j\ge0$.
Suppose that a correct process $p$ receives a $\langle \CommitLease, (-,-), j, -, - \rangle$ message.
Note that this receipt occurs in line~\ref{commitreceipt1}.
After receiving $\langle \CommitLease, (-,-), j, -, - \rangle$,
	$p$ sets $\Batch[j]$ in line~\ref{batch2},
	and then $p$ calls $\FillGaps{j-1}$ in line~\ref{gmb2}.
We claim that $p$ returns from this call.
To see this note that:
	(1) if $j \le 1$, from the code of $\FillGaps{}$, this call obviously returns;
	(2) if $j \ge 2$, by Corollary~\ref{successivebatches2} and \ref{ne-batch},
	for all $i$, $1 \le i \le j-1$,
	more than $n/2$ processes set $\Batch[i]$ to $(\Os_i,-)$ for some non-empty set $\Os_i$
	before $p$ sets $\Batch[j]$ in line~\ref{batch2},
	and therefore before $p$ calls $\FillGaps{j-1}$ in line~\ref{gmb2};
	so, by Theorem~\ref{DeSica},
	$p$ returns from this call.

By the above claim, $p$ returns from $\FillGaps{j-1}$ in line~\ref{gmb2},
	and so it calls $\ExecuteOpsUpToBatch{j}$
	in line~\ref{gmb3}.
Since the procedures $\ExecuteOpsUpToBatch{}$ and $\ExecuteBatch{}$ do not contain unbounded loops,
	$p$ returns from this call.
\qedhere~$_\text{\autoref{zozo6}}$
\end{proof}

\begin{lemma}\label{mortacci}
If a correct process $q$ calls the $\PCM()$ procedure, then this call returns.
\end{lemma}

\begin{proof}
Suppose a correct process $q$ calls the  $\PCM()$.
From the code of this procedure (lines~\ref{spcm}-\ref{epcm}),
	it is clear that $q$ could be ``stuck'' forever in $\PCM()$ only
	when it calls $\FillGaps{j-1}$ in line~\ref{gmb2},
	or when it calls $\ExecuteOpsUpToBatch{j}$ in line \ref{gmb3},
	after receiving a $\langle \CommitLease, (-,-), j, -, - \rangle$ message in line~\ref{commitreceipt1}.
By Lemma~\ref{zozo6}, these calls always return.
So $q$'s call to $\PCM()$ also returns.
\qedhere~$_\text{\autoref{mortacci}}$
\end{proof}

\begin{lemma}\label{lm6.9w}
Every correct process $q \neq \ell$ calls the $\PCM()$ procedure infinitely often. 
\end{lemma}

\begin{proof}
This follows immediately from Lemmas~\ref{lm6.8w} and~\ref{mortacci}, and lines~\ref{whileo}-\ref{whilee} of the algorithm.
\qedhere~$_\text{\autoref{lm6.9w}}$
\end{proof}

\begin{lemma}\label{lastleader}
If $\ell$ executes the loop of
	lines~\ref{est-request}-\ref{est-condition},
	\ref{mainwhile}-\ref{endwhile}, or
	\ref{prep-send}-\ref{prep-condition},
	infinitely often
	in $\LeaderWork{t}$ for some $t$, then:

\begin{enumerate}

\item\label{last-t} no process calls $\LeaderWork{t'}$ with $t' >t$, and

\item\label{tmaxito}
 every process has $\maxT \le t$ always.
\end{enumerate}

\end{lemma}

\begin{proof}
Suppose
	that $\ell$ executes the loop of
	lines~\ref{est-request}-\ref{est-condition},
	\ref{mainwhile}-\ref{endwhile}, or
	\ref{prep-send}-\ref{prep-condition},
	infinitely often in $\LeaderWork{t}$.
Thus, $\ell$ executes forever in $\LeaderWork{t}$.

\begin{enumerate}

\item Suppose, for contradiction, that some process $q$ calls $\LeaderWork{t'}$ with $t' >t$.
There are two cases:

\begin{enumerate}
\item $q=\ell$. Thus $\ell$ executes $\LeaderWork{t'}$ with $t' >t$.
Since the clock of $\ell$ is non-decreasing (Assumption~\ref{xclocks}(\ref{xcl2})),
	the code of lines~\ref{whileo}-\ref{whilee}
	implies that $\ell$ called $\LeaderWork{t'}$ \emph{after} calling $\LeaderWork{t}$.
	Since $\ell$ executes forever in the $\LeaderWork{t}$,
	this is impossible.

\item $q \neq \ell$.
Since $q$ executes $\LeaderWork{t'}$ with $t' >t$,
	$q$ calls $\ML{t'}{t'}$ and this call returns \textsc{True}.
Since $\ell$ executes the loop of
	lines~\ref{est-request}-\ref{est-condition},
	\ref{mainwhile}-\ref{endwhile}, or
	\ref{prep-send}-\ref{prep-condition},
	infinitely often,
	$\ell$ reads its $\CT$ infinitely often in line~\ref{est-condition}, \ref{taketime1}, or \ref{prep-condition}.
By Assumptions~\ref{xclocks}(\ref{xcl2}-\ref{xcl3}),
	there is a $t'' \geq t'$ such that $\ell$ gets $t''$ from its $\CT$ in line~\ref{est-condition}, \ref{taketime1}, or \ref{prep-condition},
	and then $\ell$ calls $\ML{t}{t''}$ in line~\ref{est-condition}, \ref{recheck1}, or line~\ref{prep-condition}.
Since $\ML{t'}{t'} = \textsc{true}$ at $q \neq \ell$, and $t' \in [ t, t'' ]$,
	from Theorem~\ref{leader-safety}, the call to $\ML{t}{t''}$ by $\ell$ in line~\ref{est-condition}, \ref{recheck1}, or line~\ref{prep-condition},
	returns \textsc{False}.
Thus, $\ell$ does not execute the loop of lines~\ref{est-request}-\ref{est-condition} infinitely often in $\LeaderWork{t}$, 
	since otherwise it will find $\ML{t}{\CT}$ returns \textsc{False} in line~\ref{est-condition}, exit the loop, and will not enter this loop again in $\LeaderWork{t}$.
Similarly, $\ell$ does not execute the loop of lines~\ref{mainwhile}-\ref{endwhile} infinitely often in $\LeaderWork{t}$,
	since otherwise it will find $\ML{t}{t''}$ returns \textsc{False} in
	line~\ref{recheck1} and then exit $\LeaderWork{t}$.
If $\ell$ executes the loop of lines~\ref{prep-send}-\ref{prep-condition} infinitely often,
	then it calls $\ML{t}{t''}$ with some $t'' > t'$ in line~\ref{prep-condition} 
	during a call to $\DoOps{(-,-)}{t}{j}$ for some $j$.
Since this call to $\ML{t}{t''}$ returns \textsc{False}, and
	by Lemmas~\ref{lm4} and \ref{lm2y} and the fact that $\ell$ is a correct process, 
	$\ell$ returns from this $\DoOps{(-,-,)}{t}{j}$ call.
If this call returns \textsc{Failed}, then we are done, since $\ell$ then exits $\LeaderWork{t}$ in line~\ref{doopsfailed} or line~\ref{doops2failed}.
If not, then $\ell$ continues to execute lines~\ref{taketime1} and \ref{recheck1} (whether the $\DoOps{(-,-,)}{t}{j}$ call is made in line~\ref{first-doops} or line~\ref{second-doops}).
In line~\ref{taketime1}, $\ell$ reads $\hat t$ from its clock such that $\hat t \geq t'' > t' \geq t$
	(Assumption~\ref{xclocks}(\ref{xcl2})).
Thus, the call to $\ML{t}{\hat t}$ in line~\ref{recheck1} returns \textsc{False},
	and $\ell$ then exits $\LeaderWork{t}$.
So in all cases, $\ell$ exits $\LeaderWork{t}$ --- a contradiction.
\end{enumerate}

Thus, no process calls $\LeaderWork{t'}$ with $t' >t$.

\item Suppose, for contradiction, that some process $q$ has $\maxT = t' > t$ at some time.
Since $t' > t \ge 0$, $t'$ is not the initial value $-1$ of $\maxT$.
From the way $q$ maintains $\maxT$ (line~\ref{incTmax}),
	it is clear that $q$ received an $\langle \EstRequest, t' \rangle$ message from some process $r$.
Since $r$ sends $\langle \EstRequest, t' \rangle$, $r$ previously called $\LeaderWork{t'}$.
Since $t' > t$, this contradicts the first part of the lemma (that we proved above).
So $\maxT \le t$ always at $q$.
\qedhere~$_\text{\autoref{lastleader}}$
\end{enumerate}
\end{proof}

\begin{lemma}\label{lm1}
$\ell$ does not execute the loop of lines~\ref{est-request}-\ref{est-condition} forever.
\end{lemma}

\begin{proof}
Suppose, for contradiction, that $\ell$ executes the loop of lines~\ref{est-request}-\ref{est-condition} forever.
Suppose that $\ell$ does so in the execution of $\LeaderWork{t}$ for some $t$.
Consider an arbitrary correct process $q \neq \ell$.

Since $\ell$ executes the loop of lines~\ref{est-request}-\ref{est-condition} forever,
	it sends $\langle \EstRequest,t \rangle$ to $q \neq \ell$ infinitely many times in line~\ref{est-request}.
Since the communication link between any two correct processes is fair (Assumption~\ref{fl}),
	and, by Lemma~\ref{lm6.9w},
	$q \neq \ell$ calls the $\PCM()$ procedure infinitely often,
	$q$ receives $\langle \EstRequest,t \rangle$
	infinitely often from $\ell$ in line~\ref{a2b}.
Therefore,
	$q$ sends $\langle \EstReply, t, \Ops,ts,k,-\rangle$ infinitely often to $\ell$ in line~\ref{a2e}.
Since the communication link between $q$ and $\ell$ is fair,
	$\ell$ eventually receives a $\langle \EstReply, t, \Ops,ts,k,-\rangle$ from $q$ in line~\ref{estrep1},
	and so $\ell$ eventually adds $q$ to $\replied[t]$ in line~\ref{estrep2}.
Recall that $q$ is an arbitrary correct process different from $\ell$.
Thus, there is a time after which $\replied[t]$ contains all the correct processes
	that are not $\ell$.
Since there are at least $\lfloor n/2 \rfloor$ such processes,
	there is a time after which $| \replied[t] | \ge \lfloor n/2 \rfloor$ at $\ell$.
So the exit condition of the loop of
	lines~\ref{est-request}-\ref{est-condition} is eventually satisfied,
	and $\ell$ exits this loop --- a contradiction.
\qedhere~$_\text{\autoref{lm1}}$
\end{proof}

\begin{lemma}\label{lm7w}
$\ell$ does not execute the loop of lines~\ref{prep-send}-\ref{prep-condition} forever.
\end{lemma}

\begin{proof}
Suppose, for contradiction, that $\ell$ executes the loop of lines~\ref{prep-send}-\ref{prep-condition} forever.
This occurs in the execution of some $\DoOps{(\Os,-)}{t}{j}$ in $\LeaderWork{t}$ for some $t \ge 0$.
Consider an arbitrary correct process $q \neq \ell$.
By Lemma~\ref{lastleader}(\ref{tmaxito}), process $q$ has $\maxT \le t$ always (*).

\begin{claim}\label{clamonito2}
Process $q$ has $(ts,k) \le (t,j)$ always.
\end{claim}

\begin{proof}
Suppose, for contradiction, that at some time $q$ has $(ts,k) = (t',j') > (t,j)$.
Since $t' \ge t \ge 0$, $(t',j')$ is not the initial value $(-1,0)$ of $(ts,k)$ at $q$.
Thus $q$ previously accepted a tuple $(\Os',t',j')$ for some $\Os'$.
So, by Observation~\ref{leaderfirst}, some process $r$ previously executed
	$\DoOps{(\Os',-)}{t'}{j'}$ in $\LeaderWork{t'}$.
By Lemma~\ref{lastleader}(\ref{last-t}),
	$t' \le t$.
Since $(t',j') > (t,j)$, it must be that $t' = 	t$ and $j'>j$.
Since $t' = t$, processes $\ell$ and $r$ became leader at the same local time $t$,
	by Lemma~\ref{unique-LeaderWork}, $r = \ell$.
Thus process $\ell$ called $\DoOps{(\Os',-)}{t}{j'}$ in $\LeaderWork{t}$ with $j' > j$.
By Corollary~\ref{T-k-DoopsCreduced},
	$\ell$ called $\DoOps{(\Os',-)}{t}{j'}$ after calling $\DoOps{(\Os,-)}{t}{j}$ --- contradicting the fact that
	$\ell$ executes forever in the loop of lines~\ref{prep-send}-\ref{prep-condition}
	of $\DoOps{(\Os,-)}{t}{j}$ in $\LeaderWork{t}$.
So $q$ has $(ts,k) \le (t,j)$ always.
\qedhere~$_\text{\autoref{clamonito2}}$
\end{proof}

\begin{claim}\label{clamonito0}
Process $q$ receives $\langle \Prepare,(\Os,-),t,j,-\rangle$
	infinitely often from $\ell$.
\end{claim}

\begin{proof}
Since $\ell$ executes the loop of lines~\ref{prep-send}-\ref{prep-condition} forever,
	it sends $\langle \Prepare,(\Os,-),t,j,-\rangle$ to $q \neq \ell$ infinitely many times in line~\ref{prep-send}.
Since the communication link between any two correct processes is fair (Assumption~\ref{fl}),
	and, by Lemma~\ref{lm6.9w},
	$q \neq \ell$ calls the $\PCM()$ procedure infinitely often,
	$q$ receives $\langle \Prepare,(\Os,-),t,j,- \rangle$
	infinitely often from $\ell$ in line~\ref{a3b}.
\qedhere~$_\text{\autoref{clamonito0}}$
\end{proof}

\begin{claim}\label{clamonito3}
Process $q$ eventually accepts $(\Os,t,j)$, and it does not accept any tuple thereafter.
\end{claim}

\begin{proof}
Suppose, for contradiction, that $q$ never accepts $(\Os,t,j)$.
By Claim~\ref{clamonito0}, $q$ receives \linebreak$\langle \Prepare,(\Os,\s),t,j,-\rangle$
	infinitely often from $\ell$.
Consider the first time that $q$ receives this message in line~\ref{a3b}.
Since $q$ does not accept $(\Os,t,j)$
	the guard in line~\ref{acceptcheck3},
	is not satisfied.
So $q$ has $\maxT > t$ or $(ts,k) \ge (t,j)$ in line~\ref{acceptcheck3}.
By (*) and Claim~\ref{clamonito2}, $q$ has $\maxT \le t$ and $(ts,k) \le (t,j)$ always.
Therefore $q$ has $(t,j) = (ts,k)$ in line~\ref{acceptcheck3}.
Since $t \ge 0$, $(t,j)$ is not the initial value $(-1,0)$ of $(ts,k)$ at $q$.
Thus $q$ previously accepted a tuple $(\Os',t,j)$ for some $\Os'$.
From Observation~\ref{leaderfirst}, $\ell$ executed
	$\DoOps{(\Os',-)}{t}{j}$ in $\LeaderWork{t}$.
By Lemma~\ref{doops-simplecase}, $\Os' = \Os$.
So $q$ accepted $(\Os,t,j)$ --- a contradiction.

Thus, $q$ eventually accepts $(\Os,t,j)$,
	and sets $(\Ops,ts,k)$ to $(\Os,t,j)$.
We claim that $q$ does not accept any tuple thereafter.
Suppose, for contradiction, that $q$ accepts some tuple $(\Os',t',j')$ after accepting $(\Os,t,j)$.
By Corollary~\ref{T-k-increase}, $(t',j') > (t,j)$.
Note that after $q$ accepts $(\Os',t',j')$, it has $(ts,k) = (t',j')$,
	so $q$ now has  $(ts,k) = (t',j') >  (t,j)$ --- a contradiction to Claim~\ref{clamonito2}.
\qedhere~$_\text{\autoref{clamonito3}}$
\end{proof}

From Claim~\ref{clamonito3},
	there is a real time after which $q$ has $(\Ops,ts,k) = (\Os,t,j)$ forever.
Moreover, by Claim~\ref{clamonito0},
	$q$~receives $\langle \Prepare,(\Os,-),t,j,-\rangle$
	infinitely often from $\ell$.
Therefore,
	$q$ sends $\langle \PACK,t,j \rangle$ infinitely often to~$\ell$ in line~\ref{sendPack}.
Since the communication link between $q$ and $\ell$ is fair,
	$\ell$ eventually receives a $\langle \PACK,t,j \rangle$ from $q$ in line~\ref{Rack1},
	and so $\ell$ eventually adds $q$ to $\PACKED[t,j]$ in line~\ref{Rack2}.
Recall that $q$ is an arbitrary correct process different from $\ell$.
Thus, there is a real time after which $\PACKED[t,j]$ contains all the correct processes
	that are not $\ell$.
Since there are at least $\lfloor n/2 \rfloor$ such processes,
	there is a real time after which $| \PACKED[t,j] | \ge \lfloor n/2 \rfloor$ at $\ell$.
So the exit condition of the loop of
	lines~\ref{prep-send}-\ref{prep-condition} is eventually satisfied,
	and $\ell$ exits this loop --- a contradiction.
\qedhere~$_\text{\autoref{lm7w}}$
\end{proof}

\begin{lemma}\label{lm6}
If $\ell$ calls $\DoOps{(-,-)}{-}{-}$
	then this call
	returns.
\end{lemma}

\begin{proof}
This is immediate from Lemmas~\ref{lm7w}, \ref{lm4}, and \ref{lm2y}, and the code of $\DoOps{(-,-)}{-}{-}$.
\qedhere~$_\text{\autoref{lm6}}$
\end{proof}

\begin{lemma}\label{ctmax}
$\ell$ has $\maxT < \Tz$ always.
\end{lemma}

\begin{proof}
Suppose, for contradiction, that $\ell$ has $\maxT = t' \ge \Tz$.
Since $\Tz \ge 0$ and initially $\maxT = -1$, $t'$ is not the initial value of $\maxT$.
Since $\ell$ updates $\maxT$ only in line~\ref{incTmax},
	it is clear that $\ell$ received a $\langle \EstRequest, t' \rangle$ message from some process $q$ in line~\ref{a2b}.
Note that $q \neq \ell$ because $\ell$ never sends $\langle \EstRequest, - \rangle$ messages to itself.
Furthermore, $q$ sent $\langle \EstRequest, t' \rangle$ in line~\ref{est-request} of $\LeaderWork{t'}$.
Since $q$ calls $\LeaderWork{t'}$, by Lemma~\ref{lm7},  $t' < \Tz$ --- a contradiction.
\qedhere~$_\text{\autoref{ctmax}}$
\end{proof}

\begin{lemma}\label{lm8}
For all $t \ge \Tz$, if $\ell$ calls $\DoOps{(-,-)}{t}{-}$, then this call
	returns \textsc{Done}.
\end{lemma}

\begin{proof}
Suppose that $\ell$ calls $\DoOps{(\Os,\s)}{t}{j}$, for some $\Os$, $\s$, $j$, and $t \ge \Tz$.
By Lemma~\ref{lm6}, this call returns.
Note that lines \ref{acceptcheck1}, 
	\ref{ack-condition}, and \ref{done}
	are the only return statements of $\DoOps{(\Os,\s)}{t}{j}$.
When $\ell$ executes line~\ref{acceptcheck1} of $\DoOps{(\Os,\s)}{t}{j}$, by Lemma~\ref{ctmax},
	$\ell$ has $\maxT < \Tz$.
Since $t \ge \Tz$, $\ell$ has $t > \maxT$ so it does not return in line~\ref{acceptcheck1}. 
When $\ell$ executes line~\ref{prep-condition} of $\DoOps{(\Os,\s)}{t}{j}$,
	$\CT_\ell$ is at least $t$, and hence at least $\Tz$.
So when $\ell$ calls $\ML{t}{\CT}$ in line~\ref{prep-condition} of $\DoOps{(\Os,\s)}{t}{j}$,
	by Lemma~\ref{lm6.1}(\ref{leader1}), these calls return \textsc{True}.
Thus, if $\ell$ executes line~\ref{prep-condition2},
	then it first found  $| \PACKED[t,j] | \ge  \lfloor n/2 \rfloor$ in line~\ref{prep-condition}.
Since line~\ref{Rack2} is the only place where $\ell$ modifies $| \PACKED[t,j] |$, it is clear that $| \PACKED[t,j] |$ contains a non-decreasing set of processes.
So if $\ell$ executes line~\ref{prep-condition2} of $\DoOps{(\Os,\s)}{t}{j}$, it has $| \PACKED[t,j] | \ge  \lfloor n/2 \rfloor$ and does not return in this line.
Therefore, $\DoOps{(\Os,\s)}{t}{j}$ returns \textsc{Done} in line~\ref{done}.
\qedhere~$_\text{\autoref{lm8}}$
\end{proof}

\begin{lemma}\label{lm9}
For all $t \ge \Tz$, if $\ell$ calls $\LeaderWork{t}$ then this call does not return.
\end{lemma}

\begin{proof}
Suppose $\ell$ calls $\LeaderWork{t}$ with $t \ge \Tz$.
Note that this call can return only in lines~\ref{est-condition2}, \ref{Egathercrumbs},
	\ref{doopsfailed}, \ref{recheck1}, and~\ref{doops2failed}.
We now prove that the $\LeaderWork{t}$ call does not return in any of these lines.

Since $t \ge \Tz$, if $\ell$ calls $\DoOps{(-,-)}{t}{-}$ in lines~\ref{first-doops} or~\ref{second-doops},
	then, by Lemma~\ref{lm8}, 
	this call returns \textsc{Done}.
Thus, the $\LeaderWork{t}$ call does not return in line~\ref{doopsfailed} or~\ref{doops2failed}.

When $\ell$ executes line~\ref{est-condition} of $\LeaderWork{t}$,
	$\CT$ is at least $t$ (Assumption~\ref{xclocks}(\ref{xcl2})), 
	and hence at least $\Tz$.
So, by Lemma~\ref{lm6.1}(\ref{leader1}), 
	the calls to $\ML{t}{\CT_\ell}$ in line~\ref{est-condition} return \textsc{True}.
Thus, if $\ell$ executes line~\ref{est-condition2} of $\LeaderWork{t}$,
	it must have previously found $| \replied[t] | \ge \lfloor n/2 \rfloor$ in line~\ref{est-condition}.
Since $\ell$ modifies $| \replied[t] |$ only in line~\ref{estrep2}, 
	$| \replied[t] |$ contains a non-decreasing set of processes.
So $\ell$ has $| \replied[t] | \ge \lfloor n/2 \rfloor$ when it executes line~\ref{est-condition2}, and hence it does not return in line~\ref{est-condition2}.

When $\ell$ executes line~\ref{taketime1} of $\LeaderWork{t}$
	(i.e., when $\ell$ executes ``$t' := \CT$''),
	$\ell$ gets $t'$ such that $t' \ge t \ge \Tz$
	(Assumption~\ref{xclocks}(\ref{xcl2})).
So when $\ell$ calls $\ML{t}{t'}$ in line~\ref{recheck1} of $\LeaderWork{t}$,
	by Lemma~\ref{lm6.1}(\ref{leader1}), these calls return \textsc{True}.
Thus, $\LeaderWork{t}$ does not return in line~\ref{recheck1}.

It remains to show that the $\LeaderWork{t}$ call by $\ell$ does not return in line~\ref{Egathercrumbs}.
Suppose, for contradiction, that this $\LeaderWork{t}$ call returns in line~\ref{Egathercrumbs}.
Thus, $\ell$ has $ts^* \ge t$ in line~\ref{Egathercrumbs}.
Since $t \ge \Tz \ge 0$, we have $ts^* \ge \Tz \ge 0$,
	and so $\ell$ selected a tuple $(\Ops^*,ts^*,k^*) \neq (\emptyset,-1,0)$ in line~\ref{selection}.
Thus, by Lemma~\ref{lmx3},
	some process $q^*$ previously accepted a tuple $(\Ops^*,ts^*,k^*)$.
By Observation~\ref{leaderfirst}, a process $r$ that became leader at time~$ts^*$,
	i.e., a process that called $\LeaderWork{ts^*}$,
	previously accepted $(\Ops^*,ts^*,k^*)$.
Since $r$ calls $\LeaderWork{ts^*}$ with $ts^* \ge \Tz$,
	by Lemma~\ref{lm7},
	process $r = \ell$.	
So $\ell$ accepted the tuple $(\Ops^*,ts^*,k^*)$ in $\LeaderWork{ts^*}$
	before selecting $(\Ops^*,ts^*,k^*)$ in line~\ref{selection} in $\LeaderWork{t}$.
From the code of $\LeaderWork{t}$, it is clear that $\ell$ does not accept any tuple
	between calling $\LeaderWork{t}$
	and selecting $(\Ops^*,ts^*,k^*)$ in line~\ref{selection} in $\LeaderWork{t}$.
Thus $\ell$ accepted the tuple $(\Ops^*,ts^*,k^*)$ in $\LeaderWork{ts^*}$ \emph{before} it called $\LeaderWork{t}$.
So $\ell$ called $\LeaderWork{ts^*}$ before calling $\LeaderWork{t}$.
By Lemma~\ref{monoLW}, $ts^* < t$ --- a contradiction.
\qedhere~$_\text{\autoref{lm9}}$
\end{proof}

There is a real time after which $\ell$ executes forever in the $\LeaderWork{t}$ procedure. More precisely:

\begin{theorem}\label{lm10}
There is a local time $t$ such that $\ell$ calls $\LeaderWork{t}$ and this call does not return.
Moreover, $\ell$~executes the while loop of lines~\ref{mainwhile}-\ref{endwhile}
	infinitely often in this execution of $\LeaderWork{t}$.
\end{theorem}

\begin{proof}
Consider the while loop of \textsc{Thread 2}, i.e., lines~\ref{whileo}-\ref{whilee}.

\begin{claim}\label{clamox}
$\ell$ executes a finite number of iterations of the while loop of lines~\ref{whileo}-\ref{whilee}.
\end{claim}

\begin{proof}
Suppose, for contradiction, that $\ell$ executes an infinite number of iterations of this loop.
In each iteration of this loop, $\ell$ reads $\CT_{\ell}$ in line~\ref{gt},
	and, by Assumptions~\ref{xclocks}(\ref{xcl2}-\ref{xcl3}),
	the value that $\ell$ gets from $\CT_{\ell}$ eventually exceeds $\Tz$.
Consider the first iteration where $\ell$ gets $t \ge \Tz$ in line~\ref{gt} of this loop.
Process $\ell$ then calls $\ML{t}{t}$ with $t \ge \Tz$ in line~\ref{check},
	and by Lemma~\ref{lm6.1}(\ref{leader1}), this call returns \textsc{True}.
Thus, $\ell$ calls $\LeaderWork{t}$ with $t \ge \Tz$ in line~\ref{check}.
By Lemma~\ref{lm9} this call does not return --- a contradiction.
\qedhere~$_\text{\autoref{clamox}}$
\end{proof}

By Lemma~\ref{mortacci}, whenever $\ell$ calls $\PCM()$ in line~\ref{whilee},
	this call returns.
Thus, from Claim~\ref{clamox}, the code of lines~\ref{whileo}-\ref{whilee},
	and the fact that $\ell$ is a correct process,
	it is clear that
	there is a local time $t$ such that $\ell$ calls $\LeaderWork{t}$ and this call does not return.
	
Now consider the call of $\LeaderWork{t}$ that does not return.
Since $\ell$ is correct, we note that:
	by Observation~\ref{finite-first-lw-wait}, $\ell$ completes the wait statement in line~\ref{wait-lease-expire};
	by~Lemma~\ref{lm1},
	$\ell$ exits the loop of lines~\ref{est-request}-\ref{est-condition};
	by Lemma~\ref{lm2},
	$\ell$ returns from the call of $\FillGaps{k^* -2}$
	in line~\ref{FG};
	and
	by Lemma~\ref{lm6},
	$\ell$ returns from the call of $\DoOps{(-,-)}{-}{-}$ in line~\ref{first-doops}.
Thus $\ell$ reaches line~\ref{mainwhile} of the
	``\textbf{while} \textsc{True} \textbf{do}'' loop of lines~\ref{mainwhile}-\ref{endwhile}.
Since $\ell$ executes forever in $\LeaderWork{t}$,
	$\ell$ never returns in lines~\ref{recheck1} or \ref{doops2failed} of this while loop.
Moreover, by Lemma~\ref{lm6},
	$\ell$ returns from every call of $\DoOps{(-,-)}{-}{-}$ in line~\ref{second-doops}.
Since $\ell$ is correct, it is now clear that $\ell$ executes
	infinitely many iterations of the while loop of lines~\ref{mainwhile}-\ref{endwhile}.
\qedhere~$_\text{\autoref{lm10}}$
\end{proof}

\begin{lemma}\label{lastleader2}
If $\ell$ executes in $\LeaderWork{t}$ for some $t$ forever, then:

\begin{enumerate}

\item\label{last-t2} no process calls $\LeaderWork{t'}$ with $t' >t$, and

\item\label{tmaxito2}
 every process has $\maxT \le t$ always.
\end{enumerate}

\end{lemma}

\begin{proof}
Suppose $\ell$ executes in $\LeaderWork{t}$ forever.
By Theorem~\ref{lm10}, $\ell$~executes the while loop of lines~\ref{mainwhile}-\ref{endwhile}
	infinitely often in $\LeaderWork{t}$.
By Lemma~\ref{lastleader}, no process calls $\LeaderWork{t'}$ with $t' >t$, and
	every process has $\maxT \le t$ always.
\qedhere~$_\text{\autoref{lastleader2}}$
\end{proof}

\begin{lemma}\label{zozo4}
For all $\hat{k} \ge 0$, if a process locks a tuple $(-, -, \hat{k})$
	then there is a real time after which $\ell$ has $k \ge \hat{k}$.
\end{lemma}

\begin{proof}
Suppose a process $r$ locks some tuple $(\hat{\Os}, \hat{t}, \hat{k})$.
By Observation~\ref{aboutlocking},
	$r$ locks $(\hat{\Os}, \hat{t}, \hat{k})$ in $\LeaderWork{\hat{t}}$.
By Theorem~\ref{lm10}, there is a local time $t$ such that $\ell$ executes 
	$\LeaderWork{t}$ forever.
By Lemma~\ref{lastleader2}(\ref{last-t2}), $\hat{t} \le t$.
There are two cases:

\begin{enumerate}

	\item $t = \hat{t}$. 
	Thus processes $\ell$ and $r$ became leader at the same local time $t$,
	and, by Lemma~\ref{unique-LeaderWork}, $r = \ell$.
	So $\ell$ locks $(\hat{\Os}, t , \hat{k})$ in $\LeaderWork{t}$.
	Therefore $\ell$ calls $\DoOps{(\hat{\Os},-)}{t}{\hat{k}}$ in $\LeaderWork{t}$,
		and $\ell$ sets $k$ to $\hat{k}$ in line~\ref{leader-accept} of $\DoOps{(\hat{\Os},-)}{t}{\hat{k}}$ at some real time $\tau$.
	After real time $\tau$, process $\ell$ can change its variable $k$ only by calling $\DoOps{(-,-)}{t}{k+1}$ in the
		while loop of lines~\ref{mainwhile}-\ref{endwhile} of $\LeaderWork{t}$,
		and this call just increments the value of $k$ by one (in line~\ref{leader-accept} of $\DoOps{(-,-)}{t}{k+1}$).
	Thus, process $\ell$ has $k \ge \hat{k}$ after real time~$\tau$.
	
	\item $t > \hat{t}$.
	Note that $\ell$ calls $\DoOps{(\Ops^*,0)}{t}{k^*}$ in line~\ref{first-doops} of $\LeaderWork{t}$,
		and $\ell$ sets $k$ to $k^*$ in line~\ref{leader-accept} of $\DoOps{(\Ops^*,0)}{t}{k^*}$ at some real time $\tau$.
	As we argued in case 1 above, this implies that process $\ell$ has $k \ge k^*$ after time~$\tau$.
	Since $r$ locks $(\hat{\Os}, \hat{t}, \hat{k})$
		and $\ell$ accepts $(\Ops^*, t , k^*)$ with $t > \hat{t}$,
		by Theorem~\ref{generalcase1a}(\ref{t1}), $k^* \ge \hat{k}$.
	Thus, process $\ell$ has $k \ge k^* \ge \hat{k}$ after real time~$\tau$.
	
\end{enumerate}
So in all cases there is a real time after which $\ell$ has $k \ge \hat{k}$.
\qedhere~$_\text{\autoref{zozo4}}$
\end{proof}

\begin{assumption}\label{finite-LRP}
The lease renewal period $\Ptwo$ is positive and finite.
\end{assumption}

\begin{lemma}\label{zozox1}
If there is a real time after which $\ell$ has $k \ge \hat{k}$,
	then $\ell$ sends infinitely many\linebreak
	$\langle \CommitLease, (-,-), j, -, - \rangle$ messages such that $j \ge \hat{k}$ to all processes $p \neq \ell$.
\end{lemma}

\begin{proof}
Suppose, for contradiction, that there is a real time $\tau_1$ such that,
	from real time $\tau_1$ on, $\ell$ has $k \ge \hat{k}$, 
	but $\ell$ does not send $\langle \CommitLease, (-,-), j, -, - \rangle$ messages with $j \ge \hat{k}$ to all processes $p \neq \ell$
	.
By Theorem~\ref{lm10}, there is a local time $t$ such that $\ell$ calls $\LeaderWork{t}$ and it does not return, 
	and $\ell$ executes the while loop of lines~\ref{mainwhile}-\ref{endwhile} infinitely often in this execution of $\LeaderWork{t}$.
Let $\tau_2$ be the real time when $\ell$ enters the while loop of lines~\ref{mainwhile}-\ref{endwhile} in $\LeaderWork{t}$,
	and let $\tau_3 = \max(\tau_1, \tau_2)$.

\begin{claim}\label{no-doops}
$\ell$ does not call $\DoOps{(-,-)}{-}{-}$ from time $\tau_3$ on.
\end{claim}
\begin{proof}
Suppose, for contradiction, that $\ell$ calls $\DoOps{(-,-)}{t'}{j'}$,
	for some $t'$ and $j'$,
	at some real time $\tau \geq \tau_3$.
Since $\tau\ge\tau_3\ge\tau_2$, $\ell$ is in $\LeaderWork{t}$, so $t'=t$;
	and $\ell$ makes this call in line~\ref{second-doops} of $\LeaderWork{t}$,
	so the call is of form $\DoOps{(-,-)}{t}{k+1}$.
Since $\tau \geq \tau_3 \geq \tau_1$,
	the value of $k$ is at least $\hat k$ at real time $\tau$,
	so we have $j' = k + 1 > \hat k$.
Since $\ell$ executes the while loop infinitely often in $\LeaderWork{t}$,
	this call to $\DoOps{(-,-)}{t}{j'}$ must return \textsc{Done}.
Note that before this call returns \textsc{Done} in line~\ref{done}, 
	$\ell$ sends a $\langle \CommitLease, (-,-), j', -, - \rangle$ message to all processes $p \neq \ell$ in line~\ref{sendcommit},
	which contradicts the assumption that
	$\ell$ does not send $\langle \CommitLease, (-,-), j, -, - \rangle$ messages
	with $j \geq \hat k$ from real time $\tau_1$ on.
\qedhere~$_\text{\autoref{no-doops}}$
\end{proof}
Note that it is possible that $\ell$ is executing the $\DO$ procedure at real time $\tau_3$.
We now define $\tau_4$ to be the earliest real time $\geq \tau_3$
	such that $\ell$ is executing line~\ref{mainwhile}.
Since $\ell$ executes the while loop of $\LeaderWork{t}$ infinitely often, 
	it always returns from calls to the $\DO$ procedure,
	so $\tau_4$ exists.
Since $\tau_4 \geq \tau_3$, by the definition of $\tau_4$ and Claim~\ref{no-doops}, 
	$\ell$ is never inside the $\DO$ procedure from real time $\tau_4$ on (*).
\begin{claim}\label{no-new-nst}
$\ell$ does not set its $\NextSendTime$ variable from real time $\tau_4$ on.
\end{claim}
\begin{proof}
Suppose, for contradiction, that 
	$\ell$ sets $\NextSendTime$ at some real time $\tau \geq \tau_4$.
Since $\tau \geq t_4$, by (*),
	this must happen in line~\ref{nst}.
($\NextSendTime$ is set only in lines~\ref{nst} and~\ref{nst2},
	and the latter is inside $\DO$.)
Note that just before line~\ref{nst}, 
	$\ell$ sent $\langle \CommitLease, -, k, -, - \rangle$ messages to all processes $p \neq \ell$.
Since this happens after real time $\tau_4 \ge \tau_3 \ge \tau_1$, 
	$\ell$ has $k \geq \hat k$, which contradicts the assumption about $\tau_1$.
\qedhere$_\text{\autoref{no-new-nst}}$ 
\end{proof}

Now consider the last time $\ell$ sets $\NextSendTime$ before real time $\tau_4$ 
	($\ell$ must set $\NextSendTime$ at least once before real time $\tau_4$
	since it finished a call to $\DO$ in line~\ref{first-doops},
	and it set $\NextSendTime$ in line~\ref{nst2}).
This can happen in two places, i.e., line~\ref{nst} and~\ref{nst2}.
By Assumptions~\ref{xclocks}(\ref{xcl1}), \ref{finite-LRP}, and~\ref{promiseRange}, 
	$\ell$ sets $\NextSendTime$ to some finite value $t_n$.
By Claim~\ref{no-new-nst}, 
	$\ell$ does not update $\NextSendTime$ from time $\tau_4$ on, 
	so $\ell$ has $\NextSendTime = t_n$ from time $\tau_4$ on.
Since $\ell$ executes the while loop in $\LeaderWork{t}$ infinitely often,
	consider the first iteration of the while loop after time $\tau_4$
	when $\ell$'s local clock has value at least $t_n$
	(this happens by Assumptions~\ref{xclocks}(\ref{xcl2}-\ref{xcl3})), 
	and $\ell$ gets $t' = \CT \geq t_n$ in line~\ref{taketime1}.
Thus, $\ell$ finds $t' \geq \NextSendTime$ in line~\ref{checksendtime} and
	continues to execute line~\ref{sendcommit2}.
Since this is after real time $\tau_1$, $\ell$ has $k = j \geq \hat k$ for some $j$
	in line~\ref{sendcommit2}.
So $\ell$ sends $\langle \CommitLease, (-,-), j, -, - \rangle$ messages
	with $j \geq \hat k$ to all processes $p \neq \ell$ after real time $\tau_1$ --- a contradiction.
\qedhere~$_\text{\autoref{zozox1}}$
\end{proof}

\begin{lemma}\label{zozox2}
If there is a real time after which $\ell$ has $k \ge \hat{k}$,
	then for every correct process $p \neq \ell$
	there is a $j \ge \hat{k}$ such that:
\begin{enumerate}
\item $p$ calls $\FillGaps{j-1}$ in line~\ref{gmb2} and this call returns, and
\item $p$ calls $\ExecuteOpsUpToBatch{j}$ in line~\ref{gmb3} and this call returns.
\end{enumerate}	
\end{lemma}

\begin{proof}
Suppose there is a real time after which $\ell$ has $k \ge \hat{k}$.
Let $p$ be any correct process other than~$\ell$.
By Lemma~\ref{zozox1},
	$\ell$ sends infinitely many 
	$\langle \CommitLease, (-,-), j, -, - \rangle$ messages such that $j \ge \hat{k}$ to $p$.
Since the communication link between the two correct processes $p$ and $\ell$ is fair
	(Assumption~\ref{fl}),
	$p$ eventually receives some $\langle \CommitLease, (-,-), j, -, - \rangle$ with $j \ge \hat{k}$ from $\ell$.
The result now follows from Lemma~\ref{zozo6}.
\qedhere~$_\text{\autoref{zozox2}}$
\end{proof}

Note that a process $p$ modifies $\Reply{\operation}$
	only in line~\ref{apply-op-1} of $\ExecuteBatch{}$;
	since the replies of the $\Apply$ function are not $\bot$,	
	it is clear that $p$ never sets $\Reply{\operation}$ to $\bot$ in line~\ref{apply-op-1}.\footnote{Recall that $\Apply$
	is the state transition function of the replicated object implemented by the algorithm.}
Therefore:

\begin{observation}\label{replybot}
If a process $p$ has $\Reply{\operation} \neq \bot$ for some $\operation$ at some real time $\tau$,
	then $p$ has
	$\Reply{\operation} \neq \bot$ at all real times $\tau' \ge \tau$.
\end{observation}

\begin{lemma}\label{EBreply}
Suppose that a correct process $p$ has $\Batch[j]= (\Os_j,-)$ for some non-empty set $\Os_j$ at some real time~$\tau$.
If $p$ calls $\ExecuteBatch{j}$ at some real time $\tau' \ge \tau$,
	then this call returns, and when it does and thereafter,
	$p$ has $\Reply{\operation} \neq \bot$ for every $\operation \in \Os_j$.
\end{lemma}

\begin{proof}
Suppose a correct process $p$ has $\Batch[j]= (\Os_j,-)$ for some non-empty set $\Os_j$
	at real time~$\tau$, and $p$ calls $\ExecuteBatch{j}$ at some real time $\tau' \ge \tau$.
By Corollary~\ref{finalsafetyac}, $p$ has $\Batch[j]= (\Os_j,-)$ during the entire execution of $\ExecuteBatch{j}$.
From the code of $\ExecuteBatch{j}$ and by Observation~\ref{replybot},
	it is clear that $p$ exits the loop of lines~\ref{apply-op}-\ref{set-takesEffect},
	and when it does and thereafter, $p$ has $\Reply{\operation} \neq \bot$ for every $\operation \in \Os_j$.
\qedhere~$_\text{\autoref{EBreply}}$
\end{proof}

\begin{lemma}\label{EUBreply}
Suppose that
	a correct process $p$ has $\Batch[i]= (\Os_i,-)$
	for some non-empty set $\Os_i$ for all $i$, $1 \le i \le j$, at some real time~$\tau$.
If $p$ calls $\EUB{j}$ at some real time $\tau' \ge \tau$,
	then this call returns, 
	and when it returns and thereafter,
	$p$ has $\Reply{\operation} \neq \bot$ 
	for every $\operation \in \bigcup\limits_{i=1}^{j} \Os_i$.
	
\end{lemma}

\begin{proof}
Suppose that
	a correct process $p$ has $\Batch[i]= (\Os_i,-)$ for some non-empty set $\Os_i$ for all $i$, $1 \le i \le j$, at real time~$\tau$,
	and $p$ calls $\EUB{j}$ at time $\tau' \ge \tau$.
By Corollary~\ref{finalsafetyac}, $p$ has $\Batch[i]= (\Os_i, -)$
	for all $i$, $1 \le i \le j$, during the entire execution of $\EUB{j}$.
	
Let $j_0$ be the value of $\MBD$ when $p$ executes line~\ref{EUTB} for the first time after it calls \linebreak$\EUB{j}$.
From the for loop of lines~\ref{EUTB}-\ref{addB1},
	it is clear that $p$ executes $\ExecuteBatch{i}$ for every $i$, $j_0+1 \le i \le j$.
Furthermore, by Lemma~\ref{MBD-L},
	$p$ executed the following events before calling $\EUB{j}$:
	for all $i$, $1 \le i \le j_0$, $p$ set $\Batch[i]$ to $(\Os'_i,-)$ for some non-empty set $\Os'_i$, and then it executed $\ExecuteBatch{i}$.
Note that by Corollary~\ref{finalsafetyac}, for all $i$, $1 \le i \le j_0$, $\Os'_i = \Os_i$.

So before exiting the for loop of lines~\ref{EUTB}-\ref{updateMBD1},
	(i) $p$ executes $\ExecuteBatch{i}$ for every $i$, $1 \le i \le j$,
	and (ii)~$p$~has $\Batch[i] = (\Os_i,-)$ for some non-empty set $\Os_i$ before and during the execution of each $\ExecuteBatch{i}$.
Thus, by Lemma~\ref{EBreply},
	this loop exits, and 
	when $p$ exits this loop and thereafter,
	$\Reply{\operation} \neq \bot$ for every $\operation \in \bigcup\limits_{i=1}^{j} \Os_i$.
\qedhere~$_\text{\autoref{EUBreply}}$
\end{proof}

\begin{lemma}\label{zozox3}
If there is a real time after which $\ell$ has $k \ge \hat{k}$,
	then for every correct process $p$ there is a real time after which:
\begin{enumerate}
\item\label{oopla1} for all $i$, $1 \le i \le \hat{k}$, process $p$ has $\Batch[i] = (\Os_i,-)$ for some non-empty set $\Os_i$, and
\item\label{oopla2} for every $\operation \in \bigcup\limits_{i=1}^{\hat{k}} \Os_i$, process $p$ has $\Reply{\operation} \neq \bot$.
\end{enumerate}	
\end{lemma}

\begin{proof}
Note that the lemma trivially holds for $\hat{k} < 1$.
Henceforth we assume that $\hat{k} \ge 1$.
Suppose there is a real time after which $\ell$ has $k \ge \hat{k} \ge 1$.
Let $p$ be any correct process.
There are two cases:

\begin{enumerate}[(a)]
\item $p \neq \ell$.
	By Lemma~\ref{zozox2}, there is a $j \ge \hat{k}$ such that
	$p$ calls $\ExecuteOpsUpToBatch{j}$ in line~\ref{gmb3} and this call returns.
Thus by Lemma~\ref{aboutEUB}, before $p$ calls $\ExecuteOpsUpToBatch{j}$ and at all times thereafter, the following holds:
	for all $i$, $1 \le i \le j$, 
	there is a non-empty set $\Os_i$ such that
	$p$ has $\Batch[i] = (\Os_i,-)$.
So, by Lemma~\ref{EUBreply},
	when $p$ returns from $\ExecuteOpsUpToBatch{j}$ and thereafter,
	$p$ has $\Reply{\operation} \neq \bot$ for every $\operation \in \bigcup\limits_{i=1}^{j} \Os_i$. 
	Since $j \ge \hat{k}$, 
	there is a real time after which:
\begin{enumerate}
\item for all $i$, $1 \le i \le \hat{k}$, there is a non-empty set $\Os_i$ such that process $p$ has $\Batch[i] = (\Os_i,-)$, and
\item for every $\operation \in \bigcup\limits_{i=1}^{\hat{k}} \Os_i$, process $p$ has $\Reply{\operation} \neq \bot$.
\end{enumerate}

\item $p = \ell$.
By Theorem~\ref{lm10}, there is a real time after which $\ell$ executes 
	the while loop of lines~\ref{mainwhile}-\ref{endwhile}
	of $\LeaderWork{t}$ forever.
Note that before entering the while loop of lines~\ref{mainwhile}-\ref{endwhile} in $\LeaderWork{t}$:
	$\ell$ completed a call to $\DoOps{(\Ops^*,-)}{t}{k^*}$ in line~\ref{first-doops}.
		In this call to $\DoOps{(\Ops^*,-)}{t}{k^*}$,
		process $\ell$ set $\Batch[k^*]$
		to $(\Os_{k^*},-) = (\Ops^*,-)$ in line~\ref{batch1},
		and $\ell$ completed a call to $\EUB{k^*}$ in line~\ref{ExecuteB2}.
	
By Lemmas~\ref{aboutEUB} and~\ref{EUBreply},
	when $\ell$ returns from $\ExecuteOpsUpToBatch{k^*}$ and thereafter,
$\ell$ has:

	\begin{enumerate}[(I)]
	\item for all $i$, $1 \le i \le k^*$, $\Batch[i] = (\Os_i,-)$ for some non-empty set $\Os_i$, and
	\item $\Reply{\operation} \neq \bot$ for every $\operation \in \bigcup\limits_{i=1}^{k^*} \Os_i$.
	\end{enumerate}

From the above, it is clear that if $\hat{k} \le k^*$, then parts (\ref{oopla1}) and (\ref{oopla2}) of the lemma hold. 

Now assume that $\hat{k} > k^*$.
Since $\ell$ executes $\LeaderWork{t}$ forever,
	and a process does not execute $\PCM{()}$ concurrently with $\LeaderWork{}$,
	$\ell$ sets variable $k$ only in line~\ref{leader-accept}
	during the execution of $\LeaderWork{t}$.
Consider a time when $\ell$ first sets $k$ to $j$ for some $j \geq \hat k$ in line~\ref{leader-accept} during the execution of $\LeaderWork{t}$
	(such time exists since $\ell$ has $k = k^* < \hat k$ before entering the while loop of lines~\ref{mainwhile}-\ref{endwhile} of $\LeaderWork{t}$).
After $\ell$ sets $k$ to $j$ in line~\ref{leader-accept},
	it continues to call $\EUB{j}$ in line~\ref{ExecuteB2}.
Since $j \ge \hat{k}$, the lemma then follows from Lemmas~\ref{aboutEUB} and~\ref{EUBreply}.
\qedhere~$_\text{\autoref{zozox3}}$
\end{enumerate}
\end{proof}

\begin{lemma}\label{donata}
If a process $p$ has some operation $\op \in \OpsDone$ at some real time $\tau$,
	then there is a $j \ge 1$ and a set $\Os_j$ that contains
	$\op$ such that
	$p$ has $\Batch[j] = (\Os_j,-)$ at all real times $\tau' \ge \tau$.
\end{lemma}

\begin{proof}
Suppose a process $p$ has an operation $\op \in \OpsDone$ at real time $\tau$.
Since $\OpsDone$ is initialized to~$\emptyset$ at $p$,
	process $p$ added $\op$ to $\OpsDone$ by real time $\tau$. 
Since $p$ modifies $\OpsDone$ only in line~\ref{addB1}
	by executing the statement
	``$\OpsDone := \OpsDone \cup \BatchOps{i}$'',
	it is clear that $p$ added $\op$ to $\OpsDone$
	such that $p$ has $\Batch[j] = (\Os_j,-)$ for some $j \geq 1$ and some $\Os_j$ that contains $\op$ at some real time $\tau' \leq \tau$
	($j \neq 0$ since, by Corollary~\ref{ne-batch} and the fact that the initial value of $\BatchOps{0}$ is $\emptyset$, $\BatchOps{0}$ remains $\emptyset$ forever).
Since $p$ has $\Batch[j] = (\Os_j,-)$ for some $\Os_j \neq \emptyset$ at real time $\tau'$,
	by Corollary~\ref{finalsafetyac}, $p$ has $\Batch[j] = (\Os_j,-)$ at all real times $\tau'' \ge \tau'$,
	and hence at all real times $\tau'' \ge \tau$.
\qedhere~$_\text{\autoref{donata}}$
\end{proof}

\begin{lemma}\label{lm11}
No correct process executes the \emph{periodically send-until loop} of lines~\ref{rmw-send}-\ref{rmw-until} forever.
\end{lemma}

\begin{proof}
Suppose, for contradiction, that some correct process $p$ executes the loop of lines~\ref{rmw-send}-\ref{rmw-until} forever.
Let $\operation = (o, (p,\cntr))$ be the operation that $p$ has in line~\ref{rmw4},
	just before entering the periodically send-until loop.
Since $p$ is correct, by Assumption~\ref{Omega}, there is a time
	after which
	if $p$ calls $leader()$, this call returns~$\ell$.
Thus, since $p$ executes the loop of lines~\ref{rmw-send}-\ref{rmw-until} forever,
	$p$ sends $\langle \OpRequest, \operation \rangle$ to $\ell$ infinitely often.
Since the communication link between the two correct processes $p$ and $\ell$ is fair
	(Assumption~\ref{fl}),
	this implies that $\ell$ receives $\langle \OpRequest, \operation \rangle$ infinitely often from $p$ in line~\ref{rr1}.
	
Consider the variables $\OpsRequested$ and $\OpsDone$ of $\ell$.
By Observation~\ref{opsdoneismonotonic},
	each one contains a non-decreasing set of operations.
	
\begin{claim}\label{lm11c}
There is a real time after which $\ell$ has $\operation \in \OpsDone$.
\end{claim}
\begin{proof}
Suppose, for contradiction, that $\operation$ is never in $\OpsDone$.
When $\ell$ first receives \linebreak
	$\langle \OpRequest, \operation \rangle$ from $p$ in line~\ref{rr1},
	it adds $\operation$ to its set $\OpsRequested$ in line~\ref{rr2}.
Since $\OpsRequested$ is non-decreasing, and $\operation$ is never in $\OpsDone$,
	from now on $\ell$ has 
	$\operation \in \allowbreak \OpsRequested - \OpsDone$.

By Theorem~\ref{lm10},
	there is a local time $t$ such that
	(a) $\ell$ calls $\LeaderWork{t}$,
	(b) this call does not return, and
	(c) $\ell$ executes the while loop of lines~\ref{mainwhile}-\ref{endwhile} infinitely often in $\LeaderWork{t}$.
Note that in line~\ref{nextops} of this while loop, $\ell$ sets $\NextOps$ to $\OpsRequested - \OpsDone$.

Since there is a real time after which $\ell$ has $\operation \in \OpsRequested - \OpsDone$,
	$\ell$ executes the while loop of lines~\ref{mainwhile}-\ref{endwhile} infinitely often in $\LeaderWork{t}$
	with $\operation \in \NextOps$.
Consider the first such iteration.
Since $\operation \in \NextOps \neq \emptyset$ in line~\ref{nextops},
	$\ell$ calls $\DoOps{(\NextOps,-)}{t}{j}$ for some $j$ in line~\ref{second-doops}.
Note that this call
	returns \textsc{Done}
	(because if it returned \textsc{Failed},
	then $\ell$ would exit $\LeaderWork{t}$ in line~\ref{doops2failed}, but $\ell$ does not exit $\LeaderWork{t}$).
Since $\DoOps{(\NextOps,-)}{t}{j}$ returns \textsc{Done},
	process $\ell$ 
	sets $\Batch[j]$ to $(\NextOps,-)$ in line~\ref{batch1} 
	and calls $\EUB{j}$ in line~\ref{ExecuteB2}.
When $\ell$ returns from $\EUB{j}$,
	it executed ``$\OpsDone := \OpsDone \cup \BatchOps{j}$''
	(line~\ref{addB1}),
	and by Corollary~\ref{finalsafetyac},
	$\BatchOps{j} = \NextOps$.
This implies that
	$\ell$ has $\operation \in \OpsDone$
	after line~\ref{ExecuteB2},
	contradicting that $\operation$ is never in $\OpsDone$.
\qedhere~$_\text{\autoref{lm11c}}$
\end{proof}

By Claim~\ref{lm11c}, $\ell$ has $\operation \in \OpsDone$ at some real time $\tau$.
So, by Lemma~\ref{donata},
	there is a $j \ge 1$ and a set $\Os_j$
	such that $\operation \in \Os_j$ and $\ell$ has $\Batch[j] = (\Os_j,-)$ at time $\tau$.
Thus, by Lemma~\ref{IamOutOfLableNames},
	some process locked a tuple $(\Os_j, -, j)$.
So, by Lemma~\ref{zozo4}, there is a real time after which $\ell$ has $k \ge j$.
Therefore, by Lemma~\ref{zozox3}, there is a real time after which:
\begin{enumerate}
\item $p$ has $\Batch[j] = (\Os'_j,-)$  for some non-empty set $\Os'_j$, and
\item $p$ has $\Reply{\op} \neq \bot$ for every $\op \in \Os'_j$.
\end{enumerate}
Since $\ell$ has $\Batch[j] = (\Os_j,-)$ for some non-empty set $\Os_j$ and $p$ has $\Batch[j] = (\Os'_j,-)$ for some non-empty set $\Os'_j$,
	by Theorem~\ref{finalsafetyx}, $\Os_j = \Os'_j$.
So, since $\operation \in \Os_j$, 
	there is a real time after which process $p$ has $\Reply{\operation} \neq \bot$.
Thus $p$ eventually exits the while loop of lines~\ref{rmw-send}-\ref{rmw-until} --- a contradiction.
\qedhere~$_\text{\autoref{lm11}}$
\end{proof}

We now show that no correct process executes the wait statement in line~\ref{rmw-wait-promise} forever.

\begin{definition}\label{lock-with-promise}
A process \emph{locks a tuple $(\Os,t,j)$ with promise $\s$}
	if it locks the tuple during a call to $\DoOps{(\Os,\s)}{t}{j}$.
If some process locks a tuple with promise $\s$,
	we say that \emph{the tuple is locked with promise $\s$}.
\end{definition}

\begin{observation}\label{sets-promise-when-locking}
If a process locks a tuple $(\Os,t,j)$ with promise $\s$,
	then it sets $\Batch[j]$ to $(\Os,\s)$ in line~\ref{batch1}.
\end{observation}

\begin{lemma}\label{batch-promise-locked}
For $j\geq 0$, if a process sets $\Batch[j]$ to $(\Os,\s)$ at real time $\tau$,
	then some process locks a tuple $(\Os,-,j)$ with promise $\s$ by real time $\tau$.
\end{lemma}
\begin{proof}
Suppose, for contradiction, that there is a process $p$ that sets $Batch[j]$ to $(\Os, \s)$
	for some $j, \Os$ and $\s$ at real time $\tau$
	such that no process locks a tuple $(\Os, -, j)$ with promise $\s$ by real time $\tau$.
Without loss of generality, suppose that $p$ setting $Batch[j]$ to $(\Os, \s)$
	is the first time when
	any process sets $Batch[j]$ to $(\Os, \s)$~(*).
There are several cases, depending on where $p$ sets $Batch[j]$ to $(\Os, \s)$.

\begin{enumerate}
\item $p$ sets $\Batch[j]$ to $(\Os, \s)$ in line~\ref{batch1}.
	by Definition~\ref{lock-with-promise},
	$p$ locks a tuple $(\Os, -, j)$ with promise $\s$ at the same real time
	when $p$ sets $\Batch[j]$ to $(\Os, \s)$
	--- a contradiction to (*).

\item $p$ sets $\Batch[j]$ to $(\Os, \s)$ in line~\ref{batch2}.
Thus, $p$ received a $\langle \CommitLease, (\Os, \s), j, -, - \rangle$ message
	from some process $q$.
From the code
	the first such message
	was sent by $q$ in line~\ref{sl2} during the execution of
	$\DoOps{(\Os,\s)}{-}{j}$.
Before sending that message $q$ had executed line~\ref{batch1}
	and set $\Batch[j]$ to $(\Os,\s)$ ---
	a contradiction to~(*).

\item $p$ sets $\Batch[j]$ to $(\Os, \s)$ in line~\ref{batch3}.
Line~\ref{sendbatch} is the only place where a $\langle \MyBatch, j, (\Os, \s) \rangle$ message is sent.
From the code of lines~\ref{gb}-\ref{sendbatch},
	some process $q$ has $\Batch[j] = (\Os, \s) \neq (\emptyset, \infty)$ before
	sending a $\langle \MyBatch, j, (\Os, \s) \rangle$ to $p$
	for some $j > 0$.
So $q$ must previously set $\Batch[j]$ to $(\Os, \s)$
	--- a contradiction to (*).

\item $p$ sets $\Batch[j]$ to $(\Os, \s)$ in line~\ref{setBatch4}.
From the code of lines~\ref{estrep1}-\ref{setBatch4} and lines~\ref{a2b}-\ref{a2e},
	some process $q$ sent a $\langle \EstReply, t, \EstTuple,ts,j+1, Batch[j] \rangle$ message to $p$,
	and $q$ has $\Batch[j] = (\Os, \s)$ when sending this message.
Note that $q$ has $k = j + 1 > 0$,
	by Lemma~\ref{acceptsB}, $q$ previously set $\Batch[j]$.
So $q$ must have set $\Batch[j] = (\Os, \s)$ before sending the message
	$\langle \EstReply, t, \EstTuple,ts,j+1, Batch[j] \rangle$ to $p$
	--- a contradiction to (*).

\item $p$ sets $\Batch[j]$ to $(\Os, \s)$ in line~\ref{setBatch3}.
From the code of lines~\ref{a3b}-\ref{setBatch3} and \DO,
	it is clear that some process $q$ sent a $\langle \Prepare,-,-,j + 1, (\Os, \s) \rangle$
	message to $p$ in line~\ref{prep-send}.
Note that $q$ accepts a tuple $(-, -, j + 1)$ in line~\ref{leader-accept}
	before sending this $\Prepare$ message.
By Lemma~\ref{acceptsB}, $q$ previously set $\Batch[j]$.
So $q$ must have set $\Batch[j] = (\Os, \s)$
	before sending the $\Prepare$ message to $p$	
	--- a contradiction to (*).
\qedhere~$_\text{\autoref{batch-promise-locked}}$
\end{enumerate}
\end{proof}

\begin{observation}\label{finite-locked-promise}
If a process locks a tuple with promise $\s$,
	then $\s$ is finite.
\end{observation}

Lemma~\ref{batch-promise-locked} and Observation~\ref{finite-locked-promise} imply the following:

\begin{corollary}\label{finite-batch-promise}
For $j \geq 0$, if a process sets $\Batch[j]$ to $(-,\s)$,
	then $\s$ is finite.
\end{corollary}

\begin{observation}\label{first-do-no-promise}
If a process locks a tuple with promise $\s$ 
	during a call to $\DO$ made in line~\ref{first-doops},
	then $\s = 0$.
\end{observation}

The above observation implies the following:

\begin{corollary}\label{second-do-yes-promise}
If a process locks a tuple with promise $\s > 0$,
	then it does so during a call to $\DO$ made in line~\ref{second-doops}.
\end{corollary}

\begin{lemma}\label{second-do-increasing-j}
If a tuple $(-,t,j)$ is locked and
	some process calls $\DoOps{(-,-)}{t'}{j'}$ in line~\ref{second-doops} with some $t' > t$,
	then $j' > j$.
\end{lemma}
\begin{proof}
Suppose a tuple $(-,t,j)$ is locked and 
	some process $p$ calls $\DoOps{(-,-)}{t'}{j'}$
	in line~\ref{second-doops} with some $t' > t$.
Then, $p$ previously called $\DoOps{(\Ops^*,0)}{t'}{k^*}$ in line~\ref{first-doops},
	and this call returned \textsc{Done} (since $p$ continues to execute line~\ref{second-doops}).
Thus, during the call to $\DoOps{(\Ops^*,0)}{t'}{k^*}$,
	$p$ accepted the tuple $(\Ops^*, t', k^*)$ in line~\ref{leader-accepts}.
By Theorem~\ref{generalcase1a},
	$k^* \geq j$.
Since $p$ calls $\DoOps{(-,-)}{t'}{j'}$ after $\DoOps{(\Ops^*,0)}{t'}{k^*}$,
	by Corollary~\ref{T-k-DoopsCreduced},
	$j' > k^* \geq j$.
\qedhere~$_\text{\autoref{second-do-increasing-j}}$
\end{proof}

\begin{lemma}\label{uniq-second-do}
If tuples $(-,t,j)$ and $(-,t',j')$ are locked during calls to $\DO$ made in line~\ref{second-doops}
	and $t \neq t'$,
	then $j \neq j'$.
\end{lemma}
\begin{proof}
Suppose tuples $(-,t,j)$ and $(-,t',j')$ are locked 
	during calls to $\DO$ made in line~\ref{second-doops}
	such that $t \neq t'$.
Without loss of generality assume $t < t'$.
By Lemma~\ref{second-do-increasing-j},
	$j < j'$.
\qedhere~$_\text{\autoref{second-do-increasing-j}}$
\end{proof}

\begin{lemma}\label{uniq-positive-promise-locking}
Suppose tuples $(-,t,j)$ and $(-,t',j)$ are locked
	with promises $s$ and $s'$ respectively
	during calls to $\DO$ made in line~\ref{second-doops}.
Then $t' = t$ and $s' = s$.
\end{lemma}
\begin{proof}
Suppose tuples $(-,t,j)$ and $(-,t',j)$ are locked
	with promises $s$ and $s'$ respectively
	during calls to $\DO$ made in line~\ref{second-doops}.
By definition, 
	the two tuples are locked in calls to 
	$\DoOps{(-,s)}{t}{j}$ and $\DoOps{(-,s')}{t'}{j}$ respectively.
By Lemma~\ref{uniq-second-do}, 
	$t' = t$.
By Corollary~\ref{T-k-DoopsCreduced},
	these two $\DO$ calls are made in $\LeaderWork{t}$.
By Lemma~\ref{unique-LeaderWork},
	these two $\DO$ calls are made by the same process,
	and by Corollary~\ref{T-k-DoopsCreduced}, 
	these two calls are the same call.
So $s' = s$.
\qedhere~$_\text{\autoref{uniq-positive-promise-locking}}$
\end{proof}

Lemma~\ref{batch-promise-locked}, 
	Corollary~\ref{second-do-yes-promise} and
	Lemma~\ref{uniq-positive-promise-locking}
	imply the following:
\begin{corollary}\label{uniq-positive-promise-batch}
For $j \geq 0$, if processes $p$ and $p'$ sets $Batch[j]$ to $(-,s)$ and $(-,s')$ respectively 
	such that $s > 0$ and $s' > 0$,
	then $s' = s$.
\end{corollary}

\begin{observation}\label{set-batch-and-promise}
When a process sets $\BatchOps{j}$, it also sets $\BatchPromise{j}$.
\end{observation}

\begin{lemma}\label{batch-promise-upperbound}
If a process $p$ sets $\BatchPromise{j}$ to some $s > 0$, then
	$p$ has $\BatchPromise{j} \leq \s$ thereafter.
\end{lemma}
\begin{proof}
Suppose that some process $p$ sets $\BatchPromise{j}$ to some $s > 0$,
	and $p$ later sets $\BatchPromise{j}$ to some $s'$.
If $s' \le 0$, then we have $s' < s$. 
Henceforth we assume $s' > 0$.
By Corollary~\ref{uniq-positive-promise-batch}, $s' = s$.
So if $\s > 0$,
	$p$ has $\BatchPromise{j} \leq \s$ after it sets $\BatchPromise{j}$ to $\s$.
\qedhere~$_\text{\autoref{batch-promise-upperbound}}$
\end{proof}

\begin{lemma}\label{promise-infinity}
If a process $p$ sets $\TakesEffect{\operation}$ to some $\s$, then
\begin{enumerate}
	\item $s \neq \infty$, and
	\item If $s > 0$, then
		$p$ has $\TakesEffect{\operation} \leq \s$ thereafter.
\end{enumerate}
\end{lemma}
\begin{proof}
Suppose a process $p$ sets $\TakesEffect{\operation}$ to some $\s$ for some RMW operation $\operation$.
Note that this happens in line~\ref{set-takesEffect} of $\ExecuteBatch{j}$ for some $j$,
	and $p$ has $\Batch[j] = (\Os_j, \s_j)$ for some $\Os_j$ that contains $\operation$ and
	some $\s_j$ in line~\ref{EB-get-ops}.
Since $(\Os_j,\s_j) \neq (\emptyset, \infty)$, 
	$p$ must previously set $\Batch[j]$ to $(\Os_j,\s_j)$.
Since $p$ sets $\TakesEffect{\operation}$ to $\BatchPromise{j}$,
	the lemma now follows from
	Corollary~\ref{finite-batch-promise} and
	Lemma~\ref{batch-promise-upperbound}.
\qedhere~$_\text{\autoref{promise-infinity}}$
\end{proof}

\begin{lemma}\label{finite-waiting-for-takes-effect}
No correct process executes the \emph{wait statement} of line~\ref{rmw-wait-promise} forever.
\end{lemma}
\begin{proof}
Suppose, for contradiction, that a correct process $p$ executes the \emph{wait} statement of line~\ref{rmw-wait-promise} forever.
Let $\operation$ be the operation that $p$ has in line~\ref{rmw4}.
Then, it is clear that $p$ found $\Reply{\operation} \neq \bot$ in line~\ref{rmw-until} before executing line~\ref{rmw-wait-promise}.
Since $p$ is correct and the only place where $\Reply{\operation}$ is set is in line~\ref{apply-op-1},
	$p$ continues to set $\TakesEffect{\operation}$ in line~\ref{set-takesEffect}.
By Lemma~\ref{promise-infinity}, $p$ sets $\TakesEffect{\operation}$ to some $\s \neq \infty$,
	and if $s > 0$, $p$ has $\TakesEffect{\operation} \leq s$ thereafter.
Thus, by Assumption~\ref{xclocks}(\ref{xcl2}-\ref{xcl3}), 
	there is a real time after which the local clock at $p$ has value at least $\TakesEffect{\operation}$, 
	so $p$ does not execute line~\ref{rmw-wait-promise} forever --- a contradiction.
\qedhere~$_\text{\autoref{finite-waiting-for-takes-effect}}$
\end{proof}

If a correct process invokes a read-modify-write operation $o$ on the distributed object,
	then $p$ eventually returns with a non-$\bot$ response.
More precisely:

\begin{theorem}\label{finally}
If a correct process $p$ invokes a \emph{read-modify-write operation} $\operation = (o, (p,\cntr))$
	then it eventually returns with some $\Reply{\operation} \neq \bot$.
\end{theorem}

\begin{proof}
This follows directly from the code of lines~\ref{rmw2}-\ref{rmw-return} of \textsc{Thread 1}, Lemma~\ref{lm11} and~\ref{finite-waiting-for-takes-effect}, and Observation~\ref{replybot}.
\qedhere~$_\text{\autoref{finally}}$
\end{proof}

\subsection{Read lease mechanism: basic properties}\label{readleaseBasicProps}

\begin{lemma}\label{same-leadership}
Suppose $p$ and $q$ call $\AL(\tp,\tpp)$ and $\AL(\tq,\tqq)$,
	and both these calls return $\textsc{True}$.
If the intervals $[\tp,\tpp]$ and $[\tq,\tqq]$ intersect,
	then $p=q$ and $\tp=\tq$.
\end{lemma}

\begin{proof}
Suppose, $p$ and $q$ call $\AL(\tp,\tpp)$ and $\AL(\tq,\tqq)$,
	both these calls return $\textsc{True}$, and the intervals
	$[\tp,\tpp]$ and $[\tq,\tqq]$ intersect.
By Theorem~\ref{leader-safety}, $p=q$.
It remains to show that $\tp=\tq$.

Suppose, for contradiction, that $\tp\ne\tq$.
Without loss of generality, assume that $\tp<\tq$.
Since the two intervals intersect, $\tpp > \tp$.
Clearly, $p$ calls $\AL(\tp,\tpp)$ in $\LW(\tp)$,
	and calls $\AL(\tq,\tqq)$ in either line~\ref{check} or in $\LW(\tq)$.
So $p$ must get $\tq$ from its clock at line~\ref{check} at some time.
Since $p$ becomes leader at local time $\tp$,
	by Assumptions~\ref{xclocks}(\ref{xcl2}),
	$p$ reads $\tq$ from its clock at line~\ref{check} after it exits from $\LW(\tp)$.
Since $p$ gets $\tpp$ from its clock inside $\LW(\tp)$,
	by Assumption~\ref{xclocks})(\ref{xcl4}),
	$p$ gets $\tq > \tpp$ from its clock at line~\ref{check}.
Therefore, the intervals $[\tp,\tpp]$ and $[\tq,\tqq]$ do not intersect --- a contradiction.
\qedhere~$_\text{\autoref{same-leadership}}$
\end{proof}

In the following, we use (local clock, real time clock) pairs to time events:

\begin{definition}
We say that an event occurs \emph{at time $(t_i,\tau_i)$ at a process $p$},
	if it occurs at $p$ at real time $\tau_i$, and $p$ has $\CT = t_i$ at real time $\tau_i$.
\end{definition}

We previously defined what it means for a process $\ell$ to become leader at \emph{local} local time $t$
	(Definition~\ref{bcmleader}).
We now extend this definition to say what it means for $\ell$ to become leader \emph{at time $(t,\tau)$},
	where $t$ is a \emph{local clock time}, and $\tau$ is a \emph{real time}.

\begin{definition}\label{bcmleader2}
A process $\ell$ {becomes leader at time $(t,\tau)$} if:
\begin{enumerate}
\item $\ell$ gets the value $t$ from its $\CT$ at real time $\tau$ in line~\ref{gt}, and
\item $\ell$ calls $\ML{t}{t}$, finds that $\ML{t}{t} = True$, and calls $\LeaderWork{t}$ in line~\ref{check}.
\end{enumerate}
\end{definition}

\begin{definition}\label{otherbcmleader2}
If a process $\ell$ becomes leader at time $(t, \tau)$, we also say that:
\begin{enumerate}
\item $\ell$ becomes leader at local time $t$, and
\item $\ell$ becomes leader at real time $\tau$.
\end{enumerate}
\end{definition}

\begin{observation}\label{LeaderWork2}
	If a process calls $\LeaderWork{t}$ then it becomes leader at time $(t,\tau)$ for some real time~$\tau$.
\end{observation}

Similarly, we previously defined what it means for a process $p$ to lock a tuple $(\Os,t,j)$ (Definition~\ref{locking}).
We now extend this definition to say what it means for $p$ to lock $(\Os,t,j)$ \emph{at time $(t',\tau')$},
	where $t'$ is a \emph{local clock time}, and $\tau'$ is a \emph{time}.

\begin{definition}\label{lockingwt}
A process $p$ \emph{locks a tuple $(\Os, t, j)$} at time $(t', \tau')$ 
	if $p$ executes $\DoOps{(\Os,-)}{t}{j}$ up to line~\ref{batch1}
	such that $\tau'$ is the real time when $p$ executes line~\ref{batch1} and $t' = \CT_p(\tau')$.
\end{definition}

\begin{definition}\label{otherlockingwt}

If a process $p$ locks $(\Os,t,j)$ at time some $(t',\tau')$, we also say that:

\begin{itemize}

\item	$p$ locks $(\Os,t,j)$ at local time $t'$.
\item	$p$ locks $(\Os,t,j)$ at real time $\tau'$.

\end{itemize}
	
\end{definition}

\begin{definition}\label{lockingwt-promise}
	A process $p$ \emph{locks a tuple $(\Os, t, j)$ with promise $\s$} at time $(t', \tau')$ 
		if $p$ executes $\DoOps{(\Os,\s)}{t}{j}$ up to line~\ref{batch1}
		such that $\tau'$ is the real time when $p$ executes line~\ref{batch1} and $t' = \CT_p(\tau')$.
\end{definition}

\begin{definition}\label{otherlockingwt-promise}

	If a process $p$ locks $(\Os,t,j)$ with promise $\s$ at time some $(t',\tau')$, we also say that:
	
	\begin{itemize}
	
	\item	$p$ locks $(\Os,t,j)$ with promise $\s$ at local time $t'$.
	\item	$p$ locks $(\Os,t,j)$ with promise $\s$ at real time $\tau'$.

	\end{itemize}
\end{definition}

\begin{definition}\label{lease-issue}

~
\begin{itemize}

\item A process $p$ \emph{issues a lease $(j,\tj)$ at some time $(\tjj, \tau'')$}
	if $p$ sets its $\lease$ variable to $(j, \tj)$ in line~\ref{setlease} or line~\ref{lease1} at real time $\tau''$
	and $\tjj = \CT_p(\tau'')$.

\item A process $p$ \emph{issues a lease $(j,\tj)$ in $\LeaderWork{t}$}
	if $p$ sets its $\lease$ variable to $(j,\tj)$ during  $p$'execution of $\LeaderWork{t}$.

\end{itemize}

\end{definition}

\begin{definition}\label{otherlease-issue}

If a process $p$ issues a lease $(j,t')$ at time $(\tjj,\tau'')$, we also say that:

\begin{enumerate}

\item	$p$ issues the lease $(j,t')$ at local time $\tjj$.
\item	$p$ issues the lease $(j,t')$ at real time $\tau''$.

\end{enumerate}

\end{definition}

\begin{observation}\label{lock-implies-lease}
If a process $p$ locks a tuple $(\Os,t,j)$ with promise $\s$ at time $(t', \tau')$,
	then it also issues a lease $(j,s)$ at time $(t', \tau')$.
\end{observation}

\begin{observation}\label{lease-in-doops-implies-lock}
If a process $p$ issues a lease $(j,\s)$ at time $(t', \tau')$ in line~\ref{lease1} in $\DoOps{(\Os,\s)}{t}{j}$,
	then it also locks a tuple $(\Os, t, j)$ with promise $\s$ at time $(t', \tau')$.
\end{observation}

If a process $p$ issues a lease $(j,\tj)$ in $\LeaderWork{t}$,
	then $p$ locked some tuple $(\Os, t,j)$ and this is
	the last tuple that $p$ locks before issuing this lease.
More precisely:

\begin{lemma}\label{lease-issue-obs}
Suppose a process $p$ issues a lease $(j,\tj)$ at real time $\tau$ in $\LeaderWork{t}$.
Then $p$ locks some tuple $(\Os,t,j)$ at some real time $\tau' \le \tau$ such that
	$p$ does not lock any tuple at real time $\hat \tau$ where $\tau' < \hat \tau \le \tau$.
\end{lemma}

\begin{proof}
Suppose $p$ issues a lease $(j,\tj)$ at real time $\tau$ in $\LeaderWork{t}$.
There are two possible cases:

\begin{enumerate}
\item Process $p$ issues the lease $(j,\tj)$ at real time $\tau$ in line~\ref{batch1} of $\LeaderWork{t}$.
Thus $p$ executes line~\ref{batch1} of \linebreak
	$\DoOps{(\Os,-)}{t}{j}$ 
	for some set~$\Os$,
	and $p$ locks $(\Os,t,j)$ at real time $\tau$ in line~\ref{batch1}.
So the result holds for $\tau' = \tau$.

\item Process $p$ issues the lease $(j,\tj)$ at real time $\tau$ in line~\ref{setlease} of $\LeaderWork{t}$.
From the code in line~\ref{setlease}, $p$ has $k =j$ at time $\tau$.
Since $p$ issues the lease in line~\ref{setlease},
	it has previously successfully completed at least one
	$\DoOps{(-, -)}{t}{-}$ in $\LeaderWork{t}$.
Let $\DoOps{(\Os,-)}{t}{j'}$ be the \emph{last} $\DoOps{(-, -)}{t}{-}$ that $p$ executes
	before issuing the lease $(j,\tj)$ in line~\ref{setlease} of $\LeaderWork{t}$.
During this execution of $\DoOps{(\Os,-)}{t}{j'}$,
	$p$ first sets its variables $(\Ops,ts,k)$ to $(\Os,t,j')$
	in line~\ref{leader-accept},
	and then it locks $(\Os,t,j')$ at some real time $\tau'$.
Since $\DoOps{(\Os,-)}{t}{j'}$ is the last $\DoOps{(-,-)}{t}{-}$
	that $p$ executes \emph{before} issuing the lease $(j,\tj)$ in line~\ref{setlease},
	$p$ still has $(\Ops,ts,k) = (\Os,t,j')$ at real time $\tau$,
	and $\tau' < \tau$.
Since $p$ has $k =j$ at time $\tau$, $j' =j$.
Moreover, since $\DoOps{(\Os,-)}{t}{j'}$ is the \emph{last} $\DoOps{(-,-)}{t}{-}$ that 
	$p$ executes before issuing the lease $(j,\tj)$ at time $\tjj$,
	$p$ does not lock any tuple at real time $\hat \tau$ such that $\tau' < \hat \tau \le \tau$.
\qedhere~$_\text{\autoref{lease-issue-obs}}$
\end{enumerate}
\end{proof}

\begin{lemma}\label{OnLH}
At each process $p$, the variable $\LH$ is a set of processes that does not contain~$p$.
\end{lemma}

\begin{proof}
Consider the variable $\LH$ at some process $p$.
Initially, $\LH$ equals to $\emptyset$.
Note that $p$ updates $\LH$ only in lines~\ref{lh1}, \ref{lh3} and~\ref{lh2} of the algorithm.
It is obvious that $p$ does not add~$p$ to $\LH$ in line~\ref{lh1}.
We claim that $p$ does not add $p$  to $\LH$ in line~\ref{lh2}.
To see this, note that in line~\ref{lh2}, $p$ sets $\LH$ to some set $\PACKED[t,j]$,
	and it is easy to see that $\PACKED[t,j]$ never contains $p$: in fact,
	$\PACKED[t,j]$
	contains processes that replied to a $\langle \Prepare,-,t,j,-\rangle$ message that they received from $p$,
	but $p$ does not send any $\langle \Prepare,-,-,-,-\rangle$ message to itself.
Finally we claim that $p$ does not add $p$  to $\LH$ in line~\ref{lh3}.
To see this, note that:
	(1) $p$ adds $q$ to $\LH$ in line~\ref{lh3} only if
	it receives a  $\langle \RequestLease \rangle$ from $q$,
	(2) $q$ sends a $\langle \RequestLease \rangle$ to $p$ only
	if it receives a $\langle \CommitLease,-,-,-,- \rangle$ message from $p$ in lines~\ref{lg0}-\ref{lg3}, and 
	(3) $p$ never sends a $\langle \CommitLease,-,-,-,- \rangle$ to itself (see lines~\ref{sl1} and \ref{sl2});
	so $p$ never sends a $\langle \RequestLease \rangle$ to itself.
Since initially $p \not \in \LH$, and $p$ does not add $p$ to $\LH$ in lines~\ref{lh1}, \ref{lh3} and~\ref{lh2},
	$\LH$ never contains $p$.
\qedhere~$_\text{\autoref{OnLH}}$
\end{proof}

\begin{lemma}\label{removingfromLH}
Suppose a process $p$ has $q \in \LH$ at real time $\tau_1$ and
	$q \not \in \LH$ at real time $\tau_2 > \tau_1$
	during the execution of $\LeaderWork{t}$ for some $t$.
If real time $\tau_1$ is after the real time when $p$ executes line~\ref{lh1} in $\LeaderWork{t}$,
	then there exists a real time $\hat \tau$, where $\tau_1 < \hat \tau \le \tau_2$,
	such that all of the following hold:

\begin{enumerate}[\noindent(1)]
\item\label{remv0} $p$ executes line~\ref{lh2} at real time $\hat \tau$,
\item\label{remv1} $p$ has $q \in \LH$ just before line~\ref{lh2},
\item\label{remv2} $p$ has $q \not \in \LH$ just after line~\ref{lh2},
\item\label{remv4} if $p$ executes this line~\ref{lh2} in $\DoOps{(-,s)}{t}{-}$ for some $s$ such that
	$p$ finds $s < \leasetime + \LP$ in line~\ref{LHvsAcks}, 
	then at real time $\hat \tau$, $p$ has $\CT_p \geq \leasetime + \LP$,
	where $\leasetime$ is evaluated by $p$ in line~\ref{wait-alg1}.
\end{enumerate}

\end{lemma}

\begin{proof}
Suppose a process $p$ has $q \in \LH$ at real time $\tau_1$,
	where real time $\tau_1$ is after $p$ executes line~\ref{lh1}, 
	and $q \not \in \LH$ at real time $\tau_2 > \tau_1$
	during the execution of $\LeaderWork{t}$.
Let $\hat \tau$ be the smallest real time greater than $\tau_1$
	such that $p$ has $q \not \in \LH$ at real time~$\hat \tau$.
Clearly, $\tau_1 < \hat \tau \le \tau_2$.
Note that the statements in line~\ref{lh1},~\ref{lh2} and~\ref{lh3} are the only ones that modify the content of $\LH$ at $p$.
Since (i)~real time $\tau_1$ is after when $p$ executes line~\ref{lh1},
	(ii)~the statement in line~\ref{lh3} can only \emph{add} processes to $\LH$, and
	(iii)~real time $\tau_2$ is during $p$'s execution of $\LeaderWork{t}$ 
	so line~\ref{lh1} is not executed between time $\tau_1$ and $\tau_2$,
	$p$ executed line~\ref{lh2} at real time $\hat \tau$ and 
	this execution results in $q \not \in \LH$.
By definition of $\hat \tau$, $q \in \LH$ just before the execution of line~\ref{lh2} at real time $\hat \tau$.
Thus Parts~(\ref{remv0}), (\ref{remv1}) and (\ref{remv2}) of the lemma hold.

Now suppose that $p$ executes line~\ref{lh2} during the execution of $\DoOps{(-,s)}{t}{j}$
	for some $j$ and $s$
	such that $p$ finds $s < \leasetime + \LP$ in line~\ref{LHvsAcks}.
Since $q \in \LH$ just before line~\ref{lh2}, 
	$p$ also has $q \in \LH$ when it executes line~\ref{LHvsAcks}.
Note that in line~\ref{lh2}, $p$ sets $\LH$ to a set $\PACKED[t,j]$.
Since $q \not \in \LH$ just after line~\ref{lh2}, then $q \not \in \PACKED[t,j]$ in line~\ref{lh2}.
Since $\PACKED[t,j]$ is non-decreasing
	(processes are never removed from $\PACKED[t,j]$)
	it must be that $q \not \in \PACKED[t,j]$ also in line~\ref{LHvsAcks}.
Thus, when $p$ executes line~\ref{LHvsAcks}, it has
	$q \in \LH$ and $q \not \in \PACKED[t,j]$,
	so
	$\LeaseHolders \subseteq \PACKED[t,j]$ does \emph{not} hold.

Therefore $p$ executes the \textbf{wait} statement of line~\ref{wait-alg1}.
When $p$ completes this wait, it has
	$\CT_p \geq \leasetime + \LP$.
Since $\CT_p$ is non-decreasing, 
	when $p$ executes line~\ref{lh2} at real time $\hat \tau$ after line~\ref{wait-alg1}, 
	$p$ still has $\CT_p \geq \leasetime + \LP$.
Thus Part (\ref{remv4}) of the lemma also holds.
\qedhere~$_\text{\autoref{removingfromLH}}$
\end{proof}

\begin{lemma}\label{LimpliesP}
Suppose a process $p$ locks a tuple $(\Os_i,t_i,i)$ with promise $s_i$ at real time $\tau$.
If $p$ has $q \in \LH$ at real time $\tau$
	then from real time $\tau$ on the following holds at $q$:
	\begin{enumerate}
		\item $\PendingBatchOps{i} = \Os_i$,
		\item $\PendingBatchPromise{i} = \s_i$ or $0$, and
		\item $\MaxPendingIndex \ge i$.
	\end{enumerate}
\end{lemma}

\begin{proof}
Suppose $p$ locks $(\Os_i, t_i ,i)$ 
	with promise $\s_i$ at real time $\tau$,
	and $p$ has $q \in \LH$ at real time $\tau$.
By Lemma~\ref{OnLH}, $p \neq q$.
Note that at real time $\tau$,
	$p$~is in line~\ref{batch1} of the $\DoOps{(\Os_i,\s_i)}{t_i}{i}$ procedure.
Since $p$ has $q \in \LH$ in line~\ref{batch1},
	and $p$ set $\LH$ to $\PACKED[t,i]$ in line~\ref{lh2},
	$p$ has $q \in \PACKED[t,i]$ in line~\ref{lh2}.
So $p$ has $q \in \PACKED[t,i]$ by real time $\tau$.
Thus, $q$ sent a $\langle \PACK,t,i \rangle$ to $p$ in line~\ref{sendPack} by real time $\tau$.

\begin{claim}\label{umpadumpa1}
By real time $\tau$:

\begin{enumerate}[(1)]
	\item\label{h} $q$ accepted $(\Os_i,t_i,i)$ in line~\ref{client-accept},
	\item\label{j} $q$ set $\PendingOps[i]$ to $(\Os_i,\s_i)$ in line~\ref{setPB}, and
	\item\label{k} $q$ set $\MaxPendingIndex$ to $\max (\MaxPendingIndex, i)$ in line~\ref{setMPB}.
\end{enumerate}
\end{claim}

\begin{proof}
Since $q$ sent a $\langle \PACK,t,i \rangle$ message to $p$ by real time $\tau$ in line~\ref{sendPack},
	it is clear that $q$ previously
	received a $\langle \Prepare,(\Os'_i,\s'_i),t_i,i,-\rangle$ message for some $\Os'_i$ and $\s'_i$ from $p$ in line~\ref{a3b},
	and that $q$ has $(\Ops,ts,k) = (\Os'_i,t_i,i)$ in line~\ref{sendPack}.
We claim that $(\Os'_i,\s'_i) = (\Os_i, \s_i)$.
To see this, note that $p$ sent $\langle \Prepare,(\Os'_i,\s'_i),t_i,i,-\rangle$ during an execution of $\DoOps{(\Os'_i, \s')}{t_i}{i}$.
Since $p$ calls both $\DoOps{(\Os_i, \s_i)}{t_i}{i}$ and $\DoOps{(\Os'_i, \s'_i)}{t_i}{i}$, by Lemma~\ref{doops-simplecase}, $(\Os'_i, \s'_i) = (\Os_i,\s_i)$.
So, $q$ has $(\Ops,ts,k) = (\Os_i, t_i,i)$ in line~\ref{sendPack} by real time $\tau$.
Since $p$ became leader at local time $t_i$, $t_i \neq -1$.
Thus $(\Os_i, t_i,i)$ is not the initial value of $(\Ops,ts,k)$ at~$q$.
Therefore $q$ accepted $(\Os_i, t_i,i)$ 
	before sending $\langle \PACK,t,i \rangle$ to $p$.
Note that only a process that becomes leader at local time $t_i$, i.e., only process $p$,
	can accept $(\Os_i, t_i,i)$ in line~\ref{leader-accept} of $\DoOps{(\Os_i, \s_i)}{t_i}{i}$.
Thus, since $q \neq p$, process $q$ accepted $(\Os_i, t_i,i)$ in line~\ref{client-accept}.
From the code of lines~\ref{client-accept}-\ref{sendPack},
	after accepting $(\Os_i, t_i,i)$ in line~\ref{client-accept},
	$q$ set $\PendingOps[i]$ to $(\Os_i,\s_i)$ in line~\ref{setPB} and
	$\MaxPendingIndex$ to $\max (\MaxPendingIndex, i)$ in line~\ref{setMPB},
	and then it sent $\langle \PACK,t_i,i \rangle$ to $p$ by real time $\tau$ in line~\ref{sendPack}.
\qedhere~$_\text{\autoref{umpadumpa1}}$
\end{proof}

Now suppose that after $q$ sets $\PendingOps[i]$ to $(\Os_i, \s_i)$ (i.e., after event (\ref{j}) above),
	it later resets $\PendingOps[i]$ to some $(\Os_j, \s_j)$.
We claim that $\Os_j = \Os_i$ and $\s_j = 0$.
We first show that $\Os_j = \Os_i$.
To see this, note that $q$ sets the variable $\PendingOps[i]$
	only in line~\ref{setPB}.
This implies that $q$ sets $\PendingOps[i]$ to $(\Os_j, \s_j)$ in line~\ref{setPB},
	and, just before doing so, $q$ accepts some tuple $(\Os_j, t_j ,i)$ in line~\ref{client-accept}.
Since $q$ accepted $(\Os_i, t_i ,i)$ before setting $\PendingOps[i]$ to $(\Os_i, \s_i)$,
	it is clear that $q$ accepted $(\Os_i, t_i ,i)$ before accepting $(\Os_j, t_j ,i)$.
By Lemma~\ref{T-k-increase},
	$(t_j ,i) > (t_i ,i)$, and so $t_j > t_i$.
Since $(\Os_i, t_i ,i)$ is locked and $(\Os_j, t_j ,i)$ is accepted, and $t_j > t_i$,
	by Theorem~\ref{generalcase1a}(\ref{t2}), $\Os_j = \Os_i$.

We now show that $\s_j = 0$.
Since $q$ accepts the tuple $(\Os_i, t_i ,i)$ in line~\ref{client-accept}
	and sets $\PendingOps[i]$ to $(\Os_i,\s_j)$ in line~\ref{setPB},
	it must received a $\langle \Prepare,(\Os_i,\s_j),t_j,i,- \rangle$ message
	sent by some process $r$ during a call to $\DoOps{(\Os_i,\s_j)}{t_j}{i}$.
Note that this must be the first $\DO$ call made by $r$ in $\LeaderWork{t_j}$,
	since otherwise, $r$ must have successfully completed
	a call to $\DoOps{(-,-)}{t_j}{i-1}$
	during which it accepted the tuple $(-,-,i-1)$ 
	--- a contradiction to Theorem~\ref{generalcase1a}(\ref{t1}).
Since $\DoOps{(\Os_i,\s_j)}{t_j}{i}$ is the first $\DO$ call made by $r$ in $\LeaderWork{t_j}$,
	$r$ does so in line~\ref{first-doops} and it is clear that $\s_j = 0$.

The claim that we just proved implies that $q$ has 
	$\PendingBatchOps{i} = \Os_i$ and \linebreak
	$\PendingBatchPromise{i} = \s_i$ or $0$
	from real time $\tau$ on.

Note the statement $\MaxPendingIndex := \max (\MaxPendingIndex, k)$ of line~\ref{setMPB} is the only one that changes the variable
	$\MaxPendingIndex$, thus the value of $\MaxPendingIndex$ is non-decreasing.
So after $q$ sets $\MaxPendingIndex$ to $\max (\MaxPendingIndex, i) \ge i$,
	i.e., after the event (\ref{k}) above that occurs by time~$\ti$,
	$\MaxPendingIndex \ge i$ forever.
\qedhere~$_\text{\autoref{LimpliesP}}$
\end{proof}

\begin{lemma}\label{LeaseIssue}
Suppose a process $q$ has $\lease = (j,\tj) \neq (0,-\infty)$ at real time $\tau$.
Then there is some process $r$ and a real time $\tau' \le \tau$ such that 
\begin{enumerate}
	\item $r$ issues the lease $(j,\tj)$ at real time $\tau'$, and
	\item if $r \neq q$ then $r$ has $q \in \LH$ at real time $\tau'$.
\end{enumerate}
\end{lemma}

\begin{proof}
Suppose process $q$ has $\lease = (j,\tj) \neq (0,-\infty)$ at real time $\tau$,
	so $(j,\tj)$ is not the initial value of $\lease$ at $q$.
Thus $q$ sets its $\lease$ to $(j,\tj)$ in line~\ref{setlease},~\ref{lease1} or~\ref{lg1},
	at some real time $\hat \tau \le \tau$.
If $q$ sets $\lease$ to $(j,\tj)$ in line~\ref{setlease} at time $\hat \tau$,
	then, by definition, $q$ issues the lease $(j,\tj)$ at time $\tau' = \hat \tau \leq \tau$,
	so $q = r$ in this case.
If $q$ sets $\lease$ to $(j,\tj)$ in line~\ref{lease1} at time $\hat \tau$,
	then, similarly, $q$ issues the lease $(j,\tj)$ at time $\tau' = \hat \tau \leq \tau$,
	and $q = r$ in this case.
Now, if $q$ sets $\lease$ to $(j, \tj)$ in line~\ref{lg1} at time $\hat \tau$,
	then $q$ previously received a $\langle \CommitLease, -, -, \lease', \LeaseHolders' \rangle$ message
	with $\lease' =  (j,\tj)$ and $q \in \LH'$ from some process $r \neq q$
	($r \neq q$ because no process sends a $\langle \CommitLease, -, -, -, - \rangle$ message to itself).
Note that $r$ sent this message in line~\ref{sl1} or line~\ref{sl2}.
If $r$ sent this message in line~\ref{sl1}, 
	then it issued the lease at time $\tau' \leq \hat \tau$ in line~\ref{setlease}.
If $r$ sent this message in line~\ref{sl2},
	then it issued the lease at time $\tau' \leq \hat \tau$ in line~\ref{lease1}.
For both cases, $r$ had $q \in \LH$ when it sent this $\RenewLease$ message to $q$.
Since $r$ does not modify $\LH$ in lines~\ref{setlease}-\ref{sl1} or in lines~\ref{lease1}-\ref{sl2},
	$r$ has $q \in \LH$ in line~\ref{setlease} or in line~\ref{lease1},
	so $r$ has $q \in \LH$ at time~$\tau'$.
\qedhere~$_\text{\autoref{LeaseIssue}}$
\end{proof}

\begin{lemma}\label{lease-expire-after-wait}
Suppose a process p executes $\LeaderWork{t}$
	and completes the \textbf{wait} statement in line~\ref{wait-lease-expire} at real time $\tau$.
Then for all leases $(j, t')$ issued in $\LeaderWork{t''}$ where $t'' < t$,
	$t' + \LP \leq \CT_p(\tau)$, i.e, all such leases are expired at process $p$ at real time $\tau$.
\end{lemma}
\begin{proof}
Suppose that a process $p$ calls $\LeaderWork{t}$ and 
	completes the \textbf{wait} statement in line~\ref{wait-lease-expire}
	at real time $\tau$.
Since $p$ gets $t$ from its $\CT$ in line~\ref{gt} and
	$p$ executes line~\ref{wait-lease-expire} after line~\ref{gt},
	$\CT_p(\tau) \geq t + \LP + \PP$.\footnote{Recall that $\PP$ is the parameter we called $\Pthree$ in Sections~\ref{sec-intro} and~\ref{sec-algo}.}
Now suppose a process $q$ issues a lease $(j, \tj)$ in $\LeaderWork{t''}$ and $t'' < t$.
There are two cases depending on where $q$ issues the lease:
\begin{enumerate}
\item $q$ issues this lease in line~\ref{setlease}.
From the code of lines~\ref{taketime1}-\ref{setlease},
	$q$ first got $\tj$ from its $\CT$ in line~\ref{taketime1},
	evaluated $\ML{t''}{\tj}$ to \textsc{True} in line~\ref{recheck1}
	and then issued the lease $(j, \tj)$ in line~\ref{setlease}.
We claim that $\tj < t$. 
Suppose, for contradiction, that $\tj \geq t$.
Since $p$ calls $\LeaderWork{t}$, 
	it calls $\ML{t}{t}$ in line~\ref{check} and this call returns True.
Since $t'' < t \leq \tj$, $[t,t]$ intersects $[t'', \tj]$.
Thus, by Lemma~\ref{same-leadership},
	we have $t = t''$, which contradicts the assumption that $t'' < t$.
Thus, $\tj + \LP < t + \LP \leq t + \LP + \PP \leq \CT_p(\tau)$.

\item $q$ issues this lease in line~\ref{lease1}.
Suppose that $q$ issues this lease in $\DoOps{(-,\tj)}{t''}{j}$.

If $q$ calls $\DoOps{(-,\tj)}{t''}{j}$ in line~\ref{first-doops},
	then $\tj = 0$.
Since $t''<t$, we have that $t>0$ and so
	$\tj + \LP = \LP < t + \LP + \PP \leq \CT_p(\tau)$.

Suppose $q$ calls $\DoOps{(-,\tj)}{t''}{j}$ in line~\ref{second-doops}.
Then, from the code in lines~\ref{taketime1}-\ref{second-doops},
	$q$ records $t^*$ from its $\CT$ in line~\ref{taketime1},
	calls $\ML{t''}{t^*}$ in line~\ref{recheck1}, which returns \textsc{True},
	and $q$ calls $\DoOps{(\Os,t')}{t''}{j}$ in line~\ref{second-doops},
	where $t' = t^* + \PP$.
We claim that $t^* < t$.
Suppose, for contradiction, that $t^* \ge t$.
Then, since $t'' < t$, the intervals $[t'', t^*]$ and $[t, t]$ intersect;
	since $q$ calls $\ML{t''}{t^*}$, $p$ calls $\ML{t}{t}$ and
	both these calls return \textsc{True}, 
	by Lemma~\ref{same-leadership}, $t'' = t$, 
	contradicting the fact that $t'' < t$.
Thus, $t^* < t$.
Therefore, $\tj + \LP = t^* + \LP + \PP < t + \LP + \PP \leq \CT_p(\tau)$.
\qedhere~$_\text{\autoref{lease-expire-after-wait}}$
\end{enumerate}
\end{proof}

\begin{lemma}\label{leasemonotonicity}
If a process has $\lease = (i,-)$ and later it has $\lease = (j,-)$, then $i \le j$.
\end{lemma}

\begin{proof}
Suppose that a process $q$ changes its $\lease$ variable from $(i,-)$ to $(j,-)$.
Note that $q$ sets its $\lease$ variable in only three places:
	 in line~\ref{lg1}, line~\ref{setlease}, or line~\ref{lease1} of a
	 $\DoOps{-}{-}{-}$ that $q$ called in line~\ref{first-doops} or line~\ref{second-doops}.

We now consider each one of these four cases:

\begin{enumerate}

\item \emph{Process $q$ sets $\lease$ to $(j,-)$ in line~\ref{lg1}.}
	Then the guard of line~\ref{lg1} ensures that
	$(j,-)$ is greater than its previous lease $(i,-)$ so, $ j \ge i$.
	
\item \emph{Process $q$ sets $\lease$ to $(j,-)$ in line~\ref{setlease}.}
	So $j$ is the value of $q$'s variable $k$ in line~\ref{setlease}.
From the code of $\LeaderWork{}$, it is clear that
	the last time that $q$ sets its $\lease$ before setting it to $(k,-) = (j,-)$ in line~\ref{setlease}
	is when $q$ previously issued a lease $(k,-)$
	in line~\ref{setlease} or in line~\ref{lease1} of a $\DoOps{(-,-)}{-}{-}$
	that $q$ called in line~\ref{first-doops} or line~\ref{second-doops}.
So just before it sets $\lease$ to $(j,-)$ in line~\ref{setlease},
	$q$ had $\lease = (i,-)$ with $i=j=k$.

\item \emph{Process $q$ sets $\lease$ to $(j,-)$ in line~\ref{lease1} of a $\DoOps{(-,-)}{-}{-}$ that $q$ calls in line~\ref{second-doops}.}
Note that this call is of the form $\DoOps{(-,-)}{-}{k+1}$ and $j= k+1$.
From the code of $\LeaderWork{}$, it is clear that
	the last time that $q$ sets its $\lease$ before setting it to $(j,-)$ in $\DoOps{(-,-)}{-}{k+1}$
	is when $q$ previously issued a lease $(k,-)$
	in line~\ref{setlease} or in line~\ref{lease1} of the previous $\DoOps{(-,-)}{-}{k}$ call.
So just before $q$ sets $\lease$ to $(j,-)$ in $\DoOps{(-,-)}{-}{k+1}$,
	$q$ had a $\lease = (i,-)$ with $i=k < k+1=j$.

\item \emph{Process $q$ sets $\lease$ to $(j,-)$ in line~\ref{lease1} of a $\DoOps{(-,-)}{-}{j}$ that $q$ calls in line~\ref{first-doops}.}
Note that this is the first $\DoOps{(-,-)}{-}{j}$ by $q$ in some $\LeaderWork{}$.
So the following sequence events must have occurred, in this chronological order, at process $q$:

	(a) $q$ became leader at some time $(t_j, \tau_j)$,
	
	(b) $q$ called $\LeaderWork{t}$,
	
	(c) $q$ called $\DoOps{(-,-)}{t}{j}$ in line~\ref{first-doops} of $\LeaderWork{t}$,
	
	(d) $q$ accepted $(-,{t},{j})$ in line~\ref{leader-accept} of this $\DoOps{(-,-)}{t}{j}$, and
	
	(e) $q$ issued the lease $(j,-)$ at some time $(t_j', \tau_j')$ in line~\ref{lease1} of this $\DoOps{(-,-)}{t}{j}$.

Note that from the real time $\tau_j$ when $q$ became leader up to but not including the real time $\tau_j'$
	when $q$ issues the lease $(j,-)$, $q$ does not modify its variable $\lease$.
Since $q$ has $\lease = (i,-)$ just before real time $\tau_j'$,
	$q$ must have $\lease = (i,-)$ at real time $\tau_j$ when $q$ became leader.

By Lemma~\ref{ne-doops}, $j \ge 0$.
If $i = 0$ then clearly $i \leq j$.
So, suppose $i > 0$.
Therefore, $\lease = (i,-) \neq (0,-\infty)$, i.e., $(i,-)$ is not the initial value of the variable $\lease$ at $q$.
Since $q$ has $\lease = (i,-)$ at real time $\tau_j$,
	by Lemma~\ref{LeaseIssue},
	some process $r$ issues the lease $(i,-)$ at some real time $\tau \leq \tau_j$.
By clock Assumptions~\ref{xclocks}(\ref{xcl2}) and~(\ref{xcl5}),
	this occurs while $r$ is executing $\LeaderWork{t_r}$
	for some $t_r \leq t$.
We claim that $t_r < t$.
Suppose for contradiction that $t_r = t$.
Then, since $r$ and $q$ both call $\LeaderWork{t}$,
	by Lemma~\ref{unique-LeaderWork}, $r=q$.
Since $q$ holds the lease $(i,-)$ issued by itself in $\LeaderWork{t}$
	when it became leader at local time $t$,
	$q$ calls $\LeaderWork{t}$ at least twice,
	which contradicts Corollary~\ref{LeaderWorkUnique}.
So $t_r < t$.
By Lemma~\ref{lease-issue-obs},
	$r$ previously locks some tuple $(-,t_r,i)$.
Since $r$ locks $(-, t_r,i)$ and $q$ accepts $(-,{t},{j})$ with $t_r < t$,
	by Theorem~\ref{generalcase1a}, $j \ge i$.

So in all cases we have $i \leq j$, as wanted.
\qedhere~$_\text{\autoref{leasemonotonicity}}$
\end{enumerate}
\end{proof}

From Definition~\ref{lease-issue} and Lemma~\ref{leasemonotonicity}, we have:
\begin{corollary}\label{lease-issue-order}
If a process $p$ issues a lease $(i,-)$ and later it issues a lease $(j,-)$,
	then $i \leq j$.
\end{corollary}

Next we prove that the lease times of the leases issued
	during a single execution of $\LW$ increase.
More precisely:

\begin{lemma}\label{increasing-lease-time-in-LeaderWork}
If process $p$ issues lease $(i,t_i)$ and later issues lease $(j,t_j)$
	in the same $\LeaderWork{t}$ for some $t$,
	then $t_i<t_j$.
\end{lemma}
\begin{proof}
Suppose process $p$ issues lease $(i,t_i)$ at real time $\tau_i$ 
	and later issues lease $(j,t_j)$ at real time $\tau_j$
	in the same $\LeaderWork{t}$.
So $\tau_i < \tau_j$.
We will prove that if these are consecutive leases issued by $p$
	(i.e. if $p$ issues no lease at any real time $\tau$ such that $\tau_i < \tau < \tau_j$),
	then $t_i \leq t_j$,
	and if $i = j$, then $t_i < t_j$.
Note that $p$ does not make another $\DO$ call between real times $\tau_i$ and $\tau_j$,
	since otherwise $p$ would issue a lease in line~\ref{lease1}
	and this contradicts the fact that
	$(i,t_i)$ and $(j,t_j)$ are consecutive leases issued by $p$.
 
Then by induction it follows that the lemma holds even for non-consecutive leases.

There are two places where $p$ issues leases: line~\ref{lease1} (the first lease issued for a given batch) 
	and line~\ref{setlease} (the renewal of a lease for a given batch).
There are four cases for the two leases under consideration.

\begin{case}
\item \emph{$p$ issues both leases $(i,t_i)$ and $(j,t_j)$ in line~\ref{lease1}}.
	Then the two leases must be issued by $p$ in two consecutive $\DO$ calls.
	By Corollary~\ref{T-k-DoopsCreduced}, $j = i + 1 > i$, 
		so it suffices to show that $t_i \le t_j$.
	If $p$ issued the lease $(i, t_i)$ in during a call to $\DO$ made in line~\ref{first-doops},
		then $t_i = 0$, 
		and it is clear that $t_i \le t_j$.
	Now suppose $p$ issued both leases 
		in calls to $\DO$ made in line~\ref{second-doops}.
	From the code of lines~\ref{taketime1}-\ref{second-doops},
		it is clear that the following events happened at $p$:
		\begin{enumerate}
			\item $p$ gets $t_i^c$ from its clock in line~\ref{taketime1},
			\item $p$ issues the lease $(i,t_i)$ in line~\ref{lease1} such that $t_i = t_i^c + \PP$,
			\item $p$ gets $t_j^c$ from its clock in line~\ref{taketime1}, and
			\item $p$ issues the lease $(j,t_j)$ in line~\ref{lease1} such that $t_j = t_j^c + \PP$
		\end{enumerate}
		in this order.
	Since local clocks are non-decreasing
	and in fact increase between successive readings
	(Assumptions~\ref{xclocks}(\ref{xcl2}) and~(\ref{xcl4})),
		$t_i^c < t_j^c$,
		so $t_i < t_j$ as wanted.

\item \emph{$p$ issues lease $(i,t_i)$ at real time $\tau_i$ in line~\ref{lease1} and
		lease $(j,t_j)$ at real time $\tau_j$ in line~\ref{setlease}}.
	Thus, $p$ issued $(i,t_i)$ while executing $\DoOps{(-,t_i)}{t}{i}$.
	
	Since, during the execution of $\LeaderWork{t}$,
		$p$ updates its variable $k$ only in line~\ref{leader-accept} in $\DO$,
		and it does not make another $\DO$ call between these two lease issueings,
		it does not modify its variable $k$ between real times $\tau_i$ to $\tau_j$.
	So $i = j$, and we now show that $t_i < t_j$.
	First we see that in line~\ref{nst2} of $\DoOps{(-,t_i)}{t}{i}$,
		$p$ sets $\NextSendTime$ to $t_i + \Ptwo$.
	From the code in lines~\ref{taketime1}-\ref{setlease},
		$p$ gets $t_j$ from its $\CT$ in line~\ref{taketime1},
		finds that $t_j \geq \NextSendTime$ in line~\ref{checksendtime},
		and then sets $\lease=(j,t_j)$ in line~\ref{setlease} at real time $\tau_j$.
	Since $\NextSendTime$ is changed only immediately after a lease is issued (line~\ref{nst} and line~\ref{nst2}),
		and there is no lease issued between real times $\tau_i$ to $\tau_j$,	
		$\NextSendTime$ is equal to $t_i + \Ptwo$ 
		when $p$ finds that $t_j \geq \NextSendTime$ in line~\ref{checksendtime},
		$t_j \geq t_i + \Ptwo$.
	By Assumption~\ref{finite-LRP}, $\Ptwo > 0$,	
		so we have $t_i < t_j$ as wanted.

\item \emph{$p$ issues lease $(i,t_i)$ at real time $\tau_i$ in line~\ref{setlease} and lease $(j,t_j)$ at real time $\tau_j$ in line~\ref{lease1}}.
	Thus, $p$ issues the lease $(j,t_j)$ 
		during a call to $\DoOps{(-,t_j)}{t}{j}$ in line~\ref{second-doops}.
	So $p$ has $k = i$ from real time $\tau_i$ when it issues the lease $(i,t_i)$
		to when it calls $\DoOps{(-,t_j)}{t}{k+1} = \DoOps{(-,t_j)}{t}{j}$ in line~\ref{second-doops}.
	Thus, we have $i = k < k + 1 = j$.
	We now show that $t_i \leq t_j$.
	From the code of lines~\ref{taketime1}-\ref{setlease},
		it is clear that $p$ gets $t_i$ from its clock in line~\ref{taketime1}
		and then issues the lease $(i,t_i)$ in line~\ref{setlease} at real time $\tau_i$.
	From the code of lines~\ref{taketime1}-\ref{second-doops},
		it is clear that $p$ gets some $t_j^c$ from its clock in line~\ref{taketime1}
		and then calls $\DoOps{(-,t_j)}{t}{j}$ in line~\ref{second-doops}
		such that $t_j = t_j^c + \PP$.
	Since $p$ issues the lease $(j,t_j)$ in line~\ref{lease1} in $\DoOps{(-,t_j)}{t}{j}$ 
		after it issues the lease $(i,t_i)$ in line~\ref{setlease},
		$p$ calls $\DoOps{(-,t_j)}{t}{j}$ 
		after it issues the lease $(i,t_i)$ in line~\ref{setlease}.
	So $p$ gets $t_j^c$ from its clock in line~\ref{taketime1} 
		at the same real time or after it gets $t_i$ from its clock.
	Since local clocks are non-decreasing
	and in fact increase between successive readings
	(Assumptions~\ref{xclocks}(\ref{xcl2}) and~(\ref{xcl4})),	
		$t_i < t_j^c$.
	By Assumption~\ref{promiseRange},
		$t_i < t_j^c + \PP = t_j$.
	So we have $t_i < t_j$ as wanted.

\item \emph{$p$ issues both lease $(i,t_i)$ and $(j,t_j)$ in line~\ref{setlease}}.
	Thus, it is clear that $p$ does not modify its variable $k$
		between real times $\tau_i$ and $\tau_j$.
	From the code of line~\ref{setlease},
		we have $i = k = j$.
	We now show that $t_i < t_j$.
	Since $p$ issues the lease $(i,t_i)$ in line~\ref{setlease}
		before it issues the lease $(j,t_j)$ in the same line,
		the following events occur at $p$:
	\begin{enumerate}
		\item $p$ gets $t_i$ from its clock in line~\ref{taketime1},
		\item $p$ issues the lease $(i,t_i)$ in line~\ref{setlease} at real time $\tau_i$,
		\item $p$ gets $t_j$ from its clock in line~\ref{taketime1}, and
		\item $p$ issues the lease $(j,t_j)$ in line~\ref{setlease} at real time $\tau_j$
	\end{enumerate}
	in this order.
	By Assumptions~\ref{xclocks}(\ref{xcl2}) and~(\ref{xcl4}),
		$t_i < t_j$ as wanted.
\qedhere~$_\text{\autoref{increasing-lease-time-in-LeaderWork}}$

\end{case}
\end{proof}

We now show that if a process locks batch $i$, then any process
	that holds a valid lease for an earlier batch $j$ must be notified about batch $i$. More precisely:

\begin{lemma}\label{biggiebiggie}
Suppose a process $q$ has $\lease = (j,\tqq)$
	and a process $p \neq q$ locks a tuple $(\Os_i, t ,i)$
	with promise $\s_i$ at time~$(\tpp, \tapp)$.
If $i > j$, $\tpp < \tqq + \LP$ and $\s_i < \tqq + \LP$,
	then from real time $\tapp$ on the following hold at $q$:
	\begin{enumerate}
		\item $\PendingBatchOps{i} = \Os_i$,
		\item $\PendingBatchPromise{i} = \s_i$ or $0$, and
		\item $\MaxPendingIndex \ge i$.
	\end{enumerate}
\end{lemma}
\begin{proof}
Suppose $q$ has $\lease = (j,\tqq)$, and
	$p \neq q$ locks $(\Os_i, t,i)$ at time $(\tpp,\tapp)$
	such that $i>j$, $\tpp < \tqq + \LP$ and $\s_i < \tqq + \LP$.

Since $0 \le \tpp < \tqq + \LP$ and $\LP = \Pone$,
	$\tqq \neq -\infty$.
So $q$ has $\lease = (j, \tqq) \neq (0,-\infty)$.
By Lemma~\ref{LeaseIssue},
	some process issues the lease $(j, \tqq)$.
We first show that $p$ is the \emph{unique} process that issues the lease $(j,\tqq)$ and 
	it does so in $\LeaderWork{t}$.
By Definition~\ref{locking}, 
	$p$ locks $(\Os_i, t,i)$ with promise $\s_i$ at time $(\tpp, \tapp)$ 
	during the execution of $\DoOps{(\Os_i, \s_i)}{t}{i}$,
	thus $p$ completes the wait statement in line~\ref{wait-lease-expire} by real time $\tapp$.
By Lemma~\ref{lease-expire-after-wait},
	if a process $r$ issues the lease $(j, \tqq)$ in $\LeaderWork{t_r}$ where $t_r < t$, 
	then $\tqq + \LP \leq \CT_p(\tapp) = \tpp$,
	which contradicts the assumption that $\tpp < \tqq + \LP$,
	so $r$ must issue the lease $(j, \tqq)$ in $\LeaderWork{t_r}$ where $t_r \geq t$.
Suppose that $t_r > t$;
	by Lemma~\ref{lease-issue-obs},
	$r$ locks a tuple $(\Os_j,t_r,j)$ no later than issuing this lease.
By Observation~\ref{acceptedfirst},
	$r$ accepts the tuple $(\Os_j,t_r,j)$ before it locks the tuple.
By Theorem~\ref{generalcase1a} and the fact that $p$ locks $(\Os_i,t,i)$,
	$j \geq i$, which contradicts the assumption that $i > j$.
Therefore, $r$ issues the lease $(j,\tqq)$ during the execution of $\LeaderWork{t}$,
	and by Lemma~\ref{unique-LeaderWork}, $r = p$.

Since $p$ is the unique process that issues the lease $(j, \tqq)$
	and it does so in $\LeaderWork{t}$,
	by Lemma~\ref{LeaseIssue},
	there is a real time $\taqq$ when $p$ issues the lease $(j, \tqq)$
	during the execution of $\LeaderWork{t}$
	and $p$ has $q \in \LH$ at time $\taqq$.
By Observation~\ref{lock-implies-lease},
	when $p$ locks the tuple $(\Os_i, t ,i)$ with promise $\s_i$ at time~$(\tpp,\tapp)$,
	it also issues the lease $(i, \s_i)$ at time $(\tpp,\tapp)$.
By Corollary~\ref{lease-issue-order} and the fact that $i > j$, 
	$p$ issues the lease $(j,\tqq)$ at real time $\taqq$ before it issues the lease $(i,\s_i)$ at real time $\tapp$,
	so $\taqq < \tapp$.
\begin{claim}\label{q-in-LH}
$p$ has $q \in \LH$ at real time $\tapp$.
\end{claim}
\begin{proof}
Suppose, for contradiction, that $p$ has $q \not \in \LH$ at real time $\tapp$.
Since $p$ issues $(j,\tj)$ at real time $\taqq$, it is at line~\ref{setlease} or line~\ref{lease1} at real time $\taqq$,
	which is after the real time when $p$ executes line~\ref{wait-lease-expire} in $\LeaderWork{t}$.
Since $p$ has $q \in \LH$ at real time $\taqq$ after line~\ref{wait-lease-expire} and
	$p$ has $q \notin \LH$ at real time $\tapp > \taqq$ in the same $\LeaderWork{t}$, 
	by Lemma~\ref{removingfromLH},
	there is a real time $\hat \tau$ such that:
\begin{enumerate}[(a)]
	\item $\taqq < \hat \tau \leq \tapp$,
	\item $p$ executes line~\ref{lh2} at time $\hat \tau$, and
	\item\label{cond-c}if $p$ executes this line~\ref{lh2} in $\DoOps{(\Os,s)}{t}{\hat j}$ for some $s$
		such that $p$ finds $s < \leasetime + \LP$ in line~\ref{LHvsAcks}, 
		then at time $\hat \tau$, $p$ has $\CT_p \geq \leasetime + \LP$,
		where $\leasetime$ is evaluated by $p$ in line~\ref{wait-alg1}.
\end{enumerate}
Since $p$ issues $(j,\tqq)$ at time $\taqq < \hat \tau$,
	it is clear that $p$ sets $\lease$ to $(j,\tqq)$
	before it calls $\DoOps{(\Os,\s)}{t}{\hat j}$.
Thus, by Lemma~\ref{increasing-lease-time-in-LeaderWork}, 
	when $p$ evaluates $\leasetime$ in line~\ref{LHvsAcks} in $\DoOps{(\Os,s)}{t}{\hat j}$, 
	it will find $\tqq \leq \leasetime$.
Since $p$ is in $\DoOps{(\Os,\s)}{t}{\hat j}$ at real time $\hat \tau$ and 
	$p$ is in $\DoOps{(\Os_i, \s_i)}{t}{i}$ at real time $\taqq \geq \hat \tau$,
	either these two $\DO$ calls are the same call or
	$p$ calls $\DoOps{(\Os,\s)}{t}{\hat j}$ before it calls $\DoOps{(\Os_i, \s_i)}{t}{i}$.
In the first case, we have $\s = \s_i$.
In the second case, the $\DoOps{(\Os,\s)}{t}{\hat j}$ call must return \textsc{Done},
	otherwise $p$ will exit $\LeaderWork{t}$ and, by Observation~\ref{monoLW},
	$p$ will not call $\LeaderWork{t}$ again, 
	and hence $p$ will not call $\DoOps{(\Os_i, \s_i)}{t}{i}$.
Since $p$ calls $\DoOps{(\Os,\s)}{t}{\hat j}$ before it calls $\DoOps{(\Os_i, \s_i)}{t}{i}$,
	by Corollary~\ref{T-k-DoopsCreduced}, $\hat j < i$.
Since the $\DoOps{(\Os,\s)}{t}{\hat j}$ call returns \textsc{Done}, 
	$p$ locks $(\Os,t,\hat j)$ and 
	issues lease $(\hat j, \s)$ in line~\ref{batch1}.
Note that when $p$ locks $(\Os_i, t, i)$ with promise $\s_i$,
	it also issues a lease $(\s_i, i)$
	(Observation~\ref{lock-implies-lease}).
Since $p$ issues leases $(\hat j, \s)$ and $(i, \s_i)$ in the same $\LeaderWork{t}$ and $\hat j < i$,
	by Lemma~\ref{increasing-lease-time-in-LeaderWork},
	$\s \le \s_i$.
Therefore, in both cases, we have $\s \leq \s_i$.
Thus, by the assumption that $\s_i < \tqq + \LP$,
	we have $\s \leq \s_i < \tqq + \LP \leq \leasetime + \LP$,
	so $p$ finds $\s < \leasetime + \LP$ in line~\ref{LHvsAcks} in $\DoOps{(\Os,\s)}{t}{\hat j}$.
Thus, by~(\ref{cond-c}), at real time $\hat \tau$,
	$p$ has $\CT_p \geq \leasetime + \LP \geq \tqq + \LP$,
	where $\leasetime$ is evaluated in line~\ref{wait-alg1} in $\DoOps{(\Os,\s)}{t}{\hat j}$.

Since $\taqq \geq \hat \tau$, and local clocks are monotonically increasing,
	we have $\tpp = \CT_p(\tapp) \geq \CT_p(\hat \tau) \geq \tqq + \LP$,
	which contradicts the initial assumption that $\tpp < \tqq + \LP$.
\qedhere~$_\text{\autoref{q-in-LH}}$
\end{proof}

Since $p$ locks a tuple $(\Os_i, t ,i)$ at time~$\tapp$
	and it has $q \in \LH$ at time $\tapp$,
	then, by Lemma~\ref{LimpliesP},
	from time $\taqq$ on the following holds at $q$:
	\begin{enumerate}
		\item $\PendingBatchOps{i} = \Os_i$,
		\item $\PendingBatchPromise{i} = \s_i$ or $0$, and
		\item $\MaxPendingIndex \ge i$.
	\qedhere~$_\text{\autoref{biggiebiggie}}$
	\end{enumerate}
\end{proof}

\subsection{Read lease mechanism: linearizabilty}\label{sec:linearizability}

In this section we prove that the object that the algorithm implements is linearizable with respect to its type $\mathcal{T}$.

Fix an arbitrary execution $E$ of the algorithm.
$E$ is a sequence that records the steps executed by the processes
	as they invoke operations on the object and receive responses to these operations
	by following the algorithm in Figure~\ref{ObjectAlgo-alg1-code},
	in the order in which these steps occur.

We say that an operation $\op$ \emph{appears in $E$} if
	some process assigns $\op$ to the variable $\operation$
	in line~\ref{rmw4} or~\ref{read2}.
That assignment is the \emph{invocation} of $\op$ in $E$.
The \emph{end} and the \emph{response} of an operation $\op$ that appears in $E$
	are defined as follows:
If $\op$ is a RMW operation invoked by process $p$ in line~\ref{rmw4},
	the end of $\op$ is the subsequent execution of line~\ref{rmw-return} by $p$
	(if it occurs);
	and the response of $\op$ in $E$ is the value returned in that line.
If $\op$ is a read operation invoked by process $p$ in line~\ref{read2},
	the end of $\op$ is the subsequent execution of line~\ref{read-respond} by $p$
	(if it occurs);
	and the response of $\op$ in $E$ is the value of variable $\reply$ returned in that line.
If the end of $\op$ occurs, then we say that $\op$ is \emph{complete} in $E$.

\begin{definition}\label{GBatch-def}
For all $j\in\N$, let
$$\GBatch_j=\begin{cases}
O,			&\text{if some process locks $(O,-,j)$}\\
\emptyset,	&\text{otherwise}
\end{cases}$$
\end{definition}
$\GBatch_j$ is well defined because, 
	by Theorem~\ref{generalcase2},
	if process $p$ locks $(O,-,j)$ and process $p'$ locks $(O',-,j)$,
	then $O=O'$.
Clearly, $\GBatch_j$ is a set of RMW operations.

By Lemma~\ref{IamOutOfLableNames},
\begin{corollary}\label{local-is-global-2}
For all $j\in\N$, if a process sets $\Batch[j]:=(O, -)$, then $O= \GBatch_j$.
\end{corollary}

By Theorem~\ref{intersect1},
\begin{corollary}\label{intersect-global}
For all $i,j\in\N$, if $i\ne j$ then $\GBatch_i\cap\GBatch_j=\emptyset$.
\end{corollary}

\begin{definition}\label{GPromise-def}
For all $j\in\N$, let
$$\GPromise_j=\begin{cases}
	s,			&\text{if a tuple $(-,-,j)$ is locked with promise $s$ in a call to $\DO$ made in line~\ref{second-doops}}\\
	0, 			&\parbox{15cm}{if a tuple $(-,-,j)$ is locked, and no process locks  $(-,-,j)$ in a call to $\DO$ made in line~\ref{second-doops}}\\
	\infty,	&\text{otherwise}
\end{cases}$$
\end{definition}
$\GPromise_j$ is well defined because,
	by Lemma~\ref{uniq-positive-promise-locking},
	if tuples $(-,-,j)$ and $(-,-,j)$ are locked with promise $s$ and $s'$ respectively
		during calls to $\DO$ made in line~\ref{second-doops},
	then $s = s'$.
\begin{observation}\label{lock-imply-finite-GPromise}
If a tuple of the form $(-,-,j)$ is locked, 
	then $\GPromise_j < \infty$.
\end{observation}
\begin{lemma}\label{local-less-than-global-promise}
For all $j\in\N$, if a process sets $\Batch[j]$ to $(-,\s)$, 
	then $\s \leq \GPromise_j$.
\end{lemma}
\begin{proof}
Suppose that some process $p$ sets $\Batch[j]$ to $(-, \s)$ for some $j$.
Then, by Lemma~\ref{batch-promise-locked},
	a tuple of the form $(-, -, j)$ is locked with promise $s$.
If this locking happens in a $\DO$ called in line~\ref{first-doops},
	then $s = 0 \leq \GPromise_j$ and the lemma holds.
If this locking happens in a $\DO$ called in line~\ref{second-doops},
	then by Definition~\ref{GPromise-def} and Lemma~\ref{uniq-positive-promise-locking},
	$s = \GPromise_j$.
\qedhere~$_\text{\autoref{local-less-than-global-promise}}$
\end{proof}

Given the execution $E$, we now define a subset $L$ of the operations that appear in $E$,
	called the \emph{linearized operations} of $E$;
	this consists of a set of RMW operations $\RMWOps$ and a set of read operations $\ReadOps$.
\begin{definition}\label{L-def}
Let
\begin{align*}
&	\RMWOps = \cup_{i\in\N} \GBatch_i \\
&	\ReadOps = \{\op:~\op~\text{is a read operation that appears in $E$ and is complete in $E$}\} \\
&	L = \RMWOps \cup \ReadOps
\end{align*}
\end{definition}

\begin{lemma}\label{complete-RMW-ops}
If $\op$ is a complete RMW operation in $E$,
	then there exist unique $i,j$ such that
	$\op$ is the $i$-th operation in $\GBatch_j$ (in ID order).\footnote{Recall that each operation $\op=(o,(p,\cntr))$ consists of $\op.\TYPE = o$ and a unique ID $\op.\ID=(p,\cntr)$,
	where $p$ is the process that invokes the operation and $\cntr$ is a sequence number.}
Furthermore,
	the process that invokes $\op$ set $\TakesEffect{\op}$
	in line~\ref{set-takesEffect} in the $i$-th iteration,
	and hence completed the $i$-th iteration
	of the loop in lines~\ref{apply-op}-\ref{set-takesEffect} during a call to $\EB(j)$
	before the end of $\op$ (the execution of line~\ref{rmw-return}).
\end{lemma}

\begin{proof}
Let $\op$ be a complete RMW operation in $E$.
Thus, the process $p$ that invokes $\op$ found $\CT \geq \TakesEffect{\op}$ in line~\ref{rmw-wait-promise} before the the end of $\op$ in line~\ref{rmw-return}.
Since initially $\TakesEffect{\op}=\infty$,
	$p$ must have assigned a non-$\infty$ value to $\TakesEffect{\op}$ in line~\ref{set-takesEffect}
	(the only place where $\TakesEffect{\op}$ is assigned a value after initialization).
This happens during $p$'s execution of $\EB(j)$, for some $j\in\N$.
By line~\ref{EB-get-ops},
	there is some $i$ such that $\op$ is the $i$-th operation (in ID order)
	in the set $O_j$ contained in $\BatchOps{j}$.
Since initially $\Batch[j]=(\emptyset,\infty)$,
	$p$ must have previously set $\Batch[j]$ to $(\Os_j, \s_j)$
	where $\Os_j \neq \emptyset$. 
By Corollary~\ref{local-is-global-2},
	$O_j=\GBatch_j$.
Thus, $\op$ is the $i$-th operation in $\GBatch_j$ (in ID order).
By Corollary~\ref{intersect-global}, for all $j'\ne j$, $\op\notin\GBatch_{j'}$.
So, there are unique $i,j$ such that
	$\op$ is the $i$-th operation (in ID order) in $\GBatch_j$.
Since $p$ set $\TakesEffect{\op}$ in line~\ref{set-takesEffect},
	it completed the $i$-th iteration of the loop in that line.
\qedhere~$_\text{\autoref{complete-RMW-ops}}$
\end{proof}

\begin{lemma}\label{complete-ops-in-L}
Every complete operation in $E$ is in $L$.
\end{lemma}

\begin{proof}
If $\op$ is a complete read operation, it is in $L$ by definition.
If $\op$ is a complete RMW operation,
	by Lemma~\ref{complete-RMW-ops},
	there is some $j$ such that $\op\in\GBatch_j$.
Therefore, $\op\in\RMWOps$, and so $\op\in L$.
\qedhere~$_\text{\autoref{complete-ops-in-L}}$
\end{proof}

\begin{definition}\label{earliest-realtime-clock-val}
For $t \geq 0$, we define $\GReal(t)$ to be the earliest real time when
	some process' local lock has value at least~$t$.
\end{definition}

Next we define the real time when a batch~$j$ takes effect.
Intuitively this is the earliest real time when
	a process can read the state of the object
	after the operations in batch~$j$ have been applied.

\begin{definition}\label{batch-takes-effect-def}
For any $j\in\N$ we say that
	\emph{batch $j$ takes effect at real time $\tau_j$} if and only if
	some tuple $(\GBatch_j,-,j)$ is locked and
	$\tau_j=\max(
			\min\{\tau:~\text{some process}~p~\text{locks}~(\GBatch_j,-,j)~\text{at real time}~\tau\}, 
			\GReal(\GPromise_j))$.
\end{definition}

\begin{lemma}\label{read-issued-lease}
Let $\op\in\ReadOps$ be an operation invoked by process $p$, and
	let $(k^*,t^*)$ be the value of variable $\lease$ that $p$ records
	when it executes line~\ref{getlease} in the last iteration of the loop
	in lines~\ref{getvalidlease-start}--\ref{getvalidlease-end}
	during the execution of $\op$.
Then some process issued lease $(k^*,t^*)$.
\end{lemma}

\begin{proof}
Let $\op\in\ReadOps$ be an operation invoked by process $p$.
Let $t'$ be the local time that $p$ records when it executes line~\ref{getleasetime}
	and $(k^*,t^*)$ be the value of $\lease$ that $p$ records
	when it executes line~\ref{getlease}
	in the last iteration of the loop
	in lines~\ref{getvalidlease-start}--\ref{getvalidlease-end}
	during the execution of $\op$.
By the exit condition in line~\ref{getvalidlease-end} and the fact that $t'\ge0$
	(Assumption~\ref{xclocks}(\ref{xcl1}),
	$t^* > -\infty$;
	so the value $(k^*,t^*)$ that $p$ found in $\lease$
	is not the initial value $(0,-\infty)$ of that variable.
By Lemma~\ref{LeaseIssue}, some process issues the lease $(k^*,t^*)$.
\qedhere~$_\text{\autoref{read-issued-lease}}$
\end{proof}

\begin{lemma}\label{batch-zero}
	$\Batch[0]$ equals to $(\emptyset,0)$ at all processes at all real times.
\end{lemma}
\begin{proof}
Since the initial value of $\Batch[0]$ is $(\emptyset, 0)$,
	we only need to prove that if some process sets $\Batch[0]$,
	it sets it to the same value.
Suppose that some process $p$ sets $\Batch[0]$ to $(\Os, \s)$.
Then, by Corollary~\ref{ne-batch},
	$\Os = \emptyset$.
It remains to show that $\s = 0$.
From Lemma~\ref{batch-promise-locked},
	some process locks a tuple $(\emptyset, t, 0)$ with promise $\s$ for some $t$.
This happens during a call to $\DoOps{(\emptyset,\s)}{t}{0}$.
From Lemma~\ref{ne-doops} and Corollary~\ref{T-k-increase},
	this call to $\DoOps{(\emptyset,\s)}{t}{0}$ must be made in line~\ref{first-doops},
	so $\s = 0$.
\qedhere~$_\text{\autoref{batch-zero}}$
\end{proof}

The next lemma states that
	only batches that take effect are used to determine the response of read operations.

\begin{lemma}\label{batch-read-takes-effect} 
Let $\op\in\ReadOps$ be an operation invoked by process $p$, and
	let $\hat k$ be the value that $p$ computed
	in lines~\ref{get-k-hat-1}--\ref{get-k-hat-5} during the execution of $\op$.
Then: 
\begin{enumerate}
\item\label{k-hat-has-choice} If $p$ computes $\hat k$ in line~\ref{get-k-hat-then},
	then it finds the set $\{j ~|~ 0 \leq j \leq k^*$ \textbf{and} $\BatchPromise{j} \leq t'\}$ to be non-empty.
\item\label{k-hat-is-valid} $(-,-,\hat k)$ is locked and 
	there is a $\hat \tau$ such that batch $\hat k$ takes effect at real time $\hat \tau$.
\end{enumerate}
\end{lemma}

\begin{proof}
Let 
\begin{itemize}
	\item $p$ be a process executing a operation $\op\in\ReadOps$,
	\item $\hat k$ be the value that $p$ computes
		in lines~\ref{get-k-hat-1}--\ref{get-k-hat-5} during the execution of $\op$,
	\item $t'$ be the value of $\CT_p$ that $p$ recorded when it executed line~\ref{getleasetime} in the last iteration of the loop in lines~\ref{getvalidlease-start}--\ref{getvalidlease-end}, and
	\item $(k^*,t^*)$ be the value of $\lease$ that $p$ recorded when it executed line~\ref{getlease} in the last iteration of the same loop.
\end{itemize}

Since $p$ continues to compute $\hat k$ in lines~\ref{get-k-hat-1}-\ref{get-k-hat-5},
	it found $t' < t^* + \LP$ in line~\ref{getvalidlease-end}.
So $(k^*, t^*)$  is not the initial value $(0,\infty)$ of $\lease$ at process $p$,
	and $p$ must have set $\lease$ to $(k^*,t^*)$.
By Lemma~\ref{leasemonotonicity} and the fact that the initial value of $\lease.\textit{batch}$ is $0$, 
	$k^* \geq 0$.

We will first show $(\ref{k-hat-has-choice})$.
By Lemma~\ref{batch-zero}, 
	$\BatchPromise{0} = 0$ at process $p$.
Since $p$ gets $t'$ from its clock, $t' \geq 0$.
Thus, when $p$ executes line~\ref{get-k-hat-then},
	it finds $k^* \geq 0$ and $t' \geq \BatchPromise{0} = 0$,
	so (\ref{k-hat-has-choice}) holds.

By (\ref{k-hat-has-choice}), 
	the fact that $k^* \geq 0$, and 
	from the code of lines~\ref{get-k-hat-1}-\ref{get-k-hat-5},
	it is clear that the value of $\hat k$ that $p$ computes is at least 0. 
Now we claim that $p$ sets $\Batch[\hat k]$ to some pair $(\Os,\s) \neq (\emptyset, \infty)$.

There are two cases depending on the value of $\hat k$:

\begin{case}
\item $0 \leq \hat k \leq k^*$.
Since $p$ sets $\lease$ to $(k^*, t^*)$, the claim follows from Lemma~\ref{set-batch-before-lease}.

\item $\hat k > k^*$.
It is clear that in this case, $p$ computes $\hat k$ in lines~\ref{get-k-hat-else}-\ref{get-k-hat-5}.
Since $\op \in \ReadOps$, $\op$ is a complete read operation.
So $p$ must find $\Batch[\hat{k}] \neq (\emptyset,\infty)$ in line~\ref{FG3} before $\op$ ends in line~\ref{read-respond}.
Since $(\emptyset,\infty)$ is the initial value of $\Batch[\hat{k}]$, 
	$p$ must set $\Batch[\hat{k}]$ to some pair $(\Os,\s)\neq (\emptyset,\infty)$.
\end{case}

By Observation~\ref{batch-promise-locked},
	some process locks $(\Os,-,\hat{k})$ with promise $s$.
By Observation~\ref{lock-imply-finite-GPromise},
	$\GPromise_{\hat k} < \infty$.
Thus, by Definition~\ref{batch-takes-effect-def}, 
	there is a $\hat \tau$ such that batch $\hat k$ takes effect at real time $\hat \tau$.
\qedhere~$_\text{\autoref{batch-read-takes-effect}}$
\end{proof}

Next we define the real time when an operation $\op\in\RMWOps$ takes effect.
By Corollary~\ref{intersect-global},
	there is a unique batch $j$ such that $\op\in\GBatch_j$;
	and since $\GBatch_j$ is not empty, 
	there is a real time when the tuple $(\GBatch_j, -, j)$ is locked.
By Observation~\ref{lock-imply-finite-GPromise},
	$\GPromise_j$ is finite,
	so there is a real time at which
	batch $j$ takes effect.	
Thus, we have the following definition:

\begin{definition}\label{RMW-takes-effect-def}
If $\op\in\GBatch_j$, the real time $\tau_\op$ when $\op$ \emph{takes effect}
	is the real time when batch $j$ takes effect.
\end{definition}

Next we define the real time when an operation $\op\in\ReadOps$ takes effect.

\begin{definition}\label{read-takes-effect-def}
If $\op\in\ReadOps$, the real time $\tau_\op$ when $\op$ \emph{takes effect}
	is defined as follows:
Let
\begin{itemize}
\item $p$ be the process that invoked $\op$,
\item $\tau'$ be the time when $p$ executed line~\ref{getleasetime}
	in the last iteration of the loop in lines~\ref{getvalidlease-start}--\ref{getvalidlease-end}
	during the execution of $\op$,
\item $\hat{k}$ be the value that $p$ computes in lines~\ref{get-k-hat-1}--\ref{get-k-hat-5}
	during the execution of $\op$, and
\item $\hat \tau$ be the time when batch $\hat{k}$ takes effect
	($\hat \tau$ exists by Lemma~\ref{batch-read-takes-effect}(\ref{k-hat-is-valid})).
\end{itemize}
Then $\tau_\op=\max(\tau',\hat \tau)$.
\end{definition}

We will use the real times when operations take effect
	to define a sequence $\Sigma_E$ of the operations in $L$.
Intuitively, $\Sigma_E$ is the ``linearization order'' of the operations in $E$.
Notice that in Definitions~\ref{RMW-takes-effect-def} and~\ref{read-takes-effect-def},
	different operations can take effect at the real same time.
The definition below states that in $\Sigma_E$ operations appear
	in the order in which they take effect,
	with ties resolved according to specific rules.

\begin{definition}\label{EL-def}
For any operations $\op,\op'\in L$, let $\tau_\op, \tau_{\op'}$ be real times when $\op, \op'$ take effect:
\begin{itemize}
\item If $\tau_\op < \tau_{\op'}$ then $\op$ appears before $\op'$ in $\Sigma_E$.
\item If $\tau_\op = \tau_{\op'}$ and
	$\op,\op'$ are both RMW operations or are both read operations,
	then they appear in $\Sigma_E$ in the order of their IDs.
\item If $\tau_\op = \tau_{\op'}$, $\op$ is a RMW operation, and $\op'$ is a read operation,
	then $\op$ appears before $\op'$ in $\Sigma_E$.
\end{itemize}
\end{definition}

\begin{lemma}\label{distinct-locking}
For all $i,j\in\N$, if $i<j$ and the earliest real times when tuples 
	$(\GBatch_i, -, i)$ and $(\GBatch_j, -, j)$ are locked
	are $\tau_i$ and $\tau_j$ respectively,
	then $\tau_i < \tau_j$.
\end{lemma}

\begin{proof}
Let $i,j\in\N$ be such that $i<j$,
	and suppose that the earliest real times that
	tuples $(\GBatch_i, -, i)$ and $(\GBatch_j, -, j)$ are locked
	are $\tau_i$ and $\tau_j$, respectively.
So, $\tau_i$ is the earliest real time that batch $i$ is locked and
	$\tau_j$ is the earliest real time that batch $j$ is locked.
By Observation~\ref{acceptedfirst},
	if a process $p$ locks a tuple $(-,-,j)$, $p$ previously accepted $(-,-,j)$.
Since $i,j\in\N$ and $i<j$, we have that $j\ge1$,
	and so by Corollary~\ref{accepts-to-locking},
	if $p$ accepts $(-, -, j)$, 
	some process previously locked $(-,-,j-1)$.
So, by induction,
	if some process locks $(-,-,j)$, then,
	for all $j'\in\N$ such that $j'<j$,
	some process previously locked $(-,-,j')$;
	and in particular, some process previously locked $(-, -,i)$.
Thus, the earliest real time when $(\GBatch_i,-,i)$ is locked
	is before the earliest time real when $(\GBatch_j,-,j)$ is locked.
So, $\tau_i < \tau_j$, as wanted.
\qedhere~$_\text{\autoref{distinct-locking}}$
\end{proof}

\begin{lemma}\label{zero-promise-means-GPromise-expired}
If a process locks a tuple $(-,-,j)$ with promise $s = 0$ at time $(t', \tau')$,
	then $t' \geq \GPromise_j$.
\end{lemma}
\begin{proof}
Suppose that a process $p$ locks a tuple $(-,-,j)$ with promise $s = 0$ at time $(t', \tau')$.
Then, if this happens in a $\DO$ called in line~\ref{second-doops},
	then $\GPromise_j = 0$, and the lemma holds.
So we assume that this locking happens in a $\DoOps{(-, 0)}{t}{j}$ call for some $t$ and $j$
	that is called in line~\ref{first-doops}.
By Definition~\ref{GPromise-def},
	if all processes that lock a tuple of the form $(-, -, j)$ do so in calls to $\DO$ made in line~\ref{first-doops},
	then $\GPromise_j = 0$ and the lemma holds.
Suppose that there is some process $q$ that
	locks a tuple of the form $(-, -, j)$ with promise $\s'$
	in some $\DoOps{(-, \s')}{t''}{j}$ call made in line~\ref{second-doops}.
Then $\GPromise_j = \s'$.
We claim that $t'' < t$.
Since $q$ made a $\DoOps{(-, \s')}{t''}{j}$ call in line~\ref{second-doops},
	it must have previously completed a $\DoOps{(-, -)}{t''}{j'}$ in line~\ref{first-doops},
	in which it accepted a tuple of the form $(-, t'', j')$.
By Corollary~\ref{T-k-DoopsCreduced},
	$j' < j$.
Since $p$ locks a tuple of the form $(-, t, j)$ and $q$ accepts a tuple of the form $(-, t'', j')$
	such that $j' < j$,
	by Theorem~\ref{generalcase1a}(\ref{t1}),
	$t'' \leq t$.
If $t'' = t$, then by Lemma~\ref{unique-LeaderWork},
	$p = q$ and $p$ called $\DoOps{(-, 0)}{t}{j}$ and $\DoOps{(-, \s')}{t}{j}$
	in lines~\ref{first-doops} and~\ref{second-doops},
	which contradicts Corollary~\ref{T-k-DoopsCreduced}.
So the claim $t'' < t$ holds.
By definition,
	when $q$ locks the tuple $(-, -, j)$ in $\DoOps{(-, \s')}{t''}{j}$,
	it issues a lease $(j,\s')$.
The lemma then follows from Lemma~\ref{lease-expire-after-wait},
	the monotonicity of local clocks,
	and the fact that at time $(t', \tau')$
	when it locks $(-,-,j)$,
	process $p$ is after line~\ref{wait-lease-expire}.
\qedhere~$_\text{\autoref{zero-promise-means-GPromise-expired}}$
\end{proof}

\begin{lemma}\label{zero-promise-expires-everywhere}
For $j > 0$, if a process sets $\Batch[j] = (-, 0)$ at time $(t', \tau')$,
	then $t' \ge \GPromise_j$.
\end{lemma}
\begin{proof}
Suppose a process sets $\Batch[j]$ to $(-, 0)$ at time $(t', \tau')$.
By Lemma~\ref{batch-promise-locked},
	a tuple of the form $(-, -, j)$ was locked with promise 0
	by real time $\tau'$.
The lemma then follows from Lemma~\ref{zero-promise-means-GPromise-expired}.
\qedhere~$_\text{\autoref{zero-promise-expires-everywhere}}$
\end{proof}

\begin{lemma}\label{greater-than-takes-effect}
If a process finds $\CT \geq \TakesEffect{\op}$ in line~\ref{rmw-wait-promise}
	at local time $t'$,
	then $op \in \GBatch_j$ for some $j$ and $t' \geq \GPromise_j$.
\end{lemma}
\begin{proof}
Suppose that a process $p$ finds $\CT \geq \TakesEffect{\op}$ in line~\ref{rmw-wait-promise}
	at some local time $t'$.
Since the initial value of $\TakesEffect{\op}$ is $\infty$,
	$p$ must previously set $\TakesEffect{\op}$ to some non-$\infty$ value.
This happens during $p$'s execution of $\EB(j)$ for some $j\in\N$.
From the code in line~\ref{EB-get-ops},
	there is some $i$ such that $\op$ is the $i$-th operation (in ID order)
	in the set $O_j$ contained in $\BatchOps{j}$.
By Lemma~\ref{batch-zero}, $j > 0$.
Since initially $\Batch[j]=(\emptyset,\infty)$,
	$p$ must have previously set $\Batch[j]$ to $(\Os_j, -)$
	where $\Os_j \neq \emptyset$. 
By Corollary~\ref{local-is-global-2},
	$O_j=\GBatch_j$ and hence $\op \in \GBatch_j$.
Note that line~\ref{set-takesEffect} is the only place where $\TakesEffect{\op}$ is set,
	and $p$ sets it to $\BatchPromise{j}$.
Suppose that the last value that $p$ 
	previously set to $\Batch[j]$ before line~\ref{set-takesEffect} is $(-, \s_j)$.
By Lemma~\ref{batch-promise-locked},
	some tuple of the form $(-, -, j)$ was locked with promise $\s_j$
	by the real time when $p$ sets $\Batch[j]$.
If this locking happens in time $(t'', \tau'')$ in a call to $\DO$ made in line~\ref{first-doops},
	then $\s_j = 0$.
By Lemma~\ref{zero-promise-expires-everywhere},
	$t'' \geq \GPromise_j$.
By clock Assumptions~\ref{xclocks}(\ref{xcl2}) and~\ref{xcl5}),
	when $p$ finds $\CT \geq \TakesEffect{\op}$ in line`\ref{rmw-wait-promise},
	it has $t' = \CT \geq t'' \geq \GPromise_j$.
If this locking happens in a call to $\DO$ made in line~\ref{second-doops},
	then $\s_j = \GPromise_j$ and $p$ found at local time $t'$ that
	$t' = \CT \geq  \TakesEffect{\op} = \GPromise_j$.
\qedhere~$_\text{\autoref{greater-than-takes-effect}}$
\end{proof}

\begin{lemma}\label{doops-zero}
If a process calls $\DoOps{(-, \s)}{t}{0}$,
	then this call is made in line~\ref{first-doops} and $\s = 0$.
\end{lemma}
\begin{proof}
Suppose a process $p$ makes a call to $\DoOps{(-, \s)}{t}{0}$.
By Lemma~\ref{ne-doops}(1) and Corollary~\ref{T-k-DoopsCreduced},
	this call must be made in line~\ref{first-doops}.
From the code in line~\ref{first-doops},
	$\s = 0$.
\qedhere~$_\text{\autoref{doops-zero}}$
\end{proof}

\begin{lemma}\label{greater-than-batch-promise}
If a process finds $\CT \geq \BatchPromise{j}$ in line~\ref{wait-promise-2}
	at local time $t'$, then $t' \geq \GPromise_j$.
\end{lemma}
\begin{proof}
Suppose that some process $p$ finds $\CT \geq \BatchPromise{j}$ in line~\ref{wait-promise-2}
	at local time $t'$.
So $j$ is the value of $\hat k$
	that $p$ computes in lines~\ref{get-k-hat-1}--\ref{get-k-hat-5},
	and by Lemma~\ref{batch-read-takes-effect}(2)
	a tuple of the form $(-, -, j)$ was locked.
If $j = 0$, then by Lemma~\ref{doops-zero} and the definition of locking,
	a tuple of the form $(-, -, 0)$ must be locked with promise 0.
So $\GPromise_j = \GPromise_0 = 0$, and hence $t' \geq \GPromise_j$ holds.
Henceforth we assume that $j > 0$.
Since the initial value of $\Batch[j]$ is $(\emptyset, \infty)$,
	$p$ must have previously set $\Batch[j]$.
Consider the last time $p$ sets $\Batch[j]$ before
	$p$ finds $\CT \geq \BatchPromise{j}$ in line~\ref{wait-promise-2}.
Suppose that $p$ sets $\Batch[j]$ to $(-, \s_j)$.
By Lemma~\ref{batch-promise-locked},
	some process previously locked a tuple of the form $(-, -, j)$ with promise $\s_j$
	at some time $(t'', \tau'')$.
This must happen during a $\DoOps{(-, \s_j)}{-}{j}$ call.
If this call is made in line~\ref{first-doops},
	then by Lemma~\ref{zero-promise-expires-everywhere},
	$t'' \geq \GPromise_j$.
By clock Assumptions~\ref{xclocks}(\ref{xcl2}) and (\ref{xcl5}),
	when $p$ finds $\CT \geq \BatchPromise{j}$ in line~\ref{wait-promise-2},
	it has $t' = \CT \geq t'' \geq \GPromise_j$.
If this call is made in line~\ref{second-doops},
	then $\GPromise_j = \s_j$ and 
	then $p$ finds $t' \geq \BatchPromise{j} = \s_j = \GPromise_j$ in line~\ref{wait-promise-2}.
So in all cases we have $t'\ge\GPromise_j$, as wanted.
\qedhere~$_\text{\autoref{greater-than-batch-promise}}$
\end{proof}

\begin{lemma}\label{greater-than-batch-promise-2}
If a process finds $t' \geq \BatchPromise{j}$ in line~\ref{get-k-hat-then}
	at some local time $t''$, then $t'' \geq \GPromise_j$.
\end{lemma}
\begin{proof}
The proof for this lemma is almost identical to the proof in the above lemma.
Suppose that some process $p$ finds $t' \geq \BatchPromise{j}$ in line~\ref{get-k-hat-then}.
We first show that a tuple of the form $(-, -, 0)$ was previously locked,
	so $\GPromise_0$ is not infinite.
Since the initial value of $\lease$ is $(0, -\infty)$,
	$p$ must previously set its $\lease$ variable to some $(k^*, t^*)$
	before it exists the loop in lines~\ref{getvalidlease-start}-\ref{getvalidlease-end}.
By Lemma~\ref{leasemonotonicity},
	$k^* \geq 0$.
By Lemma~\ref{lease-issue-obs},
	a tuple of the form $(-, -, k^*)$
	was previously locked.
By Observation~\ref{acceptedfirst},
	if a process $p$ locks a tuple $(-,-,k^*)$,
	$p$ previously accepted $(-,-,k^*)$.
By Corollary~\ref{accepts-to-locking},
	if $p$ accepts $(-, -, k^*)$ such that $k^* > 0$, 
	some process previously locked $(-,-,k^*-1)$.
So, by induction,
	some process previously locked $(-, -, 0)$.
This locking must happen in some $\DoOps{(-, -)}{-}{0}$,
	and by Lemma~\ref{doops-zero},
	if a process calls $\DoOps{(-, -)}{-}{0}$,
	it must do so in line~\ref{first-doops}.
So $\GPromise_0 = 0$,
	and $t'' \geq \GPromise_0$ holds.
Henceforth we assume $j > 0$.
Since the initial value of $\Batch[j]$ is $(\emptyset, \infty)$,
	$p$ must have previously set $\Batch[j]$.
Consider the last time $p$ sets $\Batch[j]$ before
	$p$ finds $t' \geq \BatchPromise{j}$ in line~\ref{get-k-hat-then}.
Suppose that $p$ sets $\Batch[j]$ to $(-, \s_j)$.
By Lemma~\ref{batch-promise-locked},
	some process previously locked a tuple of the form $(-, -, j)$ with promise $\s_j$
	at time $(t_j, \tau_j)$.
This must happen during a $\DoOps{(-, \s_j)}{-}{j}$ call.
There are two cases depending on where this $\DoOps{(-, \s_j)}{-}{j}$ call is made:
If this call is made in line~\ref{first-doops},
	then by Lemma~\ref{zero-promise-expires-everywhere},
	$t_j \geq \GPromise_j$.
By clock Assumptions~\ref{xclocks}(\ref{xcl2}) and (\ref{xcl5}),
	when $p$ finds $t' \geq \BatchPromise{j}$ in line~\ref{get-k-hat-then},
	its local time $t'' \geq t_j \geq \GPromise_j$.
If this call is made in line~\ref{second-doops},
	then $\GPromise_j = \s_j$ and
	by monotonicity of local clocks,
	$p$ has $t'' \geq t' \geq \BatchPromise{j} = \s_j = \GPromise_j$ in line~\ref{wait-promise-2}.
\qedhere~$_\text{\autoref{greater-than-batch-promise-2}}$
\end{proof}

The next lemma states that the sequence $\Sigma_E$
	preserves the order of non-concurrent operations in $E$.
	
\begin{lemma}\label{EL-respects-order}
Let $\op_1,\op_2\in L$ be operations such that
	$\op_1$ ends before $\op_2$ is invoked in $E$.
Then $\op_1$ appears before $\op_2$ in $\Sigma_E$.
\end{lemma}

\begin{proof}
It suffices to prove that for each $\op\in L$,
	$\op$ takes effect at real time $\tau_\op$ such that
	$\tau_\op$ is a real time during the execution of $\op$ in $E$,
	i.e., the interval between the real times when $\op$ is invoked
	and the time when $\op$ ends.
(In what follows, we take $\infty$ to be the ``real time''
	when an incomplete operation in $\RMWOps$ ``ends''.)
There are two cases, depending on whether $\op$ is a RMW operation or a read operation.

\begin{case}

\item $\op\in\RMWOps$.
Let $j$ be the (unique) non-negative integer such that $\op\in\GBatch_j$.
Let $\tau_j$ be the earliest real time at which 
	a process locks the tuple $(\GBatch_j,-,j)$.
By Definition~\ref{RMW-takes-effect-def},
	$\tau_\op = \max (\tau_j, \GPromise_j)$.
Recall that for the tuple $(\GBatch_j,-,j)$ to be locked,
	some process calls $\DoOps{(\GBatch_j,-)}{-}{j}$.
We have,
\begin{align*}
	&\text{real time when $p$ invokes $\op$}\\
\le\>\>	&\text{earliest real time when $p$ sends $\langle\OpRequest,\op\rangle$ (line~\ref{rmw-send})}\\
\le\>\>	&\text{earliest real time when any process receives $\langle\OpRequest,\op\rangle$ (line~\ref{rr1})}\\
\le\>\>	&\text{earliest real time when any process adds $\op$ to $\OpsRequested$ (line~\ref{rr2})}\\
\le\>\>	&\text{earliest real time when any process adds $\op$ to $\NextOps$ (line~\ref{nextops})}\\
\le\>\>	&\text{earliest real time when any process calls $\DoOps{(\NextOps,-)}{-}{-}$ with $\op\in\NextOps$}\\
\le\>\>	&\text{earliest real time when any process calls $\DoOps{(\GBatch_j,-)}{-}{j}$}\\
\le\>\>	&\text{earliest real time when any process locks a tuple $(\GBatch_j,-,j)$}\\
=\>\>	&\tau_j\\
=\>\>	&\text{earliest real time when any process sets $\Batch[j]=(\GBatch_j,-)$ (line~\ref{batch1})}\\
\le\>\>	&\text{earliest real time when any process calls $\EB(j)$}\\
\le\>\>	&\text{earliest real time when any process sets $\reply(\op)\ne\bot$  in line~\ref{apply-op} of $\EB(j)$}\\
\le\>\>	&\text{real time when $\op$ ends (line~\ref{rmw-return}).}
\end{align*}

By Lemma~\ref{greater-than-takes-effect}, 
	when $p$ finds $\CT \geq \TakesEffect{\op}$ in line~\ref{rmw-wait-promise} at time $(t', \tau')$,
	$t' \geq \GPromise_j$.
So $\tau' \geq \GReal(t') \geq \GReal(\GPromise_j)$,
	and so $\GReal(\GPromise_j)\le\text{real time when $\op$ ends}$.
Thus we have 
\begin{align*}
	\text{real time when $p$ invokes $\op$}
&\le\>\> \tau_j\\
&\le\>\>  \max (\tau_j, \GReal(\GPromise_j))\\
&=\>\>	 \tau_\op\\
&\le\>\>	\text{real time when $\op$ ends (line~\ref{rmw-return}).}
\end{align*}

\item
$\op\in\ReadOps$.
Let $\tau'$ be the real time when the process $p$ that invokes $\op$ executes line~\ref{getleasetime}
	for the last time in the loop of lines~\ref{getvalidlease-start}--\ref{getvalidlease-end}
	during the execution of $\op$,
	$\hat{k}$ be the value that $p$ computes in lines~\ref{get-k-hat-1}--\ref{get-k-hat-5}
	during the execution of $\op$,
	$\tau_{\hat k}$ be the earliest real time when any process locks a tuple $(\GBatch_{\hat{k}}, -,\hat k)$,
	and $\hat{\tau}$ be the real time when batch $\hat{k}$ takes effect
	($\tau_{\hat{k}}$ and $\hat \tau$ exist, by Lemma~\ref{batch-read-takes-effect}(\ref{k-hat-is-valid})).
By Definition~\ref{RMW-takes-effect-def}, $\hat \tau=\max(\tau_{\hat k}, \GReal(\GPromise_{\hat k}))$.

By Definition~\ref{read-takes-effect-def}, $\tau_{\op}=\max(\tau',\hat \tau)$.
If $\tau'\ge \hat \tau$, then $\tau_\op=\tau'$
	and $\tau'$ by definition is a real time during the execution of $\op$ in~$E$.
If $\tau'<\hat \tau$, then $\tau_\op = \hat \tau$
	and we must show that $\hat \tau$ is a real time during the execution of $\op$ in~$E$.
Since $\tau'<\hat \tau$ and $\tau'$ is a real time
	after $\op$ is invoked in~$E$,
	it is clear that $\hat \tau$ is after $\op$ is invoked in~$E$.
It remains to show that $\hat \tau$ is before $\op$ ends in~$E$, 
	i.e. $\tau_{\hat k}$ and $\GReal(\GPromise_{\hat k})$ are before $\op$ ends in~$E$.
(Since $\op\in\ReadOps$, $\op$ ends in $E$ --- see Definition~\ref{L-def}.)

We first prove that $\tau_{\hat k}$ is before when $\op$ ends.
Since $\op \in \ReadOps$, $\op$ is a complete read operation.
By Lemma~\ref{batch-read-takes-effect},
	a tuple of the form $(-, -, \hat k)$ was locked.
Since $p$ exits the loop in lines~\ref{getvalidlease-start}-\ref{getvalidlease-end},
	and the initial value of $\lease$ is $(0, -\infty)$,
	$p$ must have previously set $\lease$.
By Lemma~\ref{leasemonotonicity},
	$p$ sets $\lease$ to some $(k^*, t^*)$ such that $k^* \ge 0$
	before $p$ exits the loop in lines~\ref{getvalidlease-start}-\ref{getvalidlease-end}.
By Lemma~\ref{lease-issue-obs},
	a tuple of the form $(-, -, k^*)$ was locked by the real time when this lease was issued.
By Observation~\ref{acceptedfirst},
	if a process $q$ locks a tuple $(-,-,k^*)$,
	$q$ previously accepted $(-,-,k^*)$.
By Corollary~\ref{accepts-to-locking},
	if $k^* > 0$ and $q$ accepts $(-, -, k^*)$,
	then some process previously locked $(-,-,k^*-1)$.
So, by induction,
	if some process locks $(-,-,k^*)$, then,
	for all $j\in\N$ such that $j<k^*$,
	some process previously locked $(-,-,j')$;
	and in particular, some process previously locked $(-, -,0)$.
Thus, if $\hat k = 0$,
	then real time $\tau_{\hat k}$ is before the real time when $\op$ ends.
We now consider the case when $\hat k > 0$.
Since $p$ finds $\CT_p \geq \BatchPromise{\hat k}$ in line~\ref{wait-promise-2} 
	before $\op$ ends in line~\ref{read-respond}
	and the initial value of $\Batch[\hat k]$ is $(\emptyset,\infty)$,
	$p$ must set $\Batch[\hat k] \neq (\emptyset,\infty)$ before $\op$ ends.
By Lemma~\ref{IamOutOfLableNames},
	some process locks $(\GBatch_{\hat k}, -, \hat{k})$ by
	the real time when $p$ sets $\Batch[\hat k]$.
Thus,
\begin{align*}
\tau_{\hat k} &= \text{earliest real time when any process locks $(\GBatch_{\hat k},-,\hat k)$}\\
		&\le \text{earliest real time when any process sets $\Batch[\hat k]\neq(\emptyset, \infty)$}\\
		&\le \text{earliest real time when $p$ sets $\Batch[\hat k]\neq(\emptyset, \infty)$}\\
		&\le \text{real time when $p$ finds $\CT_p \geq \BatchPromise{\hat k}$ in line~\ref{wait-promise-2}} \\ 
		&\le \text{real time when $\op$ ends.}
\end{align*}

Now we prove that $\GReal(\GPromise_{\hat k}) \leq$ the real time when $\op$ ends.
By Lemma~\ref{greater-than-batch-promise},
	when $p$ finds $\CT_p \geq \BatchPromise{\hat k}$ in line~\ref{wait-promise-2}
	at time $(t'', \tau'')$ before $\op$ ends,
	$t'' = \CT_p \geq \GPromise_{\hat k}$,
	and hence $\tau'' \geq \GReal(\GPromise_{\hat k})$.
So $\GReal(\GPromise_{\hat k})$ is before the real time when $\op$ ends,
	and hence $\hat \tau$ is before the real time when $\op$ ends.
\qedhere~$_\text{\autoref{EL-respects-order}}$
\end{case}
\end{proof}

\begin{lemma}\label{distinct-LPs}
For all $i,j\in\N$, if $i<j$ and batches $i,j$ take effect at real times $\tau_i,\tau_j$,
	respectively, then $\tau_i<\tau_j$.
\end{lemma}
\begin{proof}
Let $i,j\in\N$ be such that $i<j$,
	and batches $i,j$ take effect at real times $\tau_i,\tau_j$.
Suppose that the earliest real times that tuples
	$(\GBatch_i, -, i)$ and $(\GBatch_j, -, j)$ 
	are locked are $\tau'_i,\tau'_j$, respectively.
By Definition~\ref{batch-takes-effect-def},
	$\tau_i = \max(\tau'_i, \GReal(\GPromise_i))$ and $\tau_j = \max(\tau'_j, \GReal(\GPromise_j))$.
By Lemma~\ref{distinct-locking}, $\tau'_i < \tau'_j$.
If $\GPromise_i = 0$,
	since local clocks have non-negative values,
	then $\GReal(\GPromise_i) \leq \tau'_i$.
So $\tau_i = \max(\tau'_i, \GReal(\GPromise_i)) = \tau'_i < \tau'_j \leq \max(\tau'_j, \GReal(\GPromise_j)) = \tau_j$,
	and we are done.
Henceforth we assume that $\GPromise_i > 0$.
Then by Definition~\ref{GPromise-def},
	some process $p$ locks a tuple $(-,-,i)$ with promise $\GPromise_i > 0$
	during a call to $\DoOps{(-,\GPromise_i)}{t}{i}$
	that is made in line~\ref{second-doops}
	for some $t$.
Suppose that the earliest real time when batch $j$ is locked 
	is when some process $q$ locks it in $\DoOps{(-,-)}{t'}{j}$.
Since $q$ accepts $(-,-,j)$ in line~\ref{leader-accept} of $\DoOps{(-,-)}{t'}{j}$
	and $p$ locks $(-,-,i)$ with $i < j$,
	$t' \geq t$.
There are two cases:
\begin{case}
\item $t' = t$. Then by Lemma~\ref{unique-LeaderWork}, $p = q$.
	By Corollary~\ref{T-k-DoopsCreduced},
		$p$ calls $\DoOps{(-,-)}{t}{i}$ before it calls $\DoOps{(-,-)}{t}{j}$.
	So $p$ calls $\DoOps{(-,-)}{t}{j}$ in line~\ref{second-doops}.
	Since $p$ locks $(-,-,j)$ during $\DoOps{(-,-)}{t}{j}$,
		by Definition~\ref{GPromise-def},
		this $\DO$ call is $\DoOps{(-,\GPromise_j)}{t}{j}$.
	Since $p$ issues leases $(i,\GPromise_i)$ in line~\ref{lease1} 
		and $(j,\GPromise_j)$ in the same $\LeaderWork{t}$,
		by Lemma~\ref{increasing-lease-time-in-LeaderWork},
		$\GPromise_i < \GPromise_j$.
	So $\tau_i = \max(\tau'_i, \GReal(\GPromise_i)) < \max(\tau'_j, \GReal(\GPromise_j)) = \tau_j$.

\item $t' > t$. Since $p$ locks $(-,-,i)$ in $\DoOps{(-,\GPromise_i)}{t}{i}$,
	it issues lease $(i,\GPromise_i)$ in line~\ref{lease1}.
	Thus, by Lemma~\ref{lease-expire-after-wait},
		$q$ completes the wait statement in line~\ref{wait-lease-expire} at some time $(\hat t, \hat \tau)$
		such that $\hat t \geq \GPromise_i + \LP$.
	Thus $\tau'_j > \hat \tau \geq \GReal(\GPromise_i)$.
Since $\tau'_i<t'_j$ we have that $\tau_i = \max(\tau'_i, \GReal(\GPromise_i)) < \tau'_j \leq \max(\tau'_j, \GReal(\GPromise_j)) = \tau_j$.
\qedhere~$_\text{\autoref{distinct-LPs}}$
\end{case}
\end{proof}

As a consequence of Lemma~\ref{distinct-LPs},
	the sequence $\Sigma_E$ consists of alternating (possibly empty) sequences of read operations
	and (non-empty) sequences or RMW operations,
	where every sequence of RMW operations consists of the operations of a batch.
That is (recall that batch~0 contains no operations),
$$\Sigma_E = \underbrace{\hat{\op}_0^1\hat{\op}_0^2\,\ldots\,\hat{\op}_0^{n_0}}_{\text{reads}}\;\underbrace{\op_1^1\op_1^2\,\ldots\,\op_1^{m_1}}_{\text{batch 1}}\;\underbrace{\hat{\op}_1^1\hat{\op}_1^2\,\ldots\,\hat{\op}_1^{n_1}}_{\text{reads}}\;\underbrace{\op_2^1\op_2^2\,\ldots\,\op_2^{m_2}}_{\text{batch 2}}\;\underbrace{\hat{\op}_2^1\hat{\op}_2^2\,\ldots\,\hat{\op}_2^{n_2}}_{\text{reads}}\;\cdots$$
	where,
	for $j\ge0$ and $n_j\ge0$, $\hat{\op}_j^i$, $1\le i\le n_j$, is a read operation; and
	for $j\ge1$ and $m_j\ge1$, $\op_j^i$, $1\le i\le m_j$, is the $i$-th operation in $\GBatch_j$
	(in ID order).

Now suppose the operations in $L$ are applied to the object sequentially,
	in the order in which they appear in $\Sigma_E$.
We define notation for the responses of the operations,
	and the states through which the object transitions,
	in this sequential execution.
Informally, if operations are applied in the order they appear in $\Sigma_E$, then
\begin{itemize}
\item $\rho_j^i$ is the response of $\op_j^i$;
\item $\hat{\rho}_j^i$ is the response of $\hat{\op}_j^i$;
\item $\sigma_0$ is the initial state of the object;
\item $\sigma_j^i$, for $1\le i\le m_j$,
	is the state of the object after operation $\op_j^i$ is applied; and
\item $\sigma_j=\sigma_j^{m_j}$
	(i.e., $\sigma_j$ is the state of the object after all the operations in the $j$-th batch
	have been applied).
\end{itemize}
(Read operations do not change the state of the object,
	and so we need only consider the state after each RMW operation.)

We now give the precise definition of $\sigma_j$, $\sigma_j^i$, $\rho_j^i$ and $\hat\rho_j^i$.
Recall that $\Apply$ is the state transition function of this object:
	if $\sigma$ is a state of the object and $o$ is an operation applied to the object, then 
	$\Apply(\sigma,o)$ returns a pair $(\sigma',r)$ where $\sigma'$ is the new state of the object,
	and $r$ is the response of the object.
We denote $\sigma'$ by $\Apply(\sigma,o).\sstate$ and
	$r$ by $\Apply(\sigma,o).\RESPONSE$.

We define,
\begin{alignat*}{2}
\sigma_0 &= \text{(the initial state of the object)}\\
\sigma_j^i &=
	\begin{cases}
		\Apply(\sigma_{j-1},\op_j^i.\TYPE).\sstate,	&\text{if $i=1$}\\
		\Apply(\sigma_j^{i-1},\op_j^i.\TYPE).\sstate,	&\text{if $1<i\le m_j$}\\
	\end{cases},
	&&\qquad\text{for $j\ge1$}\\
\sigma_j &= \sigma_j^{m_j},&&\qquad\text{for $j\ge1$}\\
\rho_j^i &=
	\begin{cases}
		\Apply(\sigma_{j-1},\op_j^i.\TYPE).\RESPONSE,	&\text{if $i=1$}\\
		\Apply(\sigma_j^{i-1},\op_j^i.\TYPE).\RESPONSE,	&\text{if $1<i\le m_j$}\\
	\end{cases},
	&&\qquad\text{for $j\ge1$}\\
\hat\rho_j^i &= \Apply(\sigma_j,\hat\op_j^i.\TYPE).\RESPONSE,&&\qquad\text{for $j\ge0$ and $1\le i\le n_j$}.
\end{alignat*}

$\Sigma_E$ is just a sequence of the operations in $L$, not an execution,
	so there is no \emph{a priori} meaning to ``the response of $\op$ in $\Sigma_E$''.
It is convenient to define this as follows:
\begin{definition}\label{returns-in-E-L}
For each operation $\op\in L$, \emph{the response of $\op$ in $\Sigma_E$ is
	$\rho_j^i$} if $\op=\op_j^i$, and it is $\hat{\rho}_j^i$ if $\op=\hat{\op}_j^i$.
\end{definition}

\begin{lemma}\label{EB-loop-invariant}
For all $j\ge1$, suppose that when a process $p$ calls $\EB(j)$,
	it has $\sstate[j-1]=\sigma_{j-1}$ in line~\ref{EB-start}.
For all $i$, $1\le i\le m_j$,
	if $p$ completes the $i$-th iteration of the loop in lines~\ref{apply-op}-\ref{set-takesEffect} of $\EB(j)$,
	then, when it does, $\sigma=\sigma_j^i$ and $\reply(\op_j^i)=\rho_j^i$.
Moreover, $p$  has $\reply(\op_j^i)=\rho_j^i$ thereafter.
\end{lemma}
\begin{proof}
By Lemma~\ref{before-EB-1} and Corollary~\ref{local-is-global-2},
	before $p$ calls $\EB(j)$ it has $\BatchOps{j}=\GBatch_j$.
So, when $p$ executes line~\ref{EB-get-ops}, it finds $\BatchOps{j}=\GBatch_j$,
	and so $m=|\GBatch_j|=m_j$ and $\op^i=\op_j^i$ (the $i$-th operation in $\GBatch_j$).
By assumption, $p$ has $\sstate[j-1]=\sigma_{j-1}$ in line~\ref{EB-start},
	so $\sigma$ is assigned value $\sstate[j-1]=\sigma_{j-1}$ in this line.
Then, by a straightforward induction on $i$, we can prove that 
	$p$ sets $\sigma=\sigma_j^i$ and $\reply(\op_j^i)=\rho_j^i$ in line~\ref{apply-op-1}
	in the $i$-th iteration of the loop in lines~\ref{apply-op}-\ref{set-takesEffect}
	and has $\sigma=\sigma_j^i$ when it completes the $i$-th iteration
	(since $\sigma$ is a local variable and $p$ does not modify $\sigma$ in line~\ref{set-takesEffect}).
By Lemma~\ref{complete-RMW-ops}, there exist unique $i,j$ such that 
	$\op_j^i$ is the $i$-th operation in $\GBatch_j$, 
	so $p$ sets $\reply(\op_j^i)$ only in line~\ref{apply-op-1} in the $i$-th iteration of the loop
	in lines~\ref{apply-op}-\ref{set-takesEffect} of $\EB(j)$.
Therefore, after $p$ sets $\reply(\op_j^i)=\rho_j^i$, it remains equal to $\rho_j^i$.
\qedhere~$_\text{\autoref{EB-loop-invariant}}$
\end{proof}

\begin{lemma}\label{EB0}
(a)~If a process $p$ executes $\EB(0)$ then
	the body of the loop in lines~\ref{apply-op}-\ref{set-takesEffect} is not executed.
(b)~The value of $\sstate[0]$ at process $p$ is always equal to $\sigma_0$
	(the initial state of the object).
\end{lemma}

\begin{proof}
By Corollary~\ref{ne-batch} and the fact that the initial value of $\BatchOps{0}$ is $\emptyset$,
	when $p$ calls $\EB(0)$, $p$ has $\BatchOps{0}=\emptyset$ and therefore
	the body of the loop in line~\ref{apply-op} is not executed
	($m$, the number of operations in $\BatchOps{0}$, is zero).
This proves part~(a) of the lemma.

Variable $\sstate[0]$ is initialized to $\sigma_0$.
By inspection of the code, this variable can only be assigned a value in line~\ref{update-state}
	in an execution of $\EB(0)$.
So, consider any execution of $\EB(0)$ by process $p$.
When $\EB(0)$ starts, $\sstate[-1]=\sigma_0$.
This is because $\sstate[-1]$ is initialized to $\sigma_0$,
	and is never changed
	($\sstate[i]$ is assigned only in $\EB(i)$, which is called only with $i\ge0$).
By part~(a) of the lemma,
	the body of the loop in lines~\ref{apply-op}-\ref{set-takesEffect} is not executed.
Thus, when $p$ reaches line~\ref{update-state},
	the value of variable $\sigma$ is still equal to
	the value it was assigned in line~\ref{EB-start},
	i.e., $\sstate[-1]=\sigma_0$, and so
	in line~\ref{update-state}, $p$ sets $state[0]=\sigma_0$.
Therefore, $\sstate[0]=\sigma_0$ always.
This proves part~(b) of the lemma.
\qedhere~$_\text{\autoref{EB0}}$
\end{proof}

\begin{lemma}\label{after-EB}
For all $j\ge0$, if process $p$ calls $\EB(j)$, then
\begin{enumerate}[\noindent(a)]
\item for every $i$, $1\le i\le m_j$,
	if $p$ completes the $i$-th iteration of the loop in lines~\ref{apply-op}-\ref{set-takesEffect} of $\EB(j)$,
	then, when it does,
	$\sigma=\sigma_j^i$ and $\reply(\op_j^i)=\rho_j^i$. 
	Moreover, $p$ has $\reply(\op_j^i)=\rho_j^i$ thereafter; and
\item if $p$'s call to $\EB(j)$ completes,
	then, when it does and thereafter,
	$\sstate[j]=\sigma_j$. 
\end{enumerate}
\end{lemma}

\begin{proof}
By induction on $j$.

\smallskip\noindent
\textsc{Basis.}\enspace $j=0$.
By Lemma~\ref{EB0}(a), the body of the loop in lines~\ref{apply-op}-\ref{set-takesEffect} of $\EB(0)$
	is not executed, so part~(a) of this lemma for $j=0$ holds vacuously.
Part~(b) of this lemma for $j=0$ follows directly by Lemma~\ref{EB0}(b).

\smallskip\noindent
\textsc{Induction Step.}\enspace
Consider any integer $j\ge1$.
Suppose the lemma holds for $j-1$; we will prove that it also holds for $j$.
Suppose that $p$ calls $\EB(j)$.

We first claim that
\begin{equation}\tag{*}\label{random-claim}
	\text{$p$ has $\sstate[j-1]=\sigma_{j-1}$ in line~\ref{EB-start} when it executes $\EB(j)$.}
\end{equation}
For $j=1$, (\ref{random-claim}) follows immediately by Lemma~\ref{EB0}(b).
If $j\ge2$, by Corollary~\ref{before-EB-2},
	when $p$ calls $\EB(j)$, it has previously completed a call to $\EB(j-1)$.
By part~(b) of the induction hypothesis,
	when $p$'s call to $\EB(j-1)$ ends and thereafter, $\sstate[j-1]=\sigma_{j-1}$.
So, this is still true when $p$ executes line~\ref{EB-start} in $\EB(j)$,
	and (\ref{random-claim}) holds for $j\ge2$.
By Lemma~\ref{EB-loop-invariant} and (\ref{random-claim}), part~(a) of the lemma holds for $j$.

For part~(b), suppose that $p$'s call to $\EB(j)$ completes.
By Lemma~\ref{before-EB-1} and Corollary~\ref{local-is-global-2},
	before $p$ calls $\EB(j)$, it has $\BatchOps{j}=\GBatch_j$.
Therefore, when $p$ executes line~\ref{EB-get-ops},
	$m=|\GBatch_j|=m_j$.
Since $p$'s call to $\EB(j)$ completes,
	$p$ completed the loop in line~\ref{apply-op}.
Since $m=m_j$,
	by part~(a) of the lemma, when $p$ completes the loop in line~\ref{apply-op},
	$\sigma=\sigma_j^{m_j}=\sigma_j$.
So, after $p$ executes line~\ref{update-state},
	$\sstate[j]=\sigma_j$.
Thus, since $p$ assigns $\sstate[j]$ only in line~\ref{update-state} of $\EB(j)$,
	it remains equal to $\sigma_j$ thereafter,
	and part~(b) of the lemma also holds for $j$.
\qedhere~$_\text{\autoref{after-EB}}$
\end{proof}

\begin{theorem}\label{RMW-linearizability}
For each $\op\in\RMWOps$ that is complete in $E$,
	the response of $\op$ in $E$ is the same as in~$\Sigma_E$.
\end{theorem}

\begin{proof}
Let $\op\in\RMWOps$ be complete in $E$, and
	let $p$ be the process that invokes $\op$ in $E$.
Since $\op$ is a complete RMW operation in $E$,
	$p$ returns some value $v=\Reply{\op}$ (line~\ref{rmw-return}).
By Lemma~\ref{complete-RMW-ops},
	there exist unique $i,j$ such that
	$\op$ is the $i$-th operation $\op^i_j$ in $\GBatch_j$ (in ID order),
	and $p$ completed the $i$-th iteration of the loop in lines~\ref{apply-op}-\ref{set-takesEffect}
	in $\EB(j)$ before $\op$ ends (line~\ref{rmw-return}).
By Lemma~\ref{after-EB}(a),
	$p$ has $\Reply{\op}=\rho_j^i$
	when it completes the $i$-th iteration of the loop in lines~\ref{apply-op}-\ref{set-takesEffect}
	during a call to $\EB(j)$ and thereafter.
Therefore, the response of $\op$ in $E$ is $\rho_j^i$.
By definition, however,
	$\rho_j^i$ is the response of $\op_j^i$ in $\Sigma_E$.
So the response of $\op$ in $E$ is the same as in $\Sigma_E$, as wanted.
\qedhere~$_\text{\autoref{RMW-linearizability}}$
\end{proof}

Recall that the variable $\lease$ in each process stores a pair
	$(\lease.\textit{batch},\lease.\textit{start})$.

\begin{lemma}\label{mono-lease}
If process $p$ locks $(\GBatch_i,t_i,i)$ at real time $\tau_i$,
	then, from real time $\tau_i$ on, $p$ has $\lease.\textit{batch}\ge i$.
\end{lemma}

\begin{proof}
Suppose process $p$ locks $(\GBatch_i, t_i,i)$ at time $\tau_i$.
By Observation~\ref{lock-implies-lease},
	$p$ issues a lease of the form $(i,-)$ at real time $\tau_i$,
	so $p$ sets $\lease$ to $(i,-)$ at real time $\tau_i$.
The lemma now follows from Lemma~\ref{leasemonotonicity}.
\qedhere~$_\text{\autoref{mono-lease}}$
\end{proof}

\begin{theorem}\label{Read-linearizability}
For each $\op\in\ReadOps$, the response of $\op$ in $E$ is the same as in $\Sigma_E$.
\end{theorem}

\begin{proof}
Let $\op\in\ReadOps$, and
	let $v$ be the response of $\op$ in $E$.
(Recall that, by definition, every $\op\in\ReadOps$ is complete,
	and therefore has a response, in $E$.)
We want to prove that $v$ is also the response of $\op$ in $\Sigma_E$.

Let $q$ be the process that invokes $\op$, and let
\begin{itemize}
\item $\tau'$ be the real time when $q$ executed line~\ref{getleasetime}
	in the last iteration of the loop in lines~\ref{getvalidlease-start}--\ref{getvalidlease-end}
	during the execution of $\op$,
	and $t'$ be the local time that $q$ obtained then from its clock;
\item $(k^*,t^*)$ be the value of $\lease$ that $q$ recorded when it executed line~\ref{getlease}
	in the last iteration of the same loop;
\item $u$ be the value of $\MPB$ that $q$ records in line~\ref{get-MPB} if $q$ executes line~\ref{get-MPB}
	during the execution of $\op$; and
\item $\hat k$ be the value that $q$ computes in lines~\ref{get-k-hat-1}--\ref{get-k-hat-5}
	during the execution of $\op$.
\end{itemize}

\begin{claim}\label{read-linearizability-eqn1}
$v = \Apply(\sigma_{\hat k}, op.\TYPE).\RESPONSE$.
\end{claim}
\begin{proof}
We prove that, when $q$ reaches line~\ref{client-read-end},
	$\sstate[\hat k]=\sigma_{\hat k}$.
If $\hat k = 0$, $\sstate[\hat k]$ always has value $\sigma_0$ by Lemma~\ref{EB0}(b).
If $\hat k > 0$, consider $q$'s call to $\EUTB(\hat k)$ in line~\ref{fill-gaps-to-k-hat}.
If $\MBD\ge\hat k$ when this call is made,
	by Lemma~\ref{MBD-L},
	$q$ has previously executed $\EB(\hat k)$.
If $\MBD<\hat k$ when the call to $\EUTB(\hat k)$ is made,
	before the call ends, $q$ executes $\EB(\hat k)$.
Either way, by the time $q$ reaches line~\ref{client-read-end},
	it has executed $\EB(\hat k)$.
So, by Lemma~\ref{after-EB}(b),
	when $q$ reaches line~\ref{client-read-end},
	$\sstate[\hat k]=\sigma_{\hat k}$.
In line~\ref{client-read-end} $q$ computes $\reply$ to be
	the response of $\op.\TYPE$ when applied to state $\sigma_{\hat k}$.
Since this is the value $v$ that $\op$ returns (line~\ref{read-respond}),
	$v = \Apply(\sigma_{\hat k}, op.\TYPE).\RESPONSE$.
\qedhere~$_\text{\autoref{read-linearizability-eqn1}}$
\end{proof}
We must show that $v$ is also the value that $\op$ returns in $\Sigma_E$.

Recall that $\op$ takes effect at real time $\tau_\op=\max(\tau',\hat \tau)$,
	where $\hat \tau$ is the real time when batch $\hat k$ takes effect
	(see Definition~\ref{read-takes-effect-def}).
There are two cases, depending on whether $\tau'<\hat \tau$ or $\tau'\ge\hat \tau$.

\begin{case}
\item $\tau'<\hat \tau$, hence $\tau_\op = \hat \tau$.
In this case, by the definition of $\Sigma_E$ (see Definition~\ref{EL-def}),
	$\op$ appears in $\Sigma_E$ after batch $\hat k$
	and before batch $\hat{k}+1$ (if it exists).
That is, $\op=\hat{\op}_{\hat k}^r$, for some $r$, $1\le r\le n_{\hat k}$.
Thus, the response of $\op$ in $\Sigma_E$ is
	the response of $\op.\TYPE$ when applied to state $\sigma_{\hat k}$.
By Claim~\ref{read-linearizability-eqn1}, this is equal to~$v$.
So, the response of $\op$ in $\Sigma_E$ is $v$, as wanted.

\item $\tau'\ge\hat \tau$, hence $\tau_\op =\tau'$.
Let
\begin{equation}\label{i-hat-def}
\hat{\imath} = \max \{i ~|~ \text{batch $i$ takes effect at some real time $\tau'_i \in [\hat \tau,\tau']$} \}
\end{equation}
$\hat \imath$ is well-defined because at least batch $\hat k$ takes effect during $[\hat \tau, \tau']$.
In this case, by the definition of $\Sigma_E$ (see Definition~\ref{EL-def}),
	$\op$ appears in $\Sigma_E$ after batch $\hat\imath$
	and before batch $\hat\imath+1$ (if it exists).
That is, $\op=\hat{\op}_{\hat\imath}^r$, for some $r$, $1\le r\le n_{\hat\imath}$.
Thus, the response of $\op$ in $\Sigma_E$ is
	the response of $\op.\TYPE$ when applied to state $\sigma_{\hat\imath}$.
By Claim~\ref{read-linearizability-eqn1}, it remains to show that
	that the response of $\op.\TYPE$ when applied to state $\sigma_{\hat\imath}$
	is the same as when applied to state $\sigma_{\hat k}$.
To this end, we first prove the following

\begin{claim}\label{sonofbiggiebiggie}
If some batch $i$ takes effect at a real time $\tapp$ such that
	$\hat \tau < \tapp \le \tau'$,
	then $\op.\TYPE$ 
	does not conflict with any operation in $\GBatch_i$.
\end{claim}

\begin{proof}
Since $\GBatch_0=\emptyset$, the claim is vacuously true for $i=0$.
Henceforth we assume that $i>0$.
Suppose, for contradiction, that
(A)~batch $i$ takes effect at real time $\tapp$ such that
	$\hat \tau < \tapp \le \tau'$, but
(B)~$\op.\TYPE$ conflicts with some operation in $\GBatch_i$.
Let $\tau_i$ be the earliest real time when a tuple $(\GBatch,-,i)$ is locked.
Let $p$ be the process that locks $(\GBatch,-,i)$,
	and $(t_i,\tau_i)$ be the time of that locking.
By Definition~\ref{batch-takes-effect-def},
	$\tapp = \max(\tap, \GReal(\GPromise_i))$.
Similarly, suppose the earliest real time when a tuple $(\GBatch_{\hat k}, -, \hat{k})$ is locked is $\tau_{\hat k}$, 
	then $\hat \tau = \max (\tau_{\hat k}, \GReal(\pkh))$.

Since batch $\hat k$ and batch $i$ take effect at time $\hat \tau$ and $\tapp$, respectively, 
	and $\hat \tau < \tapp$, by Lemma~\ref{distinct-LPs}, 
\begin{equation}\label{claim-3-eqn-1}
\hat{k} < i.
\end{equation}

\begin{subclaim}\label{i-set-in-else}
$i > k^*$
\end{subclaim}

\begin{proof}
Suppose by contradiction that $i \le k^*$.
By~(\ref{claim-3-eqn-1}), $\hat k<i\le k^*$,
	so $q$ sets $\hat k$ in line~\ref{get-k-hat-then}
	(otherwise, $q$ would set $\hat k$ in lines~\ref{get-k-hat-else}-\ref{get-k-hat-5} to a value at least $k^*$).
By Lemma~\ref{set-batch-before-lease},
	when $q$ has $\lease=(k^*, t^*)$ in line~\ref{getlease},
	it has previously set $\Batch[j']$ for all $j'$, $1 \leq j' \leq k^*$,
	and in particular it has previously set $\Batch[i]$.

Since $i\le k^*$, in line~\ref{get-k-hat-then}
	$q$ compares $t'$ to $\BatchPromise{i}=\s_i$, for some $\s_i$.
By Lemma~\ref{local-less-than-global-promise}, $\s_i\le\GPromise_i$.
Since $\tau' \geq \tapp = \max(\tap, \GReal(\GPromise_i))$,
	at real time $\tau'$
	some process's local clock has value at least $\GPromise_i$.
By Assumptions~\ref{xclocks}(\ref{xcl2}) and~(\ref{xcl5}),
	$q$ reads $t' \geq \GPromise_i$ at real time $\tau'$
	when it executes line~\ref{getleasetime}
	during the last iteration of the loop in lines~\ref{getvalidlease-start}-\ref{getvalidlease-end}.
Since $\GPromise_i\ge\s_i$, $q$ finds $t'\ge\s_i=\BatchPromise{i}$
	in line~\ref{get-k-hat-then},
	and sets $\hat k\ge i$, contradicting~(\ref{claim-3-eqn-1}).
\qedhere~$_\text{Subclaim \autoref{i-set-in-else}}$
\end{proof}

Recall that $p$ is the process that locks $\GBatch_i,-,i)$ at time $(t_i,\tau_i)$.

\begin{subclaim}\label{p-is-not-q}
$p \neq q$.
\end{subclaim}

\begin{proof}
Suppose by contradiction that $p=q$.
So $q$ locks $(\GBatch_i,-,i)$ at real time $\tau_i$.
By Lemma~\ref{mono-lease},
	$q$ has $\leasebatch \geq i$ from real time $\tap$ on.
Since $q$ finds $\lease=(k^*,t^*)$ in line~\ref{getlease} after real time $\tau'$,
	and therefore after time $\tau_i$
	(since $\tau' \geq \tapp \geq \tau_i$),
	$q$ has $k^* \geq i$,
	contradicting Subclaim~\ref{i-set-in-else}.
\qedhere~$_\text{Subclaim \autoref{p-is-not-q}}$
\end{proof}

From the exit condition of the loop in lines~\ref{getvalidlease-start}--\ref{getvalidlease-end},
	$t' < t^* + \LP$.
Recall that at real time $\tau'$,
	$q$ gets $t'$ from its local clock,
	and that $\tap$ is the earliest real time when a tuple $(\GBatch_i, -, i)$ is locked.
Since $\tau'_i = \max(\tau_i, \GReal(\GPromise_i))\leq \tau'$,
	$\tau_i \leq \tau'$ and $\GReal(\GPromise_i) \leq \tau'$.
By Assumptions~\ref{xclocks}(\ref{xcl2}) and (\ref{xcl5}) and the definition of $\GReal$,
	$t_i \leq t' < t^* + \LP$ and
	$\GPromise_i \leq t' < t^* + \LP$.
Therefore the following hold:
\begin{align}\label{yet-another-eqn}
&\text{$q$ has $\lease = (k^*,t^*)$},\notag \\
&\text{$p \neq q$ locks a tuple $(\GBatch_i, - ,i)$ at time~$(\tp, \tap)$},\notag \\
&i > k^*,\\
&\tp < t^* + \LP,~\text{and}\notag \\
&\GPromise_i < t^* + \LP\notag.
\end{align}
Thus, by Lemma~\ref{biggiebiggie} and Corollary~\ref{local-is-global-2},
	from real time $\tap$ on the following hold at $q$:
	\begin{align}\label{another-eqn}
	&\PendingBatchOps{i} = \GBatch_i, \notag\\
	&\PendingBatchPromise{i} = \s_i~\text{or}~0, and \\
	&\MaxPendingIndex \ge i.\notag
	\end{align}

\begin{subclaim}\label{forqed}
$t' \geq t^*$.
\end{subclaim}
\begin{proof}
Suppose the lease $(k^*,t^*)$ held by $q$ is issued by some process $r$ in $\LeaderWork{t_r}$.
Recall that $\tau_i$ is the earliest real time
	when a tuple $(\GBatch_i, -, i)$ is locked,
	and that process $p$ locked it at time $(t_i,\tau_i)$.
Suppose that this locking happens while $p$ was in $\LW(t)$, for some $t$,
	so the tuple it locked was $(\GBatch_i, t, i)$.
Since $r$ issued the lease $(k^*, t^*)$,
	by Lemma~\ref{lease-issue-obs},
	$r$ previously locked a tuple of the form $(-, t_r, k^*)$,
	and by Observation~\ref{acceptedfirst},
	$r$ previously accepted this tuple.
Since $r$ accepts the tuple $(-, t_r, k^*)$ and
	the tuple $(\GBatch_i, t, i)$ is locked in $\LW(t)$ with $i > k^*$
	(Subclaim~\ref{i-set-in-else}),
	by Theorem~\ref{generalcase1a}(\ref{t1}),
	$t_r \leq t$.
We claim that $t_r = t$.
Suppose for contradiction that $t_r < t$.
Then, before process $p$
	locks the tuple $(\GBatch_i, t, i)$ in $\LW(t)$ at time $(t_i, \tau_i)$,
	$p$ completes the wait statement in line~\ref{wait-lease-expire},
	and by Lemma~\ref{lease-expire-after-wait} and the monotonicity of local clocks,
	$t_i \geq t^* + \LP$ ---
	contradicting that $q$ finds $t_i<t^*+\LP$ in line~\ref{getvalidlease-end}
	(see (\ref{yet-another-eqn})).
So $t_r = t$,
	and by Lemma~\ref{unique-LeaderWork},
	$p = r$.
Since $r$ locks the tuples $(\GBatch_i, t_r, i)$ and $(-, t_r, k^*)$ such that $k^* < i$,
	by definition of locking and Corollary~\ref{T-k-DoopsCreduced},
	$r$ locks $(\GBatch_i, t_r, i)$ in a call to $\DO$ made in line~\ref{second-doops}.
By Definition~\ref{GPromise-def},
	this $\DO$ call is $\DoOps{(\GBatch_i, \GPromise_i)}{t_r}{i}$
	and $r$ issues the lease $(i, \GPromise_i)$ in this call.
Since $r$ issues leases $(k^*, t^*)$ and $(i, \GPromise_i)$ in $\LW(t_r)$
	and $i > k^*$,
	by Lemma~\ref{increasing-lease-time-in-LeaderWork},
	$\GPromise_i \geq t^*$.
Since $\tau' \geq \tau'_i \geq \GReal(\GPromise_i)$,
	by Assumptions~\ref{xclocks}(\ref{xcl2}) and~(\ref{xcl5})
	$t' \geq t^*$.
\qedhere~$_\text{Subclaim \autoref{forqed}}$
\end{proof}

Since $t' \geq t^*$, 
	$q$ enters the \emph{else} clause in lines~\ref{get-k-hat-3}-\ref{get-k-hat-5} to compute $\hat k$ during the execution of $\op$.
Note that $q$ sets $u$ to $\MaxPendingIndex$ in line~\ref{get-MPB} after real time $\tau' \ge \tap$.
So by (\ref{another-eqn})
	$q$ has $u \ge i$ in lines~\ref{get-MPB}--\ref{get-k-hat-5}.
Since $i > k^*$, $q$ has $u \ge i > k^*$ in lines~\ref{get-k-hat-else}--\ref{get-k-hat-5}.
Since $q$ computes $\hat{k}$ in lines~\ref{get-k-hat-else}-\ref{get-k-hat-5}
	after real time $\tau' \ge \tapp$,
	by (\ref{another-eqn}) it also has
	$\PendingBatchOps{i} = \GBatch_i$ and
	$\PendingBatchPromise{i} = \s_i$ or $0$
	in these lines.
Since $\s_i \leq \GPromise_i \leq t'$ and $0 \leq t'$,
	$q$ has $\PendingBatchPromise{i} \leq t'$ in these lines.
By (B), $\op.\TYPE$ conflicts with some operation in $\GBatch_i$.
Thus, when $q$ computes $\hat{k}$ in lines~\ref{get-k-hat-else}--\ref{get-k-hat-5},
it has $k^* < i \le u$, $\op.\TYPE$ conflicts with an operation in $\PendingBatchOps{i}$
	and $\PendingBatchPromise{i} \leq t'$,
	so $q$~computes $\hat{k} \ge i$,
	contradicting~(\ref{claim-3-eqn-1}).\qedhere~$_\text{\autoref{sonofbiggiebiggie}}$
\end{proof}
Recall that batch $\hat{k}$ takes effect at real time $\hat \tau$,
	and (by (\ref{i-hat-def}))
	batch $\hat\imath$ takes effect at some real time $\tau_{\hat{\imath}}\ge\hat \tau$.
By Lemma~\ref{distinct-LPs},
	$\hat{\imath} \ge \hat{k}$.
By Claim~\ref{sonofbiggiebiggie}, $\op.\TYPE$ does not conflict with
	any operation in any batch $i$ that takes effect
	at some real time $\tapp$ such that $\hat \tau < \tapp \le \tau'$, and therefore,
	by Lemma~\ref{distinct-LPs},
	with any operation in any batch $i$ such that $\hat k<i\le\hat\imath$.
Thus, by the definition of conflicting operations (see Section~\ref{objects-ops}),
	the response of $\op.\TYPE$ is the same when applied to $\sigma_{\hat k}$
	as when applied to $\sigma_{\hat\imath}$, as wanted.
\qedhere~$_\text{\autoref{Read-linearizability}}$
\end{case}
\end{proof}

By Lemma~\ref{complete-ops-in-L} and
	Theorems~\ref{RMW-linearizability} and~\ref{Read-linearizability},
	every operation that is complete in $E$ has the same response in $E$ as in $\Sigma_E$.
By Lemma~\ref{EL-respects-order},
	$\Sigma_E$ respects the order of non-concurrent operations in $E$.
Therefore,
	
\begin{theorem}\label{linearizability}
The algorithm in Figure~\ref{ObjectAlgo-alg1-code} implements a linearizable object of type $\mathcal{T}$.
\end{theorem}

\subsection{Read lease mechanism: liveness of reads.}\label{sec:read-liveness}

We first make one simplifying assumption that communication links are eventually FIFO.
More precisely:

\begin{assumption}\label{FIFO}
There is a real time $\tauf$ after which if a process $p$ sends a message $m$ and then $m'$ to a process $q$,
	and $q$ receives $m'$, then $q$ receives $m$ before $m'$.
\end{assumption}

We can enforce this by using sequence numbers, and postpone the receipt of messages that are out of order messages for up to $\delta$ local time units.
This does not increase the message delays to beyond $\delta$ in Assumption~\ref{message-delay}.

\begin{lemma}\label{sofia}
	$\ell$ updates $\NextSendTime$ infinitely often 
	during the execution of $\LeaderWork{t}$.
\end{lemma}

\begin{proof}
Suppose, for contradiction, that $\ell$ updates $\NextSendTime$ only a finite number of times
	during the execution of $\LeaderWork{t}$.
Then there is a real time $\tau$ after which $\NextSendTime$ does not change.
By Theorem~\ref{lm10},
	there is a real time after which
	process $\ell$~executes in the while loop of lines~\ref{mainwhile}-\ref{endwhile} in
	$\LeaderWork{t}$ forever.
Since $\ell$ executes infinitely many iterations of this loop,
	by Assumptions~\ref{xclocks}(\ref{xcl2}-\ref{xcl3}),
	there is a real time after $\tau$ such that
	the local clock of $\ell$ has value at least $\NextSendTime$.
Hence, $\ell$ finds that the condition $t' \ge \NextSendTime$ in line~\ref{checksendtime} 
	is satisfied in some iteration of the while loop.
So $\ell$ updates $\NextSendTime$ in line~\ref{nst} after real time~$\tau$ --- a contradiction.
\qedhere~$_\text{\autoref{sofia}}$
\end{proof}

\begin{corollary}\label{sofialoren}
$\ell$ sends a $\langle \CommitLease, (-,-), -, -, - \rangle$ message to every process $p \neq \ell $ infinitely often during the execution
	of $\LeaderWork{t}$.
\end{corollary}

\begin{proof}
By Lemma~\ref{sofia}, $\ell$ updates $\NextSendTime$ infinitely often
	during the execution of $\LeaderWork{t}$.
Note that $\ell$ updates $\NextSendTime$ only in line~\ref{setlease}
	of $\LeaderWork{t}$ or line~\ref{nst2} of $\DO$, and 
	$\ell$ sends a $\langle \CommitLease, (-,-), -, -, - \rangle$ message
	to every process $p \neq \ell$ just before it updates $\NextOps$
	in line~\ref{sl1} or \ref{sl2}.
\qedhere~$_\text{\autoref{sofialoren}}$
\end{proof}

\begin{lemma}\label{no-gap-after-batch-j}
There is a $j_0$ such that for all $j \geq j_0$,
	if a process $p$ receives a $\CM{j}$ message,
	then, for all $i$ such that $1 \leq i < j$,
	process $p$ previously has $\Batch[i] \neq (\emptyset,\infty)$.
\end{lemma}
\begin{proof}
By Lemmas~\ref{lm6.8w} and~\ref{lm6.9w},
	there is a real time after which no process $p \neq \ell$ executes inside $\LW$,
	so there is a $j_1$ such that if a $\CM{j}$ message is sent with $j \geq j_1$,
	it is sent by $\ell$.
By Theorem~\ref{lm10},
	there is a real time after which
	process $\ell$~executes the while loop of lines~\ref{mainwhile}-\ref{endwhile}
	infinitely often in some execution of $\LW(t)$.
So there is $j_2$ such that if $\ell$ sends a $\CM{j}$ message with $j \geq j_2$,
	then it is sent in the while loop of lines~\ref{mainwhile}-\ref{endwhile}
	in $\LW(t)$.
Since only a finite number of $\CM{-}$ messages were sent before real time $\tau_f$,
	there is a $j_3$ such that if a $\CM{j}$ message is sent with $j \geq j_3$,
	then it is sent after real time $\tau_f$.
Let $j_0 = \max(j_1, j_2, j_3 + 1)$,
	and consider any $\CM{j}$ message that $p$ receives with $j\ge j_0$.
Since $j\ge j_1$ this message is sent by $\ell$ during its execution of $\LW(t)$.
There are two places where $\ell$ could have sent this message.

\begin{case}
\item $\ell$ sends a $\CM{j}$ message in line~\ref{sl2}.
Since $j \geq j_0\ge j_2$,
	$\ell$ sends this $\CM{j}$ message in a call to $\DoOps{(-, -)}{t}{j}$
	made in line~\ref{second-doops}.
From the code of $\LW$,
	$\ell$ successfully completed a call to $\DoOps{(-, -)}{t}{j-1}$ before making this $\DO$ call.
Note that $\ell$ sent a $\CM{j-1}$ message to $p$ in $\DoOps{(-, -)}{t}{j-1}$.
Since $j\ge j_0\ge j_3+1$, we have that $j - 1 \geq j_3$, and so
	$p$ received a $\CM{j-1}$ message before it receives the $\CM{j}$ message.
From the code of lines~\ref{lg0}-\ref{fill-lease-gap} and
	the code of lines~\ref{totoz8}-\ref{totoz12},
	it is clear that $p$ sets $\Batch[i]$ to some non-$(\emptyset, \infty)$ value 
	for all $i$, $1 \leq i \leq j-1$, after receiving this $\CM{j-1}$ message,
	which is before it receives the $\CM{j}$ message.

\item $\ell$ sends a $\CM{j}$ message in line~\ref{sl1}.
By Lemma~\ref{lease-issue-obs},
	$\ell$ previously locked a tuple of form $(-, t, j)$.
Note that this happens in a $\DoOps{(-, -)}{t}{j}$ call
	in which $\ell$ sends a $\CM{j}$ message to $p$ in line~\ref{sl2},
	and we are done by Case~1.
\qedhere~$_\text{\autoref{no-gap-after-batch-j}}$
\end{case}
\end{proof}

\begin{lemma}\label{katl}
	There is a real time after which the value of the variable $k$ at $\ell$ is non-decreasing.
\end{lemma}

\begin{proof}
By Theorem~\ref{lm10}, there is a real time after which $\ell$ executes 
	the while loop of lines~\ref{mainwhile}-\ref{endwhile}
	of $\LeaderWork{t}$ forever.
In each iteration of this while loop, $\ell$ can change its variable $k$
	only by calling $\DoOps{-}{t}{k+1}$ in line~\ref{second-doops},
	and this call increments $k$ by one.
\qedhere~$_\text{\autoref{katl}}$
\end{proof}

\begin{lemma}\label{no-gap-after-time}
For each correct process $p$,
	there is a real time after which if $p$ receives a $\CM{j}$ message,
	then, for $1 \leq i < j$,
	process $p$ previously has $\Batch[i] \neq (\emptyset,\infty)$.
\end{lemma}
\begin{proof}
By Lemmas~\ref{lm6.8w} and~\ref{lm6.9w},
	there is a real time after no process $p \neq \ell$ executes inside $\LW()$.
By Theorem~\ref{lm10},
	there is a real time after which
	process $\ell$~executes the while loop of lines~\ref{mainwhile}-\ref{endwhile}
	infinitely often in some execution of $\LeaderWork{t}$.
	So there is a real time $\tau$ after which if a $\CM{j}$ message is received,
	then this message is sent by $\ell$
	in the while loop of lines~\ref{mainwhile}-\ref{endwhile} in $\LW(t)$.	
Note that $\ell$ does not send a $\CM{-}$ message to itself,
	so there is a real time after which $\ell$ does not receive $\CM{-}$ messages,
	and hence the lemma holds vacuously for $\ell$.
Henceforth we consider correct processes other than $\ell$.
There are two cases depending on if the variable $k$ grows unbounded at $\ell$:
\begin{case}
\item The variable $k$ at $\ell$ is bounded.
By Lemma~\ref{katl},
	there is a real time after which the variable $k$ at $\ell$ equals to some value $\hat k$.
By Lemma~\ref{sofialoren},
	$\ell$ sends a $\CM{-}$ message to every process $p \neq \ell$ infinitely often in $\LW(t)$.
So $\ell$ sends infinitely many $\CM{\hat k}$ messages to every process $p \neq \ell$.
Consider any correct process $p \neq \ell$.
Let $\tau'$ be the real time when $p$ receives the second $\CM{\hat k}$ message.
Let $\hat \tau = \max(\tau, \tau')$.
If $p$ receives any $\CM{j}$ message after real time $\hat \tau$,
	then $j = \hat k$ and $p$ previously received a $\CM{\hat k}$ message.
From the code of lines~\ref{lg0}-\ref{fill-lease-gap} and
	the code of lines~\ref{totoz8}-\ref{totoz12},
	it is clear that by the real time $p$ completes line~\ref{fill-lease-gap},
	$p$ has $\Batch[i]$ equal to some  $(\emptyset, \infty)$ pair
	for $1 \leq i \leq \hat{k}-1$,
	and this is before it receives the $\CM{j}$ message.

\item The variable $k$ at $\ell$ grows unbounded.
Let $j_0$ be as defined in Lemma~\ref{no-gap-after-batch-j}.
By Lemma~\ref{katl},
	there is real time after which all $\CM{j}$ messages sent have $j \geq j_0$.
The lemma then follows from Lemma~\ref{no-gap-after-batch-j}.
\qedhere~$_\text{\autoref{no-gap-after-time}}$
\end{case}
\end{proof}

Note that there is a time after which $\FG$ is called only in line~\ref{fill-lease-gap}.
Then by Lemma~\ref{no-gap-after-time}, Corollary~\ref{ne-batch}, and
	the code of lines~\ref{totoz8}-\ref{totoz12}, we have the following:
\begin{corollary}\label{constant-fill-lease-gap}
There is a real time after which if a process $p$ calls $\FillGaps{j}$ in line~\ref{fill-lease-gap},
	then this call completes in a constant number of $p$'s own steps.
\end{corollary}

In the rest of the proof we make the following simplifying assumption:
We assume that the maximum message delay $\delta$ also includes the time that the recipient of a message takes to process this message.
We use this assumption only when the message processing code consists
	of a small, constant number of steps that do not involve waiting.
More precisely:

\begin{assumption}\label{message-delay-processing} [Maximum message delay (including processing)].
There is a known constant~$\delta$ and an unknown real time $\taum$
	after which the following holds:
	For all correct processes $p$ and $q$,
	if $p$ sends a message $m$ to $q$
	then $q$ receives \textbf{and processes} $m$ within $\delta$ time units from when it was sent.
\end{assumption}

We can justify the above assumption by noting that the maximum message delay $\delta$
	guaranteed by Assumption~\ref{message-delay} in practise dwarfs the time a process
	takes to execute a small number of steps at the minimum process speed guaranteed by Assumption~\ref{process-speed}.
Note that this also holds for executing line~\ref{fill-lease-gap}
	by Corollary~\ref{constant-fill-lease-gap}.

Note that:

\begin{enumerate}[(I)]

\item\label{LM1} By Lemmas~\ref{lm6.8w} and~\ref{lm6.9w},
	there is a real time $\RTA$ after which every correct process $p \neq \ell$ executes
	the while loop of lines~\ref{whileo}-\ref{whilee}
	without calling the $\LeaderWork{}$ procedure,
	and so $p$ calls the $\PCM()$ procedure infinitely often in this while loop.

\item\label{LM2} By Theorem~\ref{lm10},
		there is a real time $\RTB$ after which
		process $\ell$~executes the while loop of lines~\ref{mainwhile}-\ref{endwhile}
		infinitely often in some execution of $\LeaderWork{t}$.

\item\label{LM3} By Assumption~\ref{message-delay-processing} 
		there is a real time $\RTC$ after which
		every message sent by $\ell$, or sent to $\ell$, is received and processed within $\delta$ units of time.
		
\end{enumerate}

\begin{definition}\label{ultimate-time1}
$\rs= \max(\RTA, \RTB, \RTC, \tauc, \tauf)$.
\end{definition}

\begin{lemma}\label{doopstimely1}
If process $\ell$ calls $\DoOps{(\Os,\s)}{t}{j}$ after real time $\rs$,
	then at most $2\delta$ units of local time elapsed
	from the instant $\ell$
		first sends a $\langle \Prepare,(\Os,\s),t,j,-\rangle$ message to all processes $p \neq \ell$ in line~\ref{prep-send},
	to the instant when
		$\PACKED[t,j] \supseteq \{ \textrm{all correct processes } q \neq \ell \}$
		first holds at $\ell$.\footnote{Since more than $n/2$ processes are correct,
	this immediately implies that
	at most $2\delta$ units of time elapsed
	from the instant $\ell$
		first sends a $\langle \Prepare,(\Os,\s),t,j,-\rangle$ message to all processes $p \neq \ell$ in line~\ref{prep-send},
		to the instant when
		$| \PACKED[t,j] | \ge  \lfloor n/2 \rfloor$
		first holds at $\ell$.}
\end{lemma}

\begin{proof}
Suppose $\ell$ calls $\DoOps{(\Os,\s)}{t}{j}$ after real time $\rs$.
Recall that after real time $\rs$,
	process $\ell$ executes forever in the while loop of lines~\ref{mainwhile}-\ref{endwhile} of $\LeaderWork{t}$.
Thus, $\ell$ calls $\DoOps{(\Os,\s)}{t}{j}$ in line~\ref{second-doops} of this loop, and this call returns $\textsc{Done}$.
In line~\ref{sendprep} of this $\DoOps{(\Os,\s)}{t}{j}$,
	process $\ell$ sends a $\langle \Prepare,(\Os,\s),t,j,-\rangle$
	to all processes $p \neq \ell$.
Let $\hat{t}$ be the value of the local clock when $\ell$ first sends this message.
Since $\ell$ sends $\langle \Prepare,(\Os,\s),t,j,-\rangle$ to all processes $p \neq \ell$
	after real time $\rs$,
	by property~\ref{LM3},
	all the correct processes $p \neq \ell$
	receive this message from $\ell$
	and process it
	by time $\hat{t} + \delta$ \emph{on $\ell$'s local clock}.
	
	\begin{claim}\label{prevnar}
	Every correct process $p \neq \ell$ sends a $\langle \PACK,t,j \rangle$ message to $\ell$
		by time $\hat{t} + \delta$ on $\ell$'s local clock.
	\end{claim}
	
	\begin{proof}
	Suppose, for contradiction, that some correct process $p \neq \ell$
	does \emph{not} send a $\langle \PACK,t,j \rangle$ message to $\ell$
		by time $\hat{t} + \delta$ on $\ell$'s local clock.
	Let $M$ be the \emph{first} $\langle \Prepare,(\Os,\s),t,j,- \rangle$ message that $p$ receives and processes from $\ell$.
	By the above, $p$ receives and processes $M$ by time $\hat{t} + \delta$ on $\ell$'s local clock.
	After $p$ received $M$ in line~\ref{a3b}, $p$ must have found
		the condition of line~\ref{acceptcheck3} to be false
		(otherwise, $p$ would have executed lines~\ref{client-accept}-\ref{setMPB}, and so it would have sent
		$\langle \PACK,t,j \rangle$ message to $\ell$ in line~\ref{sendPack}
		by time $\hat{t} + \delta$ on $\ell$'s local clock.).
	Since $p$ found that the condition of line~\ref{acceptcheck3} is false,
	there are two cases:
	
	\begin{enumerate}
	
	\item $p$ has $\maxT>t$ in line~\ref{acceptcheck3}.
	Since $\ell$ executes forever in $\LeaderWork{t}$,
		by Lemma~\ref{lastleader2}(\ref{tmaxito2}), $p$ has $\maxT \le t$ always --- a contradiction.

	\item $p$ has $(ts,k) = (t',j')$ for some $(t',j') \ge (t,j)$ in line~\ref{acceptcheck3}.
	Since $t' \ge t \ge 0$, $(t',j')$ is not the initial value $(-1,0)$ of $(ts,k)$ at $p$.
	Thus: (*) $p$ accepted a tuple $(\Os',t',j')$ for some $\Os'$,
		and $p$ accepted $(\Os',t',j')$ before receiving $M$ in line~\ref{a3b}.
	
By (*) and Observation~\ref{leaderfirst}, some process $r$ executed
	$\DoOps{(\Os',-)}{t'}{j'}$ in $\LeaderWork{t'}$.
Since $\ell$ executes forever in $\LeaderWork{t}$,
	by Lemma~\ref{lastleader2}(\ref{last-t2}), $t' \le t$.
Since $(t',j') \ge (t,j)$, it must be that $t' = t$ and $j' \ge j$.
Since $t' = t$, processes $\ell$ and $r$ became leader at the same local time $t$,
	and so, by Lemma~\ref{unique-LeaderWork}, $r = \ell$.
Thus process $\ell$ called $\DoOps{(\Os',-)}{t}{j'}$ in $\LeaderWork{t}$ with $j' \ge j$.

Since $t' = t$, by (*), $p$ accepted $(\Os',t,j')$ before receiving $M$ in line~\ref{a3b}.
Note that $p$ accepted $(\Os',t,j')$ in line~\ref{client-accept}
	($p$ cannot accept
	$(\Os',t,j')$ in line~\ref{leader-accept} of a $\DoOps{(\Os',-)}{t}{j'}$ because $p \neq \ell$,
	and so $p$ does \emph{not} execute $\LeaderWork{t}$).
Therefore: (**) $p$ received a message \mbox{$M' = \langle \Prepare,(\Os',-),t,j',- \rangle$ in line~\ref{a3b}}
	before receiving $M$ in line~\ref{a3b}.

There are two cases:

\begin{enumerate}
\item $j' = j$.
	Since $\ell$ calls $\DoOps{(\Os,\s)}{t}{j}$ and $\DoOps{(\Os',\s')}{t}{j}$,
	by Lemma~\ref{doops-simplecase}, $(\Os,\s) = (\Os',\s')$.
	By~(**), $p$ received $M' = \langle \Prepare,(\Os,\s),t,j,-\rangle$ before receiving $M$ in line~\ref{a3b}
	--- a contradiction to the definition of~$M$.

\item $j' > j$.
	By (**) $p$ received \mbox{$M' = \langle \Prepare,(\Os',-),t,j',-\rangle$ in line~\ref{a3b}}
	before receiving $M$ in line~\ref{a3b}.
	Since only a process that executes $\DoOps{(\Os',-)}{t}{j'}$
		can send a $\langle \Prepare,(\Os',-),t,j',-\rangle$ message,
		and such a process must be in $\LeaderWork{t}$,
		$M'$ was sent by $\ell$ in $\DoOps{(\Os',-)}{t}{j'}$.
	Thus $\ell$ called $\DoOps{(\Os',-)}{t}{j'}$
		before $p$ received $M'$ from $\ell$ in line~\ref{a3b},
		and so before $p$ received $M$ in line~\ref{a3b}.
	Therefore $\ell$ called $\DoOps{(\Os',-)}{t}{j'}$ by time $\hat{t} + \delta$ on $\ell$'s local clock..
	
	Since $j'>j$, by Corollary~\ref{T-k-DoopsCreduced},
		$\ell$ calls $\DoOps{(\Os,\s)}{t}{j}$ and returns from this call
		\emph{before} calling $\DoOps{(\Os',-)}{t}{j'}$.
	So $\ell$ sends $M$ to $p$ in $\DoOps{(\Os,\s)}{t}{j}$ \emph{before}
		sending $M'$ to $p$ in $\DoOps{(\Os',-)}{t}{j'}$.
	Since the communication channel from $\ell$ to $p$ is FIFO from time $\tauf$ on,
		and $\ell$ sends $M$ and $M'$ after time $t_s \ge \tf$,
		$p$ receives $M$ before receiving $M'$ --- a contradiction to (**).
	\end{enumerate}
	\end{enumerate}
Since every case leads to a contradiction, the claim holds.
	\qedhere~$_\text{\autoref{prevnar}}$
	\end{proof}
By Claim~\ref{prevnar} and property~\ref{LM3},
	$\ell$ receives and processes a $\langle \PACK,t,j \rangle$ message from every correct process $p \neq \ell$
		by time $\hat{t}+2 \delta$ on $\ell$'s local clock.
	So $\ell$ inserts every correct process $p \neq \ell$ into $\PACKED[t,j]$
		by time $\hat{t}+2 \delta$ on $\ell$'s local clock.
	Thus, $\ell$ has $\PACKED[t,j] \supseteq \{ \textrm{all correct processes } p \neq \ell \}$  
		by time $\hat{t}+2 \delta$ on $\ell$'s local clock.
\qedhere~$_\text{\autoref{doopstimely1}}$
\end{proof}

\begin{lemma}\label{cdoopstimely1}
If process $\ell$ calls $\DoOps{(\Os,\s)}{t}{j}$ after time $\rs$ then
	$\ell$'s local clock increases by at most $2\delta$
	from the instant $\ell$
		first sends a $\langle \Prepare,(\Os,\s),t,j,-\rangle$ message to all processes $p \neq \ell$ in line~\ref{prep-send},
	to the instant when
	$\ell$ completes the wait statement of line~\ref{wait2}.
\end{lemma}

\begin{proof}
Suppose $\ell$ calls $\DoOps{(\Os,\s)}{t}{j}$ after time $\rs$,
	and it first sends a $\langle \Prepare,(\Os,\s),t,j,-\rangle$ message to all processes $p \neq \ell$ in line~\ref{sendprep}
	at some local time $\hat{t}$.
By Lemma~\ref{doopstimely1}, since more than $n/2$ processes are correct,
	$\ell$ exits the repeat-until loop of lines~\ref{repeat}-\ref{endc}
	with $| \PACKED[t,j] | \ge  \lfloor n/2 \rfloor$ by local time $\hat{t} + 2\delta$.
Since $\rs \geq \RTB$, 
	$\ell$ executes the while loop of lines~\ref{mainwhile}-\ref{endwhile} infinitely often,
	it does not return in line~\ref{ack-condition}.
Note that in line~\ref{wait2}, $\ell$ waits for at most $2\delta$ local time units from local time $\hat{t}$
	it first sent the $\langle \Prepare,(\Os,\s),t,j,-\rangle$ message in line~\ref{sendprep}.
Thus $\ell$ completes the wait statement of line~\ref{wait2} by time $\hat{t} + 2\delta$.
\qedhere~$_\text{\autoref{cdoopstimely1}}$
\end{proof}

\begin{lemma}\label{LHwithNoDead}
There is a real time after which:
	(a) $\LH$ at $\ell$ contains only correct processes,
	or (b)~$\ell$ does not call $\DoOps{(-,-)}{-}{-}$.
\end{lemma}

\begin{proof}
If $\ell$ calls $\DoOps{(-,-)}{-}{-}$ only a finite number of times, then the lemma trivially holds.
Henceforth assume that $\ell$ calls $\DoOps{(-,-)}{-}{-}$ infinitely often.
By Theorem~\ref{lm10}
	process $\ell$~executes the while loop of lines~\ref{mainwhile}-\ref{endwhile}
	infinitely often in some execution of $\LeaderWork{t}$.
Thus, $\ell$ calls $\DoOps{(-,-)}{t}{-}$ infinitely often in $\LeaderWork{t}$ (and it never exits $\LeaderWork{t}$).
Let $p$ be any process that crashes.
Say that it crashes at real time $\tau$, and 
	and let $\tau'$ be the real time after which
	$\ell$ does not receive any $\langle \RequestLease \rangle$ message from $p$.
Consider the first time that
	$\ell$ calls $\DoOps{(-,-)}{t}{-}$ after real time $\max(\tau,\tau')$.
Note that this $\DoOps{(-,-)}{t}{-}$ returns $\textsc{Done}$ (because $\ell$ does not exit $\LeaderWork{t}$).
So in this $\DoOps{(-,-)}{t}{-}$
	$\ell$ sends $\langle \Prepare,(-,-),-,-,-\rangle$ to all processes except itself in line~\ref{sendprep},
	and then, in line~\ref{lh2},
	$\ell$ sets $\LeaseHolders$ to the set of processes that replied to this $\langle \Prepare,(-,-),-,-,-\rangle$ message.
Since $p$ crashed before $\ell$ called this $\DoOps{(-,-)}{-}{-}$,
	$p$ did not reply to the $\langle \Prepare,(-,-),-,-,-  \rangle$ message,
	and so $p \not \in \LH$ at $\ell$ in line~\ref{lh2}.
We claim that $\ell$ never adds $p$ to $\LH$ thereafter.
This is because:
	(1) $\ell$ does not receive any $\langle \RequestLease \rangle$ message from $p$,
		so it does not add $p$ to $\LH$ in line~\ref{lh3}, and
	(2) $\ell$ does not receive any reply to $\langle \Prepare,(-,-),-,-,- \rangle$ messages from $p$,
		so it does not add $p$ to $\LH$ in line~\ref{lh2}.
Thus, there is a real time after which $p \not \in \LH$ at $\ell$.
Since $p$ is an arbitrary process that crashed,
	there is a real time after which $\LH$ at $\ell$ contains only correct processes.
\qedhere~$_\text{\autoref{LHwithNoDead}}$
\end{proof}

Every correct process $p \neq \ell$ is in $\LH$ infinitely often at $\ell$.
More precisely:

\begin{lemma}\label{LH-has-all0}
For every correct process $p \neq \ell$, and every real time $\tau$,
	there is a real time $\tau' > \tau$ such that $p \in \LH$ at $\ell$ at real time $\tau'$.
\end{lemma}

\begin{proof}
Suppose, for contradiction, that there is a correct process $p \neq \ell$ and
	a real time $\tau$ after which \mbox{$p \not \in \LH$} at $\ell$.
By Theorem~\ref{lm10},
	there is a real time after which
	process $\ell$~executes the while loop of lines~\ref{mainwhile}-\ref{endwhile}
	infinitely often in some execution of $\LeaderWork{t}$.
 By Corollary~\ref{sofialoren},
	 $\ell$ sends a $\langle \CommitLease, -, -, \lease, \LeaseHolders \rangle$ message
	 to $p$ infinitely often during the execution of $\LeaderWork{t}$.
Let $\LM$ be the first such message
	that $\ell$ sends to $p$ after real time $\hat{\tau} = \max(\tau, \rs)$.
Note that this $\LM= \langle \CommitLease, -, -, \lease, \LeaseHolders \rangle$
	has $p \not \in \LH$ because it is sent after real time $\tau$.
Since $\LM$ is sent after real time $\rs$, by properties~\ref{LM1} and~\ref{LM3}
	$p$ eventually receives $\LM$ from $\ell$ (in line~\ref{lg0}),
Since $p \not \in \LH$, $p$ replies by sending a
	$\langle \RequestLease \rangle$ message to $\ell$ in line~\ref{lg3}.
By properties~\ref{LM2} and~\ref{LM3}, $\ell$ eventually receives this $\langle \RequestLease \rangle$ from $p$,
	and then $\ell$ adds $p$ to $\LH$ in the line~\ref{lh3}.
Since this occurs after real time~$\tau$, this contradicts the definition of $\tau$.
\qedhere~$_\text{\autoref{LH-has-all0}}$
\end{proof}

\begin{lemma}\label{LH-has-all}
For every correct process $p \neq \ell$, there is a real time after which $p \in \LH$ at $\ell$.
\end{lemma}

\begin{proof}
Suppose, for contradiction, that there is a correct process $p \neq \ell$ such that
	for every real time $\tau$,
	there is a real time $\tau' > \tau$ such that $p \not \in \LH$ at $\ell$ at real time $\tau'$.
By Lemma~\ref{LH-has-all0}, this implies that
	$\ell$ adds and removes $p$ from $\LH$ infinitely many times.
By Theorem~\ref{lm10},
	there is a real time after which
	process $\ell$~executes the while loop of lines~\ref{mainwhile}-\ref{endwhile}
	infinitely often in some execution of $\LeaderWork{t}$.
This implies that there is a real time after which
	$\ell$ can remove $p$ from $\LH$ only in line~\ref{lh2}
	during the execution of some call to $\DoOps{(-,-)}{-}{-}$.
Let $\DoOps{(\Os,\s)}{t}{j}$ be any $\DoOps{(-,-)}{-}{-}$ that
	$\ell$ calls after real time $\rs$, such that
	$\ell$ removes $p$ from $\LH$ in this $\DoOps{(-,-)}{-}{-}$:
	i.e., $\ell$ calls $\DoOps{(\Os,\s)}{t}{j}$ after real time $\rs$, and
	(i) $p \in \LH$ before $p$ executes line~\ref{lh2} of $\DoOps{(\Os,\s)}{t}{j}$,
	and
	(ii) $p \not \in \LH$ after $p$ executes line~\ref{lh2} of $\DoOps{(\Os,\s)}{t}{j}$.
	
Note that in $\DoOps{(\Os,\s)}{t}{j}$,
	process $\ell$ sends $\langle \Prepare,(\Os,\s),t,j,-\rangle$ to $p$ at some local time $\hat{t}$.
Since process $p\neq\ell$ is correct, by Lemma~\ref{doopstimely1},
	$\ell$ has $p \in \PACKED[t,j]$ by local time $\hat{t}+2 \delta$ on $\ell$'s clock.
Since $\ell$ removes $p$ in line~\ref{lh2} of $\DoOps{(\Os,\s)}{t}{j}$,
	$p \not \in \PACKED[t,j]$ in line~\ref{lh2}.
So $p \not \in \PACKED[t,j]$ during $\ell$'s execution of line~\ref{wait2}.
Since $p \in \LH$ before $p$ executes line~\ref{lh2} of $\DoOps{(\Os,\s)}{t}{j}$,
	$p \in \LH$ during $\ell$'s execution of line~\ref{wait2}.
Thus, $\LeaseHolders \subseteq \PACKED[t,j]$ does \emph{not} hold during $\ell$'s wait in line~\ref{wait2}.
So $\ell$ waits $2\delta$ units of local time (from the time it first executed line~\ref{sendprep}) in line~\ref{wait2}.
Thus $\ell$ exits the wait statement in line~\ref{wait2} at local time $\hat{t}+2 \delta$,
	and when it does so, $\ell$ has $p \in \PACKED[t,j]$.
Since $\PACKED[t,j]$ is non-decreasing, $\ell$ also has $p \in \PACKED[t,j]$ in line~\ref{lh2} --- a contradiction.
\qedhere~$_\text{\autoref{LH-has-all}}$
\end{proof}

From Lemmas~\ref{LHwithNoDead} and~\ref{LH-has-all}:

\begin{enumerate}[(I)]
\setcounter{enumi}{3}

\item\label{LM4} There is a real time $\RTD$ after which
	(a) $\LH$ at $\ell$ contains only correct processes,
	or (b)~$\ell$ does not call $\DoOps{(-,-)}{-}{-}$.

\item\label{LM5} There is a real time $\RTE$ after which
	$\LH$ at $\ell$ contains every correct process $p \neq \ell$.

\end{enumerate}

In the following, we consider the following time:

\begin{definition}\label{ultimate-time2}
$\ru= \max(\rs, \RTD, \RTE)$.
\end{definition}

\begin{definition}
A \emph{lease message} is a message of the form 
$\langle \CommitLease,(-,-),-, \lease, \LeaseHolders \rangle$.
\end{definition}

\begin{lemma}\label{doopstimely2}
If process $\ell$ calls $\DoOps{(\Os,\s)}{t}{j}$ after real time $\ru$
	then $\ell$ does not wait in line~\ref{wait-alg1}.
\end{lemma}

\begin{proof}
Suppose $\ell$ calls $\DoOps{(\Os,\s)}{t}{j}$ after real time $\ru$.
Recall that after real time $\ru$,
	process $\ell$ executes forever in the while loop of lines~\ref{mainwhile}-\ref{endwhile} of $\LeaderWork{t}$.
Thus, $\ell$ calls $\DoOps{(\Os,\s)}{t}{j}$ in line~\ref{second-doops} of this loop,
	and this call returns $\textsc{Done}$.
In line~\ref{sendprep} of this $\DoOps{(\Os,\s)}{t}{j}$,
	process $\ell$ sends a $\langle \Prepare,(\Os,\s),t,j,-\rangle$
	to all processes $p \neq \ell$.
Let $\hat{t}$ be the value of the local clock of $\ell$ when $\ell$ first sends this message.
	
By Lemma~\ref{doopstimely1},
	$\ell$ has $\{ \textrm{all correct processes } p \neq \ell \} \subseteq \PACKED[t,j]$  
	by time $\hat{t}+2 \delta$ on $\ell$'s local clock.
	We now show that $\ell$ does not wait in line~\ref{wait-alg1} of $\DoOps{(\Os,\s)}{t}{j}$.
	Suppose, for contradiction, that $\ell$ waits in line~\ref{wait-alg1}.
	Then, $\ell$ has $\neg (\LeaseHolders \subseteq \PACKED[t,j])$ in line~\ref{LHvsAcks}~(*).
	Thus $\ell$ did \emph{not} exit the wait statement of line~\ref{wait2} with $\LeaseHolders \subseteq \PACKED[t,j]$.
	So $\ell$ exits the wait statement of line~\ref{wait2} after waiting for $2\delta$
	units of local time to elapse from the moment it first executed line~\ref{sendprep}.
	Therefore when $\ell$ executes line~\ref{LHvsAcks},
		$\ell$'s local clock is at least $\hat{t}+2 \delta$,
		and so
		$\ell$ has $\{ \textrm{all correct processes } p \neq \ell \} \subseteq \PACKED[t,j]$ at this time.

We claim that when $\ell$ executes line~\ref{LHvsAcks},
	$\LH \subseteq \{ \textrm{all correct processes } p \neq \ell \}$.
This is because: (1)~$\ell$ calls $\DoOps{(\Os,\s)}{t}{j}$ after real time $\ru \ge \RTD$,
	and so, by property~\ref{LM4}, 
	$\LH$ contains \emph{only} correct processes,
	and
	(2)~by Lemma~\ref{OnLH}, $\ell \not \in \LeaseHolders$.
Thus, when $\ell$ executes line~\ref{LHvsAcks},
	$\ell$ has $\LH \subseteq \{ \textrm{all correct processes } p \neq \ell \} \subseteq \PACKED[t,j]$ --- contradicting~(*).	
\qedhere~$_\text{\autoref{doopstimely2}}$
\end{proof}

\begin{lemma}\label{Alpha123}
There are constants $\alpha_1$ and $\alpha_2$, and a real time $\tau_g \ge \tau_u$ after which
	process $\ell$ executes a full iteration of the while loop of lines~\ref{mainwhile}-\ref{endwhile} of $\LeaderWork{t}$
	in at most:
\begin{enumerate}[\noindent(1)]
\item\label{ATM1}
	$\alpha_1$ local time units, if $\ell$ does \emph{not} call the $\DoOps{(-,-)}{-}{-}$ procedure in line~\ref{second-doops}
	of this iteration.
	
\item\label{ATM2}
	$\alpha_2 + 2\delta$ local time units, if $\ell$ calls the $\DoOps{(-,-)}{-}{-}$ procedure in line~\ref{second-doops}
	of this iteration.

Moreover, $\alpha_1 \le \alpha_2 + 2\delta$ and $\ell$ is at line~\ref{mainwhile} at real time $\tau_g$.
\end{enumerate}

\end{lemma}

\begin{proof}
Each iteration of the while loop of lines~\ref{mainwhile}-\ref{endwhile}
	such that $\ell$ does \emph{not} call the $\DoOps{(-,-)}{-}{-}$ procedure in line~\ref{second-doops}
	consists of a constant number of steps by $\ell$.
By Assumption~\ref{process-speed} (and the fact that $\tau_g \ge \tau_u \ge \tauc)$,
	there is a constant $\alpha_1$
	such that $\ell$ executes these steps in at most $\alpha_1$ local time units.
So Part~(1) of the lemma holds.

Each iteration of the while loop of lines~\ref{mainwhile}-\ref{endwhile}
	such that  $\ell$ calls the $\DoOps{(-,-)}{-}{-}$ procedure in line~\ref{second-doops},
	consists of a constant number of steps by $\ell$, plus the following:
	(1) $\ell$'s execution of the periodically-until loop of lines~\ref{repeat}-\ref{endc},
	followed by $\ell$'s wait in line \ref{wait2},
	and
	(2) $\ell$'s wait in line \ref{wait-alg1}.
By Corollary~\ref{cdoopstimely1}, at most $2\delta$ local time units elapse
	from the moment $\ell$ starts executing the periodically-until loop of lines~\ref{repeat}-\ref{endc}
	to the moment $\ell$ exits the wait statement of line \ref{wait2}.
Furthermore, by Lemma~\ref{doopstimely2} (and the fact that $\tau_g \ge \tau_u$),
	$\ell$ does not wait in line \ref{wait2}.
Thus, by Assumption~\ref{process-speed},
	there is a constant $\alpha_2$
	such that
	$\ell$ takes at most $\alpha_2 + 2\delta$ local time units
	to execute an iteration of the while loop of lines~\ref{mainwhile}-\ref{endwhile} of $\LeaderWork{t}$
	that includes a call to the $\DoOps{(-,-)}{-}{-}$ procedure in line~\ref{second-doops}.
It is clear that we can chose $\alpha_2$ such that $\alpha_1 \le \alpha_2 + 2\delta$,
	and $\tau_g$ such that at real time $\tau_g$
	process $\ell$ is at the start of the loop
	in lines~\ref{mainwhile}-\ref{doops2failed}
	that it executes infinitely often.
\qedhere~$_\text{\autoref{Alpha123}}$
\end{proof}

In practice the constant $\alpha_1$ and $\alpha_2$ above are very small constants
	(they measure the time that $\ell$ takes to execute a few local steps that do not involve waiting),
	and they are negligible compared to the maximum message delay $\delta$.

\begin{definition}\label{Alpha}
Let $\alpha_0 = \alpha_1 + \alpha_2$, where $\alpha_1$ and $\alpha_2$ are specified by Lemma~\ref{Alpha123}.
\end{definition}

In the next lemma we will show that, after the system stabilizes,
	the leader sends lease messages at regular intervals.
As we will see this ensures that eventually
	all correct processes always have valid leases (Theorem~\ref{all-have-valid-lease}).

\begin{lemma}\label{lease-frequency}
For all $i \ge 0$, $\ell$ executes the following events
	in lines~\ref{setlease} and~\ref{sl1} or lines~\ref{lease1} and \ref{sl2}
	after real time~$\tau_g$:

	\begin{itemize}
	
	\item $\ev{i}{l}$ : $\ell$ sets its lease variable to $(k_i, t_i)$ for some $k_i$ and $t_i$,
	\item $\ev{i}{s}$ : $\ell$ sends the lease message $\LM_i = \langle \CommitLease, -, -, (k_i, t_i), \lh_i \rangle$
		to all $p \neq \ell$ 

	\end{itemize}

Furthermore, $\ell$ executes $\ev{i}{l}$ and $\ev{i}{s}$ at times
	$(t_i^l, \tau_i^l)$ and $(t_i^s, \tau_i^s)$, respectively,	
	such that:

	\begin{enumerate}

	\item\label{kx1} $\tau_i^l < \tau_i^s$ and $t_i^l \leq t_i^s$

	\item\label{kx2} $\lh_i$ contains every correct process $p \neq \ell$
	
	\item\label{kx3} if $i > 0$ then:
	
		\begin{enumerate}

		 \item\label{kx3-1} $\rt{i-1}{s} < \rt{i}{l}$ and $\lt{i-1}{s} \leq \lt{i}{l}$

		\item\label{kx3-2} $\lt{i}{s} \le t_{i-1} + \Ptwo + 2 \delta + \alpha_0$

		\item\label{kx3-3} $k_{i} \ge k_{i-1}$

		\item\label{kx3-4} $\ell$ does not change its $\lease$ variable between events $\ev{i-1}{l}$ and $\ev{i}{l}$
		\item\label{kx3-5} $\ell$ does not send any lease message between events $\ev{i-1}{s}$ and $\ev{i}{s}$
		 		
	\end{enumerate}	
\end{enumerate}
\end{lemma}

\begin{proof}
By induction on $j$ we now show that for all $i$ and $j$, $0 \le i \le j$, 
	process $\ell$ executes the	
	events $\ev{i}{l}$ and $\ev{i}{s}$ described in the lemma
	at some times $(\lt{i}{l}, \rt{i}{l})$ and $(\lt{i}{s}, \rt{i}{s})$, respectively,
	such that properties~\ref{kx1}-\ref{kx3} above hold.

\smallskip
\noindent
\textsc{Basis.} $j =0$ (and hence $i =0$).
By Lemma~\ref{sofia},
	$\ell$ updates $\NextSendTime$ infinitely often in the while loop of lines~\ref{mainwhile}-\ref{endwhile} of $\LeaderWork{t}$.
	Consider the first time $\ell$ updates $\NextSendTime$ after real time $\tau_g$.
Note that this can happen in lines~\ref{nst} or \ref{nst2}.
From the code, it is clear that
	just before $\ell$ updates $\NextSendTime$,
	$\ell$ executes the following events in lines~\ref{setlease} and~\ref{sl1} or lines~\ref{lease1} and \ref{sl2}:
	
	\begin{itemize}
	
	\item $\ev{0}{l}$ : $\ell$ sets its lease variable to $(k_0, t_0)$ for some $k_0$ and $t_0$ 
	\item $\ev{0}{s}$ : $\ell$ sends the lease message $\LM_0 = \langle \CommitLease, -, -, (k_0, t_0) , \lh_0 \rangle$
		to all $p \neq \ell$ 

	\end{itemize}

Clearly these two events occur after real time~$\tau_g$.
Furthermore, suppose that $\ell$ executes $\ev{0}{l}$ and $\ev{0}{s}$ at times
	$(\lt{0}{l}, \rt{0}{l})$ and $(\lt{0}{s}, \rt{0}{s})$, respectively.
Since $\ell$ executes $\ev{0}{l}$ and $\ev{0}{s}$ in this order,
	$\rt{0}{l} < \rt{0}{s}$.
By the monotonicity of the local clock of $\ell$, this implies
	$\lt{0}{l} \le \lt{0}{s}$.
Thus property~\ref{kx1} of the lemma holds.
Since $\ell$ sends $\LM_0$ after time~$\tau_g \ge \ru$,
	by the definition of $\ru$ and property~(\ref{LM5}),
	$\lh_0$ contains every correct process $p \neq \ell$;
	so property~\ref{kx2} of the lemma holds.
Since $i =0$, property~\ref{kx3} is trivially true.

\smallskip
\noindent
\textsc{Induction Step.}
Suppose that for all $i$ and $j$, such that $0 \le i \le j$,
	process $\ell$ executes the following events
	in lines~\ref{setlease} and~\ref{sl1} or lines~\ref{lease1} and \ref{sl2},
	after time $\tau_g$:

	\begin{itemize}	
	\item $\ev{i}{l}$ : $\ell$ sets its lease variable to $(k_{i}, t_i)$ for some $k_{i}$ and $t_i$,
	\item $\ev{i}{s}$ : $\ell$ sends the lease message $\LM_{i} = \langle \RenewLease, (k_{i}, t_i) , \lh_{i} \rangle$
		to all $p \neq \ell$,
	\end{itemize}
	
and $\ell$ executes $\ev{i}{l}$ and $\ev{i}{s}$ at times
	$(\lt{i}{l}, \rt{i}{l})$ and $(\lt{i}{s}, \rt{i}{s})$, respectively,
	such that \mbox{properties~\ref{kx1}-\ref{kx3}~hold.}

We now prove that the above also holds for all $i$ such that $0 \le i \le j+1$.
To do so, we show that $\ell$ executes events $\ev{j+1}{l}$, and $\ev{j+1}{s}$
	 at times $(\lt{j+1}{l}, \rt{j+1}{l})$ and $(\lt{j+1}{s}, \rt{j+1}{s})$
	 that satisfy properties~\ref{kx1}-\ref{kx3} for $i=j+1$.

By Corollary~\ref{sofialoren}, $\ell$ sends a $\langle \CommitLease, -, -, -, - \rangle$ message to all $p \neq \ell$ 
	during the execution of $\LeaderWork{t}$ infinitely many times.
Consider the \emph{first time} that
	$\ell$ sends a $\langle \CommitLease, -, -, -, - \rangle$ message to all $p \neq \ell$ \emph{after} event $\ev{j}{s}$, and let $\ev{j+1}{s}$ denote this event.
It is clear that $\ell$ executes the following
	sequence of events in lines~\ref{setlease} and~\ref{sl1} or lines~\ref{lease1} and \ref{sl2}, after real time $\tau_g$:
	
	\begin{itemize}
	\item $\ev{j+1}{l}$ : $\ell$ sets its lease variable to $(k_{j+1}, t_{j+1})$ for some $k_{j+1}$ and $t_{j+1}$, and
	\item $\ev{j+1}{s}$ : $\ell$ sends $\LM_{j+1} = \langle \CommitLease, -, -, (k_{j+1}, t_{j+1}) , \lh_{j+1} \rangle$
		for some $\lh_{j+1}$ to all $p \neq \ell$.
	\end{itemize}

Let $(\lt{j+1}{l}, \rt{j+1}{l})$ and $(\lt{j+1}{s}, \rt{j+1}{s})$ be the times when
	$\ev{j+1}{l}$ and $\ev{j+1}{s}$ occur, respectively.

We first show that property (\ref{kx3-2}) holds, i.e.,
	$\lt{j+1}{s} \le t_j + \Ptwo + 2 \delta + \alpha_0$.
We define two more events $\ev{j}{c}$ and $\ev{j+1}{c}$.
Let $\ev{j}{c}$ be the last reading of the clock by $\ell$ in line~\ref{taketime1} that occurs before $\ev{j}{s}$, 
	and similarly, 
	let $\ev{j+1}{c}$ be the last reading of the clock by $\ell$ in line~\ref{taketime1} that occurs before $\ev{j+1}{s}$.
Suppose events $\ev{j}{c}$ and $\ev{j+1}{c}$ happen
	at times $(\lt{j}{c}, \rt{j}{c})$ and $(\lt{j+1}{c}, \rt{j+1}{c})$.
Then $\ell$ reads $\lt{j}{c}$ and $\lt{j+1}{c}$ respectively from its clock
	when executing events $\ev{j}{c}$ and $\ev{j+1}{c}$.
It is clear that $\rt{j}{c} \leq \rt{j+1}{c}$, i.e.,
	$\ev{j}{c}$ happens in the same iteration of the while loop of lines~\ref{mainwhile}-\ref{endwhile} as $\ev{j+1}{c}$,
	or that $\ev{j}{c}$ happens in a previous iteration of the while loop.
By Assumption~\ref{xclocks}(\ref{xcl2}),
	$\lt{j}{c} \leq \lt{j+1}{c}$.
\begin{claim}\label{lease-at-least-clock}
$\lt{j}{c} \leq t_j$
\end{claim}
\begin{proof}
Recall that $t_j$ is the start time of the lease that is
	included in the $\CommitLease$ message that is sent during event $\ev{j}{s}$.
By definition of $\lt{j}{c}$, either $t_j = \lt{j}{c}$ (when $\ev{j}{l}$ and $\ev{j}{s}$ occur in lines~\ref{setlease} and~\ref{sl1})
	or $t_j = \lt{j}{c} + \PP$ (when $\ev{j}{l}$ and $\ev{j}{s}$ occur in lines~\ref{lease1} and~\ref{sl2}).
By Assumption~\ref{promiseRange}, 
	we have $\lt{j}{c} \leq t_j$.
\qedhere~$_\text{\autoref{lease-at-least-clock}}$
\end{proof}
After $\ell$ sends $\LM_{j}$ in line~\ref{sl1} or~\ref{sl2} at time $(\lt{j}{s}, \rt{j}{s})$ (event $\ev{j}{s}$),
	it updates $\NextSendTime : =  t_j + \Ptwo$ in line~\ref{nst} or~\ref{nst2}.
Suppose this update happens at real time $\tau_{nst}$.
Then, it is clear that $\rt{j}{s} < \tau_{nst} < \rt{j+1}{s}$.
\begin{claim}\label{nst-does-not-change}
$\ell$ does not set $\NextSendTime$ during the real time interval $(\tau_{nst}, \rt{j+1}{s}]$.
\end{claim}
\begin{proof}
Suppose, by contradiction, that $\ell$ sets $\NextSendTime$
	during the real time interval $(\tau_{nst}, \rt{j+1}{s}]$.
Then, $\ell$ would send a $\CommitLease$ message right before it updates $\NextSendTime$,
	and this sending of $\CommitLease$ messages happens between events $\ev{j}{s}$ and $\ev{j+1}{s}$,
	which contradicts the deinition of $\ev{j+1}{s}$.
\qedhere~$_\text{\autoref{nst-does-not-change}}$
\end{proof}
To show that property (\ref{kx3-2}) holds, 
	we discuss two cases depending on where $\ell$ executes event $\ev{j+1}{l}$:

\begin{case}
\item $\ell$ executes event $\ev{j+1}{l}$ in line~\ref{setlease}.
We have that $\ell$ executes $\ev{j+1}{s}$ in line~\ref{sl1}.
In this case, it is clear that 
	event $\ev{j}{c}$ occurs in an earlier iteration of the while loop of lines~\ref{mainwhile}-\ref{endwhile}
	than the iteration of the while loop in which $\ev{j+1}{c}$ occurs.
Consider the last reading the clock by $\ell$ in line~\ref{taketime1}
	before event $\ev{j+1}{c}$.
Denote this event $\ev{}{c}$.
Suppose that this event happens at time $(t^c, \tau^c)$.
Then $\ell$ reads $t^c$ from its clock when executing event $\ev{}{c}$.
So we have $\lt{j}{c} \leq t^c \leq \lt{j}{c}$.
Then we have $\lt{j}{c} \leq t^c \leq \lt{j+1}{c}$ and
			 $\rt{j}{c} < \tau^c < \rt{j+1}{c}$.
\begin{claim}\label{foo1}
$t^c < \lt{j}{c} + \Ptwo$.
\end{claim}
\begin{proof}
If $\lt{j}{c} = t^c$ the claim is trivially true.
Henceforth suppose that $\lt{j}{c} < t^c$ (so $\ev{j}{c}$ occurs before $\ev{}{c}$).
Recall that $\ell$ sets $\NextSendTime$ to $\lt{j}{c} + \Ptwo$
	at real time $\tau_{nst}$.
Since this happens in the same iteration of the while loop during which event $\ev{j}{c}$ occurs,
	$\tau_{nst} \leq \tau^c$.
After $\ell$ reads $t^c$ from its clock in line~\ref{taketime1},
	it compares $t^c$ with $\NextSendTime$ in line~\ref{checksendtime}.
Note that this comparison happens between real times $\tau_{nst}$ and $\rt{j+1}{s}$,
	by Claim~\ref{nst-does-not-change},
	$\NextSendTime$ has value $\lt{j}{c} + \Ptwo$.
We claim that $\ell$ finds $t^c < \NextSendTime$ in line~\ref{checksendtime}, 
	since otherwise, $\ell$ will send $\CommitLease$ messages in line~\ref{sl1}, 
	and this occurs between events $\ev{j}{s}$ and $\ev{j+1}{c}$, 
	which contradicts the definition of $\ev{j+1}{c}$.
So $t^c < \NextSendTime = \lt{j}{c} + \Ptwo$.
\qedhere~$_\text{\autoref{foo1}}$
\end{proof}
Between events $\ev{}{c}$ and $\ev{j+1}{s}$,
	$\ell$ executes a full iteration of the while loop from line~\ref{taketime1} to line~\ref{mainwhile},
	and an incomplete iteration of the while loop that does not call $\DO$ from line~\ref{taketime1} to line~\ref{sl1}.
By Lemma~\ref{Alpha123}, Definition~\ref{Alpha}, and Claim~\ref{foo1},
	$\lt{j+1}{s} \leq t^c +2\delta + \alpha_1 + \alpha_2 < \lt{j}{c} + \Ptwo + 2\delta + \alpha_0$. 

\item $\ell$ executes event $\ev{j+1}{s}$ in line~\ref{sl2}.
\begin{claim}\label{not-yet}
$\lt{j+1}{c} < \lt{j}{c} + \Ptwo.$
\end{claim}
\begin{proof}
Recall that $\lt{j+1}{c} \geq \lt{j}{c}$.
If $\lt{j+1}{c} = \lt{j}{c}$, then the claim follows from Claim~\ref{lease-at-least-clock}.
Henceforth we assume that $\lt{j+1}{c} < \lt{j}{c}$.
Consider when $\ell$ compares $\lt{j+1}{c}$ with $\NextSendTime$ in line~\ref{checksendtime}.
It is clear that this happens in real time interval $[\tau_{nst}, \rt{j+1}{l}]$.
By Claim~\ref{nst-does-not-change},
	$\ell$ has $\NextSendTime = \lt{j}{c} + \Ptwo$ in line~\ref{checksendtime}.
We claim that $\ell$ finds $\lt{j+1}{c} < \NextSendTime$ in line~\ref{checksendtime}
	since otherwise, $\ell$ will continue to send $\CommitLease$ messages in line~\ref{sl1},
	and this occurs between events $\ev{j}{s}$ and $\ev{j}{s+1}$, 
	which contradicts the definition of $\ev{j}{s+1}$.
Thus, we have $\lt{j+1}{c} < \NextSendTime = \lt{j}{c} + \Ptwo$.
\qedhere~$_\text{\autoref{not-yet}}$
\end{proof}
Between events $\ev{j+1}{c}$ and $\ev{j+1}{s}$, 
	$\ell$ executes an incomplete iteration of the while loop from line~\ref{taketime1} to \ref{sl2}.
By Lemma~\ref{Alpha123}, Definition~\ref{Alpha} and Claim~\ref{not-yet},
	$\lt{j+1}{s} \leq \lt{j+1}{c} + 2\delta + \alpha_2
		<  \lt{j}{c} + \Ptwo + 2\delta + \alpha_2
		< \lt{j}{c} + \Ptwo + 2\delta + \alpha_0
		\leq t_j + \Ptwo + 2\delta + \alpha_0$.
\end{case}

We now show that other properties hold.
By definition, $\ell$ executes $\ev{j+1}{l}$ and $\ev{j+1}{s}$ in this order.
So $\rt{j+1}{l} < \rt{j+1}{s}$.
By Assumption~\ref{xclocks}(\ref{xcl2}),
	$\lt{j+1}{l} \leq \lt{j+1}{s}$ and Property~\ref{kx1} holds.
Since $\ell$ sends $\LM_{j+1}$ after time~$\tau_g \ge \ru$,
	by the definition of $\ru$ and property~(\ref{LM5}),
	$\lh_{j+1}$ contains every correct process $p \neq \ell$;
	so property~\ref{kx2} of the lemma holds.
We now show that property~(\ref{kx3-1}) holds.
If $\ev{j}{c}$ and $\ev{j+1}{c}$ are the same event,
	then $\ell$ executes events $\ev{j}{l}$, $\ev{j}{s}$, $\ev{j+1}{l}$ and $\ev{j+1}{s}$ in lines~\ref{setlease}, \ref{sl1}, \ref{lease1} and \ref{sl2} respectively in this order,
	and thus $\rt{j}{s} < \rt{j+1}{l}$ and $\lt{j}{s} < \lt{j+1}{l}$ by monotonicity of local clocks.
If $\ev{j}{c}$ and $\ev{j+1}{c}$ are distinct events,
	then $\ell$ executes event $\ev{j}{s}$ before $\ev{j+1}{c}$, and event $\ev{j+1}{l}$ after $\ev{j+1}{c}$.
Thus, we still have $\rt{j}{s} < \rt{j+1}{l}$ and $\lt{j}{s} \leq \lt{j+1}{l}$.
So property~(\ref{kx3-1}) holds.
Recall that $\ell$ issues leases $(k_{j}, \lt{j}{c})$ and $(k_{j+1}, \lt{j+1}{c})$ when executing events $\ev{j}{l}$ and $\ev{j}{l+1}$.
Property~(\ref{kx3-3}) then follows from Corollary~\ref{lease-issue-order}.

From the way we defined $\ev{j+1}{l}$ and $\ev{j+1}{s}$,
	it is clear that:

\begin{itemize}
\item $\ell$ does not change its variable $\lease$ between events $\ev{j}{l}$ and $\ev{j+1}{l}$, so property~(\ref{kx3-4}) holds
\item $\ell$ does not send any lease message between events $\ev{j}{s}$ and $\ev{j+1}{s}$, so property~(\ref{kx3-5}) holds.
\qedhere~$_\text{\autoref{lease-frequency}}$
\end{itemize}

\end{proof}

\begin{definition}\label{LeasesSequence}
$\LMS$ is the infinite sequence of lease messages
	$\LM_0, \LM_1, \ldots, \LM_i , \ldots$
	that contain the leases
	$(k_0 , \lt{0}{}) ,  (k_{1} , \lt{1}{}) , \ldots , \linebreak[0] (k_{i} , \lt{i}{}) ,  \ldots$,
	respectively,
	that are sent by $\ell$ after real time $\tau_g$. 
\end{definition}

\begin{lemma}\label{Fellini}
The leases contained in the lease messages $\LM_0, \LM_1, \ldots, \LM_{i-1}, \LM_i , \ldots$
	satisfy $(k_0 , \lt{0}{}) < (k_{1} , \lt{1}{}) <  \ldots < (k_{i-1} , \lt{i-1}{}) <  (k_{i} , \lt{i}{}) < \ldots$.
\end{lemma}
\begin{proof}
Consider any two adjacent lease messages $\LM_{i - 1}$ and $\LM_{i}$
	with leases $(k_{i-1}, t_{i-1})$ and $(k_i, t_i)$.
By Lemma~\ref{leasemonotonicity},
	$k_{i-1} \leq k_i$.
If $k_{i-1} < k_i$, then we are done.
If $k_{i-1} = k_i$, the lemma then follows from Lemma~\ref{increasing-lease-time-in-LeaderWork}.
\qedhere~$_\text{\autoref{Fellini}}$
\end{proof}

\begin{lemma}\label{NoA}
There is a real time after which the only lease messages that are sent are messages in $\LMS$.
\end{lemma}

\begin{proof}
Consider any process $q \neq \ell$.
Note that $q$ sends a lease message only while executing
	 the $\LeaderWork{}$ procedure.
By Lemma~\ref{lm6.8w}, there is a real time after which
	$q$ does not execute inside the $\LeaderWork{}$ procedure.
So there is a real time after which $q$ does not send any
	lease message.
Consider process $\ell$.
By Lemma~\ref{lease-frequency}, $\ell$ eventually sends $\LM_0$, and the only lease messages that
	$\ell$ sends after $\LM_0$ are $\LM_1 , \LM_2 , \ldots , \LM_i , \ldots$.
Thus, there is a real time after which the only lease messages
	that are sent are messages in $\LMS$.
\qedhere~$_\text{\autoref{NoA}}$
\end{proof}

This immediately implies:
\begin{corollary}\label{NoB}
There is a real time after which the only lease messages that are received are messages in~$\LMS$.
\end{corollary}

A process accepts a lease message $\LM' = \langle \CommitLease, (-,-),-,\lease', \LeaseHolders' \rangle$
	if it receives this message and resets its lease to $lease'$.
More precisely,

\begin{definition}\label{accept-lease}
A process $p$ \emph{accepts a lease message $\LM' = \langle \CommitLease, (-,-),-,\lease', \LeaseHolders' \rangle$}
	at real time $\tau$
	if the following holds:

\begin{enumerate}
\item $p$ receives $\LM'$ in line~\ref{lg0},

\item $p \in \LeaseHolders'$ in line~\ref{lg1},

\item $p$ finds $\lease' > \lease$ in line~\ref{lg1}, and

\item $p$ sets $\lease := \lease'$ in line~\ref{lg2} at real time $\tau$.
\end{enumerate}

\end{definition}

\begin{lemma}\label{maspic1}
Consider any correct process $p \neq \ell$.
From real time $\ru$ on:
\begin{enumerate}
\item\label{j1}  $p$ modifies its variable $\lease$ only when it accepts a lease message, and

\item\label{j2} the value of the variable $\lease$ at $p$ is non-decreasing. 
\end{enumerate}
\end{lemma}

\begin{proof}
By the definition of $\ru$ and property~\ref{LM1},
	process $p$ does not execute in $\LeaderWork{}$ after time~$\ru$.
Thus after time~$\ru$,
	$p$ modifies its variable $\lease$ only when it accepts a lease message in lines~\ref{lg0}-\ref{lg1}.
The guard in line~\ref{lg1} ensures that $p$ does not decrease its variable $\lease$ when it accepts a lease message.
\qedhere~$_\text{\autoref{maspic1}}$
\end{proof}

Recall that in the sequence of lease messages $\LMS = \LM_0, \LM_1, \ldots, \LM_{i-1}, \LM_i , \ldots$
	sent by $\ell$ in $\LeaderWork{t}$,
	each $\LM_i$ contains a lease $(k_i , \lt{i}{})$ such that $(k_0 , \lt{0}{}) < (k_{1} , \lt{1}{}) <  \ldots < (k_{i-1} , \lt{i-1}{}) <  (k_{i} , \lt{i}{}) < \ldots$, respectively.

\begin{lemma}\label{zozo2}
If a tuple $(-, -, \hat{k})$ is locked,
	then there is a $j \ge 0$ such that $k_j \ge \hat{k}$.
\end{lemma}

\begin{proof}
Suppose a tuple $(-, -, \hat{k})$ is locked.
By Lemma~\ref{zozo4}, there is a real time after which $\ell$ has $k \ge \hat{k}$.
Note that for each $i \ge 0$, when $\ell$ sends a lease message $\LM_i  \in \LMS$
	(this occurs in line~\ref{sl1} or~\ref{sl2}),
	$\LM_i$ contains the lease $(k_i , -)$
	where $k_i$ is the \emph{current value of the variable $k$ at $\ell$}.
Since there is a real time after which $\ell$ has $k \ge \hat{k}$,
	and $\ell$ sends infinitely many messages in $\LMS$,
	it is clear that there is a $j \ge 0$ such that $\ell$ sends an $\LM_j  \in \LMS$ with a lease $(k_j , -)$
	such that $k_j \ge \hat{k}$.
\qedhere~$_\text{\autoref{zozo2}}$
\end{proof}

\begin{lemma}\label{NoD}
Every correct process $p \neq \ell$ accepts infinitely many lease messages in $\LMS$.
\end{lemma}

\begin{proof}
Suppose, for contradiction,
	that some correct process $p \neq \ell$ accepts only a finite number of lease messages in $\LMS$.
From Corollary~\ref{NoB}, $p$ accepts only a finite number of lease message that are \emph{not} in $\LMS$.
So $p$ accepts only a finite number of lease messages.
Thus, by Lemma~\ref{maspic1}(\ref{j1}),
	there is a real time after which the variable $\lease$ at $p$ does not change.
Let $(\hat{k},{\hat{t}})$ be the ``final'' value of $\lease$ at $p$, i.e.,
	there is a real time $\tau$ after which $p$ has $\lease = (\hat{k},\hat{t})$.

Consider the sequence of lease messages $\LMS = \LM_0, \LM_1, \ldots, \LM_{i-1}, \LM_i , \ldots$
	that $\ell$ sends to every $q \neq \ell$
	after time~$\ru$.
Recall that each $\LM_i$ contains a lease $(k_i , \lt{i}{})$ such that
	$(k_0 , \lt{0}{}) < (k_{1} , \lt{1}{}) <  \ldots < (k_{i-1} , \lt{i-1}{}) <  (k_{i} , \lt{i}{}) < \ldots$, respectively.
Note that $k_0 \ge 0$.

We claim that there is a $j \ge 0$ such that $k_j \ge \hat{k}$.
To see this, note that:

(a) If $\hat{k} = 0$ then $k_0 \ge \hat{k}$.

(b) If $\hat{k} \neq 0$ then, by Lemma~\ref{LeaseIssue},
	some process $r$ issued the lease $(\hat{k},\hat{t})$ while executing
	$\LeaderWork{}$;
	by Lemma~\ref{lease-issue-obs}, $r$ locks some tuple $(-, -, \hat{k})$;
	and by Lemma~\ref{zozo2}, there is a $j \ge 0$ such that $k_j \ge \hat{k}$.

Now consider the sequence of leases  $(k_j , \lt{j}{}), (k_{j+1} , \lt{j+1}{}), (k_{j+2} , \lt{j+2}{}), \ldots$ 
	contained in the lease messages $\LM_j , \LM_{j+1} , \LM_{j+2}, \ldots$.
Since $k_j \ge \hat{k}$ and $(k_j , \lt{j}{}) < (k_{j+1} , \lt{j+1}{}) < (k_{j+2} , \lt{j+2}{}), \ldots$,
	it is clear that
	there is a~$\hat{\jmath}$ such that
	for all $i \ge \hat{\jmath}$, $(k_i , \lt{i}{}) > (\hat{k},\hat{t})$.
Thus there are infinitely many lease messages in $\LMS$
	that contain a lease greater than $(\hat{k},\hat{t})$.
Consider the first time that $p$ receives
	an $\LM_i = \langle \CommitLease, -, -,(k_i , \lt{i}{}), \lh_i \rangle$
	with $(k_i , \lt{i}{}) > (\hat{k},\hat{t})$ after real time $\tau$.
By Lemma~\ref{lease-frequency},
	$p \in \lh_i$, and so
	process $p$ accepts~$\LM_i$ after real time $\tau$.
Thus $p$ sets $\lease$ to $(k_i , \lt{i}{})$ after real time $\tau$ 
	--- a contradiction to the definition of~$\tau$.
\qedhere~$_\text{\autoref{NoD}}$
\end{proof}

\begin{assumption}\label{aboutreadleases}
The read lease period $\Pone$ and the read lease renewal period $\Ptwo$ are such that
	$\Pone > 3 \delta + \alpha_0$ and
	$0< \Ptwo < \Pone- (3 \delta + \alpha_0)$.
\end{assumption}

There is a real time after which every correct process always has a valid read lease.
More precisely:

\begin{theorem}\label{all-have-valid-lease}
For every correct process $p$, there is a time $\rr$ such that
	for every real time $\tau > \rr$, the following holds at real time $\tau$ at $p$:
	$\CT < \leasetime + \Pone$.
\end{theorem}

\begin{proof}
Let $p$ be any correct process.
There are two cases:

\begin{case}

\item $p = \ell$.
By Lemma~\ref{lease-frequency}, for all $i \ge 0$,
	$\ell$ sets its lease variable to $(k_i,t_i)$ at time $(\lt{i}{l}, \rt{i}{l})$.
Let $\rr = \rt{0}{l}$, and consider any real time $\tau > \rr$.
We now show that $\CT < \leasetime + \Pone$ at real time $\tau$ at~$\ell$.

By Lemma~\ref{lease-frequency}, we have $\rt{0}{l} < \rt{1}{l} < \ldots < \rt{i}{l} < \dots$.
So, since $\tau > \rt{0}{l}$, there is an $i \ge 0$ such that $ \rt{i}{l} \le \tau < \rt{i+1}{l}$.
Suppose $\ell$ has $\CT = t_{\ell}$ at real time $\tau$.
Since $\lt{i}{l}$ and $\lt{i+1}{l}$ are the values of $\CT$ at $\ell$ at real times $\rt{i}{l}$ and $\rt{i+1}{l}$,
	by the monotonicity of local clocks (Assumption~\ref{xclocks}(\ref{xcl2})), $\lt{i}{l} \le t_{\ell} \le \lt{i+1}{l}$.
By Lemma~\ref{lease-frequency}, $\ell$ sets $\lease$ to $(k_i,t_i)$ at real time $ \rt{i}{l}$
	and does not set it again until real time $ \rt{i+1}{l}$,
	so $\ell$ has $\lease = (k_i,\lt{i}{})$ at real time $\tau$.
	
Since:

\begin{enumerate}

\item $t_{\ell} \le \lt{i+1}{l}
		  \le \lt{i+1}{s} 
		  \leq t_i + \Ptwo + 2\delta + \alpha_0$ (by Lemma~\ref{lease-frequency}(\ref{kx3-2})),

\item $\Ptwo + 2\delta + \alpha_0 < \Pone$ (by Assumption~\ref{aboutreadleases}),

\end{enumerate}
we have $t_{\ell} < t_i + \Pone$.
Since at real time $\tau$ process $\ell$ has $\CT = t_{\ell}$ and $\leasetime = t_i$,
	we have $\CT < \leasetime + \Pone$ at real time~$\tau$ at~$\ell$.
	
\item $p \neq \ell$.
From Corollary~\ref{NoB},
	there is a real time $\rp$ such that:
\begin{enumerate}[\noindent(a)]
\item $\rp > \ru$, and
\item after real time $\rp$,
	the only lease messages that $p$ accepts are messages in $\LMS$.
\end{enumerate}
By Lemma~\ref{NoD}, process $p$ accepts infinitely many messages in $\LMS$.
Let $\LM_j$ be the first message in $\LMS$~such~that:

\begin{enumerate}

\item $\ell$ \emph{sends} $\LM_j$ at some time $\rt{j}{s} > \rp$.

\item process $p$ accepts $\LM_j$.

\end{enumerate}

Let $\rr = \rt{j}{a}$ be the real time when $p$ accepts $\LM_j$.
Since $\rr \ge \rt{j}{s}$, $\rt{j}{s}  > \rp$, and $\rp > \ru$ we have:
$\rr > \rp > \ru$.

Let $\tau$ be any real time such that $\tau > \rr$.
We show that $\CT < \leasetime + \Pone$ at real time $\tau$ at~$p$.

\smallskip
Let $i = \max \{~h~|~ p$ accepts $\LM_h \in \LMS$
	during the real time interval $[\rr, \tau] \}$.\footnote{Note that this set is not empty because process $p$ accepts $\LM_j$ at time $\rr$, so the index $i$ is well-defined (and $i \ge j$).}
Let $\rt{i}{s}$ and $\rt{i}{a} \in [\rr , \tau]$ be the real times
	when $\ell$ sends $\LM_i$ and $p$ accepts~$\LM_i$, respectively.
Since $p$ accepts $\LM_i$ at real time $\rt{i}{a}$, and
	$\LM_i$ contains the lease $(k_i,t_i)$,
	process $p$ sets $\lease$ to $(k_i,t_i)$ at real time $\rt{i}{a}$.

\begin{claim}\label{maspic}
Process $p$ does not accept any lease message during the real time interval $(\rt{i}{a} , \tau]$. 

\end{claim}

\begin{proof}
Suppose, for contradiction, that $p$ accepts a lease message during the real time interval $(\rt{i}{a} , \tau]$.
Let $\LM = \langle \CommitLease,-, -, \lease, \lh \rangle$ be the first lease message 
	that $p$ accepts in interval $(\rt{i}{a} , \tau]$.
Since $p$ receives $\LM$ after real time $\rt{i}{a}$ and $\rt{i}{a} \ge \rr > \rp$,
	by the definition of $\rp$,
	$\LM$ must be in $\LMS$;
	so $\LM = \LM_h$ for some $h$.
Since $p$ accepts $\LM_i$ before accepting $\LM_h$, $i \neq h$.
Since $p$ accepts $\LM_h$ during the real time interval $(\rt{i}{a} , \tau]$,
	by the definition of $i$, we have $i > h$.
From Lemma~\ref{Fellini}, the leases
	 $(k_i,t_i)$ and $(k_h,t_h)$ contained in $\LM_i$ and $\LM_h$, respectively,
	 are such that $(k_i,t_c) > (k_h,t_h)$.
Since $\LM_h$ is the first lease message that $p$ accepts after accepting $\LM_i$,
	$p$ has $\lease = (k_i,t_i)$ just before it receives $\LM_h$.
Since $(k_i,t_c) > (k_h,t_h)$, it is clear that
	$p$ does not accept $\LM_h$ (because of the guard in line~\ref{lg2}) --- a contradiction.
\qedhere~$_\text{\autoref{maspic}}$
\end{proof}

\begin{claim}\label{maspic2}
Process $p$ has $\lease = (k_i,t_i)$ during the real time interval $[\rt{i}{a} , \tau]$.
\end{claim}

\begin{proof}
Recall that $p$ has $\lease = (k_i,t_i)$ at real time $\rt{i}{a} > \ru$.
By Lemma~\ref{maspic1}(\ref{j1}) and Claim~\ref{maspic},
	process~$p$ does not modify $\lease$ during the interval $(\rt{i}{a} , \tau]$.
Thus, $p$ has $\lease = (k_i,t_i)$ during $[\rt{i}{a} , \tau]$.
\qedhere~$_\text{\autoref{maspic2}}$
\end{proof}

\begin{claim}\label{maspic5}
Process $p$ has $\lease \le (k_i,t_i)$ during the real time interval $[\rp , \tau]$.
\end{claim}

\begin{proof}
Since $\rt{i}{a} \in [\rr , \tau]$ and $\rp < \rr$,
	we have $\rt{i}{a} \in [\rp , \tau]$.
Consider the contiguous real time intervals $[\rp , \rt{i}{a}]$ and $[\rt{i}{a} , \tau]$.
By Claim~\ref{maspic2},
	process $p$ has $\lease = (k_i,t_i)$ during $[\rt{i}{a}, \tau]$.
Since $\rp > \ru$, by Lemma~\ref{maspic1}(\ref{j2}),
	$p$ has $\lease \le (k_i,t_c)$ during $[\rp , \rt{i}{a}]$.
So $p$ has $\lease \le (k_i,t_i)$ during $[\rp , \tau]$.
\qedhere~$_\text{\autoref{maspic5}}$
\end{proof}

Claim~\ref{maspic2} immediately implies that: 
\begin{claim}\label{maspic3}
At time $\tau$, process $p$ has $\leasetime = t_i$.
\end{claim}

Suppose that at real time $\tau$,
	the local clocks of $\ell$ and $p$ are $\CT_{\ell} = t_{\ell}$ and $\CT_p= t_p$, respectively.
By Assumption~\ref{xclocks}(\ref{xcl5}),
	$t_{\ell} = t_p$.

\begin{claim}\label{enough!}
$t_{\ell} \leq t_i + \Ptwo + 3\delta + \alpha_0$.
\end{claim}

\begin{proof}
Suppose, for contradiction, that $t_{\ell} > t_i + \Ptwo + 3\delta +\alpha_0$.
By Lemma~\ref{lease-frequency}
	process $\ell$ sends a $\LM_{i+1} = \langle \CommitLease, -, -, (k_{i+1}, t_{i+1}), \lh_{i+1} \rangle$ message
	at real time $\rt{i+1}{s}$ to $p$ such that:
	
\begin{enumerate}

	\item $ \ru <\rt{i}{s} < \rt{i+1}{s}$.
	
	\item $\lt{i+1}{s} \le t_i + \Ptwo + 2\delta +\alpha_0$.

	\item $(k_i,t_i) < (k_{i+1},t_{i+1})$.

	\item $p \in \lh_{i+1}$.

\end{enumerate}

\smallskip
We now show that $p$ receives and processes $\LM_{i+1}$ during the real time interval $[\rp, \tau]$:

(a) \emph{$p$ receives $\LM_{i+1}$ after real time $\rp$.}
This is because $\ell$ sends $\LM_{i+1}$
	at real time $\rt{i+1}{s} > \rt{i}{s} \ge \rt{j}{s} > \rp$.

\smallskip
(b) \emph{$p$ processes $\LM_{i+1}$ before time $\tau$.}
To see why this holds, first note that since $\ell$ sends $\LM_{i+1}$ at local time~$\lt{i+1}{s}$,
	and this occurs after real time $\ru$, by property~\ref{LM3} and Assumption~\ref{xclocks}(\ref{xcl4}),
	$p$ receives and processes $\LM_{i+1}$ by local time $\hat{t} \le \lt{i+1}{s} + \delta$.
Since
	$\lt{i+1}{s} \le  t_i + \Ptwo + 2 \delta + \alpha_0$,
	we have $\hat{t} \le t_i + \Ptwo + 3 \delta +\alpha_0$.
By assumption $t_{\ell} > t_i + \Ptwo + 3\delta + \alpha_0$, so $\hat{t} < t_{\ell}$.
By monotonicity of local clocks,
	$p$ receives and processes $\LM_{i+1}$ before local time $t_{\ell}$.
Since $\CT_{\ell} = t_{\ell}$ at real time $\tau$, we conclude that 
	$p$ receives and processes $\LM_{i+1}$ before real time $\tau$.
	
\medskip

Since:
\begin{enumerate}

\item $p$ receives and processes $\LM_{i+1}$ during interval $[\rp, \tau]$,
\item $p$ has $\lease \le (k_i,t_i)$ during interval $[\rp , \tau]$ (Claim~\ref{maspic5}), and
\item the lease $(k_{i+1},t_{i+1})$ and the set $\lh_{i+1}$ in $\LM_{i+1}$ are such that
	$(k_{i+1},t_{i+1}) > (k_i,t_i)$ and $p \in \lh_{i+1}$,
\end{enumerate}
process $p$ accepts $\LM_{i+1}$ and sets its $\lease$ variable to $(k_{i+1},t_{i+1})$
	during the real time interval $[\rp,\tau]$
	--- a contradiction to Claim~\ref{maspic5}.
\qedhere~$_\text{\autoref{enough!}}$
\end{proof}

By Claim~\ref{enough!}, $t_{\ell} \le t_i + \Ptwo + 3\delta +\alpha_0$.
By Assumption~\ref{aboutreadleases}, $\Ptwo + 3\delta + \alpha_0 + < \Pone$.
So $t_{\ell} < t_i + \Pone$.
Since, at real time $\tau$, process $p$ has $\CT = t_{\ell}$
	and, by Claim~\ref{maspic3}, 
	$p$ has $\leasetime = t_i$
	at real time $\tau$,
we conclude that $p$ has $\CT < \leasetime + \Pone$ at real time $\tau$.
\qedhere~$_\text{\autoref{all-have-valid-lease}}$
\end{case}
\end{proof}

The previous theorem states that for every correct process $p$ there is a real time $\rr$
	after which $p$ has $\CT < \leasetime + \Pone$.
We now show that, after time $\rr$,
	in every read operation process $p$
	executes only one iteration of the
	repeat-until loop of lines~\ref{getvalidlease-start}-\ref{getvalidlease-end}.

\begin{theorem}\label{exitleaseloop}
Consider any correct process $p$, and let $\rr$ be the real time associated to $p$
	by Theorem~\ref{all-have-valid-lease}.
If $p$ starts the repeat-until loop of
	lines~\ref{getvalidlease-start}-\ref{getvalidlease-end}
	after real time $\rr$,
	then $p$ exits in line~\ref{getvalidlease-end} without looping.
\end{theorem}

\begin{proof}

\begin{claim}\label{non-decreasing-lease-start}
There is a real time $\tau$ after which the value of the variable $\leasetime$ at $p$ is non-decreasing.
\end{claim}
\begin{proof}
There are two cases:
\begin{case}
\item $p = \ell$.
By Lemma~\ref{lease-frequency},
	after real time $\tau_g$,
	$\ell$ issues leases $(k_0 , t_0) , (k_{1} , t_1) ,  \ldots ,  (k_{i} , t_i) ,  \ldots$.
By Lemma~\ref{increasing-lease-time-in-LeaderWork},
	the lease start times included in these leases are non-decreasing,
	i.e., $t_0 \leq t_1 \leq \dots$.
Thus, the claim holds for $\tau = \tau_g$.

\item $p \neq \ell$.
By Corollary~\ref{NoB} and Definition~\ref{accept-lease},
	there is a real time $\tau'$ after which
	the only lease messages accepted by $p$ are messages in $\LM$.
By Lemma~\ref{maspic1},
	after real time $\ru$, $p$ modifies its variable $\lease$ only when it accepts a lease message.
By Lemma~\ref{NoD},
	$p$ accepts infinitely many lease messages in $\LM$.
Let $\tau$ be the earliest real time when $p$ accepts a lease message after real time $\max(\ru, \tau')$.
Consider any real time $\hat \tau \ge \tap$,
	it is clear that at real time $\hat \tau$,
	the value of variable $\lease$ at $p$ is equal to $(k_{i}, t_i)$ that is included in $\LM_{i}$ for some $i$.
Consider the first time when $p$ modifies variable $\lease$ after real time $\hat \tau$.
Because of the condition in line~\ref{lg2}, 
	process $p$ must set it to some value $> (k_{i}, t_i)$.
By our choice of $\tap$ and by Lemma~\ref{Fellini}, 
	this happens only when $p$ accepts some lease message $\LM_{j}$ with $j > i$,
	and $p$ sets $\lease$ to $(k_{j}, t_j)$.
By the same argument as in Case 1, $t_j \geq t_i$.
So if $p$ sets its $\lease$ variable after real time $\tau$,
	then $\leasetime$ is non-decreasing.
\end{case}
\qedhere~$_\text{\autoref{non-decreasing-lease-start}}$
\end{proof}

Let~$\tau'$ be the real time when $p$ executes line~\ref{getleasetime} in this execution of the loop.
Since $\tau' > \rr$, by Theorem~\ref{all-have-valid-lease}, the following holds at real time $\tau'$ at $p$:
\begin{equation}\label{eqn1}
\setcounter{equation}{1}
\CT < \leasetime + \Pone
\end{equation}

Since $p$ sets $t' := \CT$ in line~\ref{getleasetime} at real time $\tau'$,
	$\CT = t'$ at real time $\tau'$ at $p$.
Let $(\ktau,\ttau)$ be the value of $p$'s $\lease$ variable at real time $\tau'$.
So $\leasetime = \ttau$ at real time~$\tau'$ at $p$.

From (1) we have:

\begin{equation}\label{eqn2}
t' < \ttau + \Pone
\end{equation}

Note that by Claim~\ref{non-decreasing-lease-start}, 
	when $p$ executes line~\ref{getlease}, the value of $\lease.start$ is at least $\ttau$,
	so $p$ sets $t^*$ to some value $\geq \ttau$ in line~\ref{getlease}.
Thus the following holds: 
\begin{equation}\label{eqn3}
t' < t^* + \LP
\end{equation}

Therefore when $p$ executes line~\ref{getvalidlease-end}, it finds that
	(\ref{eqn3}) holds, and so $p$ exits in line~\ref{getvalidlease-end}
	without looping.
\qedhere~$_\text{\autoref{exitleaseloop}}$
\end{proof}

\begin{lemma}\label{deus}
For all $j \ge 1$,
	if some tuple $(-,-,j)$ is accepted,
	then some tuple $(-,-,j)$ is locked.
\end{lemma}

\begin{proof}
For $j \ge 1$, consider the first time a tuple $(-,-,j)$ is accepted.
Suppose this occurs when a process $p$ accepts tuple $(O,t,j)$.
By Observation~\ref{leaderfirst},
	$p$ accepted $(O,t,j)$ in a call to $\DO((O,-),t,j)$
	while executing $\LW(t)$.
	
\begin{claim}\label{deus-claim}
$p$ called $\DO((O,-),t,j)$ in line~\ref{second-doops} of $\LW(t)$.
\end{claim}

\begin{proof}
Process $p$ calls $\DO((O,-),t,j)$ in line~\ref{first-doops} or~\ref{second-doops}.
Suppose, for contradiction, $p$ calls $\DO((O,-),t,j)$ in line~\ref{first-doops}.
From the code of $\LW()$,  it clear that
	$p$ had $(\Ops^*,ts^*,k^*) = (O,t',j)$, for some $t'$, in line~\ref{selection}.
Since $j \ge 1$, $(O,t',j) \neq (\emptyset, -1, 0)$.
By Lemma~\ref{lmx3},
	some process accepted tuple $(O,t',j)$ before $p$ executed line~\ref{selection}.
So $(O,t',j)$  was accepted before $p$ called $\DO((O,-),t,j)$ in line~\ref{first-doops},
	and therefore before $p$ accepted $(O,t,j)$
	--- a contradiction to the definition of $(O,t,j)$.
Thus $p$ calls $\DO((O,-),t,j)$ in line~\ref{second-doops}.
\qedhere~$_\text{\autoref{deus-claim}}$
\end{proof}

From the above claim and the code of $\LW()$,
	process $p$ calls $\DO$ at least once before calling $\DO((O,-),t,j)$
	in line~\ref{second-doops} of $\LW(t)$.
By Lemma~\ref{T-k-Doops},
	$p$ calls $\DO((O',-),t,j-1)$, for some $O'$
	before calling $\DO((O,-),t,j)$ in $\LW(t)$.
Since the call to $\DO((O',-),t,j-1)$ must return $\textsc{Done}$,

\begin{equation}\label{eqn-deus-ex-machina1}
\text{$p$ locks $(O',t,j-1)$.}
\end{equation}

Let $\ell$ be the final, stable leader (see Lemma~\ref{lm6.1}).
By Theorem~\ref{lm10},
	$\ell$ executes a non-terminating call to $\LW(t_\ell)$,
	for some $t_\ell$.
By Lemma~\ref{lastleader2}(\ref{last-t2}), $t_\ell\ge t$.
There are two cases:

\begin{case}
\item
$t_\ell=t$.
Thus $p$ and $\ell$ became leader at the same local time $t$,
	so they called $\ML{t}{t}$ and this call returned $\textsc{True}$.
By Theorem~\ref{leader-safety}, $p = \ell$.
So $\ell$ called $\DO((O,-),t_\ell,j)$ in $\LW(t_\ell)$.
Since this call returns $\textsc{Done}$
	(because $\LW(t_\ell)$ does not terminate),
	$\ell$ locks $(O,t_\ell,j)$.

\item
$t_\ell>t$.
During its initialization in $\LW(t_\ell)$,
	$\ell$ called $\DO((\Ops^*,0),t_\ell,k^*)$ in line~\ref{first-doops}, and
	it accepted $(\Ops^*,t_\ell,k^*)$
	in line~\ref{leader-accept} of this procedure.
Since $(O',t,j-1)$ is locked,
	tuple $(\Ops^*,t_\ell,k^*)$ is accepted, and
	$t_\ell>t$,
	by Theorem~\ref{generalcase1a}(\ref{t1}):
\begin{equation}\label{eqn-deus-ex-machina2}
k^*\ge j-1.
\end{equation}
	
After $\ell$ completes $\DO((\Ops^*,0),t_\ell,k^*)$,
	it initiates a RMW $\NoOp$ $\op$ in line~\ref{Do-a-NoOp}.
Since $\ell$ is correct, $\op$ is inserted in $\OpsRequested$
	(line~\ref{rr2}) after $\ell$'s call to $\DO((Ops^*,0),t_\ell,k^*)$
	is completed.
By Theorem~\ref{lm10},
	$\ell$~executes the while loop of lines~\ref{mainwhile}-\ref{endwhile}
	infinitely often during $\LW(t_\ell)$,
	so it will eventually execute
	$\DO((\NextOps,-),t,j')$, with $\op\in\NextOps$,
	for some $j'$ after completing $\DO((\Ops^*,0),t,k^*)$.
So, by Lemma~\ref{T-k-Doops},
	$\ell$ eventually calls $\DO((-,-),t,k^*+1)$.
During the execution of $\DO((\NextOps,-),t,k^*+1)$,
	$\ell$ locks $(-,t,k^*+1)$ and sets $\Batch[k^*+1]$ to some pair.
By Corollary~\ref{successivebatches2},
	for each~$i$, $0 \le i \le k^*$,
	some process previously set $\Batch[i]$ to some pair.
Thus, from Lemma~\ref{IamOutOfLableNames}, for each~$i$, $0 \le i \le k^*+1$,
	some tuple $(-,-,i)$ is locked.
By (\ref{eqn-deus-ex-machina2}), $j\le k^*+1$,
	and so some tuple $(-,-,j)$ is locked.
\end{case}

So, in both cases, some tuple $(-,-,j)$ is locked, as wanted.
\qedhere~$_\text{\autoref{deus}}$
\end{proof}

\begin{lemma}\label{nowaity}
No correct process waits forever in line~\ref{FG3}.
\end{lemma}

\begin{proof}
Let $p$ be any correct process.
Consider the wait statement of line~\ref{FG3}, namely:
\vspace{-1mm}
	 $$\text{\textbf{wait for} $(\mbox{for all } j, k^* < j \le \hat{k}, \Batch[j] \neq (\emptyset,\infty))$}$$

\vspace{-2mm}

From the way $p$ computes $\hat{k}$ in lines~\ref{get-k-hat-else}-\ref{get-k-hat-5}, it is clear that either:
	
	~~~~~(a) $\hat{k} = k^*$,
	where $k^*$ is the value of $\leasebatch$ in line \ref{getlease}, or
	
	~~~~~(b) $k^* < \hat{k} \le u$,
	where $u$ is the value of $\MPB$ in line~\ref{get-MPB}.

The wait condition is trivial if $\hat k = k^*$. 
Henceforth we assume that $\hat k > k^*$.
Since $k^* \geq 0$, we have $\hat k \geq 1$.
Since $p$ has $\MPB = u$ and $u \ge \hat k \geq 1$,
	$u$ is not the initial value of $\MPB$ at $p$.
Note that:
	(i) $p$ can set $\MPB$ to $u$ only
	in line~\ref{setMPB} of the algorithm (this is the only line that modifies this variable), and
	(ii) in line~\ref{setMPB}, $p$ sets ``$\MPB := \max (\MPB, i)$''
	\emph{right after} $p$ accepts some tuple $(-,-,i)$ in line~\ref{client-accept}.
Therefore $p$ accepted some tuple $(-,-,u)$  in line~\ref{client-accept}.
So, by Lemma~\ref{deus},
	some tuple $(-,-,u)$ is eventually locked.
Thus, by Lemma~\ref{zozo4}, there is a real time after which $\ell$ has $k \ge u$,
	and so, by Lemma~\ref{zozox3}(\ref{oopla1}), there is a real time after which
	$p$ has $\Batch[j] \neq (\emptyset,\infty)$ for all $j$, $1 \le j \le u$.
Since $1 \le \hat{k} \le u$ and $k^* \geq 0$, there is a real time after which
	$p$ has $\Batch[j] \neq (\emptyset,\infty)$ for all~$j$, $k^* < j \le \hat{k}$.

So in all cases, there is a real time after which
	$p$ has $\Batch[j] \neq (\emptyset,\infty)$ for all $j$, $k^* < j \le \hat{k}$.
Therefore $p$ eventually exits the
	wait statement of line~\ref{FG3}.
\qedhere~$_\text{\autoref{nowaity}}$
\end{proof}

\begin{lemma}\label{finite-wait-promise2}
No correct process waits forever in line~\ref{wait-promise-2}.
\end{lemma}
\begin{proof}
Let $p$ be any correct process. 
Suppose that $p$ executes line~\ref{wait-promise-2} for some $\hat k$.
By Lemma~\ref{batch-zero},
	$p$ has $\Batch[0] = (\emptyset, 0)$ always.
Thus the lemma holds if $\hat k = 0$.
Henceforth we assume that $\hat k > 0$.
We first show that $p$ sets $\Batch[\hat k] = (\Os, \s) \neq (\emptyset, \infty)$ for some $(\Os,\s)$
	before it executes line~\ref{wait-promise-2}.
Since $p$ finds $t' < t^* + \LP$ in line~\ref{getvalidlease-end},
	$p$ sets $\lease$ to some $(k^*,t^*) \neq (0, -\infty)$.
If $\hat k \le k^*$,
	then by Lemma~\ref{set-batch-before-lease},
	$p$ previously set $\Batch[\hat k]$ to some $(\Os,\s) \neq (\emptyset, \infty)$.
If $\hat k > k^*$,
	then $p$ computes $\hat k$ in the else clause of lines~\ref{get-k-hat-3}-\ref{get-k-hat-5},
	and $p$ sets $\Batch[\hat k]$ to some $(\Os,\s) \neq (\emptyset, \infty)$
	before it completes the wait statement in line~\ref{FG3}.
By Observation~\ref{batch-promise-locked},
	some process locks a tuple of form $(\Os,-,\hat k)$ with promise $\s$.
By Lemma~\ref{lock-imply-finite-GPromise} and Lemma~\ref{local-less-than-global-promise},
	$p$ has $\BatchPromise{\hat k} \leq \GPromise_{\hat k}$ after it sets $\Batch[\hat k]$
	and $\GPromise_{\hat k}$ is a constant non-infinity value.	
Thus, by Assumptions~\ref{xclocks}(\ref{xcl2}-\ref{xcl3}),
	$p$ eventually finds $\CT \geq \GPromise_{\hat k} \geq \BatchPromise{\hat k}$ in line~\ref{wait-promise-2}.
\qedhere~$_\text{\autoref{finite-wait-promise2}}$
\end{proof}

\begin{theorem}\label{bertolucci}
If a correct process starts executing a \emph{read} operation,
	then it eventually completes this operation.
\end{theorem}

\begin{proof}
Suppose a correct process $p$
	starts a \emph{read} operation (this occurs in line~\ref{read-invoke}).
By Theorem~\ref{exitleaseloop},
	$p$ eventually exits the loop in lines~\ref{getvalidlease-start}-\ref{getvalidlease-end}.
If $p$ executes line~\ref{FG3}, then by Lemma~\ref{nowaity},
	$p$ exits the wait statement of line~\ref{FG3}.
By Lemma~\ref{finite-wait-promise2},
	$p$ does not wait forever in line~\ref{wait-promise-2}.
By inspection of the algorithm,
	$p$'s call to $\EUTB(\hat k)$
	in line~\ref{fill-gaps-to-k-hat}
	terminates.
Thus, $p$ returns with a $\reply$ in line~\ref{read-respond}.
\qedhere~$_\text{\autoref{bertolucci}}$
\end{proof}

\subsection{Read lease mechanism: non-blocking reads}\label{readleaseNonBlockingProps}

Read operations that
	\emph{start} after some stabilization time
	satisfy some additional timeliness and liveness properties.
To state these properties precisely, we first define
	the notion of an operation that is \emph{pending} at some process at a given time.
Intuitively, an operation $o$ is pending at a process $p$,
	if $p$ is aware that some process is trying to ``commit'' a batch of operations $O$ that contains $o$,
	but $p$ does not know yet whether the commit of $O$ has succeeded.
There are three reasons why this may occur:
	(a) the committing of $O$ is still going on, or
	(b) $O$ was committed, but $p$ has not yet received a confirmation
	(i.e., it did not yet receive the corresponding $\CommitLease$ message), or
	(c) the commit of $O$ failed but $p$ does not know it yet.
The precise definition of pending operations is as follows.

\begin{definition}\label{pending1}
A non-empty set of operations $O$ is pending at a process $p$ at some real time $\tau$,
	if process $p$ has $\PendingOps[j] = (O, -)$ and $\Batch[j] = (\emptyset,\infty)$ 
	for
	some $j \ge 1$ at real time $\tau$.
\end{definition}

\begin{definition}\label{pending2}
An operation $o$ is pending at a process $p$ at some real time $\tau$,
	if $o$ is in a set of operations~$O$ that is pending at $p$ at real time $\tau$.
\end{definition}

\begin{observation}\label{lease-commit-have-same-k}
If a process sends a ${\langle \CommitLease, -,j, (-, j'), - \rangle}$ message, then $j = j'$.
\end{observation}

\begin{lemma}\label{sending-LMs}
There is a $j_0$ such that for all $j \ge j_0$ the following holds:
	if $\ell$ sends a $\CM{j}$ message, 
	then the sending of this message
	is event $\ev{i}{s}$ for some $i \geq 0$
	as defined in Lemma~\ref{lease-frequency},
	and $\CM{j}$ is $\LM_{i}$.
\end{lemma}
\begin{proof}
By Lemma~\ref{NoA},
	there is a real time $\tau$ after which the only $\CommitLease$ messages that are sent
	are messages in $\LMS$.
By definition,  after real time $\tau$,
	the sending of a $\CommitLease$ message is event $\ev{i}{s}$ for some $i \geq 0$ as defined in Lemma~\ref{lease-frequency}.
The lemma follows then from the fact that
	only a finite number of $\CommitLease$ messages are sent by real time $\tau$.
\qedhere~$_\text{\autoref{sending-LMs}}$

\end{proof}

\begin{observation}\label{new0}
For all processes $p \neq \ell$, there is a $j_0$ such that for all $j \ge j_0$ the following holds:
	if $\ell$ sends a $\CM{j}$ message to $p$,
	then $\ell$ sends this message after time $\tau_f$.
\end{observation}

\begin{lemma}\label{new1}
For all processes $p \neq \ell$, there is a $j_0$ such that for all $j \ge j_0$ the following holds:
	if $\ell$ sends a $\CM{j}$ or a $\PR{j}$ message to $p$,
	then for all $i$, $j_0 \le i < j$, $\ell$ previously sent a $\CM{i}$ message to $p$.
\end{lemma}
\begin{proof}
Let $j_0$ be as defined in Lemma~\ref{sending-LMs}.
Suppose that $\ell$ sends a $\CM{j}$ or a $\PR{j}$ message to $p$ for some $j \geq j_0$.
Note that by definition of $j_0$,
	this happens in $\DoOps{(-, -)}{j}{t_{\ell}}$
	and $\ell$ executes the loop of lines~\ref{mainwhile}-\ref{endwhile} infinitly often in $\LW({t_\ell})$.
So this call to $\DoOps{(-, -)}{j}{t_{\ell}}$ must return $\textsc{Done}$
	and it sends a $\CM{j}$ message before it returns.
The lemma then follows from Lemma~\ref{sending-LMs}.
\qedhere~$_\text{\autoref{new1}}$
\end{proof}

\begin{proof}
By Lemma~\ref{NoA},
	there is a real time $\tau$ after which
	the only $\CommitLease$ messages that are sent
	are those sent by $\ell$ during the non-terminating execution of $\LW({t})$,
	for some local time $t$ (see Theorem~\ref{lm10}).
Let $j_0$ be the batch number of the first such $\CommitLease$ message.
Suppose that $\ell$ sends a $\CM{j}$ or a $\PR{j}$ message to process $p$
	for some $j \geq j_0$.

\begin{claim}\label{new1-claim1}
If $\ell$ sends a $\PR{j}$ message to $p$,
	then after doing so $\ell$ also sends a $\CommitLease$ message to $p$
	and the first such message is a $\CM{j}$ message.
\end{claim}

\begin{proof}
Suppose $\ell$ sends a $\PR{j}$ message to $p$.
Since $j\ge j_0$, $\ell$ sends this message in line~\ref{sendprep}
	in a call to $\DoOps{(-, -)}{j}{t}$ made during the non-terminating
	execution of $\LW(t)$.
So this call to $\DoOps{(-, -)}{j}{t}$ must return $\textsc{Done}$.
By the code in lines~\ref{sendprep}-\ref{sendcommit},
	before returning,
	$\ell$ sends to $p$ a $\CM{j}$ message and no other $\CommitLease$ message.
\qedhere~$_\text{\autoref{new1-claim1}}$
\end{proof}

\begin{claim}\label{new1-claim2}
If $\ell$ sends a $\CM{j}$ message to $p$,
	then $\ell$ previously sent a \linebreak$\CM{i}$ message to $p$,
	for all $i$, $j_0\le i<j$.
\end{claim}

\begin{proof}
Suppose $\ell$ sends a $\CM{j}$ message to $p$.
Prior to sending this message (in line~\ref{sendcommit2} or~\ref{sendcommit}),
	$\ell$ issues a lease $(j,-)$ (in line~\ref{setlease} or~\ref{lease1}).
By Lemma~\ref{lease-issue-obs},
	if $\ell$ issues a lease $(j,-)$ in $\LW(t)$,
	$\ell$ previously locked $(-,j,t)$.
This can only happen while $\ell$ is executing
	a call to $\DoOps{(-, -)}{j}{t}$.
By Lemma~\ref{T-k-Doops},
	consecutive calls to $\DO$ during the execution of $\LW(t)$
	are for successive batches.
Therefore,
	if $\ell$ sends a $\CM{j}$ message to $p$,
	then $\ell$ previously sent a $\CM{i}$ message to $i$ for every $i$,
	$j_0\le i<j$.
\qedhere~$_\text{\autoref{new1-claim2}}$
\end{proof}

The lemma now follows from Claims~\ref{new1-claim1} and~\ref{new1-claim2}.
\qedhere~$_\text{\autoref{new1}}$
\end{proof}

\begin{lemma}\label{new3}
For all processes $p \neq \ell$, there is a $j_0$ such that for all $j \ge j_0$ the following holds:
	if $p$ sets $\Batch[j]$ to some pair $(\Os_j,\s_j)$,
	then $p$ previously received a $\CM{j}$ from $\ell$.
\end{lemma}
\begin{proof}
Let $p \neq \ell$.

\begin{claim}\label{new3-claim}
~
\begin{enumerate}
\item\label{cx1} There is a $j_1$ such that for all $j \ge j_1$, $p$ does not call $\FillGaps{j}$ in line~\ref{FG}.

\item\label{cx2} There is a $j_2$ such that for all $j \ge j_2$, $p$ does not set $\Batch[j]$ in line~\ref{batch1}.

\item\label{cx3} There is a $j_3$ such that for all $j \ge j_3$, $p$ does not set $\Batch[j]$ in line~\ref{setBatch4}.

\item\label{cx4} There is a $j_4$ such that for all $j \ge j_4$, no process $q \neq \ell$ sends a $\CM{j}$ or a $\PR{j}$. 

\item\label{cx5} There is a $j_5$ such that for all $j \ge j_5$, if $\ell$ sends a $\CM{j}$ then $\ell$ sends this $\CM{j}$ after real time $\tau_f$.

\item\label{cx6} There is a $j_6$ such that for all $j \ge j_6$, if $\ell$ sends a $\CM{j+1}$ or a $\PR{j+1}$ to $p$ then
	for all $i$, $j_6 \le i \le j$, $\ell$ previously sent a $\CM{i}$ to $p$.

\end{enumerate}

\end{claim}
\begin{proof}
Since $p \neq \ell$, by Lemma~\ref{lm6.8w},
	there is a real time after which $p$ does not execute inside the $\LW()$ procedure.

\begin{enumerate}
\item
Since line~\ref{FG} is in the $\LW()$ procedure,
	there is a real time after which $p$ does not call $\FillGaps{}$ in line~\ref{FG}.
	So $p$ calls $\FillGaps{}$ in line~\ref{FG} only finitely many times.
	This implies part~(\ref{cx1}) of the claim.

\item
Since $\DO((-,-),-,-)$ is called only inside $\LW()$,
	there is a real time after which $p$ does not call $\DO$.
	So $p$ executes line~\ref{batch1} of $\DO$ only finitely many times.
	This implies part~(\ref{cx2}) of the claim.

\item
Note that a process sends $\langle \EstRequest, - \rangle$
	messages only in line~\ref{est-request} of the $\LW()$ procedure.
By Lemma~\ref{lm6.8w} and Theorem~\ref{lm10},
	there is a real time after which only $\ell$ is in $\LW$.
So there is a real time after which only $\ell$ can send $\langle \EstRequest, - \rangle$ messages.
By Lemma~\ref{lm6.8w},
	there is a real time after which
	$\ell$ executes in the while loop of lines~\ref{mainwhile}-\ref{endwhile} of a $\LW()$ procedure forever.
Thus, there is a real time after which
	no process sends $\langle \EstRequest, - \rangle$ messages.
So only a finite number of such messages are received (in line~\ref{spcm}),
	and only a finite number of $\langle \EstReply, -, -,-,-,- \rangle$ are sent and received
	(in line~\ref{reply} and line~\ref{estrep1}, respectively).
Therefore, line~\ref{setBatch4} is executed only finitely many times.
This implies part~(\ref{cx3}) of the claim.

\item 
By Lemma~\ref{lm6.8w}, there is a real time after which
	no process $q \neq \ell$ executes inside the $\LW()$ procedure.
Since $\CM{-}$ and $\PR{-}$ messages are sent only in this procedure,
	part~(\ref{cx4}) of the claim holds.

\item
Process $\ell$ can send only
	a finite number of messages before real time $\tau_f$.
This implies part~(\ref{cx5}) of the claim.

\item
Lemma~\ref{new1} implies part~(\ref{cx6}) of the claim,
	where $j_6$ is the constant $j_0$ described in Lemma~\ref{new1}.
\qedhere~$_\text{\autoref{new3-claim}}$
\end{enumerate}
\end{proof}

Let $j_0 = \max (j_1,j_2, j_3, j_4, j_5, j_6)$.
Consider any $j \ge j_0$ and suppose $p$ sets $\Batch[j]$ to some pair $(\Os_j,\s_j)$.

Note that this can occur only in lines~\ref{batch1}, \ref{setBatch3}, \ref{batch2}, \ref{setBatch4}, or \ref{batch3} of the algorithm.
Since $j \ge j_2$ and $j \ge j_3$, by part~(\ref{cx2}) and~(\ref{cx3}) of Claim~\ref{new3-claim},
	$p$ does not set $\Batch[j]$ in lines~\ref{batch1} or line~\ref{setBatch4}.
We now consider each one of the remaining three cases.

\begin{enumerate}
\item $p$ sets $\Batch[j]$ to $(\Os_j,\s_j)$ in line~\ref{batch2}.
	Thus, $p$ previously received a $\CM{j}$ message
	in line~\ref{commitreceipt1}.
	
\item $p$ sets $\Batch[j]$ to $(\Os_j,\s_j)$ in line~\ref{setBatch3}.
So $p$ received some $P_{j+1} = \langle \Prepare,-,-,j+1, (\Os_j,\s_j) \rangle$ message
	in line~\ref{a3b} before setting $\Batch[j]$ to $(\Os_j,\s_j)$ in line~\ref{setBatch3}.
Since $j \ge j_4$, by part~(\ref{cx4}) of Claim~\ref{new3-claim}, 
	$P_{j+1}$ was sent by $\ell$.
Since $j \ge j_6$, by part~(\ref{cx6}) of Claim~\ref{new3-claim},
	$\ell$ sent a $C_j = \CM{j}$ to $p$ before sending $P_{j+1}$ to $p$.
Since $j \ge j_5$, by part~(\ref{cx5}) of Claim~\ref{new3-claim},
	$\ell$ sent $C_j$ after real time $\tau_f$.
Since the communication channel from $\ell$ to $p$ is FIFO from real time $\tau_f$ on
	(Assumption~\ref{FIFO}), and $\ell$ sent $C_j$ to $p$ before sending $P_{j+1}$ to $p$,
	$p$ received $C_j$ before receiving $P_{j+1}$ in line~\ref{a3b}.
	So $p$ received $C_j = \CM{j}$ before setting $\Batch[j]$ to $(\Os_j,\s_j)$ in line~\ref{setBatch3}.

\item $p$ sets $\Batch[j]$ to $(\Os_j,\s_j)$ in line~\ref{batch3}.
Thus $p$ previously received a $\langle \MyBatch, j, (\Os_j,\s_j) \rangle$ message from some process $q$ (line~\ref{batchreceipt}).
	Thus $q$ previously sent a $\langle \MyBatch, j, (\Os_j,\s_j) \rangle$ message to $p$ (line~\ref{sendbatch}).
	So $q$ previously received a $\langle \MyGaps,\Gaps \rangle$ message with $j \in \Gaps$ from $p$ (line~\ref{gb}).
	Thus, $p$ previously sent a $\langle \MyGaps,\Gaps \rangle$ message with $j \in \Gaps$ to~$q$ (line~\ref{totoz10}).
	So $p$ previously called $\FillGaps{j'}$ with $j' \ge j$.
	Note that $p$ can call $\FillGaps{j'}$ in line~\ref{gmb2} or line~\ref{FG}.
	Since $j' \ge j \ge j_1$, by part~(\ref{cx1}) of Claim~\ref{new3-claim}, $p$ does not call $\FillGaps{j'}$ in line~\ref{FG}.
	So $p$ called $\FillGaps{j'}$ in line~\ref{gmb2}.
	Thus, $p$ previously received a $C_{j'+1} = \CM{j'+1}$ message in line~\ref{commitreceipt1}.
	Note that $p$ received $C_{j'+1}$ before setting $\Batch[j]$ to $(\Os_j,\s_j)$ in line~\ref{batch3}.
	
	Since $j' \ge j \ge j_4$, by part~(\ref{cx4}) of Claim~\ref{new3-claim}, this $C_{j'+1} = \CM{j'+1}$ was sent by $\ell$.
	Since $j' \ge j \ge j_6$, by part~(\ref{cx6}) of Claim~\ref{new3-claim},
	for all $i$, $j_6 \le i \le j'$, $\ell$ sent a $\CM{i}$ to $p$ before sending $C_{j'+1}$ to $p$.
	In particular, since $j_6 \le j \le j'$, $\ell$ sent a $C_j= \CM{j}$ to $p$ before sending $C_{j'+1}$ to $p$.
	Since $j \ge j_5$, by part~(\ref{cx5}) of Claim~\ref{new3-claim},
		$\ell$ sent $C_j$ after real time $\tau_f$.
Since the communication channel from $\ell$ to $p$ is FIFO from real time $\tau_f$ on
	(Assumption~\ref{FIFO}), and $\ell$ sent $C_j$ to $p$ before sending $C_{j'+1}$ to $p$,
	$p$ received $C_j$ before receiving $C_{j'+1}$ in line~\ref{a3b}.
So $p$ received $C_j = \CM{j}$ before
	setting $\Batch[j]$ to $(\Os_j,\s_j)$ in line~\ref{batch3}.
\end{enumerate}
Therefore in all possible cases $p$ received a $\CM{j}$ message
	before setting $\Batch[j]$ to $(\Os_j,\s_j)$.
\qedhere~$_\text{\autoref{new3}}$
\end{proof}

\begin{lemma}\label{new2}
For all processes $p \neq \ell$, there is a $j_0$ such that for all $j \ge j_0$ the following holds:

	if $p$ receives a $\CM{j}$ from $\ell$,
	then for all $i$, $j_0 \le i < j$,
	$p$ previously received a $\CM{i}$ message from $\ell$.

\end{lemma}
\begin{proof}
Let $p \neq \ell$.
By Lemma~\ref{new1}, 
	there is a $j_1$ such that for all $j \ge j_1$:
	(*) if $\ell$ sends a $\CM{j}$ message to $p$,
	then for all $i$, $j_1 \le i < j$, $\ell$ previously sent a $\CM{i}$ message to $p$.
By Observation~\ref{new0},
	there is a $j_2$ such that for all $i$, $j_2 \le i$:
	(**) if $\ell$ sends a $\CM{i}$ message to $p$,
	then $\ell$ sends this message after real time $\tau_f$.
Let $j_0 = \max(j_1 , j_2)$.
Consider any $j \ge j_0$, and suppose that $p$ receives
	a message $C_j=\CM{j}$ from $\ell$. 
Since $j \ge j_0 \ge \max(j_1 , j_2)$, by (*) and (**) we have:
	for all $i$, $j_0 \le i < j$, $\ell$~sent a $\CM{i}$ message to $p$
	before sending $C_j$
	and after real time~$\tau_f$.
Since after real time~$\tau_f$, the communication channel from $\ell$ to $p$ is FIFO (Assumption~\ref{FIFO}),
	for all $i$, $j_0 \le i < j$,
	$p$ receives this $\CM{i}$ from $\ell$ before receiving $C_j$ from $\ell$.
\qedhere~$_\text{\autoref{new2}}$
\end{proof}

\begin{lemma}\label{new4}
For all processes $p \neq \ell$, there is a $j_0$ such that for all $j \ge j_0$ the following holds:

	if $p$ sets $\Batch[j]$ to some pair $(\Os_j,\s_j)$, 
	then for all $i$, $j_0 \le i < j$, $p$ previously received a $\CM{i}$ message from $\ell$.
\end{lemma}
\begin{proof}
Immediate from Lemmas~\ref{new3} and~\ref{new2}.
\qedhere~$_\text{\autoref{new4}}$
\end{proof}

\begin{lemma}\label{new5}
For all processes $p \neq \ell$, there is a $j_0$ such that for all $j \ge j_0$ the following holds:

	if $p$ sets $\Batch[j]$ to some pair $(\Os_j,\s_j)$, 
	then for all $i$, $j_0 \le i < j$, $p$ previously set $\Batch[i]$ to some pair $(\Os_i,\s_i)$.
\end{lemma}
\begin{proof}
The proof follows from Lemma~\ref{new4} and the fact that when a process $p \neq \ell$
	receives a \linebreak 
	$\langle \CommitLease, (\Os_i,\s_i), i, -, -\rangle$ message for any pair $(\Os_i, \s_i)$,
	$p$ sets $\Batch[i]$ to $(\Os_i, \s_i)$ before doing anything else (see lines~\ref{commitreceipt1} and~\ref{batch2}).
\qedhere~$_\text{\autoref{new5}}$
\end{proof}

\begin{lemma}\label{new6}
For all processes $p \neq \ell$, there is a $j_0$ such that for all $j \ge j_0$ the following holds:
	if $p$ has $\Batch[j] \neq (\emptyset,\infty)$ at some real time $\tau$, 
	then for all $i$, $j_0 \le i \le j$, $p$ has $\Batch[i] \neq (\emptyset,\infty)$ at real time $\tau$.

\end{lemma}
\begin{proof}
Let $p \neq \ell$.
By Lemma~\ref{new5},
	there is an $j_0 > 0$ such that for all $j \ge j_0$:
	(*) if $p$ sets $\Batch[j]$ to some pair $(\Os_j,\s_j)$, 
	then for all $i$, $j_0 \le i < j$, 
	$p$ previously set $\Batch[i]$ to some pair $(\Os_i,\s_j)$.
Consider any $j \ge j_0$ and suppose $p$ has $\Batch[j] \neq (\emptyset, \infty)$ at some real time $\tau$.
Since $\Batch[j]$ is initialized to $(\emptyset,\infty)$ at $p$,
	process $p$ set $\Batch[j]$ to some pair $(\Os_j,\s_j) \neq (\emptyset,\infty)$ by real time $\tau$.
Since $j \ge j_0$,
	by (*),
	for all $i$, $j_0 \le i < j$, $p$ set $\Batch[i]$ to some pair $(\Os_i, \s_i)$ before real time~$\tau$.
Since $j_0>0$, by Corollary~\ref{ne-batch},
	for all $i$, $j_0 \le i < j$, $p$ has $\Batch[i] = (\Os_i,\s_i) \neq (\emptyset,\infty)$ before real time $\tau$.
By Corollary~\ref{finalsafetyac} and the fact that $p$ has $\Batch[j] \neq (\emptyset,\infty)$ by real time $\tau$,
	we conclude that for all $i$, $j_0 \le i \le j$, $p$ has $\Batch[i] \neq (\emptyset,\infty)$ at real time $\tau$.
\qedhere~$_\text{\autoref{new6}}$
\end{proof}

\begin{lemma}\label{new7}
There is a $j_0$ such that for all $j \ge j_0$ the following holds:
	if $\ell$ has $\Batch[j] \neq (\emptyset,\infty)$ at some real time $\tau$, 
	then for all $i$, $j_0 \le i \le j$, $\ell$ has $\Batch[i] \neq (\emptyset,\infty)$ at real time $\tau$.
\end{lemma}
\begin{proof}
By Theorem~\ref{lm10}, there is a real time after which $\ell$ executes in a $\LeaderWork{t}$ for some $t$.
In this $\LeaderWork{t}$, $\ell$ first executes $\DoOps{(-, -)}{t}{k^*}$ in line~\ref{first-doops},
	and then $\ell$ iterates forever
	in the while loop of lines~\ref{mainwhile}-\ref{endwhile}.
In this loop, $\ell$ calls $\DoOps{(-, -)}{t}{-}$ a finite or infinite number of times.
From the code of $\LeaderWork{t}$ and Lemma~\ref{T-k-Doops},
	the (possibly empty) sequence of consecutive calls to $\DoOps{(-,-)}{t}{-}$ that
	$\ell$ makes in this while loop is of the form:
	$\DoOps{(-,-)}{t}{k^*}, \DoOps{(-,-)}{t}{k^*+1} , \DoOps{(-,-)}{t}{k^*+2} \ldots$

Let $j_0 = k^* +1 > 0$.
Consider any $j \ge j_0$ and
	suppose that $\ell$ has $\Batch[j] \neq (\emptyset, \infty)$ at some real time $\tau$.
We must show that 
	for all $i$, $j_0 \le i \le j$, $\ell$ has $\Batch[i] \neq (\emptyset, \infty)$ at real time $\tau$.

\begin{claim}\label{new7-claim}
$\ell$ locked some tuple $(-, t , j)$ in $\LeaderWork{t}$ by real time $\tau$.
\end{claim}

\begin{proof}
Since $\ell$ has $\Batch[j] \neq (\emptyset, \infty)$ at real time $\tau$,
	and $\Batch[j]$ is initialized to $(\emptyset, \infty)$ at $\ell$,
	$\ell$ set $\Batch[j]$ to some pair $(\Os_j, \s_j) \neq (\emptyset, \infty)$ by real time $\tau$.
Thus, 
	by Lemma~\ref{IamOutOfLableNames},
	some process $r$ locked a tuple $(\Os_j , t' , j)$ for some $t'$ by real time $\tau$.
By Observation~\ref{aboutlocking}, $r$ did so in $\LeaderWork{t'}$.
Since $\ell$ executes forever in $\LeaderWork{t}$,
	by Lemma~\ref{lastleader2}(\ref{last-t2}),
	no process calls $\LeaderWork{t'}$ with $t' > t$.
So $t' \le t$.
We will show that, in fact, $t=t'$.
Suppose, for contradiction, that $t' < t$;
	since $r$ locks $(\Os_j , t' , j)$ and $\ell$ locks some tuple $(-, t, k^*)$ in $\DoOps{(-,-)}{t}{k^*}$,
	by Theorem~\ref{generalcase1a}, $k^* \ge j$;
	so $k^* \ge j \ge j_0 = k^* +1$ --- a contradiction.
Therefore $t' = t$.
Since $r$ and $\ell$ called $\LeaderWork{t}$, they called $\ML{t}{t}$ and got $\textsc{True}$.
By Theorem~\ref{leader-safety}, $r = \ell$.
So $\ell$ locked $(\Os_j , t , j)$ in $\LeaderWork{t}$ by real time~$\tau$.
\qedhere~$_\text{\autoref{new7-claim}}$
\end{proof}

Since $\ell$ locked $(- , t , j)$ in $\LeaderWork{t}$, $\ell$ called $\DoOps{(-,-)}{t}{j}$ in $\LeaderWork{t}$.
Since $j \ge k^*+1$, by Lemma~\ref{T-k-Doops},
	$\ell$ called
	$\DoOps{(-,-)}{t}{i}$ for $i=k^*, k^*+1,\ldots,j-1$
	before calling
	$\DoOps{(-,-)}{t}{j}$ in $\LW(t)$.
Thus, for all $i$, $k^* \le i \le j-1$,
	$\ell$ set $\Batch[i]$ to some pair $(\Os_i, \s_i)$ in $\DoOps{(-,-)}{t}{i}$ before locking $(-, t , j)$ in $\DoOps{(-,-)}{t}{j}$,
	and therefore before real time $\tau$.
Since $j_0 = k^* +1$, for all $i$, $j_0 \le i \le j-1$,
	$\ell$ set $\Batch[i]$ to $(\Os_i, \s_i)$ before real time $\tau$,
	and since $i \ge j_0 \ge 1$, by Corollary~\ref{ne-batch}, $\Os_i \neq \emptyset$,
	and so $(\Os_i, \s_i) \neq (\emptyset, \infty)$.
So, by Lemma~\ref{finalsafetyac},
	for all $i$, $j_0 \le i \le j-1$,
	$\ell$ has $\Batch[i] = (\Os_i,-) \neq (\emptyset, \infty)$ at real time $\tau$.
Since $\ell$ also has $\Batch[j] \neq (\emptyset, \infty)$ at real time $\tau$,
	for all $i$, $j_0 \le i \le j$,
	$\ell$ has $\Batch[i] \neq (\emptyset,\infty)$ at real time $\tau$.
\qedhere~$_\text{\autoref{new7}}$
\end{proof}

\begin{lemma}\label{toto2}
For all processes $p$, there is a $j_0$ such that for all $j \ge 0$
	the following holds:

	if $p$ has $\Batch[j] \neq (\emptyset,\infty)$ at some real time $\tau$,
	then for all~$i$, $j_0 \le i \le j$, $p$ has $\Batch[i] \neq (\emptyset,\infty)$ at real time $\tau$.
\end{lemma}
\begin{proof}
Consider any process $p$.
Define $j_0$ to be the constant described by Lemmas~\ref{new6} if $p \neq \ell$,
		or the constant described by Lemmas~\ref{new7} if $p = \ell$.
Let $j \ge 0$ and suppose that $p$ has $\Batch[j] \neq (\emptyset,\infty)$ at some real time $\tau$.
We must show that: (*) for all~$i$, $j_0 \le i \le j$, $p$ has $\Batch[i] \neq (\emptyset,\infty)$ at real time $\tau$.
If $j < j_0$ then (*) is vacuously true; if $j \ge j_0$ then (*) follows from Lemma~\ref{new6} if $p \neq \ell$,
	and from Lemma~\ref{new7} if $p = \ell$.
\qedhere~$_\text{\autoref{toto2}}$
\end{proof}

\begin{lemma}\label{no-gaps-in-batch}\label{tognazzi2}
For all correct processes $p$, there is a real time $\rrb$ such that: for all $\tau > \rrb$,
	if $p$ has $\Batch[j] \neq (\emptyset,\infty)$ for some $j \ge 0$
	at real time $\tau$,
	then for all~$i$, $1 \le i \le j$, process $p$ has $\Batch[i] \neq (\emptyset,\infty)$ at real time~$\tau$.
\end{lemma}
\begin{proof}
Let $p$ be any correct process.
Consider the (value of the) variable $k$ of process $\ell$.
There are two cases:

\begin{enumerate}

\item \emph{$k$ is bounded.}
Thus, from Lemma~\ref{katl}, there is a real time after which $k = \hat{k}$ for some integer $\hat{k}$.
So, by Lemma~\ref{zozox3}(\ref{oopla1}):
	(*) there is a real time $\rrb$ after which
	for all~$i$, $1 \le i \le \hat{k}$, $p$ has $\Batch[i] \neq (\emptyset,\infty)$.
	
\begin{claim}\label{toto3}
For all $k' > \hat{k}$, $\Batch[k'] = (\emptyset,\infty)$ at $p$ (always).
\end{claim}

\begin{proof}
Suppose, for contradiction, that for some $k' > \hat{k}$,
	process $p$ has $\Batch[k'] \neq (\emptyset,\infty)$ at some time.
By Lemma~\ref{IamOutOfLableNames},
	a process previously locked some tuple $(-,-,k')$.
Thus, by Lemma~\ref{zozo4}, there is a real time after which $\ell$ has $k \ge k' > \hat{k}$
	--- contradicting the definition of~$\hat{k}$.
\qedhere~$_\text{\autoref{toto3}}$
\end{proof}

Suppose that $p$ has $\Batch[j] \neq (\emptyset,\infty)$
	for some $j \ge 0$ at some real time $\tau > \rrb$.
By Claim~\ref{toto3}, $j \le \hat{k}$.
Since $j \le \hat{k}$ and $\tau > \rrb$,
	by (*) we have that for all~$i$, $1 \le i \le j$,
	$p$ has $\Batch[i] \neq (\emptyset,\infty)$ at real $\tau$.

\item \emph{$k$ grows unbounded.}
By Lemma~\ref{toto2}, there is a $j_0$ such that for all $j\ge0$:
(**) if $p$ has $\Batch[j] \neq (\emptyset,\infty)$ at some real time $\tau$,
	then
	for all~$i$, $j_0 \le i \le j$, process $p$ has $\Batch[i] \neq (\emptyset,\infty)$ at real time $\tau$.
Since $k$ grows unbounded, there is a real time after which $\ell$ has $k \ge j_0$.
So , by Lemma~\ref{zozox3}(\ref{oopla1}):
	(***)~there is a real time $\rrb$ after which
	for all~$i$, $1 \le i \le j_0$, process $p$ has $\Batch[i] \neq (\emptyset,\infty)$.

Suppose that $p$ has $\Batch[j] \neq (\emptyset,\infty)$ for some $j \ge 0$ at some real time $\tau > \rrb$.
	By (**): for all~$i$, $j_0 \le i \le j$,
	process $p$ has $\Batch[i] \neq (\emptyset,\infty)$ at real time $\tau$.
Combining this with (***) we have: for all~$i$, $1 \le i \le j$,
	process $p$ has $\Batch[i] \neq (\emptyset,\infty)$ at real time $\tau$.

\end{enumerate}

So in all cases, there is a real time $\rrb$ such that
	if $p$~has $\Batch[j] \neq (\emptyset,\infty)$ at a real time $\tau > \rrb$,
	then for all~$i$, $1 \le i \le j$, $p$~has $\Batch[i] \neq (\emptyset,\infty)$ at real time $\tau$.

\qedhere~$_\text{\autoref{no-gaps-in-batch}}$
\end{proof}

\begin{lemma}\label{promise-is-great}
There is a $j_0$ such that for all $j\geq j_0$,
	if a process $p\neq\ell$ has $\PendingOps[j] = (\Os_j,\s_j)\neq (\emptyset, \infty)$ for some $\Os_j$ and $\s_j$,
	then
	\begin{enumerate}
		\item\label{stable-pending}
		$p$ has $\PendingOps[j] = (\Os_j,\s_j)$ thereafter, and
		\item\label{promise-core}
		$p$ sets $\Batch[j]$ to $(\Os_j,\s_j)$ by local time $\s_j - \Pthree + \alpha_2 + 3\delta$.\footnote{Recall that $\alpha$ is the value of the parameter $\PP$.}
	\end{enumerate}
\end{lemma}
\begin{proof}
By Claim~\ref{new3-claim}(4), there is a $j_1$ such that for all $j \geq j_1$,
	no process $q\neq\ell$ sends a $\PR{j}$ or a $\CM{j}$ message.
Since $\ell$ sends a finite number of $\PR{-}$ messages by real time $\tau_g$,
	there is a $j_2$ such that for all $j \geq j_2$,
	if $\ell$ sends a $\PR{j}$ message,
	it does so after real time $\tau_g$.
Let $j_0 = \max(j_1, j_2)$.
Suppose a process $p\neq\ell$ has $\PendingOps[j] = (\Os_j, \s_j) \neq (\emptyset, \infty)$ 
	for some $j \geq j_0$, $\Os_j$ and $\s_j$ at real time $\tau$.
Since a process sets $\PendingOps[j]$ only in line~\ref{setPB},
	$p$ previously received a $\langle \Prepare,(\Os_j,\s_j),-,j,- \rangle$ message.
Since $j\geq j_0$,
	this message is sent by $\ell$ in $\DoOps{(\Os_j,\s_j)}{t}{j}$ after real time $\tau_g$,
	and $\ell$ executes the while loop of lines~\ref{mainwhile}-\ref{endwhile} in $\LeaderWork{t}$ forever after time $\tau_g$.
If $p$ later resets $\PendingOps[j]$ to some pair $(\Os_j', \s_j')$,
	then it must receive a $\langle \Prepare,(\Os_j',\s_j'),-,j,- \rangle$ message.
Similar as above, since $j\geq j_0$, 
	this message must be sent by $\ell$ in $\DoOps{(\Os_j',\s_j')}{t}{j}$ after real time $\tau_g$.
By Corollary~\ref{T-k-DoopsCreduced}, these two $\DO$ calls are the same call,
	so $(\Os_j',\s_j') = (\Os_j,\s_j)$,
	and hence $p$ has $\PendingOps[j] = (\Os_j,\s_j)$ at all real times after $\tau$,
	so (\ref{stable-pending}) holds.

Since $\ell$ executes the while loop of lines~\ref{mainwhile}-\ref{endwhile} forever after time $\tau_g$,
	the $\DoOps{(\Os_j,\s_j)}{t}{j}$ call is made in line~\ref{second-doops}.
Consider the iteration of the while loop in which $\ell$ makes this $\DO$ call.
Suppose $\ell$ gets $t'$ from its local clock in line~\ref{taketime1} at time $(t', \tau')$.
Since $\ell$ is at line~\ref{mainwhile} at real time $\tau_g$
	and $\ell$ sends $\langle \Prepare,(\Os_j,\s_j),-,j,- \rangle$ after $\tau_g$,
	$\tau' > \tau_g$.
By Lemma~\ref{Alpha123},
	it takes at most $\alpha_2 + 2\delta$ units of local time 
	from local time $t'$ in line~\ref{taketime1} to when $\ell$ sends
	$\langle \CommitLease,(\Os_j,\s_j), j, -, - \rangle$
	to all process $q \neq \ell$
	in line~\ref{sendcommit} of $\DoOps{(\Os_j,\s_j)}{t}{j}$.
Since this happens after $\tau_g \geq \tau_3$,
	by property~\ref{LM3} and the clock synchronization Assumption~\ref{xclocks}(\ref{xcl5}),
	$p$ receives this message and sets $\Batch[j]$ to $(\Os_j,\s_j)$ by its local time $t' + \alpha_2 + 3\delta$.
From the way $\ell$ calls $\DoOps{(\Os_j,\s_j)}{t}{j}$ in line~\ref{second-doops},
	it is clear that $\s_j = t' + \Pthree$.
Thus, $p$ sets $\Batch[j]$ to $(\Os_j,\s_j)$ 
	by local time $\s_j - \Pthree + \alpha_2 + 3\delta$.
So (\ref{promise-core}) holds.
\qedhere~$_\text{\autoref{promise-is-great}}$
\end{proof}

\begin{lemma}\label{sordi}
There is a real time after which if a correct process $p$ starts executing a read operation~$o$
	(in lines~\ref{read-invoke}-\ref{read-respond}) then:

\begin{enumerate}
	\item\label{sordi1} $p$ executes lines~\ref{getvalidlease-start}-\ref{getvalidlease-end} only once.

	\item\label{sordi2} $p$ waits in line~\ref{FG3} only if $o$ conflicts with some operation $o'$ that is pending at $p$ when $p$ executes line~\ref{get-k-hat-else}.
	
	\item\label{sordi3} $p$ waits in line~\ref{FG3} only if it has $\PendingOps[\hat{k}] \neq (\emptyset,\infty)$ just before line~\ref{FG3}.
	
	\item\label{sordi4} $p$ waits in line~\ref{FG3} only if it has $\PendingBatchPromise{\hat{k}}  \leq t'$ in line~\ref{get-k-hat-5},
		where $t'$ is the value that $p$ gets from its clock in line~\ref{getleasetime}.
\end{enumerate}
\end{lemma}
\begin{proof}
Let $p$ be any correct process.
Suppose $p$ starts executing a \emph{read} operation~$o$
	(in lines~\ref{read-invoke}-\ref{read-respond}) after
	real time $\rrs = \max (\tau_r, \rrb)$, 
	where $\tau_r$ and $\rrb$
	are described in Theorem~\ref{exitleaseloop} and Lemma~\ref{no-gaps-in-batch},
	respectively.
Suppose $t'$ is the value that $p$ gets from its clock in line~\ref{getleasetime}
	during the last iteration of the loop of lines~\ref{getvalidlease-start}-\ref{getvalidlease-end}.
	
\begin{enumerate}
\item Since $p$ starts the repeat-until code
	of lines~\ref{getvalidlease-start}-\ref{getvalidlease-end} after real time $\rrs \ge \tau_r$,
	by Theorem~\ref{exitleaseloop}, $p$ exits in line~\ref{getvalidlease-end} without looping.
	In other words,
	$p$ execute lines~\ref{getvalidlease-start}-\ref{getvalidlease-end} only once.

\item Suppose $p$ waits in line~\ref{FG3}.
	Thus, there is a $j$, $1 \le j \le \hat{k}$,
		such that $p$ has $\Batch[j] = (\emptyset,\infty)$ in line~\ref{FG3}.
	Since this holds after real time $\rrs \ge \rrb$, Lemma~\ref{tognazzi2} implies
		that $p$ has $\Batch[\hat{k}] = (\emptyset,\infty)$ in line~\ref{FG3}.
	By Corollary~\ref{finalsafetyac} and Corollary~\ref{finite-batch-promise},
		$p$ also has $\Batch[\hat{k}] = (\emptyset,\infty)$ in line~\ref{get-k-hat-else}.

	From the way $p$ computes $\hat{k}$ in lines~\ref{get-k-hat-else}-\ref{get-k-hat-5}, $\hat{k} = k^*$ or $\hat{k} > k^*$.
	So there are two cases:
	
	\begin{enumerate}
	\item $\hat{k} = k^*$.
	In this case, it is clear that $p$ has $\leasebatch = k^* = \hat{k}$ in line~\ref{getlease}.
	By Lemma~\ref{set-batch-before-lease}, 
		$p$ has $\Batch[\hat k] \neq (\emptyset, \infty)$ by the real time when it sets lease to $(k^*,-)$, 
		and hence by the real time when line~\ref{getlease} is executed.
	Thus, $p$ also has $\Batch[\hat{k}] \neq (\emptyset,\infty)$ line~\ref{get-k-hat-else} --- a contradiction; so
	case (a) is not possible.
	
	\item $\hat{k} > k^*$. 
	In this case it is clear that when $p$ executes lines~\ref{get-k-hat-else}-\ref{get-k-hat-5},
	$p$ finds that
	$o$ conflicts with some operation $o'$
	in $\PendingBatchOps{\hat{k}}$,
	and that $t' \geq \PendingBatchPromise{\hat k}$.

	Since $p$ has $\Batch[\hat{k}] = (\emptyset,\infty)$ in line~\ref{get-k-hat-else},
	this means that 
		$o$ conflicts with some operation $o'$
		that is pending when $p$ executes line~\ref{get-k-hat-else}.
	Since $p$ finds that $t' \geq \PendingBatchPromise{\hat k}$ in line~\ref{get-k-hat-5},
		$p$ has $\PendingOps[\hat{k}] \neq (\emptyset,\infty)$ just before line~\ref{FG3}. 

\end{enumerate}
\end{enumerate}

The above shows that:
	(1)~$p$ executes lines~\ref{getvalidlease-start}-\ref{getvalidlease-end} only once,
	(2)~$p$ waits in line~\ref{FG3} only if
	$o$ conflicts with some operation $o'$
	that is pending at $p$ when $p$ executes line~\ref{get-k-hat-else},
	(3)~$p$ waits in line~\ref{FG3} only if it has $\PendingOps[\hat{k}] \neq (\emptyset,\infty)$ just before line~\ref{FG3}, and
	(4)~$p$ waits in line~\ref{FG3} only if it has $\PendingBatchPromise{\hat{k}} \leq t'$ in line~\ref{get-k-hat-5}.
\qedhere~$_\text{\autoref{sordi}}$
\end{proof}

\begin{lemma}\label{gafni}
There is a real time after which there are no pending operations at process $\ell$.
\end{lemma}
\begin{proof}
By Theorem~\ref{lm10}, eventually $\ell$ executes
	in the $\LeaderWork{t}$ procedure forever for some $t$.
Let $\tau_0$ the real time when $\ell$ calls $\LeaderWork{t}$,
	and let $\PBZ$ be the value of the array $\PB$ at $\ell$ at real time $\tau_0$.
We claim $\PB$ remains equal to $\PBZ$ forever after real time $\tau_0$.
More precisely:

\begin{claim}\label{toto32}
$\ell$ has $\PB = \PBZ$ at all times $\tau \ge \tau_0$.
\end{claim}

\begin{proof}
When $\ell$ starts $\LeaderWork{t}$ at real time $\tau_0$, 
	it has $\PB = \PBZ$.
Since $\ell$ modifies its array $\PB$ only in line~\ref{setPB} of the $\PCM()$ procedure,
	and $\ell$ does not execute this procedure when it is in $\LeaderWork{t}$,
	process $\ell$ does not modify its $\PB$ array in $\LeaderWork{t}$.
Since $\ell$ remains in $\LeaderWork{t}$ forever, the claim follows.
\qedhere~$_\text{\autoref{toto32}}$
\end{proof}

There are two cases:

\begin{enumerate}

\item For all $j \ge 1$, $\PBZ[j] = (\emptyset,\infty)$.
By Claim~\ref{toto32}, for all $j \ge 1$, $\PB[j] = (\emptyset,\infty)$ at $\ell$ at all real times $\tau \ge \tau_0$.
Thus, from Definitions~\ref{pending1}-\ref{pending2}, 
	there are no pending operations at~$\ell$ after real time $\tau_0$.

\item  There is a $j \ge 1$, such that $\PBZ[j] \neq (\emptyset,\infty)$.
Let $j_0 = \max \{ j ~|~ \PBZ[j] \neq (\emptyset,\infty) \}$.
(This maximum exists by Claim~\ref{toto32}, since by real time $\tau_0$
	process $\ell$ has $\PB[j]\ne(\emptyset,\infty)$
	for only a finite number of indices.)
Note that $j_0 \ge 1$, since $\PB[0]$ remains $(\emptyset,\infty)$ forever.

\begin{claim}\label{toto33}
There is a real time $\tau_1$ after which
	for all $j$, $1 \le j \le j_0$, $\Batch[j] \neq (\emptyset,\infty)$ at $\ell$.
\end{claim}

\begin{proof}
Since $\PB[j_0] = \PBZ[j_0] \neq (\emptyset,\infty)$ at real time $\tau_0$,
	it is clear from the code of lines~\ref{a3b}-\ref{setPB} that
	$\ell$ previously accepted some tuple $(-,-,j_0)$.
So, by Lemma~\ref{deus},
	some tuple $(-,-,j_0)$ is eventually locked.
By Lemma~\ref{zozo4}, there is a real time after which $\ell$ has $k \ge j_0$.
So, by Lemma~\ref{zozox3}(\ref{oopla1}), there is a real time $\tau_1$ after which
	for all $j$, $1 \le j \le j_0$,
	process $\ell$ has $\Batch[j] = (\Os_j,-)$ for some non-empty set~$\Os_j$.
\qedhere~$_\text{\autoref{toto33}}$
\end{proof}

Let $\hat \tau = \max(\tau_0, \tau_1)$.

\begin{claim}\label{toto34}
There are no pending operations at process $\ell$ after real time $\hat{\tau}$.
\end{claim}

\begin{proof}
Suppose, for contradiction, that some operation $o$ is pending at $\ell$ at some real time $\tau > \hat{\tau}$.
Thus, by Definitions~\ref{pending1}-\ref{pending2},
	there is a set of operations $\Os$ and an index $j \ge 1$ such that: (a) $o \in \Os$, and
	(b) $\PB[j] =  (\Os,-)$ and $\Batch[j] = (\emptyset,\infty)$ at $\ell$ at real time $\tau$.
Since $\ell$ has $\PB[j] =  (\Os,-) \neq (\emptyset,\infty)$ at real time  $\tau > \hat{\tau} \ge \tau_0$,
	by Claim~\ref{toto32}, $\PBZ[j] \neq (\emptyset,\infty)$.
	So, by the definition of $j_0$ and the fact that $j \ge 1$, $1 \le j \le j_0$.
Therefore, by Claim~\ref{toto33},
	$\ell$ has $\Batch[j] \neq (\emptyset,\infty)$ at real time $\tau > \hat{\tau} \ge \tau_1$ --- a contradiction.
\qedhere~$_\text{\autoref{toto34}}$
\end{proof}

\end{enumerate}

Thus, in all cases, there is a real time after which there are no pending operations at process $\ell$.
\qedhere~$_\text{\autoref{gafni}}$
\end{proof}

From Lemmas~\ref{sordi} and~\ref{gafni}, we have:
\begin{corollary}\label{leader-does-not-wait-22}
There is a real time after which process $\ell$ does not wait in line~\ref{FG3}.
\end{corollary}

\begin{theorem}\label{three-delta-promise-thm}\label{three-delta-thm}
There is a real time after which
	no correct process executes a wait statement in line~\ref{FG3}
	that lasts more than $\max(3\delta-\Pthree+\alpha_2, 0)$ local time units.
\end{theorem}
\begin{proof}
By Corollary~\ref{leader-does-not-wait-22},
	the theorem holds for $\ell$.
So we consider processes other than $\ell$.
Suppose, for contradiction, that:
\begin{align}\label{eqn-three-delta-1}
&\text{some correct process $p \neq \ell$ executes infinitely often a wait statement
in line~\ref{FG3} }\\[-2pt]
&\text{that lasts more than $\max(3\delta-\Pthree+\alpha_2, 0)$ local time units.}\notag
\end{align}
By Lemma~\ref{sordi}(\ref{sordi3}),
	there is a real time $\tau_\nb$ after which $p$ waits in line~\ref{FG3}
	only if it has $\PB[\hat k]\ne(\emptyset,\infty)$
	just before executing that line
	(the subscript ``$\nb$'' stands for ``no-blocking'').
From this and (\ref{eqn-three-delta-1}),
	$p$ executes infinitely often a wait statement that
	lasts more than $\max(3\delta-\Pthree+\alpha_2, 0)$ local time units
	\emph{and} starts after real time~$\tau_\nb$.
Let $W_i$ denote the $i$-th instance of such a wait statement
	and $\hat{k}_i$ be the value of $\hat k$ in the execution of~$W_i$.

\begin{claim}\label{distinct-pending-k-hat}
$\hat k_i$ strictly increases with $i$.
\end{claim}
\begin{proof}
It is clear that $W_i$'s are not executed concurrently.
Suppose that $p$ reads $(k^*,-)$ from its $\lease$ variable
	in line~\ref{getlease} during the last iteration of the loop of lines~\ref{getvalidlease-start}-\ref{getvalidlease-end}
	before executing $W_i$ for some $i \geq 1$.
By Lemma~\ref{set-batch-before-lease},
	$p$ has $\Batch[j]\neq (\emptyset, \infty)$ for $1\leq j \leq k^*$
	before executing $W_i$.
By Corollary~\ref{finalsafetyac}, 
	$p$ has $\Batch[j]\neq (\emptyset, \infty)$ for $1\leq j \leq k^*$ thereafter.
From the wait statement in line~\ref{FG3}, 
	when $p$ completes the execution of $W_i$,
	it has $\Batch[j] \neq (\emptyset, \infty)$ for $k^* < j \leq \hat k_i$.
Thus, by Corollary~\ref{finalsafetyac}, 
	$p$ has $\Batch[j] \neq (\emptyset, \infty)$ for all $1\leq j \leq \hat{k}_i$ thereafter.
Now consider $W_{i'}$ for some $i' > i$.
This must occur after $p$ completes $W_i$.
From the code of line~\ref{FG3},
	it must be that $\hat k_{i'} > \hat k_i$ since otherwise $p$ does not wait in this line. 
\qedhere~$_\text{\autoref{distinct-pending-k-hat}}$
\end{proof}

By Lemma~\ref{tognazzi2},
	there is a real time $\tau_\ngap$ after which,
	if $p$ has $\Batch[j]\ne(\emptyset,\infty)$ for some $j\ge0$,
	then $p$ also has $\Batch[i]\ne(\emptyset,\infty)$ for all $i$, $1\le i\le j$
	(the subscript ``$\ngap$'' stands for ``no-gaps'').
Let $j_0$ be as defined in Lemma~\ref{promise-is-great}.
By Lemma~\ref{nowaity}, 
	$p$ does not wait forever in line~\ref{FG3}.
Thus, by Claim~\ref{distinct-pending-k-hat},
	there is a $m\geq 1$ such that $\hat k_m \geq j_0$ \emph{and} 
	$W_m$ starts after time $\tau_\ngap$.

Let $t'$ be the value that $p$ gets from its clock in line~\ref{getleasetime}
	during the last iteration of the loop of lines~\ref{getvalidlease-start}-\ref{getvalidlease-end} before $W_{\hat k_m}$,
	so this is at local time $t'$.
Since $p$ has $\PendingOps[\hat k_m] = (\Os_m,\s_m)\neq (\emptyset, \infty)$
	for some pair $(\Os_m,\s_m)$ 
	before executing $W_m$, 
	by Lemma~\ref{promise-is-great}(\ref{stable-pending}),
	$p$ has $\PendingOps[\hat k_m] = (\Os_m,\s_m)$ thereafter.
Thus, for $p$ to wait in line~\ref{FG3},
	 $p$ must find in line~\ref{get-k-hat-5} that $t' \geq \PendingBatchPromise{\hat k_m} = \s_m$.
By Lemma~\ref{promise-is-great}(\ref{promise-core}),
	$p$ sets $\Batch[\hat k_m]$ to $(\Os_m,\s_m)$ by local time 
	$\s_m - \Pthree + 3\delta + \alpha_2$.
Since this happens after real time $\tau_\ngap$,
	$p$ has $\Batch[j] \neq (\emptyset,\infty)$ for all $j$, 
	$1 \leq j \leq \hat k_m$, 
	by local time $\s_m - \Pthree + 3\delta + \alpha_2$.
Since $p$ executes $W_{\hat k_m}$ in line~\ref{FG3} after local time $t'$,
	$p$ waits in line~\ref{FG3} for at most the time period from
	local time $t' \geq s_m$ to local time $\s_m - \Pthree + 3\delta + \alpha_2$.
This implies that $p$ waits in line~\ref{FG3} for at most
	$\max(3\delta-\Pthree+\alpha_2, 0)$ local time units.
\qedhere~$_\text{\autoref{three-delta-promise-thm}}$
\end{proof}

\begin{lemma}\label{stable-batch-promise}
There is a $j_0$ such that for all $j\geq j_0$,
	if a process $p \neq \ell$ sets $\Batch[j]$ to some pair $(\Os_j,\s_j)$ at real time $\tau$,
	then it has $\Batch[j] = (\Os_j,\s_j)$ at all real times $\tau' \geq \tau$.
\end{lemma}
\begin{proof}
By Theorem~\ref{lm10} and Lemma~\ref{lm6.8w},
	there is a real time after which $\ell$ executes $\LeaderWork{t}$ forever,
	and no process $p\neq\ell$ executes in $\LW()$.
This implies that there is a $j_0$ such that for all $j \geq j_0$,
	any call to $\DoOps{(-,-)}{-}{j}$ is made by $\ell$ in $\LeaderWork{t}$.
Suppose some process $p \neq \ell$ sets $\Batch[j]$ 
	to some pair $(\Os_j,\s_j)$ for some $j \geq j_0$.
By Observation~\ref{batch-promise-locked},
	some process locked a tuple $(\Os_j,-,j)$ with promise $\s_j$.
Since $j \geq j_0$, this process is $\ell$ and 
	$\ell$ does so during a call to $\DoOps{(\Os_j,\s_j)}{t}{j}$.
If $p$ later sets $\Batch[j]$ to some pair $(\Os_j',\s_j')$,
	then by the same reasoning as above, 
	$\ell$ calls $\DoOps{(\Os_j',\s_j')}{t}{j}$.
By Corollary~\ref{T-k-DoopsCreduced}, these two $\DO$ calls are the same call,
	so $(\Os_j',\s_j') = (\Os_j,\s_j)$.
\qedhere~$_\text{\autoref{stable-batch-promise}}$
\end{proof}

\begin{lemma}\label{leader-lease-larger-than-promise}
There is a real time after which if a lease $(k, t')$ is issued,
	then $t' \geq \GPromise_k$.
\end{lemma}
\begin{proof}
By Theorem~\ref{lm10} and Lemma~\ref{lm6.8w},
	there is a real time $\tau$ after which $\ell$ executes while loop of lines~\ref{mainwhile}-\ref{endwhile}
	in some $\LeaderWork{t}$ forever
	and no process $p\neq\ell$ executes in $\LW()$.
There are two cases depending on whether $\ell$ calls $\DO$ in line~\ref{second-doops}:
\begin{case}
	\item $\ell$ does not call $\DO$ in line~\ref{second-doops} in $\LW(t)$.
	Since after real time $\tau$ only $\ell$ executes in $\LW(t)$,
		there is a real time after which all leases are issued in line~\ref{setlease} in $\LW(t)$.
	Suppose $\ell$ called $\DoOps{(-, 0)}{t}{j}$ in line~\ref{first-doops} in $\LW(t)$.
	Then in this $\DO$ call,
		$\ell$ sets its variable $k$ to $j$ 
		and locks a tuple of form $(-, t, j)$.
	By Observation~\ref{lock-imply-finite-GPromise},
		$\GPromise_j \neq \infty$.
	Since $\ell$ does not call $\DO$ in line~\ref{second-doops} in $\LW(t)$
		and it does not execute in $\PCM$ while executing $\LW(t)$,
		it does not change its variable $k$,
		so $k$ remains equal to $j$ at $\ell$.
	By Assumptions~\ref{xclocks}(\ref{xcl2}) and~(\ref{xcl3}),
		there is a real time $\hat \tau$ after which if $\ell$ reads from its local clock,
		it reads value at least $\GPromise_j$.
	So all leases issued after real time $\hat \tau$ have $\leasetime \geq \GPromise_j$.
	So the lemma holds for real time $\hat \tau$.

	\item $\ell$ calls $\DO$ in $\LW(t)$.
	Let $\hat \tau$ be the real time when the first such $\DO$ call is made.
	Consider any lease $(k, t')$ issued after real time $\hat \tau$.
	There are two cases:
	\begin{subcase}
		\item $\ell$ issues $(k, t')$ in line~\ref{sl2}.
		Suppose this happens in $\DoOps{(-, \s_j)}{t}{j}$.
		Since $\ell$ executes infinitely often the while loop in $\LW(t)$,
			this call to $\DoOps{(-, \s_j)}{t}{j}$ returns $\textsc{Done}$.
		So $\ell$ locks a tuple of form $(-, t, j)$ with promise $\s_j$
			and issues the lease $(k, t')$ in line~\ref{sl2}.
		Hence $k = j$ and $t' = \s_j$.
		By Definition~\ref{GPromise-def},
			$\GPromise_j = \s_j$
			and the lemma holds in this case.
		\item $\ell$ issues $(k, t')$ in line~\ref{sl1}.
		By Lemma~\ref{lease-issue-obs},
			$\ell$ previously completed a call to $\DoOps{(-, \s_k)}{t}{k}$
			and locked a tuple of form $(-, t, k)$.
		Since this is after $\hat \tau$,
			this call to $\DoOps{(-, \s_k)}{t}{k}$ must be made in line~\ref{second-doops}.
		So by Definition~\ref{GPromise-def},
			$\GPromise_k = \s_k$.
		During this call to $\DoOps{(-, \s_k)}{t}{k}$,
		$\ell$ issued a lease $(k, \s_k)$ in line~\ref{sl2}.
		By Lemma~\ref{increasing-lease-time-in-LeaderWork},
			$t' \geq \s_k = \GPromise_k$.
\qedhere~$_\text{\autoref{leader-lease-larger-than-promise}}$
	\end{subcase}
\end{case}
\end{proof}

\begin{lemma}\label{no-wait-for-promise}
There is a real time after which no correct process waits in line~\ref{wait-promise-2}.
\end{lemma}
\begin{proof}
Let $p$ be any correct process and
	$t'$ be the value that $p$ gets from its clock in line~\ref{getleasetime}
	during the last iteration of repeat-until loop of lines~\ref{getvalidlease-start}-\ref{getvalidlease-end}.
Consider the value of $\hat k$ that $p$ computes in lines~\ref{get-k-hat-1}-\ref{get-k-hat-5}.
If $\hat k = 0$,
	then by Lemma~\ref{batch-zero},
	$\BatchPromise{0}$ remains $0$ in line~\ref{wait-promise-2},
	and the lemma holds.
Henceforth we assume $\hat k > 0$.
There are three cases:
\begin{case}
\item $p$ computes $\hat k$ in line~\ref{get-k-hat-then}.
Thus, $p$ finds	
	$\BatchPromise{\hat k} \leq t'$ in line~\ref{get-k-hat-then}.
Since the initial value of $\BatchPromise{\hat k}$ is $\infty$,
	$p$ must previously set $\Batch[\hat k]$.
By Lemmas~\ref{greater-than-batch-promise-2} and~\ref{local-less-than-global-promise}
	and monotonicity of local clocks
	(Assumption~\ref{xclocks}(\ref{xcl2})),
	$p$ has $\CT \geq \BatchPromise{\hat k}$
	when it starts executing line~\ref{wait-promise-2},
	and hence $p$ does not wait in line~\ref{wait-promise-2}.

\item $p$ computes $\hat k$ in lines~\ref{get-k-hat-else}-\ref{get-k-hat-5} \emph{and} $\hat k = k^*$.
By Lemma~\ref{leader-lease-larger-than-promise},
	there is a real time after which if a $\lease$ $(k^*,t^*)$ is issued,
	then $t^* \geq \GPromise_{k^*}$.
By Lemmas~\ref{lease-frequency} and~\ref{NoD},
	$p$ sets its $\lease$ variable infinitely often.
So there is a real time $\hat \tau$ after which,
	if $p$ has $\lease = (k^*,t^*)$,
	then $t^* \geq \GPromise_{k^*}$.
Consider any read operation started by $p$ after real time $\hat \tau$.
For $p$ to compute $\hat k$ in lines~\ref{get-k-hat-else}-\ref{get-k-hat-5},
	it must find $t' \geq t^*$ in line~\ref{get-k-hat-1}.
By the above argument,
	$t' \geq t^* \geq \GPromise_{k^*}$.
By Lemma~\ref{set-batch-before-lease} and Observation~\ref{set-batch-and-promise},
	$p$ sets $\Batch[k^*]$ to $(\Os_{k^*},\s_{k^*})$ for some non-empty set $\Os_{k^*}$
	by the real time when it sets $\lease$ to $(k^*,t^*)$.
By Lemma~\ref{local-less-than-global-promise},
	$\s_{k^*} \leq \GPromise_{k^*}$ and
	$p$ has $\BatchPromise{k^*} \leq \GPromise_{k^*}$ thereafter.
Thus, by monoticity of local clocks,
	when $p$ executes line~\ref{wait-promise-2} with $\hat k = k^*$,
	it has $\CT \geq t' \geq \GPromise_{k^*} \geq \BatchPromise{k^*}$,
	and $p$ does not wait in this line.
So there is a real time after which $p$ does not wait in line~\ref{wait-promise-2}.

\item $p$ computes $\hat k$ in lines~\ref{get-k-hat-else}-\ref{get-k-hat-5} 
	\emph{and} $\hat k > k^*$.
By Lemma~\ref{gafni},
	there is a real time $\tau_{np}$ after which there is no pending operation at $\ell$.
Thus, after time $\tau_{np}$,
	if $\ell$ computes $\hat k$ in lines~\ref{get-k-hat-else}-\ref{get-k-hat-5},
	then it must compute $\hat k$ to be $k^*$.
So after time $\tau_{np}$
	this case does not happen for process $\ell$.
Henceforth we assume $p \neq \ell$.

There are two subcases:
\begin{enumerate}
\item \emph{The value of $\leasebatch$ at $p$ is bounded.}
So there is a $k'$ and a real time $\hat \tau$ after which
	$\leasebatch = k'$ at $p$.
\begin{claim}\label{no-label-can-you-believe-it}
There is no $j > k'$ such that $\PendingOps[j] \neq (\emptyset,\infty)$ at $p$.
\end{claim}
\begin{proof}
Suppose, by contradiction, that
	$p$ has $\PendingOps[j] \neq (\emptyset, \infty)$ for some $j > k'$.
Then $p$ must have received a $\PR{j}$ message,
	and accepted some tuple $(-,-,j)$.
By Lemma~\ref{deus},
	some tuple $(-,-,j)$ is eventually locked.
By Lemma~\ref{zozo4}, 
	there is a real time after which $\ell$ has $k \geq j$.
By Lemma~\ref{lease-frequency},
	$\ell$ sends $\lease$ messages infinitely often
	where $\leasebatch$ is the value of variable $k$ at $\ell$.
So there is a real time after which
	all the $\lease$ messages sent by $\ell$ has a $\leasebatch \geq j$.
By Lemma~\ref{NoD},
	$p$ eventually accepts some $\lease$ message with $\leasebatch \geq j$.
This contradicts the fact that
	$p$ has $\leasebatch = k' < j$ at all real times after $\hat \tau$.
\qedhere~$_\text{\autoref{no-label-can-you-believe-it}}$
\end{proof}
Thus, after real time $\hat \tau$,
	if $p$ computes $\hat k$ in lines~\ref{get-k-hat-else}-\ref{get-k-hat-5}
	then $\hat k = k^*$.
So after real time $\hat \tau$,
	this case does not happen.

\item  \emph{The value of $\leasebatch$ at $p$ is unbounded.}
By Lemma~\ref{promise-is-great}, 
	there is a $j_1$ such that for all $j\geq j_1$,
	if $p$ has $\PendingOps[j] = (\Os_j,\s_j)\neq (\emptyset, \infty)$ for some $\Os_j$ and $\s_j$,
	then $p$ sets $\Batch[j]$ to $(\Os_j,\s_j)$ at some time.
Since $p\ne\ell$, by Lemma~\ref{stable-batch-promise}
	there is a $j_2$ such that for all $j\geq j_2$,
	if $p$ sets $\Batch[j]$ to some pair $(\Os_j,\s_j)$,
	then $p$ has $\Batch[j] = (\Os_j,\s_j)$ at all real times after.
Let $j_0 = \max(j_1, j_2)$.
Since the value of $\leasebatch$ at $p$ is unbounded,
	by Lemma~\ref{leasemonotonicity},
	there is a real time $\hat \tau$ after which the value of $\leasebatch$ at $p$ is at least $j_0$.
Consider when $p$ computes $\hat k$ in lines~\ref{get-k-hat-else}-\ref{get-k-hat-5}
	after real time $\hat \tau$ such that $\hat k > k^*$.
Since this happens after real time $\hat \tau$, 
	we have $\hat k > k^* \geq j_0$.
Since $\hat k > k^*$,
	$p$ finds $\PendingOps[\hat k] = (\Os_{\hat k}, \s_{\hat k}) \neq (\emptyset, \infty)$
	in line~\ref{get-k-hat-else} for some $\Os_{\hat k}$ and $\s_{\hat k}$
	such that $t' \geq \s_{\hat k}$.
Since $\hat k > j_0$,
	when $p$ completes the wait statement in line~\ref{FG3},
	it has $\Batch[\hat k] = (\Os_{\hat k}, \s_{\hat k})$ thereafter.
Thus, when $p$ starts executing line~\ref{wait-promise-2} after local time $t'$,
	it has $\CT \geq t' \geq \s_{\hat k} = \BatchPromise{\hat k}$,
	and hence it does not wait in line~\ref{wait-promise-2}.
\qedhere~$_\text{\autoref{no-wait-for-promise}}$
\end{enumerate}
\end{case}
\end{proof}

\begin{theorem}\label{granfinale1}
	There is a real time after which if a correct process $p$ starts executing a \emph{read} operation~$o$,
		$p$ completes this operation in a (small)
		constant number of its own steps,
		unless $o$ conflicts with another operation
		that is pending at $p$ when $p$ executes line~\ref{get-k-hat-else}.
\end{theorem}
\begin{proof}
This follows from Lemmas~\ref{sordi} and~\ref{no-wait-for-promise} and
	the code of lines~\ref{read-invoke}-\ref{read-respond}.
\end{proof}

\begin{theorem}\label{granfinale2}
There is a real time after which if process $\ell$ starts executing a \emph{read} operation,
	$\ell$ completes this operation in a constant number of its own steps.
\end{theorem}
\begin{proof}
This follows from Lemmas~\ref{sordi}(2), \ref{gafni} and~\ref{no-wait-for-promise}.
\qedhere~$_\text{\autoref{granfinale2}}$
\end{proof}

\begin{theorem}\label{granfinale3}
There is a real time after which if a correct $p\neq \ell$ starts executing a \emph{read} operation,
	$\ell$ completes this operation in a constant number of its own steps plus
	at most $\max(3\delta-\Pthree + \alpha_2, 0)$ units of local time.
\end{theorem}
\begin{proof}
This follows from Lemma~\ref{sordi}(\ref{sordi1}), Theorem~\ref{three-delta-promise-thm}
	and Lemma~\ref{no-wait-for-promise}.
\qedhere~$_\text{\autoref{granfinale3}}$
\end{proof}

Recall that $\alpha_2$ is a very small constant
	(which measures the time that $\ell$ takes to execute a few local steps that do not involve waiting),
	and is negligible compared to the maximum message delay $\delta$.
Thus, the maximum blocking time of a read operation is effectively $\max(3\delta-\Pthree, 0)$.

\end{document}